%% file: paper.tex
\documentclass[twocolumn,aps,prd,floatfix,preprintnumbers,a4paper,nofootinbib,superscriptaddress]{revtex4-1}
 
\usepackage{graphicx}				
\usepackage{placeins}
\usepackage{amssymb}
\usepackage{amsmath}
\usepackage[dvipsnames]{xcolor}
\usepackage{verbatim}
\usepackage{rotating}
\usepackage{comment}
\usepackage[colorlinks]{hyperref}

\definecolor{LinkColor}{rgb}{0, 0, 0.75}
\definecolor{CiteColor}{rgb}{0.75, 0, 0}
\definecolor{UrlColor}{rgb}{0, 0, 0.75}
\hypersetup{linkcolor=LinkColor}
\hypersetup{citecolor=CiteColor}
\hypersetup{urlcolor=UrlColor}

\usepackage{graphicx}
\usepackage{amsfonts}
\usepackage{dcolumn}
\usepackage{longtable}
\usepackage[caption=false]{subfig}
\usepackage{soul}
\usepackage{mathrsfs}
\usepackage{enumitem} 

\begin{document}

% Input macros (commands, shorthand and functions)
\input{./macro}

% Define affiliation shorthand
\newcommand{\KCL}{King's  College  London,  Strand,  London  WC2R  2LS,  United Kingdom}
\newcommand{\Amsterdam}{Institute for High-Energy Physics, University of Amsterdam, Science Park 904, 1098 XH Amsterdam, Netherlands}

% TITLE
\title{A radial scalar product for Kerr quasinormal modes}

% AUTHOR LIST
\author{Lionel T. London} 
\affiliation{\KCL} 
\affiliation{\Amsterdam}

% ABSTRACT
\begin{abstract}
A scalar product for quasinormal mode solutions to Teukolsky's homogeneous radial equation is presented. Evaluation of this scalar product can be performed either by direct integration, or by evaluation of a confluent hypergeometric functions. The related scalar product will be useful for better understanding analytic solutions to Teukolsky's radial equation, particularly the quasi-normal modes, their potential spatial completeness, and whether the quasi-normal mode overtone excitations may be estimated by spectral decomposition rather than fitting. With that motivation, the scalar product is applied to confluent Heun polynomials where it is used to derive their peculiar orthogonality and eigenvalue properties. A potentially new relationship is derived between the confluent Heun polynomials' scalar products and eigenvalues. Using these results, it is shown for the first time that Teukolsky's radial equation (and perhaps similar confluent Heun equations) are, in principle, exactly tri-diagonalizable. To this end, ``\textit{canonical}'' confluent Heun polynomials are conjectured. 
\end{abstract}

\maketitle

%%%
\section{Introduction}
\label{s1}
\par The aftermath of \bbh{} merger is thought to be a spinning and perturbed \bh{} that \textit{rings down} like a struck, and shimmering gong~\cite{Vishv:1970,LIGOScientific:2016aoc,TheLIGOScientific:2016src,LIGOScientific:2020ibl,LIGOScientific:2021djp,LIGOScientific:2018mvr}.
The corresponding \grad{} encodes information about the progenitor \bbh{} system (i.e. masses, spins, momenta and environment), and about the true nature of gravity~\cite{Kamaretsos:2012bs,London:2014cma,Hughes:2019zmt,Berti:2005ys,QNMTopicalReview09,Maselli:2017kvl,LivRevQNM,Evstafyeva:2022rve,LISA:2022kgy,Pacilio:2023mvk}. 
Ongoing and planned experiments endeavor to detect and learn from that radiation with increasing precision~\cite{Bailes2021,Sathyaprakash:2012jk,Berti:2005ys,Vallisneri:2014vxa,Karnesis:2022vdp,LISA:2022kgy}.
At all stages, those efforts are guided by our mathematical understanding of \bh{s}, their \gw{s}, and the special functions best suited to them~\cite{Abbott:2017oio,TheLIGOScientific:2016wfe,Misner1973,Maggiore2007,NP66,NP62,PhysRevLett.29.1114,Whiting:1989ms}.
Recently, it has become clear that there are limitations in our mathematical understanding of \textit{ringdown} radiation, and that these limitations may actively confound the interpretation of numerical and experimental data~\cite{Giesler:2019uxc,McWilliams:2018ztb,Carullo:2019flw,Ota:2019bzl,Bhagwat:2019dtm,Forteza:2020hbw,Leong:2023nuk,Khera:2023lnc,Franchini:2023eda,Nee:2023osy,Cotesta:2022pci,Baibhav:2023clw}. 
% %
% \par The purpose of this article is to introduce a framework (via a radial scalar product) that may enable a deeper practical understanding of ringdown's ``overtones'', and perhaps \gw{s} from compact binaries broadly. 
% %
% {
% The primary techincal focus of this work is the vector space properties of eigensolutions to \tk{s} equation for \qnms{s} \textit{after} physical boundary conditions have been imposed.
% }
%
\par The purpose of this article is to introduce a framework via a radial scalar-product that may offer a deeper, practical understanding of ringdown “overtones,” and potentially of gravitational waves from compact binaries more broadly.
{
The primary technical focus of this work is the vector space structure of the eigensolutions to \tk{s} equation for \qnms{}, considered \textit{after} the imposition of physical boundary conditions.
}
\par To date, the \lvk{} has confidently observed \grad{} from {$83$} \bbh{} mergers; among them, the {\rd{}} radiation of {$31$} have been compared against \gw{} signal models (theory predictions) to test for physics beyond \gr{}~\cite{LIGOScientific:2020ibl,LIGOScientific:2021djp,LIGOScientific:2018mvr,LIGOScientific:2021usb,LIGOScientific:2019fpa,LIGOScientific:2020tif,LIGOScientific:2021sio}.
Thus far, all \bbh{} remnants have been found to be consistent with spinning (i.e. Kerr) \bh{s}~\cite{LIGOScientific:2019fpa,LIGOScientific:2020tif,LIGOScientific:2021sio}.
\par Planned \gw{} detectors such as the \et{}, \ce{} and the \lisa{} are, for example, expected to detect signal to noise ratios of $\sim10^3$~(ET \& CE) and $\sim10^5$~(LISA) from GW150914-like\footnote{This example refers to \bbh{s} that have similar configurations to that of the first \gw{} detection: approximately non-spinning progenitors and having a luminosity distance of $\sim410~\mathrm{Mpc}$~\cite{TheLIGOScientific:2016wfe,LIGOScientific:2016aoc}.} \bbh{s} of $10^2~\mathrm{M}_\odot$ and $10^6~\mathrm{M}_\odot$, respectively~\cite{Berti:2016lat,Mishra:2010tp,Robson:2018ifk}.
Such detections are expected to dramatically increase the potential for \rd{} data analyses to verify proposed environmental, {dark matter} and quantum effects~\cite{Klein:2015hvg,Franchini:2023eda,Macedo_2013,Berti:2005ys,Barausse:2014pra,Zhao:2023itk}.
However, the correct interpretation of analyses involving \rd{} is contingent upon our understanding of \rd{}'s underlying mathematical structure~\cite{Berti:2003jh,Berti:2014fga,Berti:2005gp,London:2018gaq,London:2014cma,London:2020uva,London:2018nxs,Husa:2015iqa,Hamilton:2021pkf,Hamilton:2023znn,Cotesta:2020qhw,Pompili:2023tna,Ossokine:2020kjp,Hughes:2019zmt,Lagos:2022otp}.
\par Here, \textit{ringdown} will refer to linear combinations of \bh{} \qnm{s}~\cite{Press:1971ApJ,leaver85,Press:1973zz}.
While the \gw{s} from \bbh{} \textit{postmerger} will refer to all linear and nonlinear contributions~\cite{London:2014cma}, the mathematical structure of \rd{} is that of the \qnms{}.
Due to their astrophysical relevance, the \qnm{s} of Kerr \bh{s} are perhaps the most studied~\cite{Berti:2003jh,Berti:2014fga,Berti:2005gp,London:2018gaq,London:2014cma,London:2020uva,London:2018nxs,Husa:2015iqa,Hamilton:2021pkf,Hamilton:2023znn,Cotesta:2020qhw,Pompili:2023tna,Ossokine:2020kjp,Hughes:2019zmt,Ghosh:2017gfp,Yang:2012he,Yang:2012pj,Zimmerman:2015trm,Lagos:2022otp}. 
%
% Henceforth, ``\qnm{}'' will refer to those of Kerr unless specified otherwise. 
%
\par The Kerr \qnm{s} emerge within the study of \ee{} linearized around the Kerr metric~\cite{Vishv:1970,Press:1973zz,leaver85}.
Within that topic, a central result is that each single \qnm{} solution is a function of time (a damped sinusoid), times a function of spherical polar angles (a spheroidal harmonic), and times a function of radial distance (a confluent Heun function)~\cite{Press:1973zz,Dariescu:2021zve,Fiziev:2010,Hatsuda:2020iql}.
\par The time domain's sinusoidal damping is determined by discrete \qnm{} (or \rd{}) frequencies~\cite{leaver85}.
For gravitational perturbations of Kerr, \qnm{} frequencies are labeled by three integers: $\ell$, $m$ and $n$, where $\ell\ge 2$, $|m|\le\ell$ and $n\ge0$~\cite{Berti:2005ys}.
Typically, $\ell$ is referred to as a polar index, $m$ an azimuthal index, and $n$ an ``overtone'' index.
Respectively, $\ell$, $m$ and $n$ originate from the polar, azimuthal and radial dimensions, as defined in a \bh{} centered frame~\cite{leaver85}. 
\par For physically relevant scenarios, i.e. where the \qnm{} frequencies depend on $\ell$, the spheroidal harmonics have recently been shown to not only be biorthogonal\footnote{Two sequences, $\{a_i\}_{i=0}^{\infty}$ and $\{b_j\}_{j=0}^{\infty}$, are biorthogonal if $\brak{a_i}{b_j}=\delta_{ij}$~\cite{brauer1964,Christensen2003}.} with the so-called adjoint-spheroidal harmonics, but also complete~\cite{London:2020uva}. 
This conclusion is based on the fact that $\ell$ is directly related to the eigenvalues of e.g. Jacobi polynomials~\cite{Fackerell:1977}. 
As a result, the spheroidal harmonics may be thought of as a basis of special functions, capable of representing \gw{s} from arbitrary sources.
\par This is well known to be true of the spherical harmonics, which are a special case of the spheroidal harmonics. Consequently, spherical harmonics play a variety of key roles in \gw{} theory and data analysis, from the development of perturbative approximations (e.g. Post-Newtonian theory and \gw{} self force), and the storage of simulated \gw{s} (Numerical Relativity), to the evaluation of \gw{} signal models for use in parameter inference~\cite{NP66,Goldberg:1966uu,Blanchet:2013haa,TheLIGOScientific:2016pea,OSullivan:2014ywd,Ruiz:2007yx,Breuer77}.
Analogous applications for the spheroidal and adjoint-spheroidal harmonics have been proposed~\cite{London:2021P2}. 
\textit{Perhaps surprisingly, there has been comparatively little research into whether the \qnm{}'s radial functions are special functions of potentially broad use in \gw{} theory}~\cite{Green:2022htq,Lagos:2022otp,Yang:2012he,Leaver86c}.
\par In $1973$, \tk{} derived the differential equation that the radial functions must satisfy, and at the time noted that his radial equation has two regular singular points and one irregular one, meaning that it cannot be transformed into a hypergeometric-type equation for which there are many well known analytic methods of solutions~\cite{Teukolsky:1973ha}.  
This appears to have informed a branch of research which seeks to solve \tk{}'s radial equation using numerical methods, e.g.~\cite{OSullivan:2014ywd,Hughes:1999bq,Taracchini:2014zpa,Hughes:2019zmt,Islam:2022laz,Rifat:2019fkt,Campanelli:2000nc,Krivan:1996da,Vishal:2023fye}.
\par {However, in 1985 Leaver presented what was perhaps the first Frobenius-type series solution for the \qnm{'s} radial functions~\cite{leaver85,Leaver:1986JMP}.
Leaver's work on \qnm{s} intersects with topics in \gw{} self-force, in which asymptotic solutions to \tk{s} radial equation play a key role~\cite{Isoyama:2021jjd,Casals:2018eev,Sasaki:2003xr,Fujita:2004rb}.}
\par In parallel, there has been broad and growing interest in a generalization of the hypergeometric equation called the Heun equation~\cite{ronveaux1995heun,Chen:2023ese,Fiziev:2009ud,Hortacsu:2011rr,Dariescu:2021zve,MAGNUS2021105522,Bonelli:2021uvf}.
Unlike the hypergeometric equation, which has three regular singular points, the Heun equation has four~\cite{ronveaux1995heun}. 
It was perhaps first pointed out by Leaver and then Fiziev that \tk{}'s radial equation is an instance of the \textit{confluent Heun equation}, also known as a generalized spheroidal equation~\cite{Fiziev:2009ud}.
Fiziev's work appears to coincide with what is perhaps a recent and multi-disciplinary interest in Heun equations\cite{Hortacsu:2011rr}. 
For example, it has very recently been observed that a careful consideration of the confluent Heun function's asymptotic behavior allow them to be applied directly to astrophysically relevant scenarios in particle perturbation theory (e.g. \gw{s} from extreme mass ratio inspirals)~\cite{Chen:2023ese}.
\par Together, methods for evaluating the \qnm{}'s radial functions that are derived from Teukolsky, MST and Heun underlay a vast (and growing) body of research.
However, thus far, this work appears to have not fully addressed the following questions:
\begin{enumerate}[label=(\hspace{-1pt}{\emph{Q}\arabic*}\normalfont)]
    \item\label{Q1} The properties of many special functions in physics may be framed using ideas from Sturm-Liouville theory and functional analysis~\cite{ARFKEN2013401,Kristensson:2010,Courant1954,adkins2012algebra}. There, solutions to differential equations are conceptualized as eigenvectors, where the vector scalar products (i.e. symmetric nondegenerate bilinear forms~\cite{adkins2012algebra}) are defined by an integral whose integrand is \textit{weighted} such that the differential equations is formally \textit{self-adjoint}. For example, the scalar product for the \qnm{'s} angular functions can be framed as that of the Jacobi polynomials (i.e. with the Jacobi \textit{weight function}), and the Jacobi differential equation is self-adjoint with respect to that scalar product~\cite{Fackerell:1977}. {Further, the Jacobi polynomials are compatible with the spheroidal harmonic's analytic (i.e. regular and differentiable) structure~\cite{Fackerell:1977}}. In analogy: 
    \\
    \\
    \textit{Is there an appropriate scalar product for the \qnm{'s} radial functions?} 
    \item\label{Q2} Classical polynomials (i.e. those of Hermite, Laguerre and Jacobi) are known to be solutions to special cases of the hypergeometric equation~\cite{abramowitz+stegun}. It is well known that the \qnm{} angular functions are underpinned by spin-weighted spherical harmonics, which are themselves underpinned by Jacobi polynomials~\cite{Fackerell:1977}. Therefore, there is good reason to wonder whether an analogous conclusion be developed for the \qnm{}'s radial functions: 
    \\
    \\
    \textit{Are the \qnm{}'s radial functions underpinned by classical polynomials, or some generalization thereof? Equivalently, is the overtone label, i.e. $n$, directly related to the order of a complete and orthonormal polynomial sequence?} 
\end{enumerate}
\par Regarding \ref{Q1}, many studies have encountered the need to perform integrals over solutions to \tk{}'s radial equation, e.g.~\cite{Green:2022htq,Motl:2003cd,Hughes:1999bq,Leaver86c,Andersson:1996cm,Jaramillo:2020tuu}.
Perhaps most relevant to the present article, Leaver~(Ref.~\cite{Leaver86c}) studied Schwarzschild (i.e. non-spinning) \bh{s}, and used a Green's function approach to derive various spectral-type expansions, including those for \qnms{}.
Similar approaches have recently been used to study non-linear (particularly quadratic) \qnms{} of Schwarzschild, e.g.~\cite{Lagos:2022otp}.
This branch of research intersects with \ref{Q1}, however it does not directly draw from Sturm-Liouville theory due to apparent divergences in  \qnm{}-type problems that are not present in classical resonance problems\footnote{Similarly, Ref.~\cite{Price1999} implies that the \qnm{} problem may not be compatible with Sturm-Liouville theory \textit{because} \qnms{} damp in the time domain. In \sec{s3} of this article, this is shown to not be the case. {In paricular, the \qnm{} problem is non-Hermitian; however, this does not preclude the self-adjoint perspective taken in this work.}}~\cite{Leaver86c}.
Interestingly, Ref.~\cite{Leaver86c} observes that exactly such divergence may be overcome by allowing for \textit{a radial integration path that is complex valued}. 
In this article, the same idea is encountered for Kerr, and is motivated by coordinate dependence. 
\par In an adjacent branch of research, many studies have made use of what's called the ``formal adjoint'' of operators related to \qnms{} where, in effect, a weight function of $1$ is considered, e.g.~\cite{TeuPre74_3,Wald:1979a,Green:2022htq,Dias:2022oqm,Fiziev:2009ud,Jaramillo:2020tuu}. 
From that perspective, \tk{}'s radial equation is not self-adjoint due to the weight function or scalar product used.
Consequently, it appears that an investigation of {the \qnm{s}' radial functions} in the context of Sturm-Liouville theory with a {physically motivated} weight has not yet been pursued. 
\par Regarding \ref{Q2}, a number studies have attempted to use classical polynomials to represent solutions to the \qnm{s}' radial functions, e.g.~\cite{Chung:2023zdq,Chung:2023wkd,ghojogh2023eigenvalue}. 
These efforts encounter the following mathematical difficulty: 
{while \tk{'s} radial equation may be similarity transformed to be similar to the equation that governs e.g. Jacobi polynomials, the related transformations differ from those required by the physical boundary conditions~\cite{Teukolsky:1973ha,Press:1973zz,TeuPre74_3}.
%
% the asymptotic (i.e. large distance from the \bh{}) behavior of the classical polynomials is incompatible with the \qnm{}'s radial functions, specifically the constraint that \grad{} from \qnm{s} must be purely outgoing towards spatial infinity~\cite{Teukolsky:1973ha,Press:1973zz,TeuPre74_3}. 
%
In turn, this means that unlike \tk{}'s angular eigenvalue problem for \qnm{s}, \tk{}'s radial eigenvalue problem cannot be exactly represented in terms of a standard spectral problem involving classical polynomials.
Instead,} one encounters a ``generalized spectral problem''~\cite{Chung:2023zdq,ghojogh2023eigenvalue}.
%
% \par In contrast, a purely spectral formulation would result in a matrix representation of the problem that is e.g. \textit{tridiagonal}, or approximately diagonal, and has no artificial eigensolutions~\cite{Cook:2014cta,Fackerell:1977}.
% %
Although it has been recognized that \tk{}'s radial equation is a confluent Heun equation, it appears that \textit{confluent Heun polynomials} have not been used to frame the \qnm{} {radial} problem as purely spectral until the present work. 
\par In 2014, Cook and Zalutskiy studied the connection between {confluent Heun polynomails}, ``algebraically special'' \qnm{s} of nonspinning \bh{}'s, and ``total transmission'' \qnms{} in the presence of \bh{} spin~\cite{Cook:2014cta}. 
There, it was found that for special \qnm{} frequencies, the associated radial functions can be directly identified with select confluent Heun polynomials.
With \ref{Q2} in mind, there is good reason to ask whether the confluent Heun polynomials are only relevant to the total transmission \qnm{s}, or whether there exists a single formulation that connects confluent Heun polynomials to {all} \qnm{} solutions?
\smallskip
\par This article provides a first-principles development of a radial scalar product for \qnm{s}, thereby affirmatively answering \ref{Q1}. 
Towards \ref{Q2}, this article applies the scalar product to confluent Heun polynomials, where it is shown that the polynomials may be defined for \textit{any} wave frequency, including the complex valued \qnm{} frequencies. 
While a full treatment of \ref{Q2} is determined to be outside of this article's scope, significant progress is provided towards an affirmative answer. 
In particular, this article concludes by showing that \tk{}'s radial equation is, at least in principle, exactly tridiagonalizable.
To this end, the existence of potentially new {``canonical confluent Heun polynomials''} is conjectured. 
\section{Scope and organization}
\subsection{Scope}
\par This article discusses \qnm{} solutions to \tk{} radial equation in the context of isolated Kerr \bh{s}, absent any effects beyond \gr{}.
Some aspects of the discussion make minimal physical assumptions, while others are specific to \gw{} \rd{}.
{Broadly, this work studies the discrete spectra of \tk{}'s radial equation. By construction, all versions of this equation are predicated on a single \qnm{} frequency (i.e. the radial equation for \qnm{s} may only be deﬁned for a single frequency at a time).  
%
% The reader is cautioned that the author does not aspire to the language of formal mathematical rigor (e.g. the text will not include theorems and lemmas).
%
\par The underlying perspective of this work is that the structure \qnm{} frequencies is influenced by the underlying structure of the discrete spectra of the radial equation for single frequencies. This concept is outlined in \sec{s2a}, and it is encapsulated by the overtone number, $n$. It is known that many references refer to the overtone number as being heuristically related to an “apparent” order in the imaginary part of \qnm{} frequencies \cite{Berti:2005ys,Nollert:1999ji,Nollert:1998ys}. That apparent order is equally present in the \qnm{} separation constants, and in the context of this work, each \qnm{} separation constant is one of many radial eigenvalues.}
\par Most mathematical arguments make no assumption about spin weight, $s$. 
However, $s\in \{-2,+2\}$ are given particular attention due to their correspondence with \grad{}. 
Further, \tk{}'s radial equation, even when formatted for the \qnms{}, may be considered for \textit{any} frequency.
Therefore, much of this article's content is not specific to the \qnm{}'s frequencies, but rather the structure of \tk{}'s radial equation when formatted with \qnm{} boundary conditions.
The reader is cautioned that the author does not aspire to the language of formal mathematical rigor (e.g. the text will not include theorems and lemmas).
\par {This article's primary focus is a {frequency dependent} scalar product that is uniquely determined by a choice of radial coordinate and the \qnm{} boundary conditions. While it is noted in \sec{s3b} that this product is related to a range of ostensibly frequency independent products, a key feature of all such products is that they are conditioned on the frequency parameter used to separate \tk{'s} equation. In that sense, all such products are frequency dependent: some place the frequency dependence in a pre-factor in front of the \qnm{} radial functions, while others, including the one studies here, place the frequency dependence within the scalar product's weight function. This concept is expanded upon in \sec{s3b}.
}
\par Throughout this article, \qnm{} frequencies are used to evaluate numerical examples due to their relevance to \gw{} theory and experiment. 
\qnm{} frequencies for this article have been generated using Leaver's method via the publicly available \texttt{positive} and \texttt{qnm} Python packages~\cite{positive:2020,Stein:2019mop}. 
All numerical results presented have been computed with $64$ significant figures of precision via the \texttt{mpmath} package~\cite{mpmath}.
\subsection{Organization}
\par \Sec{s2} provides an overview of Kerr \qnms{} and \tk{}'s radial equation.
\Sec{s3} discusses a simplified version of \tk{}'s radial equation in the context of Sturm-Liouville theory, and uses ideas therefrom to derive a scalar product for the Kerr \qnm{s}.
\Sec{s4} focuses on two complementary numerical methods (direct integration and analytic continuation) for evaluating the scalar product.
\Sec{s6} presents the limitations of direct integration and analytic continuation, and then compares and contrasts them.
\par \Sec{s5} presents two pedagogical applications of the scalar product. 
There, a pedagogical construction of the confluent Heun polynomials is provided, along with a derivation their most relevant properties.
\Sec{s7} discusses the implications of the scalar product and confluent Heun polynomials on \tk{}'s radial equation. 
There, a class of ``canonical confluent Heun polynomials'' is proposed, and used to argue that \tk{}'s radial equation may be exactly tridiagonalizable.
\par Finally, in \sec{s8}, further implications of the scalar product and related results are discussed in the context of future work.
%
%%%
\section{\tk{}'s equation with QNM boundary conditions}
\label{s2}
\par In this section we connect the observable \gw{} polarizations with this article's topical starting point, \tk{}'s radial equation with \qnm{} boundary conditions.
Along the way, we will encounter spin weighted fields~\cite{NP66,NP62}, the definition of \qnm{s}~\cite{Nollert:1999ji,Berti:2016lat}, the importance of a scalar product and polynomials to \tk{}'s angular equation, and how \qnm{} solutions to \tk{}'s radial equation may be framed in close analogy with the angular case~\cite{leaver85}.
These ideas converge on \tk{}'s radial equation in a simplified form~\cite{Cook:2020otn,Fiziev:2010,Chen:2023ese}.
That equation sets the stage for our application of Sturm-Liouville theory~\cite{pinchover_rubinstein_2005,abramowitz+stegun,ARFKEN2013401,Courant1954}, and the development of a radial scalar product in \sec{s3}.
\subsection{From \gw{} strain to the perturbative wave quantity}
\label{s2a}
\par Any gravitational wave detector will be sensitive to the two \textit{strain} polarizations predicted by \gr{}, $h_+$ and $h_\times$~\cite{Ruiz:2007yx,Thorne:1980}.
The study of \gw{} plane waves motivates an extremely useful complex valued strain given by 
\begin{subequations} 
    \label{p1}
    \begin{align}
        \label{p1a}
        h \; &= \; h_+ \, - \, i \, h_\times \;
        \\
        \label{p1b}
             &= \; \int_{-\infty}^{t} \int_{-\infty}^{t'} \psi_4(t'',r,\theta,\phi) \; dt''   dt'  \; .
    \end{align}
\end{subequations} 
In \eqn{p1a}, $h$ is the complex valued \gw{} strain~\cite{Ruiz:2007yx}. 
We will consider it to depend on Boyer–Lindquist coordinates, where time, radius, polar angle and azimuthal angle are denotes by $t$, $r$, $\theta$ and $\phi$ respectively~\cite{BoyerLindquist:1967}.   
In \eqn{p1b}, $\psi_4$ is the $5$'th Weyl scalar in Newman-Penrose notation~\cite{Ruiz:2007yx,NP62}.
In that formalism, $\psi_4$ is a complex valued scalar field of \textit{spin weight} $s=-2$~\cite{NP66,Teukolsky:1973ha}.
Henceforth, complex valued qualities encode or modify the relative importance of $h_+$ and $h_\times$ within $h$.
Within the study of gravitationally perturbed isolated \bh{s}, two Weyl scalars are of particular interest:  $\psi_4$ ($s=-2$), and the closely related $\psi_0$ ($s=+2$).
Both encode information about \gw{s}~\cite{Teukolsky:1973ha}. 
\par For linear perturbations of Kerr, $\psi_4$ is treated in terms of a generic wave quantity, $\psi$:
\begin{align}
    \label{p2}
    \psi \; &= \; -( r - i a \cos\theta )^{-4} \; \psi_4 \;.
\end{align}
In \eqn{p2}, the quantity scaling $\psi$ is a spin coefficient~\cite{Teukolsky:1973ha}, where $a=|J|/M$ is the \bh{} spin magnitude.
We will hold that $G=c=1$, leaving $a$ to have units of the black hole mass, $M$.
\par The use of $\psi$ is that it satisfies exactly the same partial differential equation as other physically relevant wave quantities. 
In the case of the first Weyl scalar, $\psi_0$, we simply have that $\psi=\psi_0$~\cite{NP62}.
Whether $\psi$ refers to $\psi_0$ or $\psi_4$ is toggled by the spin weight of the field, $s$.
Thus for $\psi_0$ ($s=+2$) or $\psi_4$ (s=-2), $\psi$ satisfies Teukolsky's equation, which is one way of representing \ee{} linearized about the Kerr solution~\cite{Hughes:2000pf,Teukolsky:1973ha,PhysRevLett.29.1114,TeuPre74_3,Press:1973zz,Mino:1997bx}.
\subsection{Teukolsky's equation and its \qnm{} solutions}
\label{s2b}
\par If we denote the partial differential operator for \tk{}'s equation as $\LMaster$, with
\begin{align}
    \label{p3}
    \LMaster \; = \; \LMaster( s, r, \theta, \partial_t, \partial_r, \partial_\theta, \partial_\phi ) ,
\end{align}
then the schematic form of Teukolsky's equation is
\begin{align}
    \label{p4}
    \LMaster \;\; \psi (t,r,\theta,\phi) \; = \;  4\pi \Sigma T\; .
\end{align}
In \eqn{p3}, $\{\partial_t,\partial_r,\partial_\theta,\partial_\phi\}$ denote differentiation with respect to the the Boyer-Lindquist coordinates~\cite{BoyerLindquist:1967}.
An explicit definition of $\LMaster$ is provided in \apx{Apx-1}.
In \eqn{p4}, $\Sigma$ is the usual quantity that appears in the Kerr metric,
\begin{align}
    \label{p5}
    \Sigma \equiv r^2 + a^2\cos^2\theta \; ,
\end{align}
and $T$ is a source term derived from the stress-energy tensor. 
Since we are interested in isolated \bh{}s, we have that
\begin{align}
    \label{p6}
    T(t,r,\theta,\phi) \; = \; 0 \; .
\end{align}
\par {The structure of $\LMaster$~(see \ceqn{p3}) is known to be such that $\psi$ of the form
\begin{align}
    \label{p7}
    \psi \; \propto \; R(r) \, S(\theta) \, e^{-i \cw{} t} \, e^{-i m \phi}\;,
\end{align}
separates\footnote{The ansatz provided in \eqn{p7} obscures the fact that our net solution will depend on the tortoise coordinate via the retarded time~\cite{Press:1973zz,Teukolsky:1973ha}. The radial part of the retarded time is ultimately encoded within $R(r)$.} $\LMaster$ into two coupled equations: one for $R(r)$ and another for $S(\theta)$.}
In \eqn{p7}, $\cw$ is generally a complex valued frequency:
\begin{align}
    \label{p8}
    \cw \; = \; \omega \, - \, i/\tau \; .
\end{align}
In \eqn{p8}, one should think of any physical situation in which an isolated \bh{} experiences a momentary and small perturbation.
The post-merger of two sufficiently compact objects is an example of particular importance to astrophysics~\cite{Helfer:2016ljl,LIGOScientific:2021djp,LIGOScientific:2020ibl,Baker:2005vv}. 
For the simplest scenarios, the resulting \grad{} will, qualitatively, ring down~\cite{Nollert:1999ji,Press:1971ApJ,Berti:2007zu,LivRevQNM,Price1999}. 
The central frequency of that ringing will be $\omega$, and the exponential decay rate of that ringing will be $1/\tau$, where $\tau$ is strictly \textit{positive}; otherwise, \eqn{p7} would diverge at late times, and the \bh{} would be linearly unstable~\cite{Vishv:1970,Press:1973zz,Press:1971ApJ,Whiting:1989ms}.
\par In general, \grad{} will be a sum of such mode-like terms~(\ceqn{p7})
weighted by constant factors that depend on spacetime geometry and the perturber~\cite{Berti:2006:ExFacs,Kamaretsos:2012bs,London:2014cma,London:2018gaq}.
The net spatial dependence of each term in the sum defines a single \qnm{}~\cite{Berti:2007zu,LivRevQNM,Price1999}.
There will always be a period in time for which \qnm{} radiation dominates over other solutions to the homogeneous radial equation, such as power-law tails, prompt emission, and nonlinear \qnm{s}~\cite{Leaver86c,Lagos:2022otp,Andersson:1996cm}.
This inherently poorly defined region of \qnm{} dominance is colloquially referred to as {ringdown}~\cite{Press:1971ApJ,Nollert:1999ji}. 
\par Properties of \qnm{} solutions can be determined by applying the solution ansatz defined in \eqn{p7}, and then studying the resulting coupled eigenvalue problems.
These are separate radial and angular differential equations, connected by the separation constant, $A$:
\begin{subequations} 
    \label{p9}
    \begin{align}
        \label{p9a}
        \mcL{_u} \, S(u) \; &= \; - A \, S(u)
        \\
        \label{p9b}
        \mcL{_r} \, R(r) \; &= \; + A \, R(r) \; .
    \end{align}
\end{subequations} 
Only the schematic structure of \eqnsa{p9a}{p9b} are important here, but reader may find full expressions for  $\mcL{_u}$ and $\mcL{_r}$ in \apx{Apx-1}. 
Henceforth, we will refer to \eqn{p9a} as \textit{the angular equation}, and \eqn{p9b} as \textit{the radial equation}.
Respectively, \textit{angular problem} and \textit{radial problem} will refer to the solving of each eigenvalue equation.
\par In \eqn{p9}, $u=\cos(\theta)$, $\mcL{_u}$ will be referred to as the angular differential operator, and $\mcL{_r}$ will be referred to as the radial differential operator.
The angular functions, $S(u)$, are called spheroidal harmonics, and we will refer to $R(r)$ as \tk{}'s radial functions.
Both are known to be closely related to the confluent Heun functions\footnote{As noted by Leaver, they are also called generalized spheroidal wave functions~\cite{Leaver:1986JMP}.}~\cite{Fiziev:2010,Cook:2014cta,Cho:2009wf,Chen:2023ese}.  
\subsection{Lessons from \tk{}'s angular problem: the roles of the scalar product and Jacobi polynomials}
\label{s2c}
\par Although both angular and radial equations are known to be similar in form (e.g. Ref.~\cite{Leaver:1986JMP}), there are good reasons to first study the angular equation, and then apply lessons therefrom to the radial equation. 
Relative to the radial equation, the angular equation is well known to be simpler; as a result, there are many well known and highly accurate methods for computing its global\footnote{i.e. series solutions that apply across the physical domain} analytic solutions~\cite{leaver85,Cook:2014cta,OSullivan:2014ywd,Berti:2005gp,Yang:2012he}. 
Pertinent to this article is the spectral method presented in Ref.~\cite{Fackerell:1977}, and \textit{whether two key concepts therein may be applied to the radial problem}.
\par For the convenience of the reader, a modern and self-contained presentation this method is provided in \apx{Apx-3}.
There, the two key concepts are:
\begin{itemize}
    \item[(\textit{i})] The structure of the differential equation allows the development of a scalar product that is natural for the solution space. This concept motivates \ref{Q1}. 
    \item[(\textit{ii})] The structure of the differential equation points the way to problem specific special functions, namely the Jacobi polynomials. In turn, the Jacobi polynomials may be used to exactly represent analytic solutions. This concept motivates \ref{Q2}. 
\end{itemize}
\par If applicable to the radial problem, concept (\textit{i}) may enable greater understanding of \qnm{} (spatial) orthogonality, bi-orthogonality, and completeness~\cite{London:2020uva}.
Similarly, concept (\textit{ii}) may yield a representation of the radial functions that may not only be computationally efficient, but may also lend clear insight into the mathematical meaning of the so-called overtone label, $n$.
For these reasons, concepts (\textit{i}) and (\textit{ii}) will guide forthcoming discussion of the radial equation.
\subsection{\qnm{} solutions of \tk{}'s radial equation}
\label{s2d}
\par The study of \bh{} \qnm{s} is traditionally contained within the study of spacetime exterior to the event horizon~\cite{Regge:1957td,Vishv:1970,Press:1973zz}. 
In this context, \qnm{} boundary conditions are defined on two asymptotic surfaces: the event horizon and spatial infinity~\cite{Teukolsky:1973ha,Press:1973zz,TeuPre74_3}. 
This makes it computationally and algebraically useful to consider a compactified radial coordinate~\cite{leaver85,Leaver:1986JMP},
\begin{align}
    \label{p10}
	\xi \; &= \; (r-\rp)/(r-\rm) \; .
\end{align}
In \eqn{p10}, $\rp=M+\sqrt{M^2-a^2}$ is the outer Kerr horizon (i.e. the event horizon), and $\rm=M-\sqrt{M^2-a^2}$ is the inner Kerr horizon.
The compactified coordinate, $\xi$, has the effect of mapping $r=\rp$ to $\xi=0$, $r=\rm$ to $\xi\rightarrow-\infty$, and $r\rightarrow\infty$ to $\xi=1$.
\par The \qnm{} boundary conditions are that the \grad{} is purely ingoing (towards the \bh{}) at $\xi=0$ and purely outgoing near $\xi=1$. 
A derivation of these conditions is provided in e.g. Refs.\cite{Press:1973zz,Cook:2014cta}.
Since these conditions are imposed between $r=\rp$ and $r\rightarrow\infty$, we will only consider $r\in[\rp,\infty)$, and so $\xi\in[0,1)$.
\par The \qnm{} boundary conditions are asymptotic in the sense that they are restrictions on the functional form of $\psi$ near the event horizon and spatial infinity.
As used by Leaver, these boundary conditions amount to a \textit{similarity transformation} on the radial problem~\cite{ronveaux1995heun,leaver85,Leaver:1986JMP}. 
This transformation proceeds by seeking solutions to the radial problem of the form 
\begin{align}
    \label{p11}
    R(r(\xi)) \; &= \; \mu(\xi) f(\xi) \; .
\end{align}
In \eqn{p11}, $\mu(\xi)$ is a product of functional forms required by the asymptotic boundary conditions,
\begin{align}
    \label{p12}
    \mu(\xi) \; &= \; 
    e^{\frac{2 i \delta  \cw}{1-\xi }} 
    (1-\xi )^{1+2(s- i M \cw)} 
    \\ \nonumber
    &\quad\quad \times \xi ^{-(i M \cw+s) + \frac{i \left(a m-2 M^2 \cw\right)}{2 \delta }} \; .
\end{align}
The factors that make up $\mu$ are determined by structure of Frobenius (series) solutions to \eqn{p9b}.
In \eqn{p12}, the argument of the exponential, as well as the powers of $\xi$ and $1-\xi$ are selected by the physical boundary conditions~\cite{Cook:2014cta,Fiziev:2010,Leaver:1986JMP}.
In turn, \eqn{p11} strictly requires that $f(\xi)$ be non-zero at $\xi=0$ and $\xi=1$; otherwise the scalings imparted by $\mu$ (i.e. the \qnm{} boundary conditions) would be spoiled.
This requirement makes $f(\xi)$ the natural quantity for which to seek a power series solution. 
\par The application of \eqn{p11} to radial eigenproblem (\ceqn{p9b}) has the effect of transforming $\mcL_{r}$ into a new differential operator, $\mcL_\xi$,
\begin{subequations}
    \label{p13}
    \begin{align}
        \label{p13a}
        \left[ \mu(\xi)^{-1} \, \mcL{_r} \, \mu(\xi(r)) \right] f(\xi) \; &= \; \left[\mu(\xi)^{-1} A \, \mu(\xi) \right] f(\xi) \;
        \\
        \label{p13b}
        \mcL{_\xi} \, f(\xi) \; &= \; A \, f(\xi) \; .
    \end{align}
\end{subequations}
In \eqn{p13a}, we have applied the solution ansatz given by \eqn{p11} to the radial eigenproblem, \eqn{p9b}.
We have then divided by $\mu$.
Since $\mu$ is closely related to the singular exponents used in Frobenius' method, the division is {exact}~\cite{leaver85}.
The result, \eqn{p13b}, is a transformed eigenproblem that has the same eigenvalues, $A$.
It is in this sense that $\mcL_\xi$ is formally \textit{similar} to $\mcL_r$~\cite{Axler:2015,lax2002functional,Christensen2003},
\begin{align}
    \label{p13a2}
    \mcL{_\xi} \; &= \; \mu(\xi)^{-1} \, \mcL{_r} \, \mu(\xi(r)) \; .
\end{align}
While we refer to the effect of \eqnsa{p13}{p13a2} as a similarity transformation, it is sometimes referred to as an elementary power transformation, or a homotopic transformation~\cite{ronveaux1995heun,Chen:2023ese,Cho:2020tzx}.
The reader should note that \eqnsa{p13a}{p13a2} contain a coordinate transformation ($r$ to $\xi$), as well as a similarity transformation.
\par In \eqn{p13a2}, the new differential operator, $\mcL_\xi$, has the form given by \eqn{p14}.
\begin{widetext}
\begin{align}
    \label{p14}
	\mcL{_\xi}  \; &= \; (\text{C}_0+\text{C}_1 (1-\xi ))+ \left(\text{C}_2+\text{C}_3 (1-\xi )+\text{C}_4 (1-\xi )^2\right)\partial_{\xi} + \xi  (\xi -1)^2 \partial_{\xi}^2
\end{align}
\end{widetext}
There, the constants, $\text{C}_0$ to $\text{C}_4$, are defined in \eqn{p15} using the physical parameters discussed previously.
\begin{subequations} 
    \label{p15}
    \begin{align}
        \label{p15a}
        \delta \; &= \; \sqrt{ M^2 - a^2 }
    \end{align}
    \begin{align}
        \label{p15b}
        \text{C}_0 \; &= \; -2 a m \cw-2 i \cw (-\delta +M(2  s+1))
        \\ \nonumber
        & \quad \;\; +\cw^2 (\delta +M) (\delta +7 M)
        \\
        \label{p15c}
        \text{C}_1 \; &= \; 8 M^2 \cw^2+ s (4 i M \cw-1)+6 i M \cw-1
        \\ \nonumber 
        & \quad \;\; - (4 M \cw+i) \frac{\left(a m-2 M^2 \cw\right)}{\delta } \; 
    \end{align}
    \begin{align}
        \label{p15d}
        \text{C}_2 \; &= \; 4 i \delta  \cw
        \\
        \label{p15e}
        \text{C}_3 \; &= \; -2(s+1)+ 4 i \cw (M-\delta )
        \\
        \label{p15f}
        \text{C}_4\; &= \; s+3-6 i M \cw + \frac{i \left(a m-2 M^2 \cw\right)}{\delta }
    \end{align}
\end{subequations}
\par \Eqn{p14} has a few special properties.
We have started with Boyer-Lindquist coordinates, but one also arrives at \eqn{p14} if one instead begins with Kerr Ingoing or Kerr Outgoing coordinates~\cite{TeuPre74_3}. 
One also arrives at \eqn{p14} (albeit with $s\rightarrow-s$ and $A\rightarrow A-2s$) if one uses the alternative tetrad choice discussed in Ref.~\cite{TeuPre74_3}, which amounts to $t\rightarrow-t$ and $\phi\rightarrow-\phi$.
As noted in e.g. Refs.~\cite{TeuPre74_3,Whiting:1989ms}, the use of either $s=2$ or $s=-2$ results in mathematically different descriptions of the same \grad{}. 
\par \Eqn{p14} has regular singular points at $\xi\in\{-\infty,0\}$ and one irregular singular point at $\xi=1$.
Thus \eqn{p14} is an instance of the confluent Heun differential operator~\cite{ronveaux1995heun}.
This is also true in the Schwarzschild limit.
While it is sometimes said that {Ref.~\cite{Leaver:1986JMP}} treats \eqn{p14} as what is called a generalized spheroidal operator, we note here that this is a slight misnomer~\cite{ronveaux1995heun,Hortacsu:2011rr}.
\Eqn{p14} may only be transformed into a generalized spheroidal operator if one applies an asymptotic boundary condition at $r=r_-$ that is \textit{inconsistent} with \qnm{s}~\cite{leaver85,Cook:2014cta}.
\Apx{Apx-2} provides an expanded disambiguation between different forms of the confluent Heun equation and the radial equation.
\par \Eqns{p13b}{p15} are commonly used to define series solutions for $f(\xi)$. 
The requirement that these series solutions converge on $\xi\in[0,1)$ happens to be equivalent to finding eigenvalues, $A$, such that certain continued fractions are zero~\cite{leaver85,Leaver:1986JMP,Mano:1996vt}.
In particular, see Ref.\cite{Leaver:1986JMP} for a proof of convergence for series solutions, example applications of continued fractions to the radial and angular eigen-problems, and how the eigenvalues may be determined by searching for roots of continued fractions. 
\par {From the use of continued fractions, values of $A$ are well known to be discrete.}
However, the use of continued fractions has arguably obscured deeper understanding of the distribution of eigenvalues, $A$, and their related radial (labeled in $n$) and angular functions (labeled in $\ell$ and $m$). 
In search of that understanding, it appears that there has been relatively little application of Sturm-Liouville theory to \eqn{p14} in the \bh{} context~\cite{leaver85,Nollert:1999ji,Berti:2006:ExFacs,Ripley:2022ypi,Lin:2022ynv,Press:1971ApJ,Motl:2003cd,ronveaux1995heun,Zimmerman:2014aha}.
Nevertheless, since the Kerr \qnm{s}' eigenvalues are well known to be unique for most physical scenarios
    \footnote{There is some ambiguity in how one defines the eigenvalues. 
    In particular, the fact that $\text{C}_0$ is non-zero means that the eigenvalue that one starts with, $A$, need not strictly define the eigenvalue that is used in \eqn{p9a}. 
    Thus in e.g. the limit that $a=0$, where $A=\ell(\ell+1)-s(s+1)$ for every radial eigenfunction, the effective eigenvalue, $A-\text{C}_0$, may be a unique function of $\cw$.}
~(e.g. ~\cite{London:2018nxs,Cook:2020otn,Seidel:1988ue,Yang:2012he}), the \qnm{}'s themselves are necessarily linearly independent\footnote{More precisely, finite sets of \qnm{s} will be linearly independent when their eigenvalues are unique. Infinite sets of \qnm{s} will be \textit{minimal} if their eigenvalues remain distinct as some order label on the set goes to infinity~\cite{Christensen2003,London:2020uva}.}, and thus well suited to Sturm-Liouville theory~\cite{Courant1954}.
\section{Definition of a radial scalar product}
\label{s3}
\par The aim of Sturm-Liouville theory is to understand the existence and asymptotic behavior of eigenvalues and eigenfunctions~\cite{pinchover_rubinstein_2005,ARFKEN2013401,Courant1954}. 
Here, we will focus on a facet of the theory that pertains to whether there exists a \textit{scalar product} and an associated \textit{weight function}, for which $\mcL_\xi$ is \textit{self-adjoint}.
{While a scalar product may be developed under which the problem is self-adjoint (symmetric under scalar products), it is still the case the the \qnm{} radial problem is non-Hermitian; e.g. it has non-trivially complex eigenvalues.
\par The immediate aim of what follows is to derive a scalar product and related weight function such that \eqn{p14} is \textit{formally} self-adjoint. 
In other words, a function, $\tx{W}(\xi)$, is determined such that $\tx{W}(\xi) \mcL_\xi$ is in Sturm-Liouville form.
As the scalar product will ultimately be an integral over $\xi$ involving $\tx{W}(\xi)$, the behavior $\tx{W}(\xi)$ at the domain boundaries is addressed in \sec{s4}.}
\subsection{The scalar product and the weight function}
\label{s3a}
\par By scalar product, we mean a symmetric bilinear form that takes two non-singular functions\footnote{The reader should not confuse the non-singular but otherwise arbitrary function, $\mathrm{a}(\xi)$, with the \bh{} spin, $a$.}, $\mathrm{a}(\xi)$ and $\mathrm{b}(\xi)$, and performs an operation on them that results in a number~\cite{adkins2012algebra,gohberg2006indefinite}.
In less technical parlance, we simply mean a bra-ket operation that is similar to what one encounters in quantum mechanics: 
\begin{align}
    \label{p16}
    \brak{\mathrm{a}}{\mathrm{b}} \; = \; \int_{0}^{1} \, \mathrm{a}(\xi) \mathrm{b}(\xi) \; \text{W}(\xi) \; d\xi \; .
\end{align}
In \eqn{p16}, $\tx{W}(\xi)$ is a called a measure, or {weight function}. It will soon be the focus of our discussion.
\par Since there is no conjugation involved in \eqn{p16}, the scalar-product shown is not conjugate-symmetric. 
This feature distinguishes it from an inner-product, and allows analysis of $\mcL_\xi$, irrespective of its complex valued coefficients.
The analysis of immediate interest pertains whether $\tx{W}(\xi)$ may be defined such that $\mcL_\xi$ is formally {self-adjoint}~\cite{pinchover_rubinstein_2005,Courant1954}.
\par The adjoint of $\mcL_\xi$ is defined to be an operator, $\adj{\mcL_{\xi}}$, such that 
\begin{align}
    \label{p17}
    \brak{\mathrm{a}}{ \mcL{_\xi}  \mathrm{b}}  \; &= \; \brak{ \adj{\mcL{_\xi}} \mathrm{a}}{\mathrm{b}} \; .
\end{align}
Chapter $5$ Section 3 of Ref.~\cite{Courant1954} provides a detailed overview of how $\adj{\mcL_\xi}$ may be explicitly defined using integration by parts. 
In that derivation, the fact that the coefficient of the second derivative term in \eqn{p14} vanishes at $\xi=0$ and $\xi=1$ is key to Sturm-Liouville boundary conditions being satisfied~\cite{pinchover_rubinstein_2005}.
There, it is also shown that $\mcL_\xi$ is formally self-adjoint, i.e.  
\begin{align}
    \label{p18}
    \mcL{_\xi}\; &= \; \adj{\mcL{_\xi}} \; ,
\end{align}
if it also holds that
\begin{subequations}
    \label{p19}
    \begin{align}
        \label{p19a}
        \mcL_\xi \; &= \; p_0(\xi) \; + \; \tx{W}(\xi)^{-1}\,\partial_\xi \left[ \, (\tx{W}(\xi) \,p_2(\xi)) \, \times \, \partial_\xi \,  \right]
        \\
        \label{p19b}
        \; &= \; p_0(\xi) \; + \; \frac{\partial_\xi(\tx{W}(\xi)\,p_2(\xi))}{\tx{W}(\xi)}\,\partial_\xi  \;+ \;  p_2(\xi)\partial^2_\xi \; .
    \end{align}
\end{subequations}
In \eqn{p19}, $p_0(\xi)$ and $p_2(\xi)$ are the coefficient functions seen in zero'th and second derivative terms of \eqn{p14}, respectively. %
If we define $p_1$ to be the coefficient function of $\partial_\xi$ in \eqn{p14}, then
\begin{subequations}
    \label{p20}
    \begin{align}
        \label{p20a}
        p_0 \; &= \; \text{C}_0+\text{C}_1 (1-\xi ) \; ,
        \\
        \label{p20b}
        p_1 \; &= \; \text{C}_2+\text{C}_3 (1-\xi )+\text{C}_4 (1-\xi )^2
        \\
        \label{p20c}
        p_2 \; &= \; \xi  (\xi -1)^2 \; .
    \end{align}
\end{subequations}
In \eqn{p19a}, the appearance of the weight function is a direct result of \eqnsa{p17}{p18}.
To go from \eqn{p19a} to \eqn{p19b}, we have applied the product rule (a.k.a. Leibniz's rule), to $\partial_\xi \, \left[ (\tx{W} p_2) \times \partial_\xi \right]$.
For $\mcL_{\xi}$ in \eqn{p14} to be equivalent to that in \eqn{p19b}, it must be the case that coefficients of $\partial_\xi$ match,
\begin{subequations}
    \label{p21}
    \begin{align}
        \label{p21a}
        p_1 \; &= \; \tx{W}^{-1}\, \partial_\xi(\tx{W}\,p_2)
        \\
        \label{p21b}
        \tx{W}^{-1}\, \partial_\xi\tx{W} \; &= \; p_2^{-1}(p_1-\partial_\xi p_2) \; .
    \end{align}
\end{subequations}
In \eqn{p21a} we have simply equated the first derivative coefficient of \eqn{p19b} with that of \eqn{p14}.
In \eqn{p21b}, we have used Leibniz's rule to expand the right-hand-sid of \eqn{p21a}, and then we have rearranged terms to emphasize that \eqn{p21b} is a linear differential equation whose left-hand-side is equal to $\partial_\xi \ln \tx{W}$.
\par \Eqn{p21b} can be solved for $\tx{W}$ using the coefficient functions defined in \eqn{p20},
\begin{subequations}
    \label{p22}
    \begin{align}
        \label{p22a}
        \text{W}(\xi) \; &= \; \, \frac{1}{p_2(\xi)} \, \exp\left(  \int^{\xi}  \frac{p_1(\xi')}{p_2(\xi')}  \, d\xi'   \, \right)
        \\
        \label{p22b}
                        &= \; e^{\frac{\text{C}_2}{1-\xi }} \;  (1-\xi )^{-\text{C}_2-\text{C}_3-2} \;  \xi ^{\text{C}_2+\text{C}_3+\text{C}_4-1} \; .
    \end{align}
\end{subequations}
In \eqn{p22a}, the integration constant has been chosen to be zero, and in \eqn{p22b} we explicitly evaluate the integral using \eqnsa{p20b}{p20c}.
\par It will soon be useful to simplify our notation for $\mathrm{W}(\xi)$ by defining a shorthand for combinations of parameters,
\begin{subequations}
    \label{p23}
    \begin{align}
        \label{p23a}
        \text{B}_0 \; &= \; \text{C}_2+\text{C}_3+\text{C}_4-1
        \\
        \label{p23b}
        \text{B}_1 \; &= \; -\text{C}_2-\text{C}_3-2
        \\
        \label{p23c}
        \text{B}_2 \; &= \; \text{C}_2 \; .
    \end{align}
\end{subequations}
Consequently, the weight function takes a less cumbersome form,
\begin{align}
    \label{p24}
    \text{W}(\xi) \; = \;  \xi ^{\text{B}_0} (1-\xi )^{\text{B}_1} e^{\frac{\text{B}_2}{1-\xi }} \; ,
\end{align}
and the scalar product between two any two arbitrary functions, $\text{a}(\xi)$ and $\text{b}(\xi)$, may now be written as,
%
% \begin{widetext}
\begin{align}
    \label{p25}
    \brak{\mathrm{a}}{\mathrm{b}} \; = \; \int_{0}^{1} \, \mathrm{a}(\xi) \, \mathrm{b}(\xi) \;\; \xi ^{\text{B}_0} (1-\xi )^{\text{B}_1} e^{\frac{\text{B}_2}{1-\xi }} \; d\xi \; .
\end{align}
% \end{widetext}
%
\subsection{Properties and non-uniqueness of the weight function}
\label{s3b}
\par With \eqn{p25}, we arrive at one of this article's key results. 
\Eqn{p25} along with \eqnsa{p15}{p23} define a scalar product for which \tk{}'s radial equation, \eqn{p14}, is self-adjoint. 
In turn, self-adjointness, in principle, implies that as long as the eigenvalues may be defined, and that they are unique, then the eigenfunctions are orthogonal, and very likely complete~\cite{ronveaux1995heun}.
However, there are good reasons why the explicit demonstration of orthogonality and completeness are beyond the scope of this article\footnote{{In particular, one route to demonstrating completeness of the single-frequency radial functions would be to construct an isomorphism between them, and a known basis~\cite{London:2020uva,Christensen2003}.}}. 
The most pressing reason has to do with the fact that $\tx{W}(\xi)$, \textit{naively}, appears to diverge on the event horizon, and at spatial infinity.
Understanding that \textit{apparent} divergence and its consequences is preliminary to further investigation.
\par Visual inspection of \eqn{p22} points to the origin of these divergences.
In \eqn{p22a}, the factor 
\begin{align}
    \label{p27}
    p_2(\xi)^{-1} \; = \; \xi^{-1} (1-\xi)^{-2}
\end{align}
clearly creates a potential for the weight function to become singular at $\xi=0$ and $\xi=1$.
For this reason, the radial eigen-problem may be said to be of the \textit{singular}\footnote{Note that e.g. the Jacobi and Legendre polynomial problems are also singular in this way~\cite{ARFKEN2013401}.} Sturm-Liouville type~\cite{pinchover_rubinstein_2005}.
In \eqn{p22b}, we see that the net exponents of $\xi$ and $1-\xi$ may still prevent divergence if their total \textit{real part} is positive. 
Similarly, the exponential factor may diverge if the real part of $\tx{C}_2=4i\delta \cw$ is greater than zero. 
\par Thus, to determine the physical conditions in which $\tx{W}(\xi)$ diverges, one might embark on an analysis of the $\text{sign}$ of parameters within \eqn{p22b} or, equivalently, \eqn{p24}. 
While this is a somewhat cumbersome exercise, it is easy to see that the exponential factor, 
\begin{align}
    \label{p28}
    e^{\tx{C}_2/(1-\xi)} \; = \; e^{4i\delta\cw/(1-\xi)} \;,
\end{align}
will generally diverge at spatial infinity.
For that we need to recall the discussion below \eqn{p8}: Since the imaginary part of $\cw$ must be negative for time-domain stability, the real part of $4i\delta\cw/(1-\xi)$ is generally \textit{positive} for the \qnm{s},
\begin{align}
    \label{p29}
    \tx{Re}\, \frac{4i\delta\cw}{1-\xi} \; = \;\frac{4\delta}{\tau (1-\xi)}  \; > \; 0 \;. 
\end{align}
The inequality in \eqn{p29} holds because $\delta=\sqrt{M^2-a^2}>0$ and $\tau>0$.
Thus it would appear that $\tx{W}(\xi)$ will always diverge on at least one boundary.
This appearance happens to be just that. 
\par Above, we have implicitly assumed that $\xi=0$ and $\xi=1$ are approached from the real line.
While that is a physically meaningful assumption, previous studies have found it to be an unnecessary practical impediment~\cite{Leaver:1986JMP,Green:2022htq,Motl:2003cd,Mano:1996vt}.
In particular, when evaluating the scalar product, $\xi$ need not be a real valued quantity.
This point is explored in the two sections that follow. 
A kind of ``physicist's analytic continuation'' is detailed in \sec{s4a}, and then a complementary analytic continuation technique is presented in \sec{s4b}.
\smallskip
\par We conclude this section by pointing out two alternative perspectives on the weight function.
The first hinges on the fact that \tk{}'s equation allows for the definition of a weight function prior to applying the \qnm{} boundary conditions.
The second pertains to whether the weight function is unique, once the \qnm{} boundary conditions have been applied.
\par The first perspective is that the form of the weight function is closely linked to the scaling function, $\mu(\xi)$, defined in \eqn{p12}.
In particular, it is easy show (see e.g. \ceqn{d24}) that 
\begin{align}
    \label{p30}
    \tx{W}(\xi) \; &= \; \mu(\xi)^2 \, \tx{W}_0(\xi)  \; .
\end{align}
In \eqn{p31}, $\tx{W}_0$ is the radial equation's weight function \textit{before} applying the similarity transformation shown in \eqn{p14},
\begin{align}
    \label{p31}
    \tx{W}_0(\xi) \; &= \; \xi^s \, (1-\xi )^{-2 (s+1)} \; .
\end{align}
This weight function is independent of $\cw$, $m$ and $a$, and it is directly relevant for the original radial functions, $R(\xi)$~(see \ceqn{p9b}).
In essence, \eqn{p30} defines how the weight function changes under similarity transformations on the main problem.
One way to arrive at \eqn{p30} is to note that such transformations preserve the value of scalar products.
{While $\tx{W}_0$ has already been used in operator perturbation theory (e.g.~\cite{Zimmerman:2014aha}), it is not directly considered here because the corresponding differential operator does not support polynomial solutions that are compatible with the \qnm{} boundary conditions. }
\par The second perspective is that $\tx{W}(\xi)$ is not the only weight function for which $\mcL_{\xi}$ is formally self-adjoint.
It is left as an exercise for the reader to show that one can actually enforce a strictly positive and real valued weight function, 
\begin{align}
    \label{p32}
    \mathcal{W}(\xi) \; = \;  |\xi ^{\text{B}_0} (1-\xi )^{\text{B}_1} e^{\frac{\text{B}_2}{1-\xi }} | \; .
\end{align}
This exercise follows exactly the steps discussed in \eqns{p19}{p22}, but with the requirement that the weight function is wrapped within an absolute value.
{Further, one must use the fact that $\partial_\xi \mathcal{W}=\tx{sign}(\tx{W})\,\partial_\xi \tx{W}$.}
While $\mathcal{W}(\xi)$ is amenable to the direct integration approach that we will discuss shortly, it is not at all clear to the author how it may be used with the analytic continuation approach that will be discussed in \sec{s4b}. 
Further, there do not appear to be significant gains in using $\mathcal{W}(\xi)$ rather than $\tx{W}(\xi)$. 
Since complexification of the radial coordinate is also needed for $\mathcal{W}(\xi)$, scalar products will in general still be complex valued. 
$\mathcal{W}(\xi)$ is noted for potential future interest.
Henceforth, this article will focus on the evaluation of scalar products with the comparatively simpler weight function, $\tx{W}(\xi)$. 
\begin{figure*}
    \hspace{-1cm}
    \begin{tabular}{ccc}
        \includegraphics[width=0.34\textwidth]{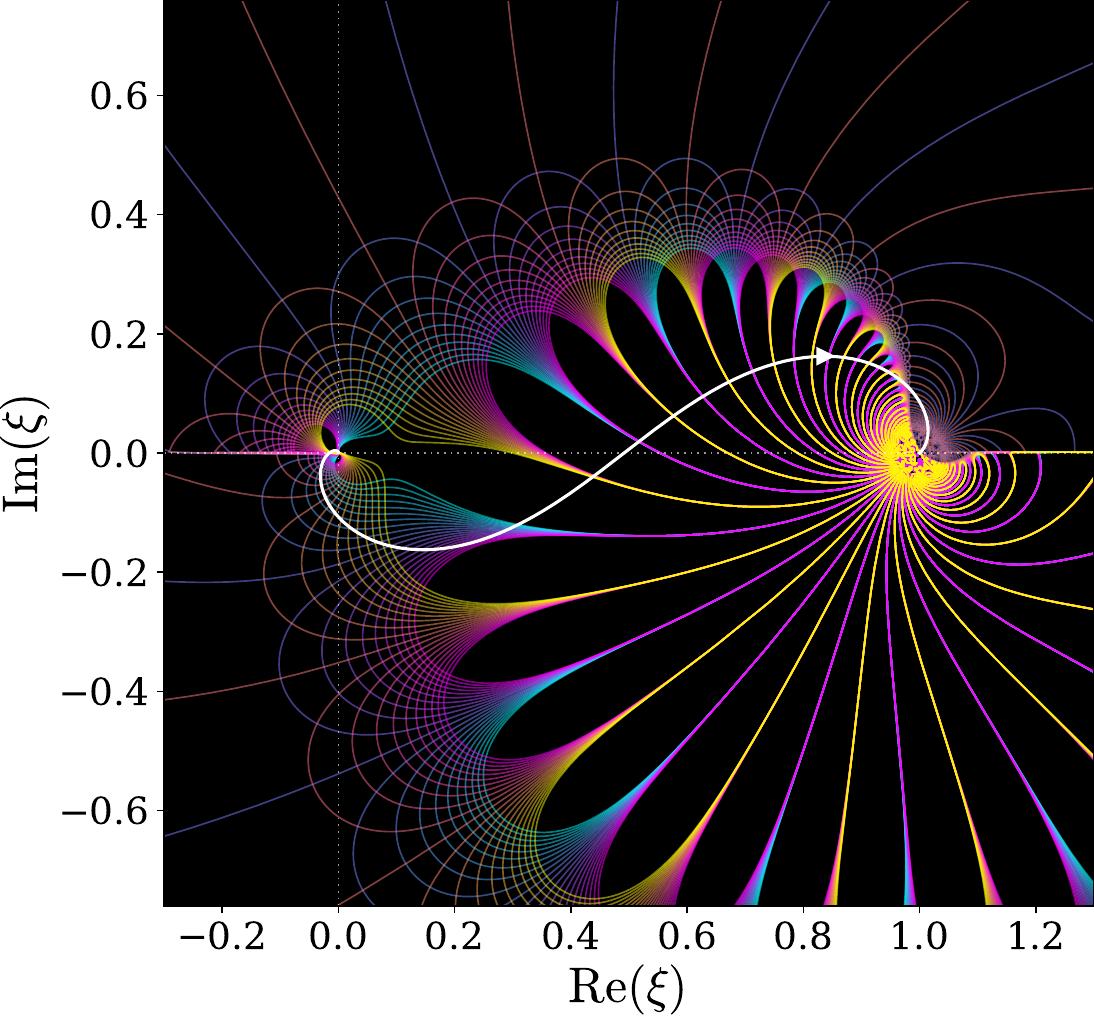}
        &
        \includegraphics[width=0.34\textwidth]{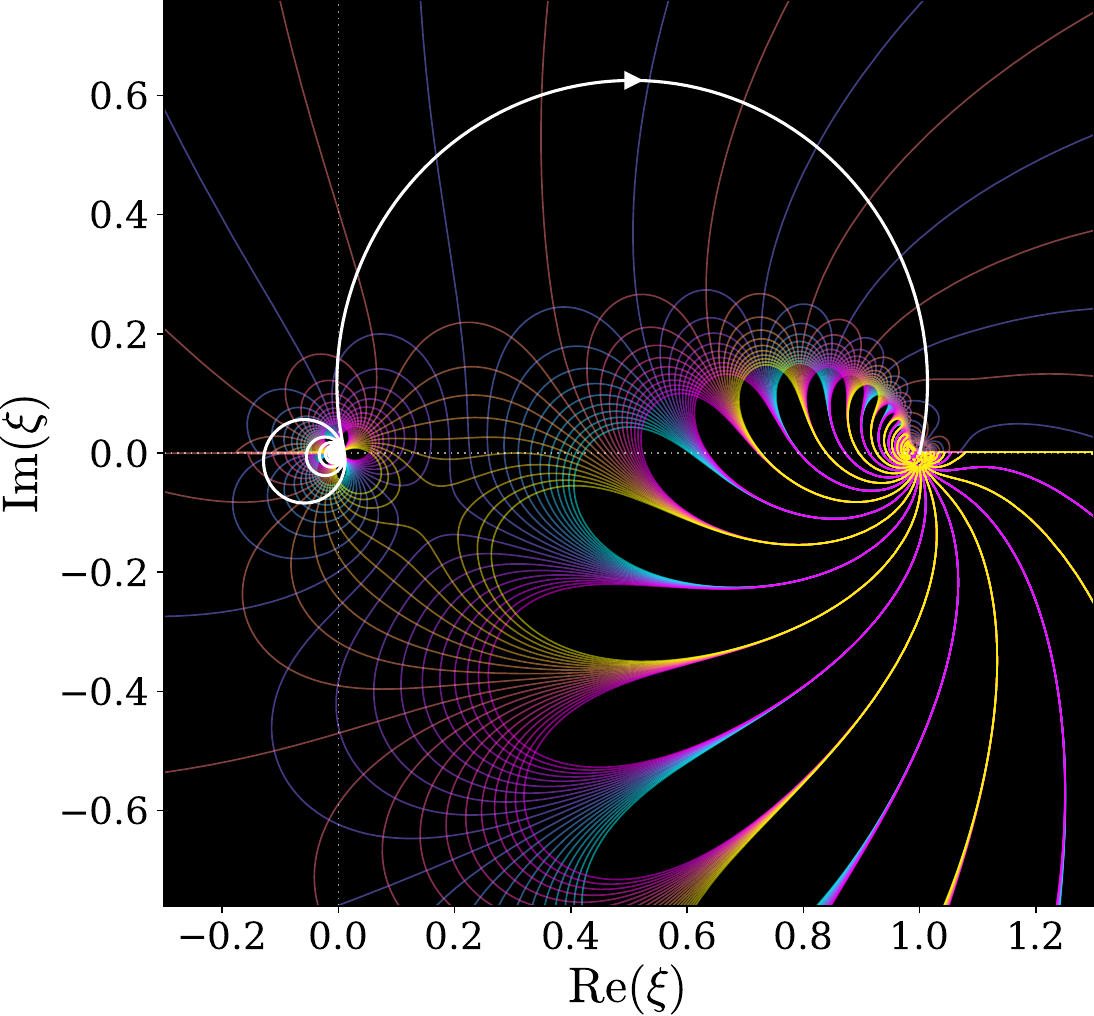}
        &
        \includegraphics[width=0.34\textwidth]{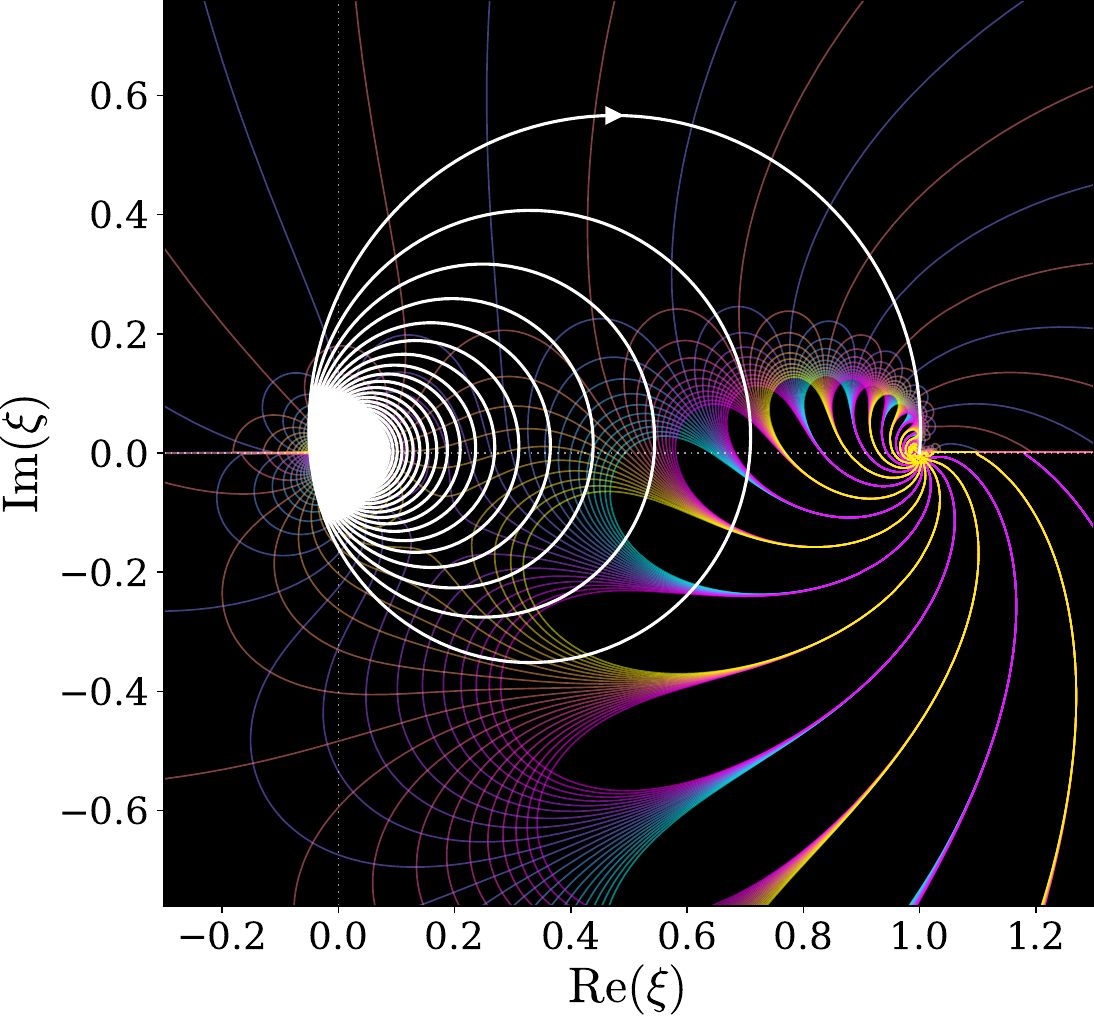}
    \end{tabular}
    \caption{Examples integration paths (white curves) against weight function contours (colored contours) for the $(s,\ell,m,n)=(-2,2,2,3)$ \qnm{} at three \bh{} spins: (left) $a/M=0.7$, $M\cw=0.4713- 0.5843i$, (center) $a/M=0.99$, $M\cw=0.8695-0.2058i$, and (right) $a/M=0.999$, $M\cw=0.9558-0.07299i$. Contours for the real and imaginary parts of the weight function, $\text{W}(\xi)=\mathrm{Re}(\text{W})+i\,\mathrm{Im}(\text{W})$, are shown. For $\mathrm{Re}(\text{W})$, colors between cyan and magenta represent dimensionless values between $-50$ and $50$ respectively. For $\mathrm{Im}(\text{W})$, colors between magenta and yellow represent values between $-50$ and $50$ respectively. Convergence of contours at spatial infinity ($\mathrm{Re}(\xi)=1$) corresponds to the divergence of the weight function. Horizontal contours along $\mathrm{Re}(\xi)<0$ and $\mathrm{Re}(\xi)>0$ are branch cuts. For left, center and right panels, coordinate parameters, $(K_0,K_2)$, are $(0.8530,23.4621)$, $(0.7105,1.4010\times 10^{-4})$ and $(0.1585,2.5088\times 10^{-4})$, respectively. 
    Dotted horizontal and vertical lines mark the respective locations of the real and imaginary axes. }
    \label{F1}
\end{figure*}
\section{Evaluation of scalar products}
\label{s4}
The scalar product developed in the previous section has been copied below,
\begin{align}
    \label{e1}
    \brak{\tx{a}}{\tx{b}} \; = \; \int_{0}^{1} \, \tx{a}(\xi) \, \tx{b}(\xi) \;\; \xi ^{\text{B}_0} (1-\xi )^{\text{B}_1} e^{\frac{\text{B}_2}{1-\xi }} \; d\xi \; .
\end{align}
\par One of the simplest ways to evaluate \eqn{e1} is direct integration (i.e. explicit numerical integration via discretization of the domain). 
However, divergence of the weight function at spatial infinity or the \bh{} horizon means that naive integration along real values of $\xi$ is bound to fail.
So other means must be considered. 
\par In \sec{s4a}, we make use of {coordinate dependence} to develop a method to \textit{directly integrate} \eqn{e1}.
In \sec{s4b}, we note that \eqn{e1} coincides with the integral representation of the Tricomi confluent hypergeometric function, also known as the Kummer function of the second kind~\cite{ARFKEN2013401,Chen:2010}.
That fact, along with the well known \textit{analytic continuation} of e.g. the Gamma function to complex inputs, is used to analytically evaluate \eqn{e1}. 
This section concludes with a brief comparison of the pros and cons of direct integration and analytic continuation. 
%
%%%
\subsection{Direct integration}
\label{s4a}
An extremely useful lesson from calculus is that, given an integral, we may change the coordinate representation of its integrand without changing the value of the integral. 
So while the integrand of \eqn{e1} is coordinate {dependent}, its value, $\brak{\tx{a}}{\tx{b}}$, is coordinate {independent}.
This concept leads us to seek a new real valued coordinate, $z$, such that the integrand of \eqn{e1} does not diverge. 
Henceforth, we will use $\xi(z)$ to refer to a complex valued integration path, and $\xi$ to refer to the quantity defined by \eqn{p10}.
\par The following definition of $\xi(z)$ has the desired effect of augmenting singular behavior at the horizon and spatial infinity,
\begin{align}
    \label{e2}
    \xi(z) \; = \; \frac{{\tx{x}_2} z^{{\tx{x}_0}}}{({\tx{x}_2}+1) z^{{\tx{x}_0}}-1} \; .
\end{align}
In \eqn{e2}, $\tx{x}_0$ and $\tx{x}_2$ are coordinate parameters that modify $\tx{B}_0$ and $\tx{B}_2$, respectively.
We will consider $\tx{Re}(x0)>0$, such that $\xi(0)=0$.
The dependence of \eqn{e2} on $\tx{x}_2$ is motivated by the complex radial coordinate used in Ref.~\cite{Mano:1996vt}.
It may also be developed by placing the Laguerre polynomial equation into a $\xi$-like compactified coordinate, and then studying that system's weight function.
The dependence of \eqn{e2} on $\tx{x}_0$ is motivated by the observation that $z^{\tx{x}_0}$ is the simplest apparent coordinate modification that impacts singular behavior arising from $\tx{B}_0$.
\par The effect of this coordinate choice is found by applying \eqn{e2} to the scalar product's integrand,
\begin{align}
    \label{e3}
    \brak{\tx{a}}{\tx{b}}  =  \int_{0}^{1}  \tx{a}(\xi(z)) \, \tx{b}(\xi(z)) \; \left[\text{W}(\xi(z)) \; \partial_{z}\xi(z)\right] \; dz \;.
\end{align}
In \eqn{e3}, we have used the fact that $d\xi = \partial_{z}\xi(z)\,dz$ to write the integral explicitly in $z$.
{According to \eqn{e2}, there may exist coordinate parameters, $\tx{x}_0$ and $\tx{x}_2$, such that $\xi(z=0)=0$, and $\xi(z=1)=1$, which would mean that the limits of integration do not change.} 
\par Within \eqn{e2}, the quantity in square brackets is of particular practical use. 
It is an \textit{effective weight function},
\begin{subequations}
    \label{e4}
    \begin{align}
        \label{e4a}
        \text{W}_{\text{eff.}}(z) \; &= \; \text{W}(\xi(z)) \; \partial_{z}\xi(z)
        \\
        \label{e4b}
        &= \; e^{\frac{\text{B}_{2} \text{x}_{2} }{z^{\text{x}_{0} }-1}} \;
        z^{ \text{x}_{0}(\text{B}_{0}  + 1) -1} \;
        \left(1-z^{\text{x}_{0} }\right)^{\text{B}_{1}}  \;
        \\ \nonumber 
        &\quad\quad \times \; \left(1-(\text{x}_{2} +1) z^{\text{x}_{0} }\right)^{-\text{B}_{0}-\text{B}_{1}-2} \; .
    \end{align}
\end{subequations}
While \eqn{e2} provides a coordinate that modifies the boundary behavior of $\tx{W}(\xi)$, it is \eqn{e4} that must be analyzed to determine the special values of $\tx{x}_0$ and $x_2$ that prevent the effective weight function from diverging at the domain boundaries. 
In essence, we have two independent conditions: (i) $\text{W}_{\text{eff.}}(0)=0$, and (ii) $\text{W}_{\text{eff.}}(1)=0$.
We also have two unknowns: $\tx{x}_0$ and $\tx{x}_1$.
Thus, $\tx{x}_0$ and $\tx{x}_1$ may be determined (up to positive constants) by careful analysis of $\text{W}_{\text{eff.}}$.
\par This analysis is somewhat tedious, but tractable.
First, one should focus on the factor of $z^{ \text{x}_{0}(\text{B}_{0}  + 1) -1}$ within $\text{W}_{\text{eff.}}$.
If $\tx{x}_0$ is chosen appropriately, then the real part of $\text{x}_{0}(\text{B}_{0}  + 1) -1$ should equal some positive constant, $\tx{K}_0$, thereby allowing $\text{W}_{\text{eff.}}$ to go to zero as $z$ goes to zero.
Since, $\tx{B}_0$ is complex valued, $\tx{x}_0$ must be also.
These ideas result in the following constraints,
\begin{subequations}
    \label{e5}
    \begin{align}
        \label{e5a}
        \tx{K}_0  &=  -1 + \left[1 + \tx{Re}(\tx{B}_0)\right] \tx{Re}(\tx{x}_0) - \tx{Im}(\tx{B}_0) \tx{Im}(\tx{x}_0) ,
        \\
        \label{e5b}
        \tx{K}_0 &> 0 \; .
    \end{align} 
\end{subequations}
If we assume that $\tx{Re}(\tx{x}_0)$ only depends on $\tx{Re}(\tx{B}_0)$, then then the result for $\tx{x}_0$ follows,
\begin{align}
    \label{e6}
    \text{x}_0 \; &= \;\text{K}_{0} \, \frac{\text{sgn}(\mathrm{Re}(\text{B}_{0})+1)}{\mathrm{Re}(\text{B}_{0})+1}
    \\ \nonumber
    &\quad \; + \; i \, \frac{ -1 \,  + \,  \text{K}_{0} \, [\text{sgn}(\;\mathrm{Re}(\text{B}_{0})+1\;)-1] }{\mathrm{Im}(\text{B}_{0})} \; .
\end{align}
Note that, in \eqn{e6}, $\tx{Re}(\tx{x}_0)>0$, as is needed for $\xi|_{z=0}$ to be zero.
\par Given, $\tx{x}_0$, one may then determine $\tx{x}_2$.
Returning to $\text{W}_{\text{eff.}}$, if its exponential factor has a negative argument as $z\rightarrow1$, then $\text{W}_{\text{eff.}}$ is assured to be regular in that limit.
To that end, one should consider $e^{\frac{\text{B}_{2} \text{x}_{2} }{z^{\text{x}_{0} }-1}}$, and seek $\tx{x}_2$ such that 
\begin{subequations}
    \label{e7}
    \begin{align}
        \label{e7a}
        \lim_{z\rightarrow1} \frac{\text{B}_{2} \text{x}_{2} }{z^{\text{x}_{0} }-1} \; &\sim \; -\frac{\tx{K}_2}{1-z} \; ,
        \\
        \label{e7b}
        \tx{K}_2 \; &> 0 \; .
    \end{align}
\end{subequations}
To evaluate the left-hand-side of \eqn{e7a}, one can use the binomial theorem\footnote{For this it is useful to work with $z'=1-z$.}; this results in a quantity with the same form as the right-hand-side of \eqn{e7a}.
Upgrading the asymptotic equivalence in \eqn{e7a} to an equality yields the following solution for $\tx{x}_2$,
\begin{align}
    \label{e8}
    \text{x}_2 \; &= \; \text{K}_{2} \;  \text{x}_{0} \, / \, \text{B}_{2} \; .
\end{align}
Note that, since $\tx{x}_0$ is complex valued, the factor of $\left(1-(\text{x}_{2} +1) z^{\text{x}_{0} }\right)^{-\text{B}_{0}-\text{B}_{1}-2}$ within $\text{W}_{\text{eff.}}$ will never encounter a root.
Thus it is of little practical concern.
\par In \eqnsa{e6}{e8} we define the values of the $\tx{x}_0$ and $\tx{x}_2$ are allowed to take such that $\text{W}_{\text{eff.}}$ is regular (actually zero) at the event horizon and spatial infinity.
In doing so, we have transferred the coordinate dependence to two new, strictly positive, parameters, $\tx{K}_0$ and $\tx{K}_2$.
What remains is the possibility that $|\xi(z)|\gg1$.
\par If $\tx{K}_0$ and $\tx{K}_2$ are not chosen such that $|\xi(z)|\lessapprox1$, then numerical evaluation of \eqn{e3} can encounter extremely large numbers. 
This numerically problematic possibility can be avoided if one determines $\tx{K}_0$ and $\tx{K}_2$ by requiring that $\xi(z)$ is contained as much as possible within a unit circle centered about $\xi=1/2$.
This may be accomplished via the following minimization,
\begin{subequations}
    \label{e9}
    \begin{align}
        \label{e9a}
        \{ \tx{K}_0,\tx{K}_2 \} = \tx{argmin}_{\tx{K}_0,\tx{K}_2} S(\tx{K}_0,\tx{K}_2,z) 
    \end{align}
where
    \begin{align}
        \label{e9b}
        S(\tx{K}_0,\tx{K}_2,z) \; &= \; |1/2-\tx{mean}_z(|\xi(z)|)|
        \\ \nonumber 
        &\quad\; + \; \tx{max}_z|\tx{Im}\,\xi(z)| \; + \; |\tx{min}_z \; \tx{Re}\, \xi(z) |\; .
    \end{align}
\end{subequations}
In \eqn{e9}, $\tx{argmin}$ outputs the values for which $S$ is minimized in $\tx{K}_0$ and $\tx{K}_2$,
$\tx{mean}_z$ computes the unweighted mean over values of $z$, $\tx{min}_z$ returns the smallest value of a quantity in $z$, and $\tx{max}_z$ returns the largest value of a quantity in $z$.
\par In practice, one has a number of remaining options for actually evaluating \eqn{e3}. 
For the evaluation of \eqn{e9}, it is found to be useful to densely sample, on log-scale, $z$ near $0$ and $1$.
The coordinate freedom associated with $\tx{x}_0$ introduces oscillations into $\xi(z)$ which must be well resolved for the accurate determination of $\tx{K}_0$ and $\tx{K}_2$.
Similarly, for the evaluation of \eqn{e3}, one will want to densely sample near the boundaries; it is found that $z=\frac{1}{2}(1+\sin(t \pi/2))$, with $t$ uniformly spaced on $(-1,1)$, is sufficient for this purpose.
For the determination of $\tx{K}_0$ and $\tx{K}_2$, and for the evaluation of \eqn{e3}, it is found that at least $10^4$ points in $z$ are needed for robust results.
With these details in hand, it is now possible to compute the integration path, $\xi(z)$, and related scalar-products, \eqn{e3}, using e.g. $5^{\text{th}}$ order spline integration~\cite{2020SciPy-NMeth}. 
\par \Fig{F1} shows three example integration paths, $\xi(z)$, set against contours of the weight function, $\tx{W}(\xi)$, in the complex plane. 
In focus is the effect of \bh{} spin $a$.
The result for $a=0$ is qualitatively similar to that for $a=0.7$.
Branch cuts are visible as overlapping horizontal contours along the real line for $\xi<0$ and $\xi>1$.
The shape of integration paths are seen to vary significantly with increasing \bh{} spin, changing most drastically near the extremal limit, $a=M$.
\par
\begin{figure}
    \hspace{-0.5cm}
    \begin{tabular}{c}
        \includegraphics[width=0.465\textwidth]{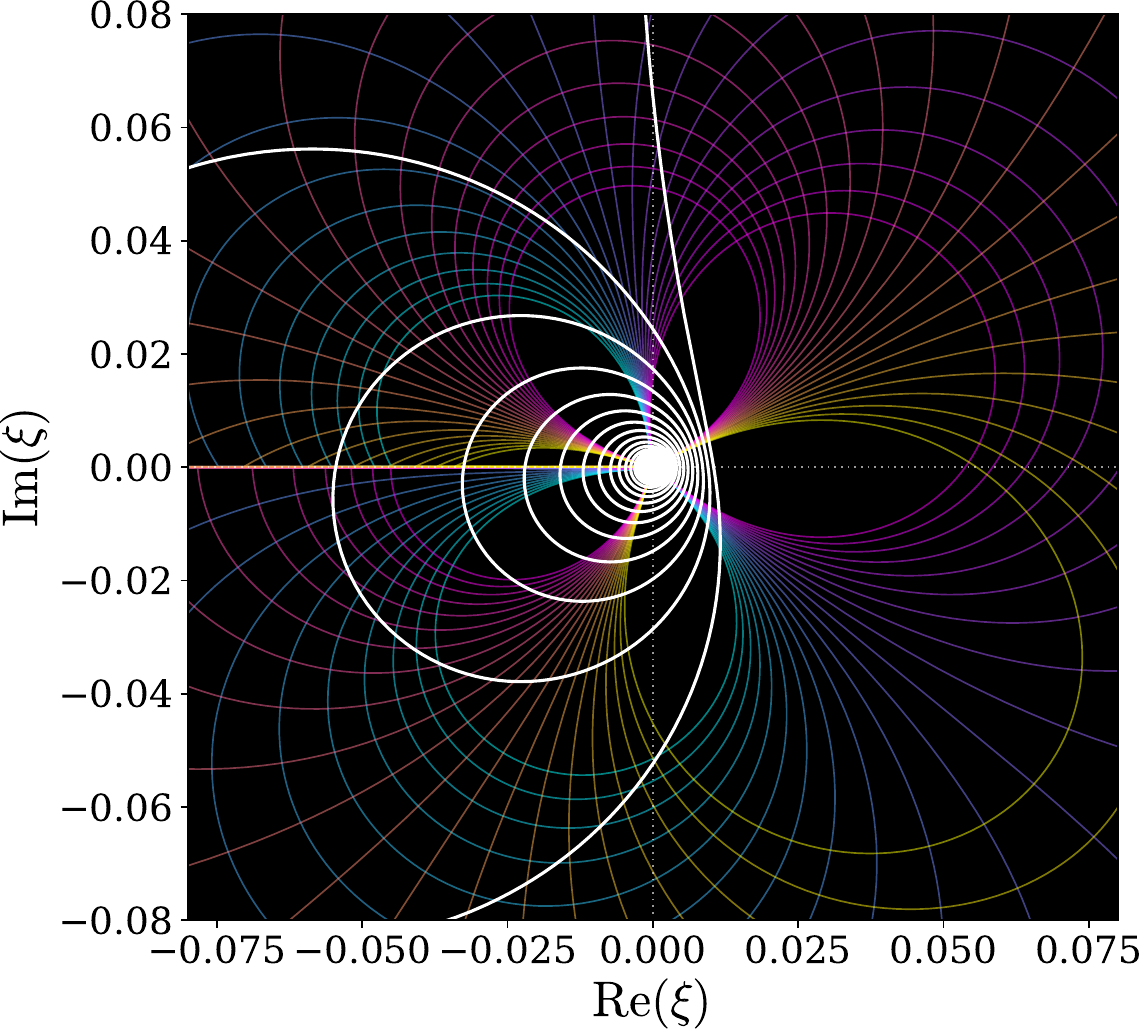}
        \\
        \includegraphics[width=0.465\textwidth]{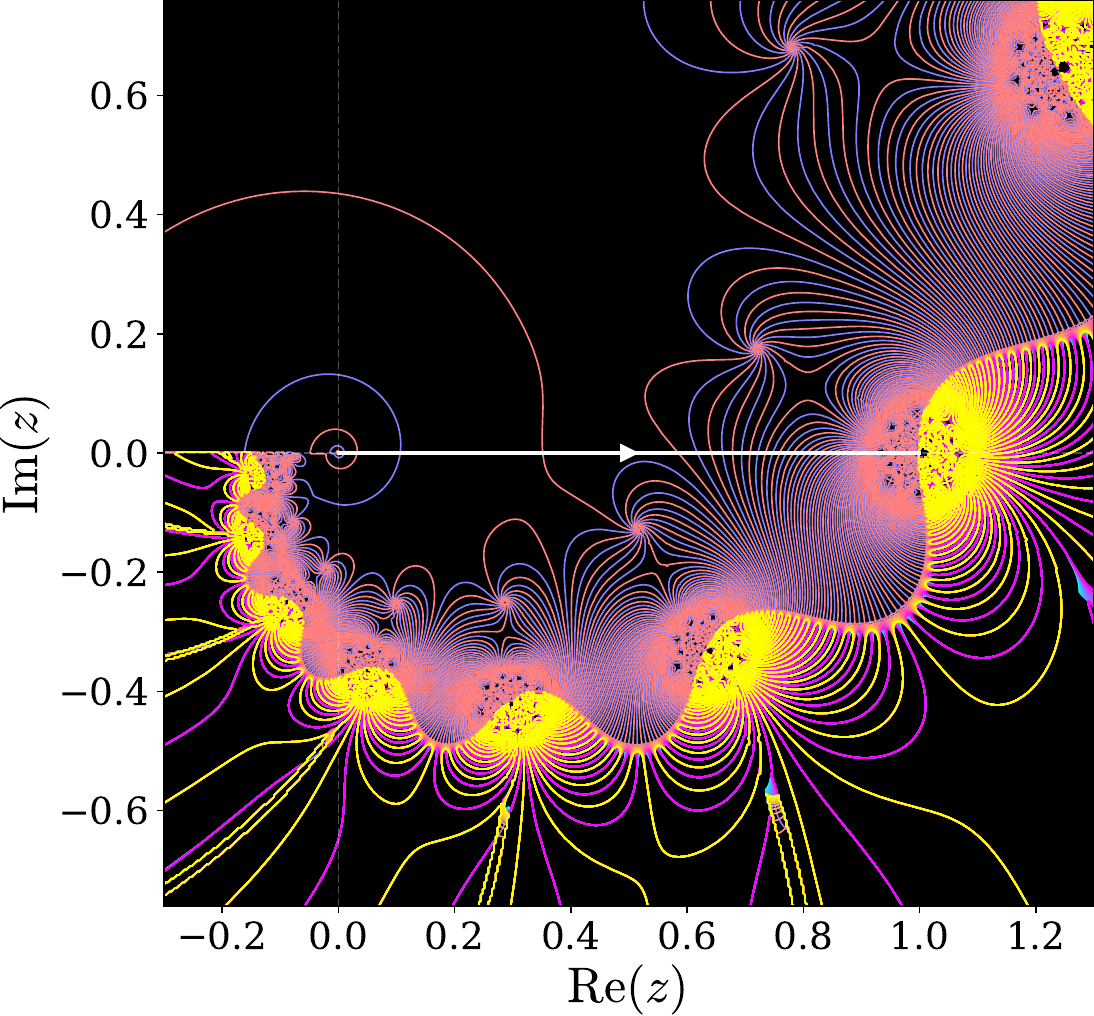}
    \end{tabular}
    \caption{Examples integration paths (white curves) against weight function contours (colored contours) for the $(s,\ell,m,n)=(-2,2,2,3)$ \qnm{} at \bh{} spin $a/M=0.99$. (top) A zoom-in of the central panel of \fig{F1} with axes centered around the \bh{} horizon at $\xi=0$. (bottom) Visualization of the coordinate transformed integration path and effective weight function for the same case. Contour formatting and coordinate parameters are identical to those in \fig{F1}.  }
    \label{F2}
\end{figure}

 For $a=0.99$ (the central panel), the integration path appears to cross the branch cut\footnote{{For complex numbers $y$, We define the principle branch of $\log(y)$ by taking $\arg(y)\in(-\pi,\pi)$ and place the branch cut on the line defined by $\Re(y)<0$. As seen in \fig{F1}, this convention results in a branch cut along $Re\xi<0$ (for powers of $\xi$), and another along $Re\xi>1$ (for powers of $1-\xi$).}} many times. 
The top panel of \fig{F2} shows this in detail. 
There, the apparent crossing of branch cuts is an artifact of presentation: 
$\xi(z)$ actually runs along all branches of the Riemann surface defined by $\tx{W}(z)$, but in \fig{F1}, only a single branch of $\tx{W}(\xi)$ is shown~\cite{kuijlaars2003orthogonality}. 
This can be understood by noting that the transformed scaler-product, \eqn{e3}, uses an integration path that follows the real line, where there are no branch cuts. 
This is shown explicitly in the bottom panel of \fig{F2}.
%
%%%
\subsection{Analytic continuation}
\label{s4b}
\par The scalar product developed in \sec{s3} is,
\begin{align}
    \label{e10}
    \brak{\tx{a}}{\tx{b}} \; = \; \int_{0}^{1} \, \tx{a}(\xi) \, \tx{b}(\xi) \;\; \xi ^{\text{B}_0} (1-\xi )^{\text{B}_1} e^{\frac{\text{B}_2}{1-\xi }} \; d\xi \; .
\end{align}
The method described in the previous section uses coordinate freedom to extend the domain upon which \eqn{e10} can be evaluated.
This is the basic purpose of the set of techniques known as \textit{analytic continuation}~\cite{Ahlfors1966,ronveaux1995heun}.
It happens that there is a more direct, and ultimately more robust, route to evaluating \eqn{e10} with analytic continuation.
\par This alternative route begins with the observation that if $\tx{a}(\xi)=\tx{b}(\xi)=1$, then the left-hand-side of \eqn{e10} coincides exactly with the integral representation of the Tricomi\footnote{The nomenclature for these functions appears to be particularly inconsistent across sources. Some readers may find Refs.~\cite{NIST:DLMF} and \cite{Chen:2010} of particular use.} confluent hypergeometric function~\cite{ARFKEN2013401,Chen:2010,abramowitz+stegun,NIST:DLMF,Leaver:1986JMP},
\begin{align}
    \label{e11}
    \int_{0}^{1} \, &\xi^{\text{B}_0} \, (1-\xi)^{\text{B}_1} \, e^{\frac{\text{B}_2}{1-\xi }}  \; d\xi
    \\ \nonumber 
    &= \; e^{\text{B}_{2}} \, \Gamma (\text{B}_{0}+1) \, U(1+\text{B}_{0},-\text{B}_{1},-\text{B}_{2}) \; .
\end{align}
In \eqn{e11}, $\Gamma(x)$ is the usual Euler-Bernoulli Gamma function, and $U(\tx{x},\tx{y},\tx{z})$ is the Tricomi confluent hypergeometric function,
\begin{align}
    \label{e12}
    U(\tx{x},\tx{y},\tx{z}) \; = \; &\frac{\Gamma(1-\tx{y})}{\Gamma(\tx{x}-\tx{y}+1)} M(\tx{x},\tx{y},\tx{z})
    \\ \nonumber 
    &\; + \;  \frac{\Gamma(\tx{y}-1)}{\Gamma(\tx{x})}\tx{z}^{1-\tx{y}}M(\tx{x}-\tx{y}+1,2-\tx{y},\tx{z}) \; .
\end{align}
Both may be evaluated for complex inputs by considering the appropriate integration contour~\cite{whittaker_watson_1996}.
Finally, in \eqn{e12}, $M(\tx{x},\tx{y},\tx{z})$ is the confluent hypergeometric function of the first kind~(See e.g. Eq. $13.2.42$ of Ref.~\cite{NIST:DLMF}).
\par The Gamma function is constructed to extend the factorial function to the complex numbers.
The confluent hypergeometric function is a product of Gamma functions and a power series in $\tx{z}$, and so it is straightforward to evaluate.
\par This parade of special functions has the following use:
Every $\tx{a}(\xi)$ and $\tx{b}(\xi)$ of interest to \eqn{e10} will be smooth, and so possesses a series expansion,
\begin{align}
    \label{e13}
    \tx{a}(\xi) \; = \; \sum_{j=0} \, \text{a}_j \, \xi^j \quad\text{  and  }\quad  \tx{b}(\xi) \; = \; \sum_{k=0} \, \text{b}_k \, \xi^k \;  .
\end{align}
This allows every scalar product to be expressed as a sum over scalar products of monomials,
\begin{align}
    \label{e14}
    \brak{\tx{a}}{\tx{b}} \; = \; \sum_{j,k} \; \text{a}_j \, \text{b}_k \, \left< \xi^{j+k} \right> \; .
\end{align}
In \eqn{e14}, $\langle \xi^p \rangle$ is the $p^\text{th}$ \textit{monomial moment} defined as,
\begin{subequations}
    \label{e15}
    \begin{align}
        \label{e15a}
        \left< \xi^{p} \right> \; &= \; \int_{0}^{1} \, \xi^{p} \, \text{W}(\xi) \; d\xi
        \\ 
        \label{e15b}
        &= \; \int_{0}^{1} \, \xi ^{\text{B}_0+p} \, (1-\xi )^{\text{B}_1} \, e^{\frac{\text{B}_2}{1-\xi }}  \; d\xi
        \\
        \label{e15c}
        &= \; e^{\text{B}_{2}} \, \Gamma (\text{B}_{0}+p+1) \, U(1+\text{B}_{0}+p,-\text{B}_{1},-\text{B}_{2}) \; . 
    \end{align}
\end{subequations}
In \eqn{e15a}, we have effectively defined $\left< \xi^{p} \right>$ to be $\brak{1}{\xi^p}$.
In \eqn{e15b}, we have applied the definition of the scalar product, and in \eqn{e15c}, we have equated the integral over $\xi$ with the product of special functions found in \eqn{e11}, but with $\tx{B}_0\rightarrow\tx{B}_0+p$.
The result is that the scalar product of any two smooth functions may be equated with a sum over confluent hypergeometric terms,
\begin{align}
    \label{e16}
    \brak{\tx{a}}{\tx{b}} \; &= \; e^{\text{B}_{2}} \, \sum_{j,k} \; \text{a}_j \, \text{b}_k \, \Gamma (\text{B}_{0}+j+k+1) \,
    \\ \nonumber
    & \hspace{2cm} \times \, U(1+\text{B}_{0}+j+k,-\text{B}_{1},-\text{B}_{2}) \; .
\end{align}
\par In practice, it is useful to precompute monomial moments using \eqn{e15}, and then compute scalar products using \eqn{e14}.
The monomial moments are defined up to an overall constant.
Henceforth, we chose to normalize all $\left< \xi^{p} \right>$ such that $\left< \xi^{1} \right>=1$,
\begin{align}
    \label{e17}
    \left< \xi^{p} \right> \; \leftarrow \; \left< \xi^{p} \right>/\left< \xi^{1} \right> \; .
\end{align}
%
%
%
%
%%%
\section{Limitations of direct integration and analytic continuation}
\label{s6}
Direct integration, analytic continuation, and their application to monomial moments and confluent Heun polynomials are all subject to various caveats and limitations. 
Here, these facets are reviewed.
At times, technical details are omitted for brevity, but particular attention is given to practical workarounds.
{While analytic continuation and direct integration are equivalent when their respective caveats are satisfied, they are otherwise found to be complementary.} 
Ultimately, analytic continuation is determined to be easier to use in practice.
\subsection{Direct Integration: disallowed \bh{} spins, and the zero frequency limit }
\label{s6a}
\par The effective weight function introduced in \sec{s4a} was
\begin{align}
    \label{l82x}
    \text{W}_{\text{eff.}}(z) \; &= \; e^{\frac{\text{B}_{2} \text{x}_{2} }{z^{\text{x}_{0} }-1}} \;
    z^{ \text{x}_{0}(\text{B}_{0}  + 1) -1} \;
    \left(1-z^{\text{x}_{0} }\right)^{\text{B}_{1}}  \;
    \\ \nonumber 
    &\quad\quad \times \; \left(1-(\text{x}_{2} +1) z^{\text{x}_{0} }\right)^{-\text{B}_{0}-\text{B}_{1}-2} \; .
\end{align}
In \sec{s4a}, a direct integration approach was introduced to evaluate scalar product using $\text{W}_{\text{eff.}}(z)$.  
This approach has two limiting features: (\textit{i}) for a given value of $\cw$, there are values of the \bh{} spin, $a$, such that an integration path cannot be constructed, and (\textit{ii}) when $\cw=0$, the weight function may diverge at $\xi=1$.
\par For any value of $\cw=\omega-i/\tau$, there may exist values of the \bh{} spin, 
\begin{align}
    \label{l83}
    a=a_*\;, 
\end{align}
such that an integration path cannot be constructed as described in \sec{s4a}.
This limitation results from the requirement that $\tx{Re}(\tx{x}_0)>0$.
\par In \sec{s4a}, the coordinate parameter $\tx{x}_0$ was introduced so that $\text{W}_{\text{eff.}}(z)$ is zero at $z=0$.
The value found for $\tx{x}_0$ to have that effect was,
\begin{align}
    \label{l84}
    \text{x}_0 \; &= \;\text{K}_{0} \, \frac{\text{sgn}(\mathrm{Re}(\text{B}_{0})+1)}{\mathrm{Re}(\text{B}_{0})+1}
    \\ 
    \nonumber % \label{l84b}
    &\quad \; + \; i \, \frac{ -1 \,  + \,  \text{K}_{0} \, [\text{sgn}(\;\mathrm{Re}(\text{B}_{0})+1\;)-1] }{\mathrm{Im}(\text{B}_{0})}
\end{align}
where $\tx{sgn}(x)$ is defined by $x=\tx{sgn}(x)\,|x|$, and 
\begin{align}
    \label{l85}
    \text{B}_0 \; =& \; -\left(\frac{2 M (\delta +M)}{\tau \delta   }+s\right) 
    \\ \nonumber
    &\hspace{0.5cm} \; + \; i \left(\frac{a m-2 M^2 \omega }{\delta }-2 M \omega \right) \; .
\end{align}
and $\tx{K}_0>0$.
As seen in \eqn{l84}, the real part of $\tx{x}_0$ is generally regular, since $\text{sgn}(\mathrm{Re}(\text{B}_{0})+1)$ is defined to be zero when $\tx{Re}(\text{B}_{0})+1$ is zero.
However, the imaginary part of $\tx{x}_0$ may diverge if $\tx{Im}(\tx{B}_0)=0$.
If we had constructed $\tx{x}_0$ differently, e.g. by allowing $\tx{Re}(\tx{x}_0)$ to depend on both $\tx{Re}(\tx{B}_0)$ and $\tx{Im}(\tx{B}_0)$, then the requirement that $\tx{Re}(\tx{x}_0)>0$ would still have resulted in a $1/\tx{Im}(\tx{B}_0)$ divergence.
\par Taking $\omega$ as known, there may exist values of the \bh{} spin, namely $a_*$, such that
\begin{subequations}
    \begin{align}
        \label{l86a}
        \text{Im}{(\text{B}_0)}|_{a=a_*} \; &= \;  0
    \end{align}
Applying the definition of $\tx{B}_0$~(see \ceqn{l85}), and recalling that $\delta=\sqrt{M^2-a^2}$ allows \eqn{l86a} to be solved for $a_*$,
    \begin{align}
        \label{l86b}
        a_* \; &= \; \frac{4 m M^2 w}{m^2+4 M^2 w^2} \; .
    \end{align}
\end{subequations}
\par For physical applications, it happens that for any \qnm{} with $(\ell,m,n)$, having frequency 
\begin{align}
    \label{l88}
    \cw(a) \; = \; \omega(a)-i/\tau(a) \; ,
\end{align}
there is a discrete set \bh{} spins where $\omega$ is such that $a=a_*$.
For a given \qnm{} $\ell$, $m$ and $n$, this set may be defined by first evaluating $\omega(a)$ using e.g. \texttt{positive.physics.qnmobj}\cite{positive:2020}, and then searching for values of $a$ such that 
\begin{align}
    \label{l89}
    |a-a_*(\cw{(a)})| \; = \; 0 \; .
\end{align}
\par For $(\ell,m,n)=(2,2,0)$, there is only one such \bh{} spin, 
\begin{align}
    \label{l90}
    a_* \; \approx \; 0.996632 \; .
\end{align}
It is instructive to contemplate the behavior of $\tx{x}_0$ when $a$ is above this critical value.
\par Towards extremal \bh{} spin, $a\rightarrow 1$, $\tx{x}_0$ typically remains regular, meaning that an integration path, $\xi(z)$, may still be constructed.
This is a result of how $\tx{x}_0$ depends on $\tx{B}_0$.
Since $\lim_{a\rightarrow 1}|\tx{Re}(\tx{B}_0)|\rightarrow \infty$, $\tx{Re}(\tx{x}_0)$ tends to zero in that same limit.
Similarly, it may be deduced that $\tx{Im}(\tx{x}_0)$ tends to zero in that limit.
Recall from \eqn{e2} that the integration path, $\xi(z)$, is not defined when $\tx{x}_0=0$.
Despite this, the reasoning above implies that $\xi(z)$ may be constructed in the near extremal limit, within the limits of numerical precision; i.e. when $\tx{x}_0$ is small, but strictly non-zero.
\par Lastly, a second coordinate parameter, $\tx{x}_2$, was introduced so that exponential divergence of the weight function was avoided~(see \csec{s4a}).
The full expression for $\tx{x}_2$ was
\begin{subequations}
\label{l91}
\begin{align}
    \label{l91a}
    \text{x}_2 \; &= \;  \frac{ \text{K}_{2}\,\text{x}_{0} }{\text{B}_2 } \; 
    \\ 
    \label{l91b}
    &= \;  \frac{ \text{K}_{2}\,\text{x}_{0} }{4i \delta \cw } \; .
\end{align}
    
\end{subequations}
In \eqn{l91b}, we have used the fact that $\tx{B}_2=4i \delta \cw$.
Since $\tx{x}_2$ is proportional to $\cw^{-1}$, it generally diverges in the zero frequency limit.
However, since only $\tx{x}_2 \,\tx{B}_2 \,=\, \text{K}_{2}\,\text{x}_{0}$ appears in the weight function~(see e.g. \ceqn{l82x}), $\text{W}_{\text{eff.}}(z)$ may still be evaluated. 
\par Further, in the zero frequency limit there is no exponential divergence of the weight function at $\xi=1$, and so there is no formal need for $\tx{x}_2$ \textit{at all} in that limit.
Seeing as 
\begin{align}
    \label{l92}
    \text{W}_{\text{eff.}}(z)|_{\cw=0} \; \propto \; \left(1-z^{\text{x}_{0} }\right)^{\text{B}_{1}} \; ,
\end{align}
$\text{W}_{\text{eff.}}$ will have a pole at $z=1$ if $\tx{Re}( \tx{B}_1 ) < 0$.
This is of no concern when $\cw\neq 0$, since $\tx{x}_2$ is designed to enforce exponential damping at $z=1$; otherwise, when $\cw=0$, it is possible to construct a different coordinate, $z'$, such that this pole is avoided.
This work-around may \textit{only} be viable when $\cw=0$; otherwise, it is found that when $\cw\neq0$, the required coordinate change necessarily introduces oscillations in $\xi(z')$ that grow exponentially as $z$ increases.
Thus, when $\cw=0$, one can only use coordinate freedom to either avoid the essential singularity at $\xi(z)=1$ (i.e. that resulting from the exponential), \textit{or }the pole at $\xi(z)=1$, but not both.
\par In this, direct integration's underlying limitation is one of mathematical perspective: it happens to be more robust to use analytic continuation via integration over closed contours, rather than singularity avoiding integration over finite paths {as has been discussed above}.
For example, Ref.~\cite{ronveaux1995heun} details how single and double Pochhammer contours may be used in similar mathematical settings. 
This approach makes use of standard theorems in complex analysis, and is ultimately equivalent to analytic continuation as outlined in \sec{s4b}.
\subsection{Analytic continuation and limitations of the Gamma function}
\label{s6b}
It was found in \sec{s4b} that scalar products may be evaluated analytically using monomial moments, $\left<\xi^p\right>$,
\begin{subequations}
    \begin{align}
        \label{l93a}
        \left< \xi^{p} \right> \; &= \; \Gamma (1+\text{B}_{0}+p) \, U(1+\text{B}_{0}+p,-\text{B}_{1},-\text{B}_{2}) \; .
    \end{align}
The qualitative behavior of the monomial moments can be described by 
    \begin{align}
        \label{l93b}
        \left< \xi^{p} \right>\; &\approx \;  \frac{ \Gamma (1+\text{B}_{0}+p) \, \Gamma(1+\text{B}_{1}) }{ \Gamma(2+\text{B}_{0}+p+\text{B}_{1}) } \;,
    \end{align}
\end{subequations}
where the leading term in the power series definition for $U(1+\text{B}_{0}+p,-\text{B}_{1},-\text{B}_{2})$ has been used to approximate it with ${ \Gamma(1+\text{B}_{1}) }/{ \Gamma(2+\text{B}_{0}+p+\text{B}_{1}) }$~\cite{NIST:DLMF}.
Since $\tx{B}_2=4i\delta\cw{}$, \eqn{l93b} is exact in the zero frequency limit, $\cw=0$, or in the extremal spin limit, $\delta=\sqrt{M^2-a^2}=0$.
\par For either \eqn{l93a} or \eqn{l93b}, $\left< \xi^{p} \right>$ will be regular when instances of the Gamma function are regular. 
Since $\Gamma(z)$ is well known to have simple poles when $z$ is a negative integer, it follows that $\left< \xi^{p} \right>$ can diverge when,
\begin{subequations}
    \label{l95}
    \begin{align}
        \label{l95a}
        1+\text{B}_{0}+p &\in \{\;z\in\mathbb{Z}\;|\;z<0\;\} \; ,
    \end{align}
and
    \begin{align}
        \label{l95b}
        1+\text{B}_{1} &\in \{\;z\in\mathbb{Z}\;|\;z<0\;\} \; .
    \end{align}
\end{subequations}
In \eqn{l95}, $\{z\in\mathbb{Z}\,|\,z<0\}$ denotes the set of all negative integers.
The practical implications of \eqn{l95} may be investigated by rewriting $\text{B}_{0}$ and $\text{B}_{1}$ in terms of physical parameters,
\begin{subequations}
    \label{l96}
    \begin{align}
        \label{l96a}
        p-s+1-\frac{2 M (\delta +M)}{\tau \, \delta }\quad\quad\quad\quad\;\;&
        \\ \nonumber
        +\; \frac{i (a m-2 M \omega  (\delta +M))}{\delta } &\;=\;-z_0 \;,
        % \frac{i \left(a m-2 M^2 \cw\right)}{\delta }-2 i M \cw-s+1+p &\;=\;-z_0 \; 
        \\
        \label{l96b}
        2 s+1-\frac{4 M}{\tau } -4 i M \omega  &\;=\;-z_1 \; .
        % -4 i M \cw+2 s+1 &\;=\;-z_1 \; .
    \end{align}
\end{subequations}
In \eqn{l96}, $z_0$ and $z_1$ are natural numbers, and we have used \eqn{p8}, \eqnsa{p15}{p23} to expand in terms of basic physical quantities.
\Eqn{l96a} is equivalent to \eqn{l95a}, and \eqn{l96b} is equivalent to \eqn{l95b}.
\par Inspection of \eqnsa{l96a}{l96b} reveals that they can only hold when $M\cw{}$ is \textit{purely imaginary}~(i.e. $\omega=0$).
This limits the physical relevance of these equations to the total transmission \qnm{s}~\cite{Cook:2014cta}.
Furthermore, \eqn{l96a} can only be true if $a=0$, thus it is only relevant to the algebraically special modes of Schwarzschild~\cite{Cook:2014cta,Berti:2003jh}.
\par Both \eqnsa{l96a}{l96b} have non-trivial implications for evaluation of the scalar product with analytic continuation.
If \eqn{l96a} holds, then there exist some monomials, $\xi^p$, such that the respective monomial moments are infinite when defined by \eqn{l93a}. 
If \eqn{l96b} holds, then no monomial moments are finite according to \eqn{l93a}.
\par These limitations are, again, a result of mathematical perspective: 
evaluation of the scalar product with Gamma and confluent hypergeometric functions may fail, depending on $\tx{Re}(\tx{B}_0)$ or $\tx{Re}(\tx{B}_1)$;
however, we saw in the previous section that the viability of direct integration only depends on $\tx{Im}(\tx{B}_0)$.
Thus the two perspectives have complementary limitations. 
\subsection{Direct Integration vs. Analytic continuation}
\label{s6c}
\par In practice, analytic continuation has the following relative attributes that may make it preferable in most circumstances.
Since analytic continuation does not require the determination of optimal coordinate parameters~(see \csec{s4a}), its implementation requires fewer steps.
In addition, analytic continuation requires monomial moments to compute scalar products, ans do does not require the storage of arrays (discretizations of $\xi(z)$) of length $10^5$ or greater~(see \csec{s4a}).
Instead, monomial moments with degrees up to twice that of the largest polynomial order of interest may be precomputed, ideally using arbitrary precision arithmetic~\cite{mpmath}, and then used for all scalar products~(\csec{s4b}).
Thus, while direct integration may serve as a valuable alternative, particularly in cases where the Gamma function cannot be evaluated, analytic continuation is expected to be structurally simpler and computationally more efficient.
\smallskip
\par All forthcoming numerical results will make use of analytic continuation with $64$ digits of precision via the public \texttt{mpmath} package~\cite{mpmath}.
\begin{figure*}
    \hspace{-1cm}
    \begin{tabular}{ccc}
        \includegraphics[width=0.34\textwidth]{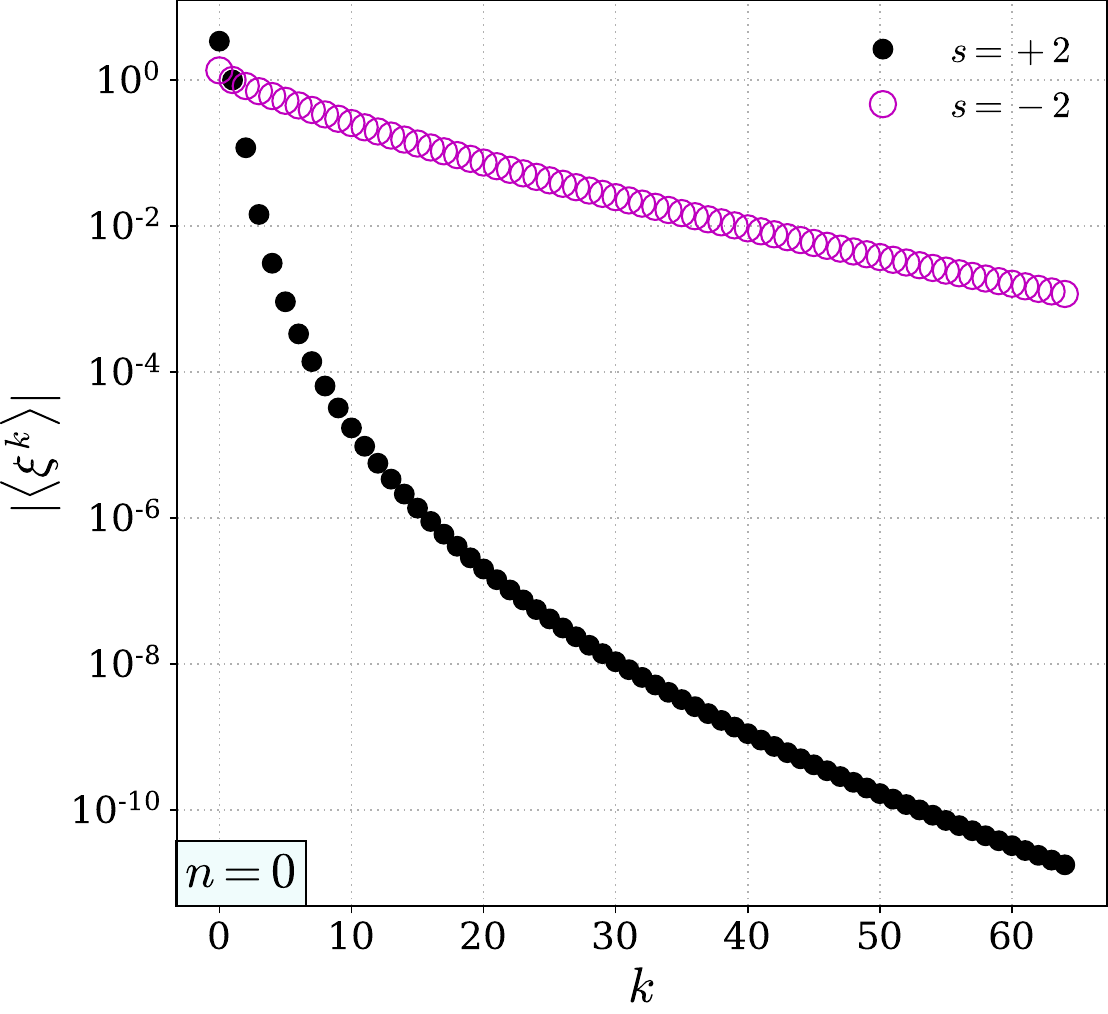}
        &
        \includegraphics[width=0.34\textwidth]{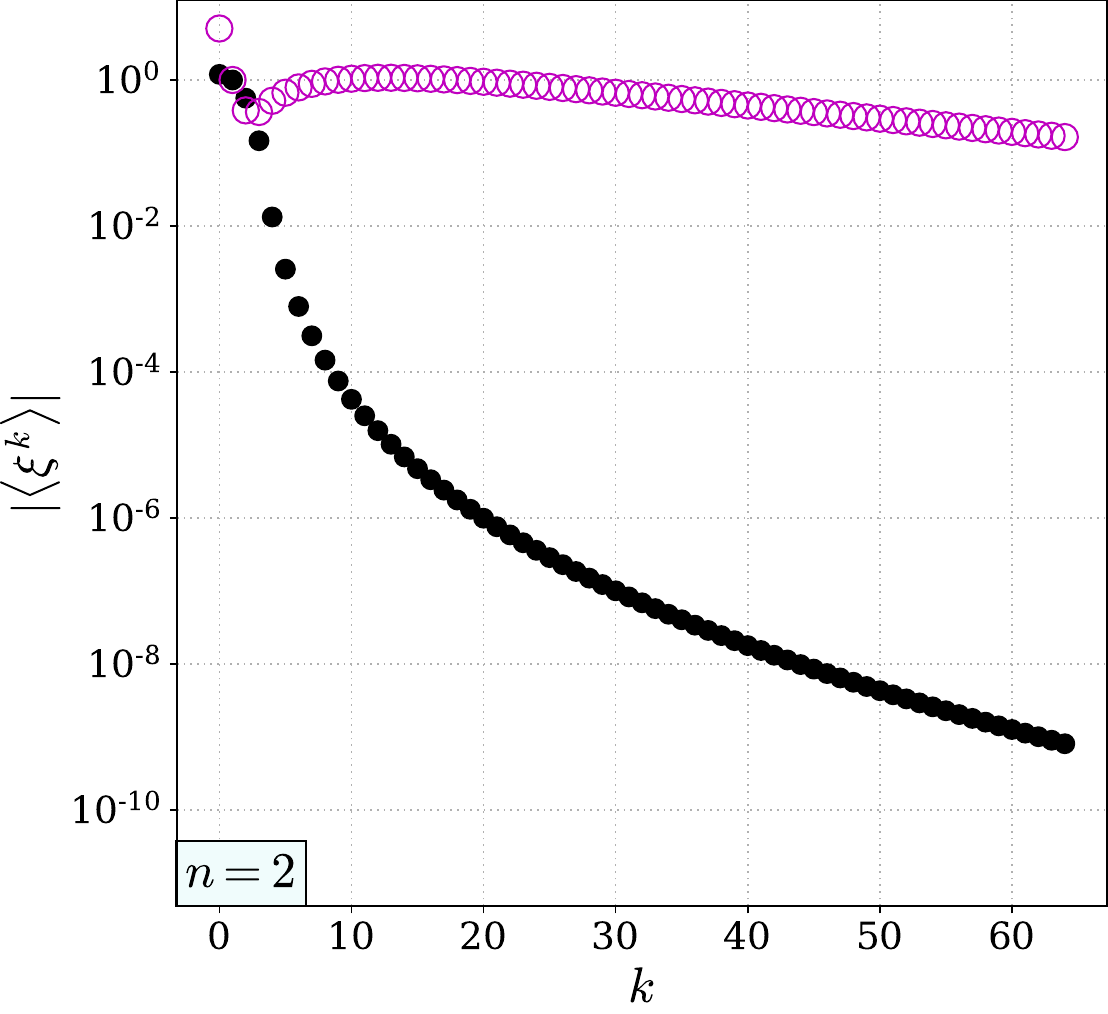}
        &
        \includegraphics[width=0.34\textwidth]{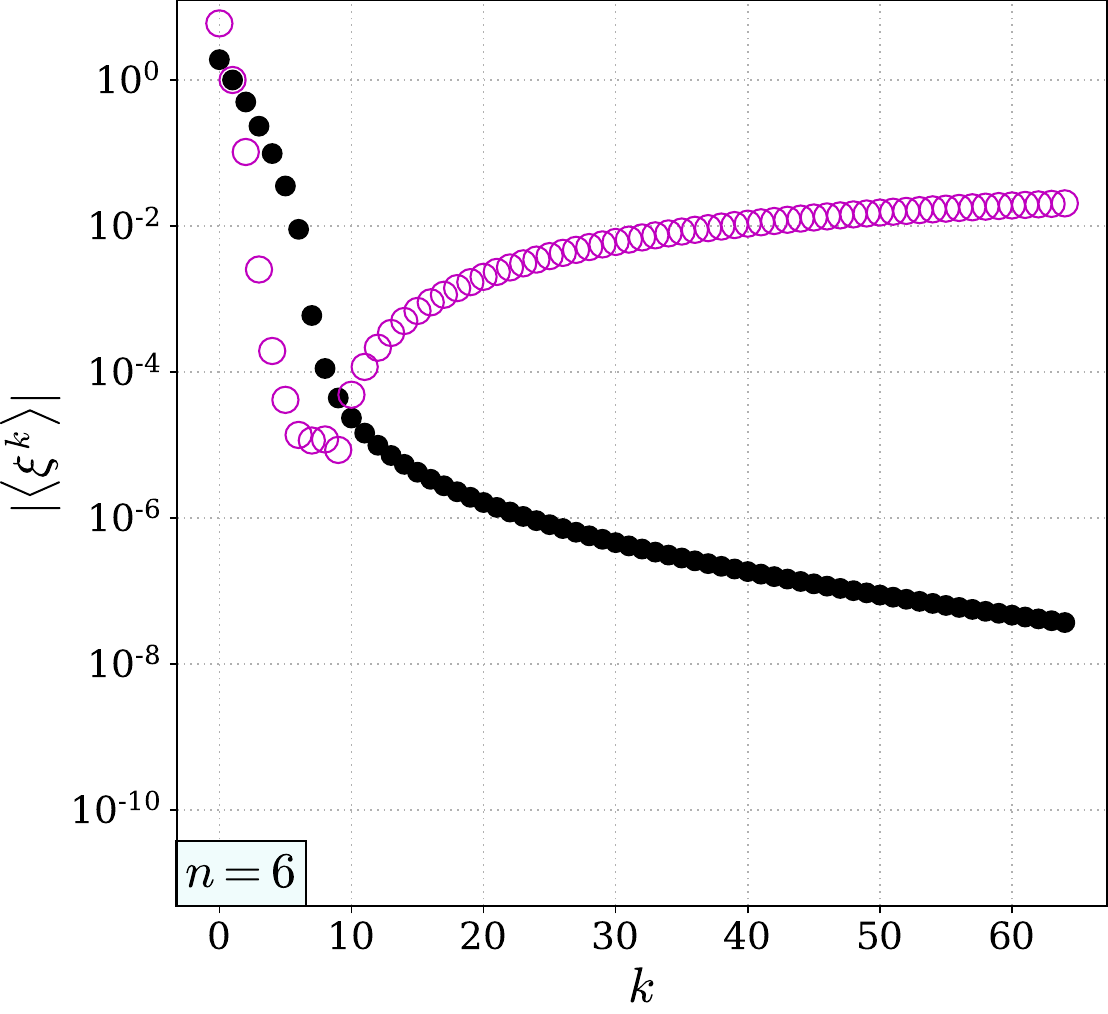}
    \end{tabular}
    \caption{
        Example distributions of absolute monomial moments, $|\langle \xi^k\rangle|$ v.s. monomial order, $k$, for the $(\ell,m)=(2,2)$ \qnm{} at \bh{} dimensionless spin of $a/M=0.7$. Left, central and right panels show results for $n$ of $0$, $2$ and $6$, respectively. The corresponding \qnm{} frequencies are $M\cw=0.5326-0.0808i$, $M\cw=0.4999-0.4123i$, and $M\cw=0.4239-1.0954i$ respectively. At focus is the effect of negating the spin weight, $s$, for increasing overtone label. For all cases shown, $|\langle \xi\rangle|=1$ has been enforced. Closed circles show points for $s=+2$, and open circles show points for $s=-2$. 
    }
    \label{F3}
\end{figure*}
%
%
%
%%%
\section{Pedagogical Applications}
\label{s5}
Previous sections provide a foundation for applying the scalar product, \eqn{e18}, to physical scenarios in which the series expansion for radial functions is either {known} (analytic continuation), or {unknown} (direct integration).
Direct integration is applicable in both scenarios.
\begin{align}
    \label{e18}
    \brak{\tx{a}}{\tx{b}} \; = \; \int_{0}^{1} \, \tx{a}(\xi) \, \tx{b}(\xi) \;\; \xi ^{\text{B}_0} (1-\xi )^{\text{B}_1} e^{\frac{\text{B}_2}{1-\xi }} \; d\xi \; .
\end{align}
Before leaping into applications to \tk{}'s radial equation, there are good reasons to first look closely at two pedagogical cases.
\par In this section, we will first apply the scalar product to the monomial moments for different \qnm{} values of the frequency $\cw$, paying particular attention to the effect of spin weight, $s$.
For a broad range of scenarios, monomial moments with $s=2$ decrease monotonically with monomial order, while this is less so the case for $s=-2$. 
We will then apply the scalar product to the confluent Heun polynomials.
\par Previously, Cook and Zalutskiy considered confluent Heun polynomials for \bh{s} in Ref.~\cite{Cook:2014cta,Fiziev:2009ud}.
There, the polynomials were defined such that a special choice $\cw{}$ resulted in $\mcL_\xi$~(\ceqn{p14}) having polynomial eigenfunctions.
These are the well known algebraically special and total transmission \qnms{}~\cite{Cook:2014cta,Andersson:1994tt,Dias:2013hn,Berti:2003jh,Whiting:1989ms,Fiziev:2009ud}. 
\par Here, we take a different perspective:
We will \textit{a priori} define $\cw{}$ to be \textit{any} \qnm{} frequency 
\footnote{In the language of Ref.~\cite{ronveaux1995heun}, we choose to focus on \texttt{Class I} of confluent Heun polynomials.}.
This requires that, in effect, we redefine the potential of $\mcL_\xi$ such that its eigenfunctions are polynomials.
{The consequences of this choice enable confluent Heun polynomials to be defined for arbitrary frequencies; in turn, this will have a significant impact on forthcoming results and discussion.}
%
% The consequences of this choice will have a significant impact on forthcoming results and discussion.
%
\par We will use the scalar product to derive the orthogonality between confluent Heun polynomials of \textit{like order}.
Lastly, we will use the scalar product to derive a relationship between the confluent Heun polynomials' eigenvalues, and a specific average of $\xi$.
This relationship provides some insight into the structure of the eigenvalues. 
%
%%%
\subsection{Monomial moments}
\label{s5a}
The scalar-product of any two smooth functions may be evaluated using a linear combination of monomial moments,
\begin{align}
    \label{e20}
    \brak{\tx{a}}{\tx{b}} \; = \; \sum_{j,k} \; \text{a}_j \, \text{b}_k \, \left< \xi^{j+k} \right> \; .
\end{align}
For many common scalar products, monomial moments $\left<\xi^{p}\right>$ decrease monotonically with $p$.
This is easily demonstrated by considering a \textit{hypothetical} constant weight function:
\begin{subequations}
    \label{e21}
    \begin{align}
        \label{e21a}
        \tx{W}(\xi)_{\text{hypothetical}} \; &= \; 1\; ,
        \\
        \label{e21b}
        \brak{\tx{a}}{\tx{b}}_{\text{hypothetical}} \; &= \; \int_{0}^{1} \, \tx{a}(\xi) \, \tx{b}(\xi) d\xi \; ,
        \\
        \label{e21c}
        \left<\xi^{p}\right> _{\text{hypothetical}}\; &= \; (p+1)^{-1}\sim p^{-1} \; .
    \end{align}
\end{subequations}
In \eqn{e21a}, a weight function of $1$ results in a scalar product equivalent to that used for the shifted Legendre polynomials, \eqn{e21b}.
In turn, this results in monomial moments that are asymptotically equivalent to one over the degree of the monomial, \eqn{e21c}. 
In practice, this means that if the sequences comprised of $\tx{a}_j$ and $\tx{b}_k$ are convergent, then so too is $\text{a}_j \, \text{b}_k \, \left< \xi^{j+k} \right>$.
We will now examine this argument in the context of the \qnm{}'s scalar product.
Our goal is to gain a basic and qualitative understanding of how monomial moments depend  on two physical parameters: $s$ and $M\cw$.
\par We previously observed that monomial moments may be computed via analytic continuation of the confluent hypergeometric function, $U$, and the gamma function, $\Gamma$,
\begin{align}
    \label{e22}
    \left< \xi^{p} \right> \; &= \; \Gamma (1+\text{B}_{0}+p) \, U(1+\text{B}_{0}+p,-\text{B}_{1},-\text{B}_{2}) \; 
\end{align}
The confluent hypergeometric function, $U$, is defined as a series expansion in its third input~\cite{NIST:DLMF}.
Thus some understanding of $\left< \xi^{p} \right>$ may be found by examining $U$ expanded in powers of $\text{B}_{2}$.
Examination of the zeroth order term happens to be sufficient for \textit{qualitative} understanding of how $\left< \xi^{p} \right>$ behaves when $p$ is large.
For precise and predictive understanding, one would have to extend the forthcoming analysis to higher orders.
This will not be done here so that the following discussion remains fairly simple.
\par To zero'th order in its third input, $U$ takes the approximate form,
\begin{align}
    \label{e23}
    U(\tx{x},\tx{y},\tx{z}) \; \approx  \; \frac{ \Gamma(1-\tx{y}) }{ \Gamma(\tx{x}-\tx{y}+1) } + \mathcal{O}(z) \; .
\end{align}
When applied to \eqn{e22}, this yields
\begin{subequations}
    \label{e24}
    \begin{align}
        \label{e24a}
        \left< \xi^{p} \right> \; &\approx \;  \frac{ \Gamma (1+\text{B}_{0}+p) \, \Gamma(1+\text{B}_{1}) }{ \Gamma(2+\text{B}_{0}+p+\text{B}_{1}) } 
        \\
        \label{e24b}
        \; &\sim \;  p^{  - 1-\text{B}_{1} } \;  \Gamma(1+\text{B}_{1}) \; .
    \end{align}
\end{subequations}
To go from \eqn{e24a} to \eqn{e24b}, we have used the fact that, as $p\rightarrow\infty$, $\Gamma(p+\alpha)\,\sim\,p^{\alpha}\,\Gamma(p)$~\cite{abramowitz+stegun}.
The physics of the situation is revealed by using \eqnsa{p23}{p15} to rewrite $\tx{B}_1$ in terms of $M$, $\cw$ and $s$,
\begin{subequations}
    \label{e25}
    \begin{align}
        \label{e25a}
        \left< \xi^{p} \right> \; &\sim \;  p^{-1-2 s+4 i M \cw} \; 
        \\ 
        \label{e25b}
        &\sim \;  p^{-1-2 s+4M/\tau} e^{4 i M \omega \, \ln(p)} \; .
    \end{align}
\end{subequations}
\par In \eqn{e25a}, we see that the previous result of $p^{-1}$ is effectively generalized to depend on the spin weight, $s$, and the dimensionless frequency, $M\cw$.
Since the net exponent of $p$ is now complex valued, it is the real part of the exponent that determines the large-$p$ behavior of $|\left< \xi^{p} \right>|$.
In \eqn{e25b}, this is shown explicitly by using $\cw=w-i/\tau$. Recall that $\tau$ is strictly positive as is needed for time domain stability.
\par The presence of $e^{4 i M \omega \, \ln(p)}$ in \eqn{e25b} means that, in general, the real and imaginary parts of $\left< \xi^{p} \right>$ will trace out a logarithmic spiral.
\par The presence of $p^{-1-2 s+4M/\tau}$ within \eqn{e25b} allows us to discern three additional features:
\begin{enumerate}
    \item[(\textit{i})] There exist some combination of $s$ and $M/\tau$ such that monomial moments either decrease, asymptote to a constant, or increase as $p$ tends towards infinity.
    \item[(\textit{ii})] Since \eqn{e25b} depends on $p^{-2 s}$, monomial moments for $s=+2$ will be much more inclined to decrease as $p\rightarrow\infty$:
    \begin{align}
        \label{e27aa}
        \left< \xi^{p} \right>|_{s=+2} \; \sim \;\ p^{-8}\, \left< \xi^{p} \right>|_{s=-2}
    \end{align}
    In \eqn{e27aa}, we have used the fact that if $\cw{}$ is a \qnm{} frequency, then $\cw|_{s=+2}=\cw|_{s=-2}$.  
    \item[(\textit{iii})] Lastly, the presence of $4M/|\tau|$ in \eqn{e25b} means that, as $\tau$ decreases, the rate of change of $\left< \xi^{p} \right>$ will typically increase.
    This is perhaps most relevant for the \qnm{} overtones, for which $|\tau|$ decreases as the overtone label, $n$, increases~\cite{Zimmerman:2015trm,Berti:2003jh}.
\end{enumerate}
\par These three qualitative features are visible in \fig{F3}:
\begin{enumerate}
    \item[(\textit{i})] In the left panel ($n=0$), and for $s=-2$, we see that the monomial moments $|\left< \xi^{p} \right>|$ decrease monotonically with $p$. However, for $s=-2$ and $n\in\{2,6\}$, moments are non-monotonic, and at $n=6$, they increase as $p$ increases.
    \item[(\textit{ii})] In each panel, we see that monomial moments for spin weight $s=+2$ are significantly depressed relative to those for $s=-2$.
    \item[(\textit{iii})] The scale of each panel is identical so that they effect of increasing overtone label (i.e. decreasing $\tau$) is easily visible for both spin weights. 
\end{enumerate}
\par The fact that absolute monomial moments, $|\left< \xi^p \right>|$, are not strictly decreasing with $p$ has implications for the numerical behavior of scalar products. 
Naively, it might seem that, if $|\left< \xi^p \right>|$ increases with $p$, then scalar products might diverge.
This happens to not be the case because the monomial moments are generally complex valued. 
\begin{figure*}[ht]
    \hspace{-1cm}
    \begin{tabular}{cc}
        \includegraphics[width=0.48\textwidth]{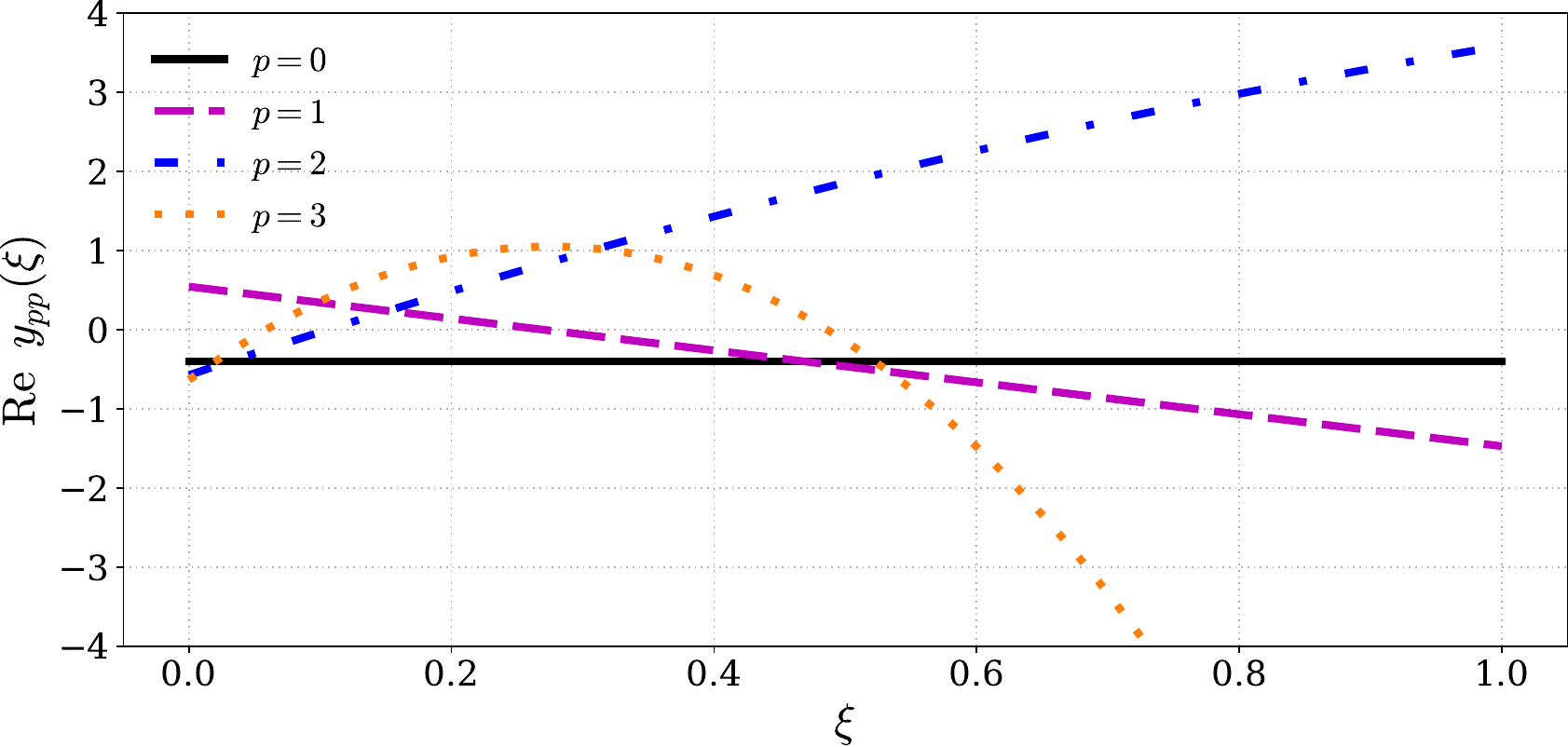}
        &
        \includegraphics[width=0.48\textwidth]{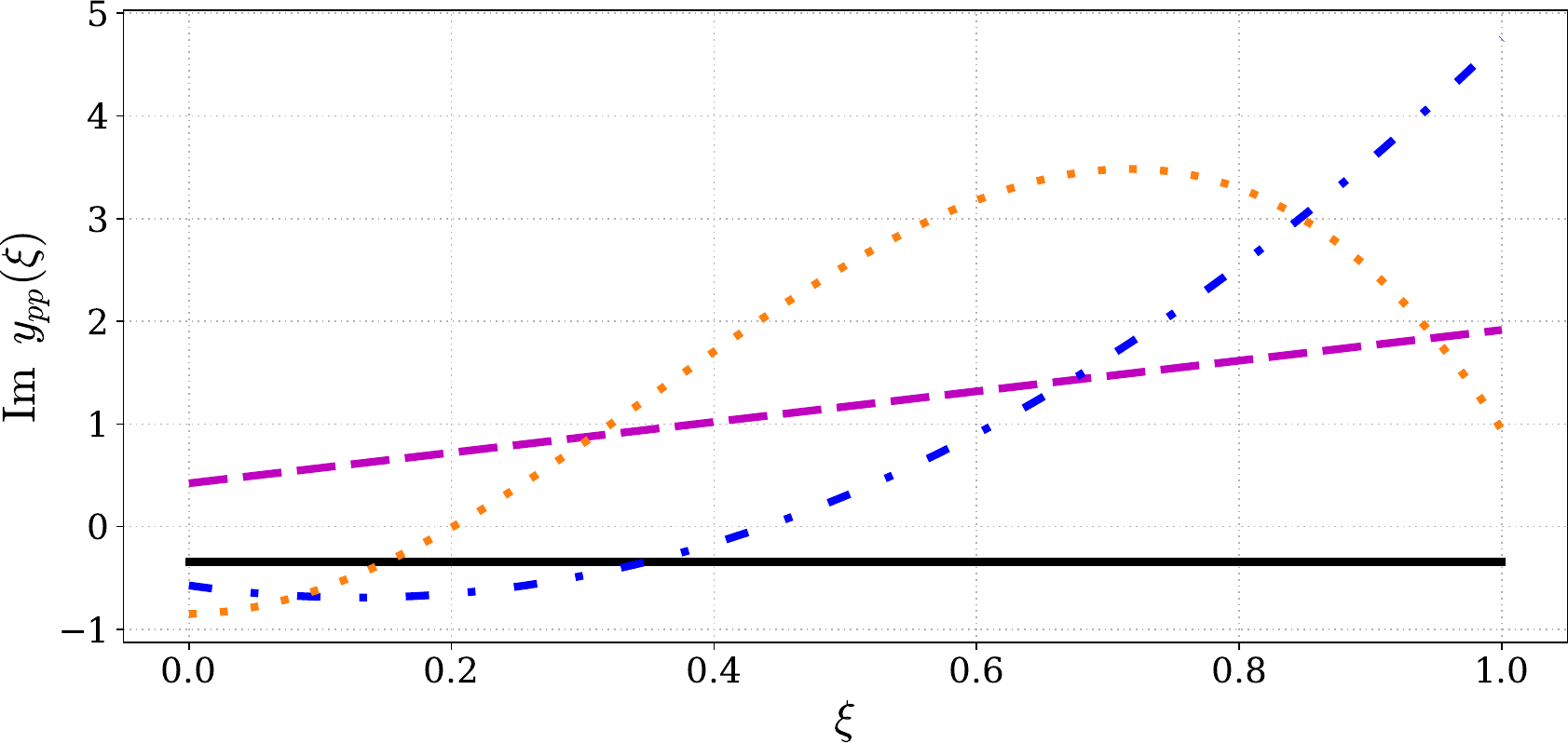}
        \\
        \includegraphics[width=0.48\textwidth]{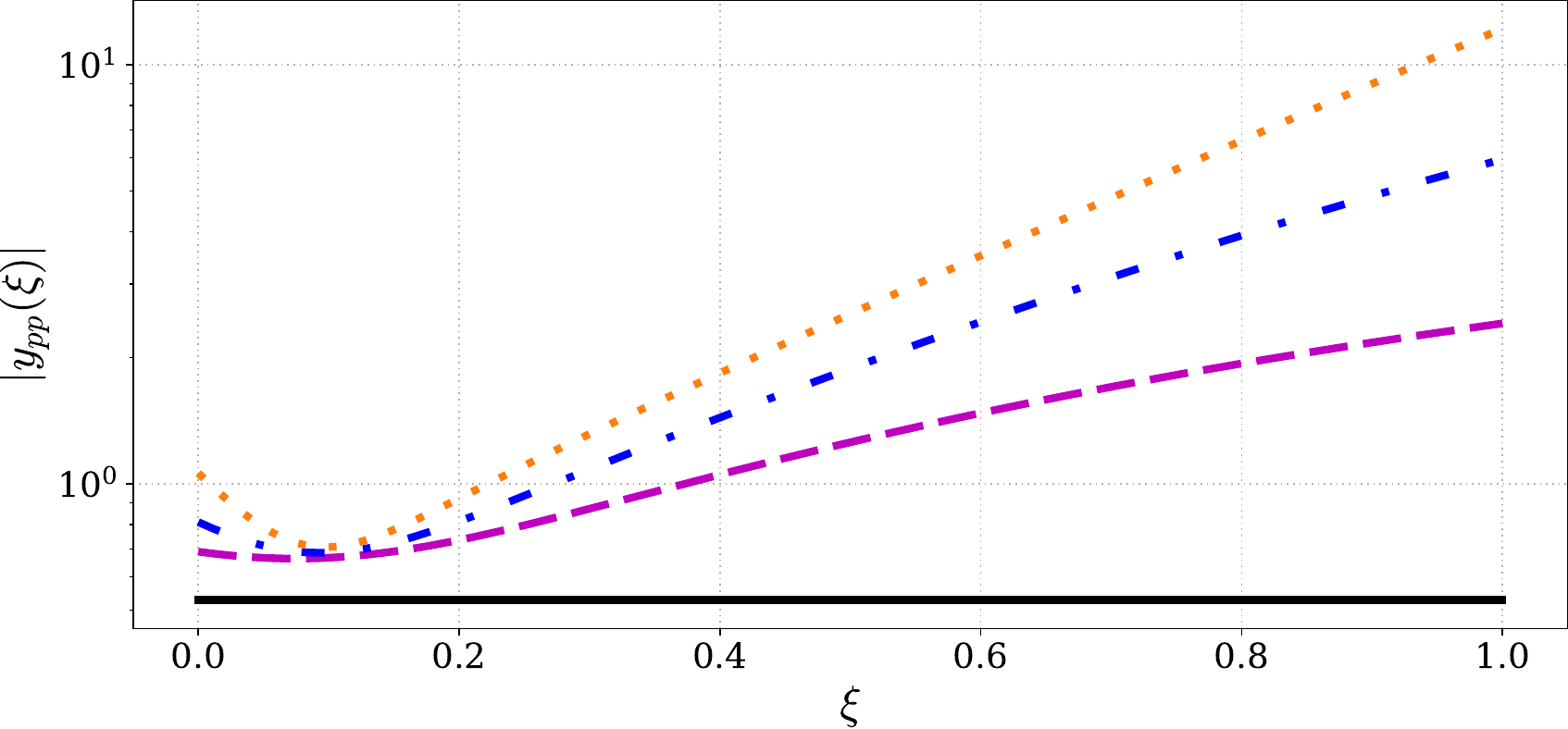}
        &
        \includegraphics[width=0.48\textwidth]{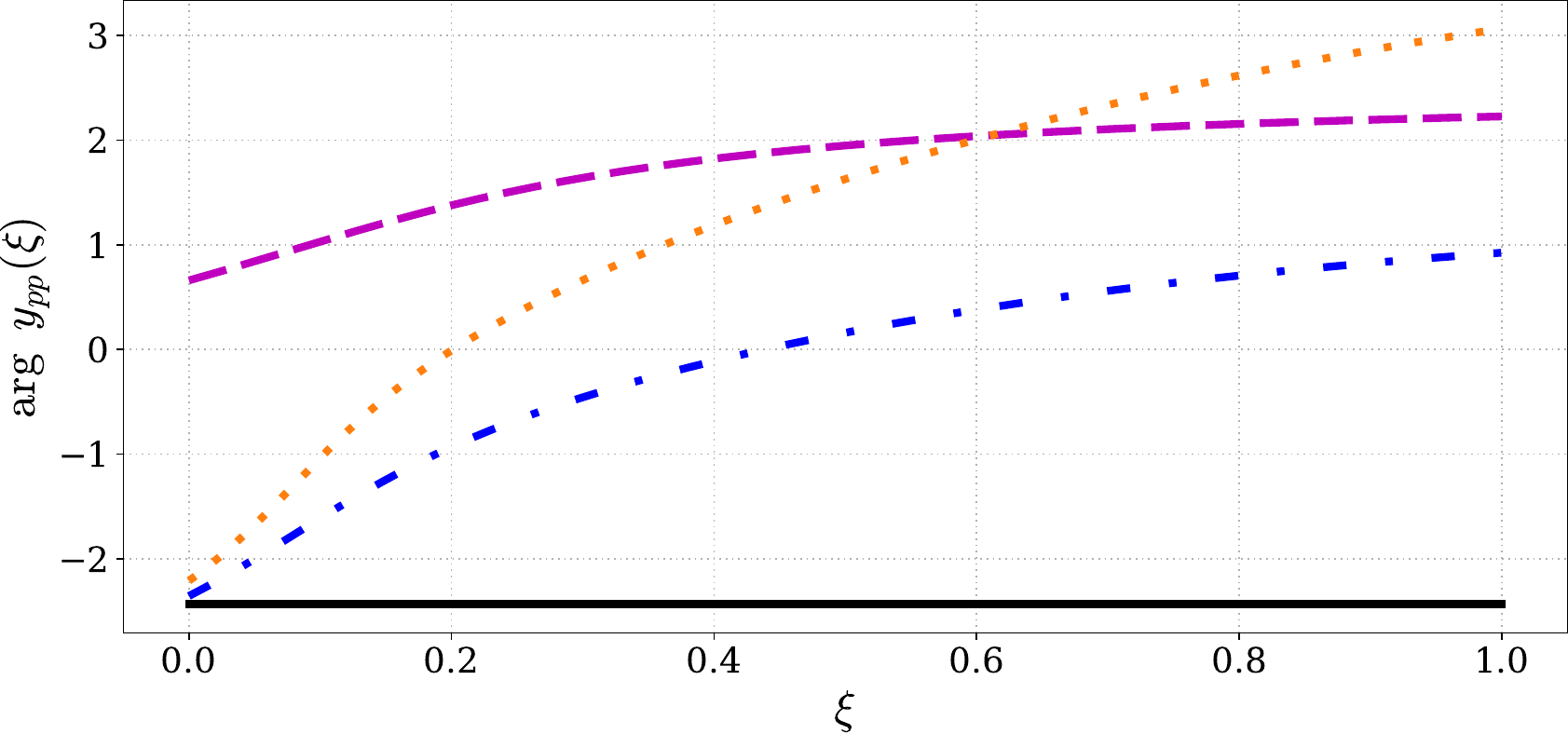}
    \end{tabular}
    
    \caption{ The first four confluent Heun polynomials for $p=k$, spin weight $s=-2$, black hole dimensionless spin $a/M=0.86$, and $\cw$ defined by the $(\ell,m,n)=(3,3,1)$ \qnm{}. The corresponding \qnm{} frequency is $M\cw=0.9840-0.2140i$. Polynomial orders $p\in \{0,1,2,3\}$ are shown as a function of the physically valued fractional radius $\xi$. The top left and right panels shows the real and imaginary parts of each polynomial, respectively. Bottom left and right panels show the corresponding polynomial amplitudes and phases, respectively.  }
    \label{F4}
\end{figure*}
\par We have previously noted that the real and imaginary parts of $\left< \xi^p \right>$ will trace out a logarithmic spiral in $p$.
This implies that linear combinations of monomial moments will, in general, have a telescoping quality: as $p$ increases, successive terms will interfere destructively, meaning that although $|\left< \xi^p \right>|$ may increase, its net contribution must be considered along with adjacent terms.
This telescoping quality behoves the use of arbitrary precision (e.g. \texttt{mpmath}~\cite{mpmath}) when evaluating moments by analytic continuation\footnote{Leaver came to the same conclusion during his analysis of series expansions in $\xi$, which he refers to as Jaff\'e-type expansion~\cite{Leaver:1986JMP}.}~\cite{Leaver:1986JMP,ronveaux1995heun}. 
However, this quality also prompts a question.
\par Since the weight function, $\tx{W}(\xi)$, allows the computation of arbitrary monomial moments, the monomials are a basis for the scalar product space defined by \eqn{e20}~\cite{adkins2012algebra,gohberg2006indefinite}.
Thus, it is fair to wonder whether there exists a related basis that is more suitable for numerical approximation?
This is the basic question that links the study of monomial moments to applied polynomial theory~\cite{chihara2011introduction}.
A complete investigation of this question is beyond the current scope.
Nevertheless, a first step towards an answer may be found by applying the scalar product to \textit{confluent Heun polynomials}~\cite{Whiting:1989ms,Cook:2014cta,Fiziev:2009,NIST:DLMF:ConfHeunPoly,MAGNUS2021105522,Fiziev:2009ud}.
%
%
%%%
\subsection{Confluent Heun polynomials}
\label{s5b}
The differential operator that results from applying \qnm{} boundary conditions to \tk{}'s radial equation was 
\begin{subequations}
    \label{e26}
    \begin{align}
        \label{e26a}
        \mcL{_\xi}  \; &= \; (\text{C}_0+\text{C}_1 (1-\xi ))
        \\
        \label{e26b}
        & + \; \left(\text{C}_2+\text{C}_3 (1-\xi )+\text{C}_4 (1-\xi )^2\right)\partial_{\xi} 
        \\
        \label{e26c}
        & + \; \xi  (\xi -1)^2 \partial_{\xi}^2 \; .
    \end{align}
\end{subequations}
Since the operator given in \eqn{e26} is of the confluent Heun type, its eigen-solutions are the aptly named confluent Heun functions~\cite{ronveaux1995heun,MAGNUS2021105522}.
\par The subset of these with terminating series solutions are, unsurprisingly, named confluent Heun polynomials~\cite{NIST:DLMF,Cook:2014cta,Fiziev:2009,ronveaux1995heun}.
\par A confluent Heun polynomial of order $p$ is typically considered to result from special values of $\tx{C}_0$ and $\tx{C}_1$~\cite{Cook:2014cta,Fiziev:2009}.
In the context of \tk{}'s radial equation, these special values have been interpreted as a constraint on $\cw$.
This is the case for the algebraically special modes of Schwarzschild and the total transmission modes of Kerr~\cite{Cook:2014cta,Berti:2003jh,Fiziev:2009ud}.
Here, a different perspective on the confluent Heun polynomial conditions for \bh{s} will be used.
This perspective is based on two observations: 
\begin{enumerate}
    \item[(\textit{i})] Classical polynomials are typically defined to be eigenfunctions of a purely differential operator (i.e. one with no $\partial_\xi^0$ term).
    For the confluent Heun operator, \eqn{e26}, this purely differential part is denoted by \eqnsa{e26b}{e26c}, and repeated below in \eqn{e27},
    \begin{align}
        \label{e27}
        \mcD{_\xi}  \; &= \; \left(\text{C}_2+\text{C}_3 (1-\xi )+\text{C}_4 (1-\xi )^2\right)\partial_{\xi}
        \\ \nonumber 
        & \hspace{.5cm} + \; \xi  (\xi -1)^2 \partial_{\xi}^2  \; .
    \end{align}
    \item[(\textit{ii})] Any attempt to construct order $p$ polynomials that are eigenfunctions of $\mathcal{D}_\xi$ requires that the ``eigenvalue'' be a \textit{linear} function of the domain variable,
    \begin{subequations}
        \label{e28}
        \begin{align}
            \label{e28a}
            \mcD{_\xi} \; \mathbf{y}_{pk}(\xi)  \; &= \; \sigma_{pk}(\xi) \; \mathbf{y}_{pk}(\xi) \;,
            \\
            \label{e28b}
            \sigma_{pk}(\xi) \; &= \; \lambda_{pk} \; + \; \mu_{p} \, \xi \; .
        \end{align}
    \end{subequations} 
    Plainly, $\sigma_{pk}(\xi)$ is not an eigenvalue in the traditional sense. 
    Instead, it is an instance of a \textit{two parameter} eigenvalue problem~\cite{ronveaux1995heun,Atkinson:2011mpe}.
    We will see that $\mu_{p}$ is uniquely determined by the order of the polynomial, meaning that it need not be conceptually separated
    from the traditional eigenvalue, $\lambda_{pk}$.
\end{enumerate}
\par In what follows, we will apply these ideas to construct confluent Heun polynomials that are valid for \textit{any} value of $\cw$ (i.e. \textit{any} \qnm{} frequency).
We will see that the operator, $\mcD{_\xi}$, is self-adjoint under the same scalar product that we have discussed previously. 
We will then use the self-adjointness of this operator to prove and numerically demonstrate the orthogonality of confluent Heun polynomials for \textit{fixed} polynomial order. 
Lastly, we will use self-adjointness to derive an expression for the confluent Heun polynomials' eigenvalues in terms of ratios of scalar products. 
\subsubsection{Construction}
\label{s5b1}
Let $\mathbf{y}_{pk}(\xi)$ be a confluent Heun polynomial of order $p$,
\begin{align}
    \label{e29}
    \mathbf{y}_{pk}(\xi) \; = \; \sum_{q=0}^{p} \, \text{a}_{pkq} \, \xi^q \; .
\end{align}
We will soon see that the label, $k$, is useful because there are will generally be $p+1$ order $p$ polynomials.
\par The polynomial coefficients, $\tx{a}_{pkq}$, are determined by applying the form above for $\mathbf{y}_{pk}(\xi)$ to the eigen-relationship given by \eqn{e28a}.
{One then asserts that the coefficients of each power in $\xi$ must be equal on both sides of \eqn{e28a}.}
The result is a three term recursion relationship or, equivalently, a discrete eigenvalue relationship.
To simplify our notation, we will henceforth drop two indices, $p$ and $k$, from $\tx{a}_{pkq}$,
\begin{align}
    \label{e30}
    \tx{a}_{q} \; \leftarrow \; \tx{a}_{pkq} \; . 
\end{align}
Thus the discrete eigenvalue relationship takes the schematic form,
\begin{align}
    \label{e31}
    \text{a}_{q-1}\, \alpha_{q} \; + \; \text{a}_{q}\, \beta_{q} \; + \; \text{a}_{q+1} \,\gamma_{q}  \; = \; \lambda_{pk}\, \text{a}_{q} \; ,
\end{align}
where
\begin{subequations}
    \begin{align}
        \label{e32}
        \alpha_{q} \; &= \; -(q-1) (\text{C}_{4}+q-2)-\mu_{p}
        \\
        \label{e33}
        \beta_{q} \; &= \;  q\, (\text{C}_{3}+2 (\text{C}_{4}+q-1))
        \\
        \label{e34}
        \gamma_{q} \; &= \; -(q+1) (\text{C}_{2}+\text{C}_{3}+\text{C}_{4}+q) \; .
    \end{align}
\end{subequations}
An initial boundary condition, 
\begin{align}
    \label{e31b}
    \text{a}_{0}\, \beta_{0} \; + \; \text{a}_{1} \,\gamma_{0}  \; = \; \lambda_{pk}\, \text{a}_{0} \; ,
\end{align}
is imposed for consistency with \eqn{e29}. In \eqn{e31b}, $\text{a}_{0}=1$ is a common initial normalization choice.
\par A necessary but insufficient condition for the three-term recursion to terminate at the $p^{\tx{th}}$ term, is simply that $\alpha_{p+1}=0$.
This is the so-called $\Delta_{p+1}=0$ condition~\cite{NIST:DLMF:ConfHeunPoly,Cook:2014cta}.
In our notation, this takes the form,
\begin{align}
    \label{e36}
    \alpha_{p+1} \; &= \; p \,(\,p+\text{C}_{4}-1)-\mu_{p} \; = \; 0 \; .
\end{align}
In turn, this can be interpreted as a constraint that allows the determination of $\mu_p$,
\begin{align}
    \label{e37}
    \mu_{p} \; &= \; p\,(\,p+\text{C}_4-1\,) \; .
\end{align}
Thus $\mu_p$ must be non-zero for the existence of non-trivial polynomial solutions to \eqn{e28a}.
\par Furthermore, if $\mu_p$ is given by \eqn{e37}, then the discrete eigenvalue relationship given by \eqn{e31} defines a finite dimensional matrix eigenvalue problem\footnote{
    Note that, along with the series' boundary conditions~(\ceqns{e31b}{e36}), it is the finite nature of the system of equations that justifies the eigenproblem interpretation. In contrast, non-polynomial solutions to the radial problem will not satisfy \eqn{e36}, meaning that e.g. a continued fraction based method, and a related boundary condition at the infinite end of the series are required~\cite{Leaver:1986JMP,leaver85}.
}~\cite{ronveaux1995heun,NIST:DLMF:ConfHeunPoly,Cook:2014cta}.
The matrix of interest, $\hat{\mathbf{A}}_p$, is asymmetric and tridiagonal,
\begin{equation}
    \label{e38}
    \hat{\mathbf{A}}_p \; = \; \left(\begin{array}{ccccc}
    \beta_0 & \gamma_0 & 0 & \cdots & 0 \\
    \alpha_1 & \beta_1 & \gamma_1 & \ddots & 0 \\
    0 & \alpha_2 & \beta_2 & \ddots & \vdots \\
    \vdots & \ddots & \ddots & \ddots & \gamma_{p-1} \\
    0 & 0 & \cdots & \alpha_p & \beta_p
    \end{array}\right) \; .
\end{equation}
By construction, the eigenvectors of $\hat{\mathbf{A}}_p$ are the polynomial coefficients, and the eigenvalues of $\hat{\mathbf{A}}_p$ are the polynomial eigenvalues,
\begin{subequations}
    \label{e39}
    \begin{align}
        \label{e39a}
        \{ \text{a}_{pkq} \}_{q=0}^{p} \; &= \; {k}^\text{th}\text{ Eigenvector of } \hat{\mathbf{A}}_p \; ,
        \\
        \label{e39b}
        \lambda_{pk} \; &= \; {k}^\text{th}\text{ Eigenvalue of } \hat{\mathbf{A}}_p \; .
    \end{align}
\end{subequations}
Since $\hat{\mathbf{A}}_p$ is a $(p+1) \times (p+1)$ matrix, there will be $p+1$ eigenvalues, $\lambda_{pk}$.
If the eigenvalues are unique, then we may impose an order on $\mathbf{y}_{pk}$, whereby 
\begin{align}
    \label{e39c}
    k \in \{0,1,2,...,p\} \; .
\end{align}
In general, the $p+1$ values of $\lambda_{pk}$ will correspond to the roots of $\hat{\mathbf{A}}_p$'s characteristic polynomial.
Lacking a deeper understanding how the roots of all such polynomials are distributed, the following ordering will henceforth be used
\begin{align}
    \label{e39d}
    \tx{Im}\,\lambda_{p0} \, < \, \tx{Im}\,\lambda_{p1} \, < \, \tx{Im}\,\lambda_{p2} \, < \, ... \, < \, \tx{Im}\,\lambda_{pp} \; .
\end{align}
In this sense, for a given order $p$, there is a \textit{multiplex} comprised of $p+1$ solutions, where each solution is labeled by the \textit{multiplex index}, $k$.
\subsubsection{Orthogonality between polynomials of like order}
\label{s5b2}
\par If each $\lambda_{pk}$ is unique, then each corresponding polynomial, $\mathbf{y}_{pk}$, will be linearly independent~\cite{Axler:2015}.
Since $\mcL_\xi=\mathcal{D}_\xi + (\tx{C}_0+(1-\xi)\tx{C}_1)$ is self-adjoint with respect to the scalar product, \eqn{e20}, $\mathcal{D}_\xi$ is also self-adjoint,
\begin{align}
    \label{e40}
    \mcD{_\xi} \; &= \;  \adj{\mcD{_\xi}} \; .
\end{align}
Here, we will use the self-adjointness of $\mathcal{D}_\xi$ to show that when $\lambda_{pk}$ are unique, $\mathbf{y}_{pk}$ must be orthogonal.
To that end we will use a standard argument that is based on the behavior of $\brak{\mathbf{y}_{p'k'}}{\mathcal{D}_\xi \mathbf{y}_{pk} }$ when $(p',k')$ swaps places with $(p,k)$.
Along the way, we will encounter a twist that results from the fact that $\mathbf{y}_{pk}$ satisfy a non-standard eigenvalue equation, \eqn{e28a}.
\par Let us begin by analyzing the matrix whose elements are $\brak{\mathbf{y}_{p'k'}}{\mathcal{D}_\xi \mathbf{y}_{pk} }$.
Since $\mathcal{D}_\xi$ is self-adjoint with respect to the scalar product~(\ceqn{p25}), we have that  
\begin{align}
    \label{e41}
    \brak{\mathbf{y}_{p'k'}} { \mcD{_\xi} \, \mathbf{y}_{pk}}  \; &= \;  \brak{\mcD{_\xi}\,\mathbf{y}_{p'k'}}{\mathbf{y}_{pk}} \; .
\end{align}
Applying the eigenvalue relationship,
\begin{align}
    \label{e42}
    \mcD{_\xi} \; \mathbf{y}_{pk}(\xi)  \; &= \; (\lambda_{pk} \; + \; \mu_{p} \, \xi) \; \mathbf{y}_{pk}(\xi)
\end{align}
to both sides of \eqn{e41} yields 
\begin{align}
    \label{e43}
    \lambda_{pk}&\brak{\mathbf{y}_{p'k'}}{\mathbf{y}_{pk}} \; + \; \mu_{p} \,\bra{\mathbf{y}_{p'k'}} \xi \; \ket{\mathbf{y}_{pk}}
    \\ \nonumber 
    &= \; \lambda_{p'k'}\brak{\mathbf{y}_{p'k'}}{\mathbf{y}_{pk}} \; + \; \mu_{p'} \,\bra{\mathbf{y}_{p'k'}} \xi \; \ket{\mathbf{y}_{pk}} \; .
\end{align}
A useful rearrangement of \eqn{e43} is given by \eqn{e44},
\begin{widetext}
    \begin{align}
        \label{e44}
        (\lambda_{pk}-\lambda_{p'k'})\,&\brak{\mathbf{y}_{p'k'}}{\mathbf{y}_{pk}} 
        \; = \; (\mu_{p'}-\mu_{p}) \,\bra{\mathbf{y}_{p'k'}} \xi \; \ket{\mathbf{y}_{pk}} \; .
    \end{align}
\end{widetext}
The aforementioned \textit{twist} is that, unlike in single parameter eigenvalue problems, the right-hand-side of \eqn{e44} is non-zero.
Nevertheless, the standard orthogonality argument is recovered by restricting \eqn{e44} to polynomials of like order, 
\begin{align}
    \label{e45}
    p\;&=\;p' \; .
\end{align}
The resulting version of \eqn{e44} is,
\begin{align}
    \label{e46}
    (\lambda_{pk}-\lambda_{pk'})\,\brak{\mathbf{y}_{pk'}}{\mathbf{y}_{pk}}  \; &= \; 0 \;.
\end{align}
If $\lambda_{pk}$ are distinct for different $k$, then \eqn{e46} can only hold if
\begin{align}
    \label{e47}
    \brak{\mathbf{y}_{pk'}}{\mathbf{y}_{pk}}  \; &= \; \eta_{pk}^2 \,  \delta_{k'k} \; . 
\end{align}
In \eqn{e47}, $\eta_{pk}$ is an inverse normalization constant. 
Henceforth, we will work with the \textit{normalized confluent Heun polynomials},
\begin{align}
    \label{e48}
    {y}_{pk}  \; &= \;  \eta_{pk}^{-1} \;\mathbf{y}_{pk}  \; ,
\end{align}
where,
\begin{align}
    \label{e49}
    \brak{{y}_{pk'}}{{y}_{pk}}  \; &= \; \delta_{k'k} \; .
\end{align}
\begin{figure*}[htb]
    \hspace{-1cm}
    \begin{tabular}{ccc}
        \includegraphics[width=0.34\textwidth]{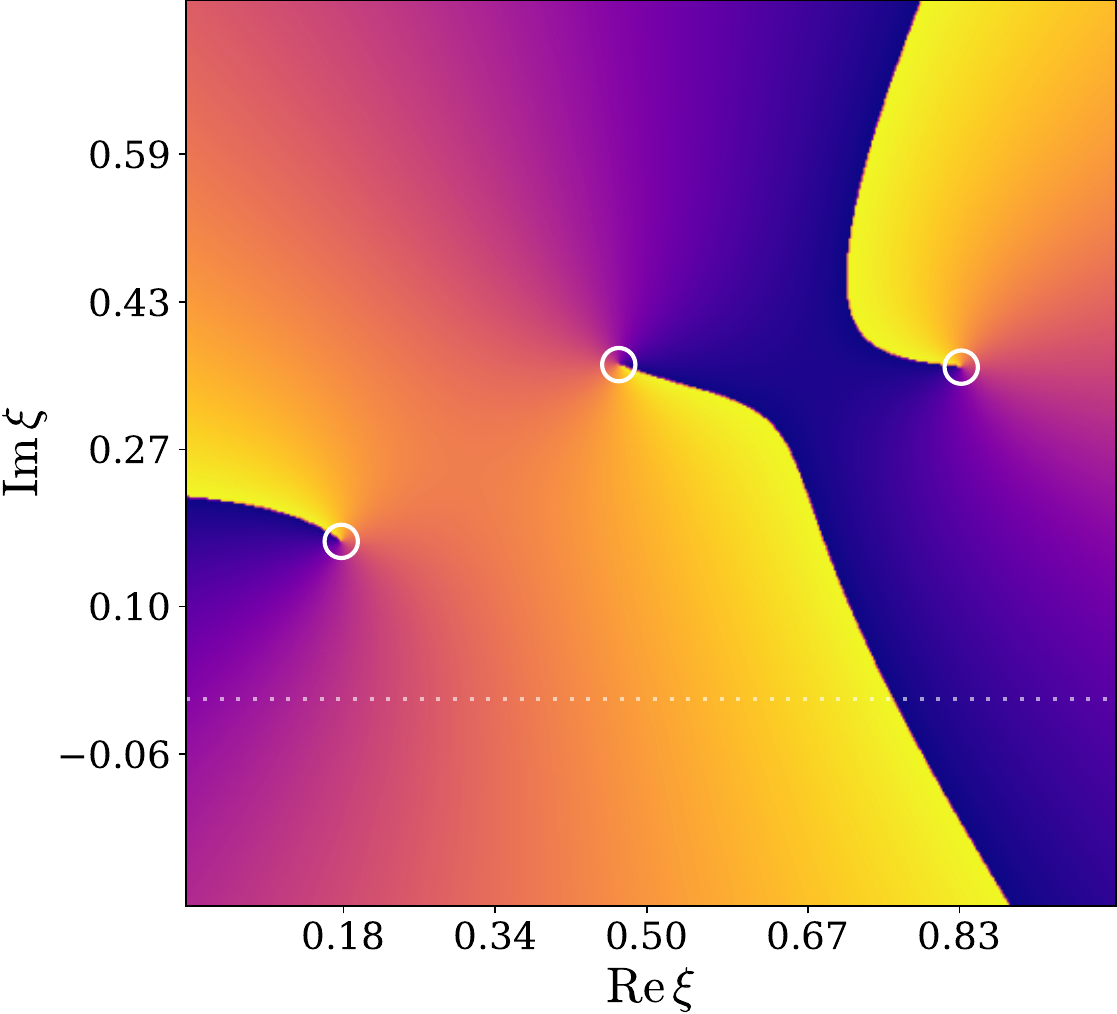}
        &
        \includegraphics[width=0.34\textwidth]{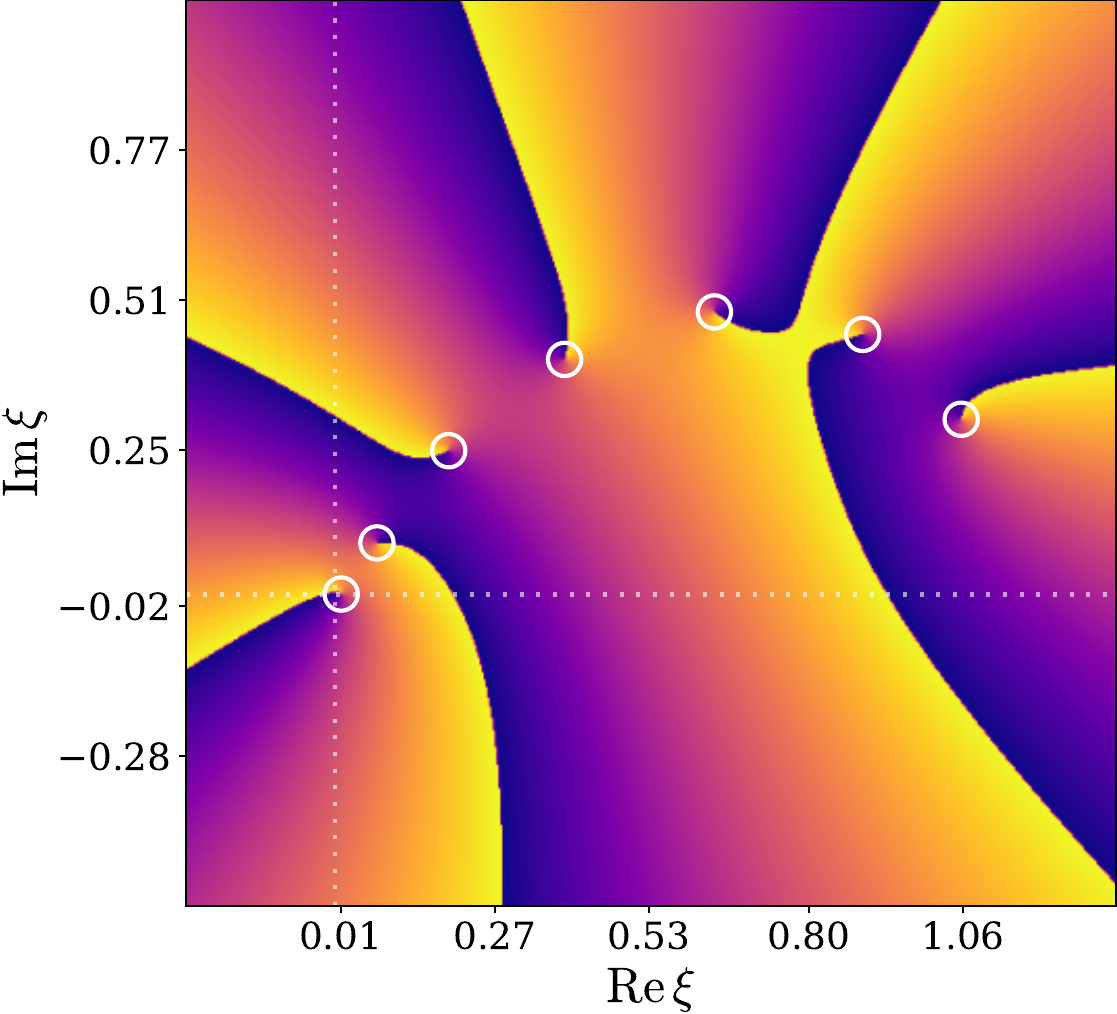}
        &
        \includegraphics[width=0.34\textwidth]{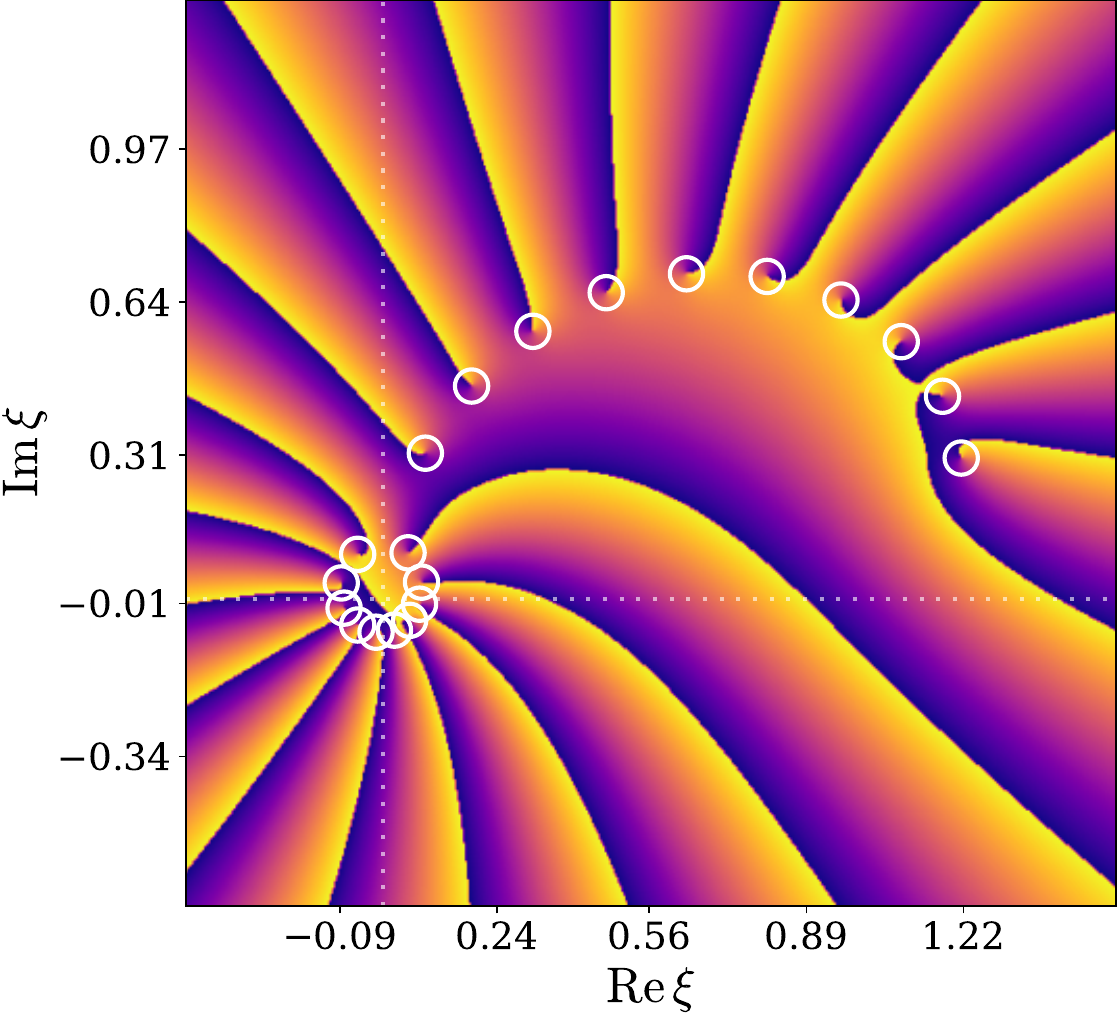}
    \end{tabular}
    \caption{
        Distribution of roots for confluent Heun polynomials, $y_{pk}(\xi)$, for three cases, all with $s=-2$, $a/M=0.7$, and $\ell=m=2$. The three cases differ in polynomial order, $p$, and the frequency, $\cw$, used: (left) $y_{33}$, the $n=0$ \qnm{} frequency, $M\cw=0.5326-0.0808i$, was used, (center) $y_{77}$, the $n=0$ \qnm{} frequency, $M\cw=0.4713-0.5843i$, was used, and (right) $y_{20,20}$, the $n=12$ \qnm{} frequency, $M\cw=0.4155-2.5050i$, was used.
        Polynomial roots are marked by white open circles. Roots were calculated using \texttt{numpy.roots}~\cite{Horm:1999ma,2020NumPy-Array}.
        Values of polynomial phase, $\arg(y_{pk})$, are shaded from blue to yellow, indicating values between $-\pi$ and $\pi$, respectively.
        Dotted horizontal and vertical lines mark the respective locations of the real and imaginary axes. 
    }
    \label{F7}
\end{figure*}
\subsubsection{A scalar product expression for eigenvalues}
\label{s5b3}
For polynomials of unlike order, \eqn{e44} may be recast as a relationship between polynomial eigenvalues.
This is done by solving for $\lambda_{pk}$,
\begin{align}
    \label{e50}
    \lambda_{pk} 
        \; = \; \lambda_{p'k'} \, + \, (\mu_{p'}-\mu_{p}) \,\frac{\bra{{y}_{p'k'}} \xi \; \ket{{y}_{pk}}}{ \brak{{y}_{p'k'}}{{y}_{pk}} } \; .
\end{align}
\Eqn{e50} applies to all $(p,k)$ and $(p',k')$, where $p\neq p'$.
An instructive special case is that of
\begin{align}
    \label{e51}
    p'=0 \; .
\end{align}
This corresponds to ${y}_{00}$, which must be constant.
If ${y}_{00}$ is a constant, then for the eigenvalue relation~(\ceqn{e42}) to hold, it must be true that   
\begin{align}
    \label{e52}
    \lambda_{00}\;&=\;\mu_{0}\;=\;0 \; .
\end{align}
Applying \eqn{e52} to \eqn{e50} gives 
\begin{align}
    \label{e53}
    \lambda_{pk}\, 
    \; &= \; \mu_{p} \, \frac{\bra{y_{00}} \xi \; \ket{y_{pk}}}{\brak{y_{00}}{y_{pk}}} \; .
\end{align}
Recalling that $\mu_{p}=p\,(\,p+\text{C}_4-1\,)$, \eqn{e53} is equivalent to 
\begin{align}
    \label{e54}
    \lambda_{pk}\, 
    \; &= \; p\,(\,p+\text{C}_4-1\,) \, \frac{\bra{y_{00}} \xi \; \ket{y_{pk}}}{\brak{y_{00}}{y_{pk}}} \; .
\end{align}
If $y_{pk}$ are known, then \eqn{e54} allows the computation of the related eigenvalues via the scalar product. {In most instances this will not be needed to define the eigenvalues; however, \eqns{e50}{e54} are presented to provide a way to check consistency between evaluations of the scalar product, and the underlying confluent Heun system.} 
\subsubsection{Numerical examples}
\label{s5b4}
Numerical computation of the confluent Heun polynomials may proceed by first computing $\tx{C}_2$, $\tx{C}_3$ and $\tx{C}_4$.
Then, for a chosen polynomial order, one would compute $\hat{\mathbf{A}}_p$~(\ceqn{e38}), so that its eigenvalues and vectors may be determined. 
Since scalar products between polynomials is a weighted sum over monomial moments~(\ceqn{e20}), it is useful to precompute and store all of the monomial moments relevant to a given polynomial order.
This allows all polynomial scalar products to be evaluated without explicit integration~(\ceqn{e15}).
From this perspective, numerical evaluation of confluent Heun polynomials may be accomplished on-the-fly via the appropriate sum over monomials~(\ceqn{e29}).
Here, this procedure is used to provide select numerical examples.   
\par \Fig{F4} shows example confluent Heun polynomials of orders $p\in\{0,1,2,3\}$.
Only the polynomials for which $p=k$ are shown; others have qualitatively similar behavior.
As expected, the real and imaginary parts of each case behave as polynomials in $\xi$. 
However, since each polynomial has complex value coefficients, the polynomials' amplitudes and phases are non-polynomial functions of $\xi$.
Similarly, since the polynomials are complex valued, we should not expect the roots of their respective real and imaginary parts to be simultaneous. 
For example, the top left panel of \fig{F4} shows $\tx{Re}\,y_{11}$ having a root near $\xi=0.5$, but $\tx{Im}\,y_{11}$ having no root at all on the real line.
\par \Fig{F7} shows the distribution of polynomial roots for three cases.
In focus is the distribution of roots in the complex plane as polynomial order and overtone label increase.
As expected, roots are typically not real valued. 
For large $n$ and large polynomial orders, roots are found to form two families, one situated around the origin, and another dispersed across the upper-right quadrant.
This is seen in the right panel mof \fig{F7}.
In all panels of \fig{F7}, each root is seen along with its associated phase boundary.
Any smooth curve, e.g. $\xi(z)$, may be defined to have a continuous phase by adding the appropriate integer factor of $2\pi$ at each phase boundary.
\par
\begin{figure}[ht]
    
    \begin{tabular}{cc}
        \hspace{-0.3cm} \includegraphics[width=0.23\textwidth]{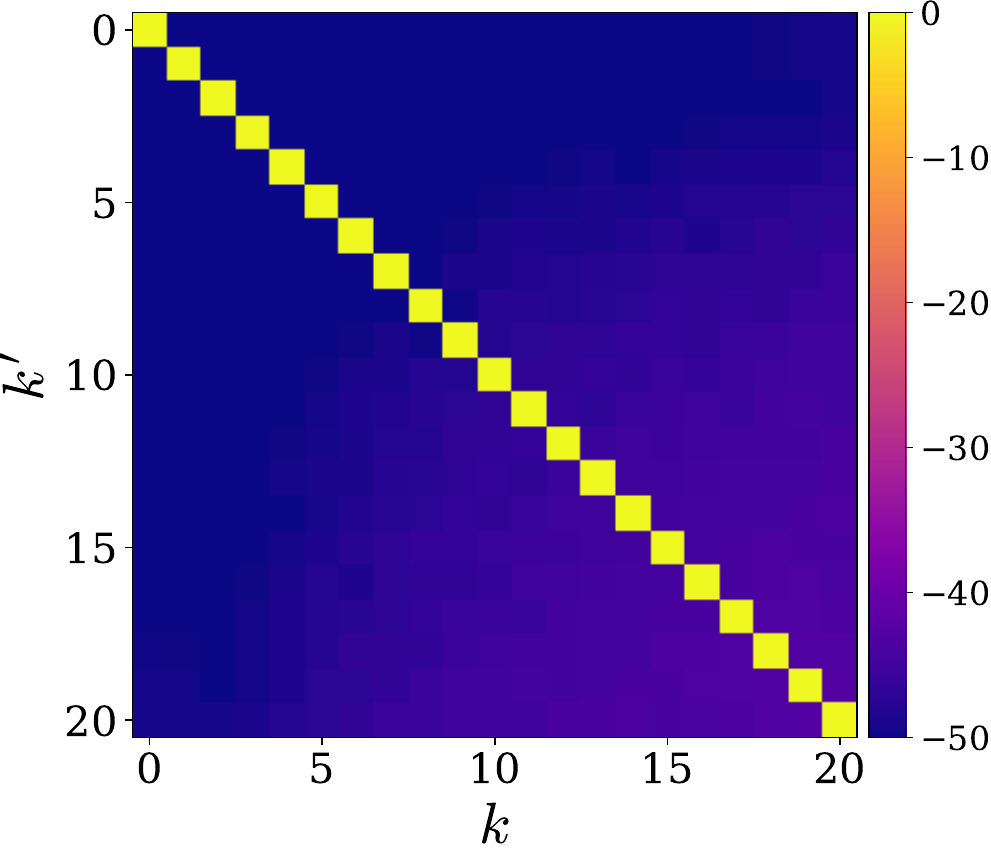}
        &
        \includegraphics[width=0.23\textwidth]{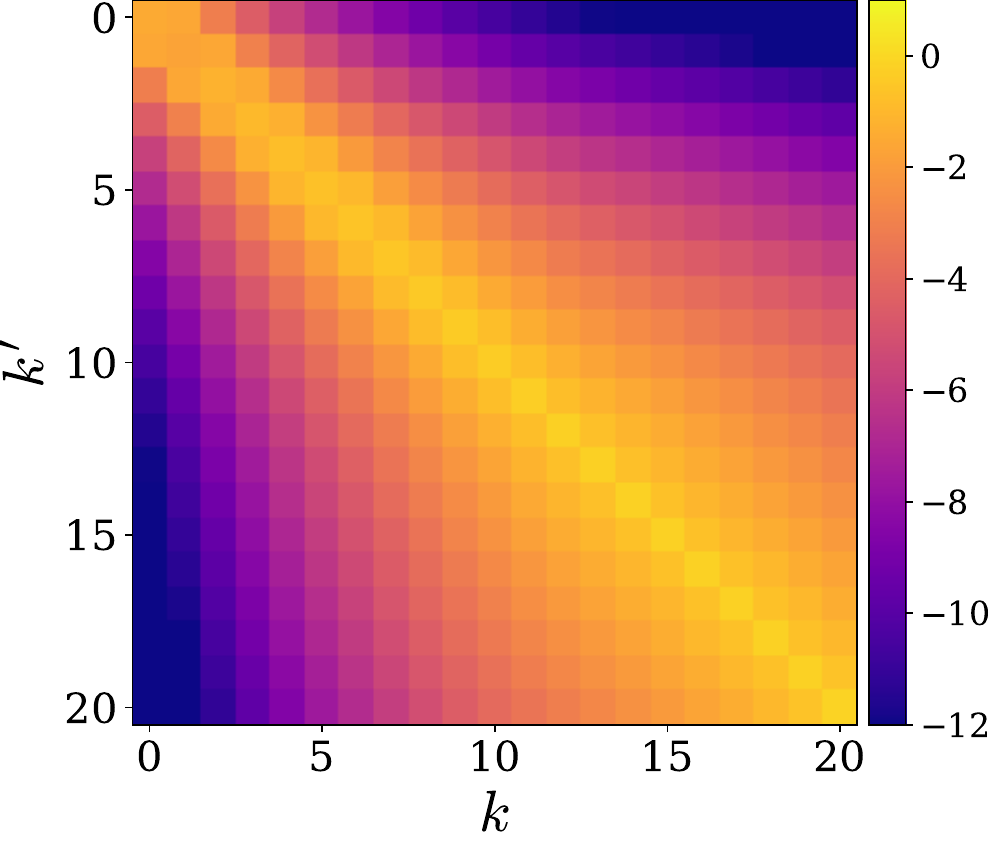}
        \\
        \hspace{-0.3cm} \includegraphics[width=0.23\textwidth]{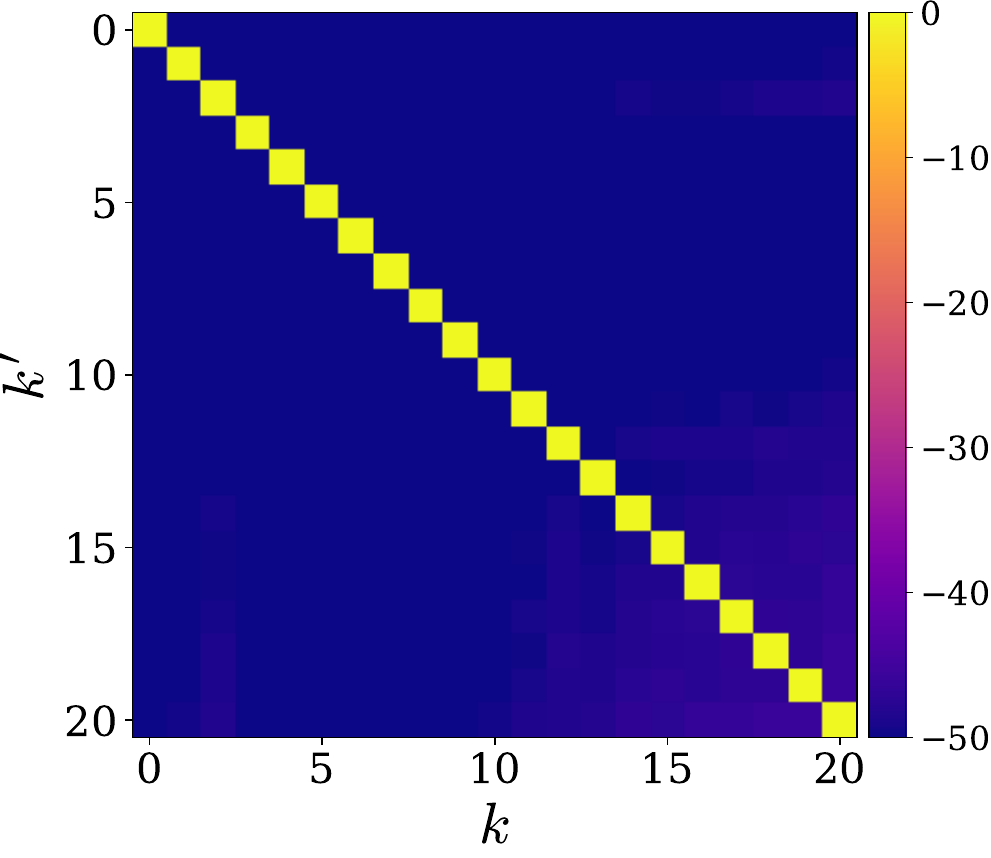}
        &
        \includegraphics[width=0.23\textwidth]{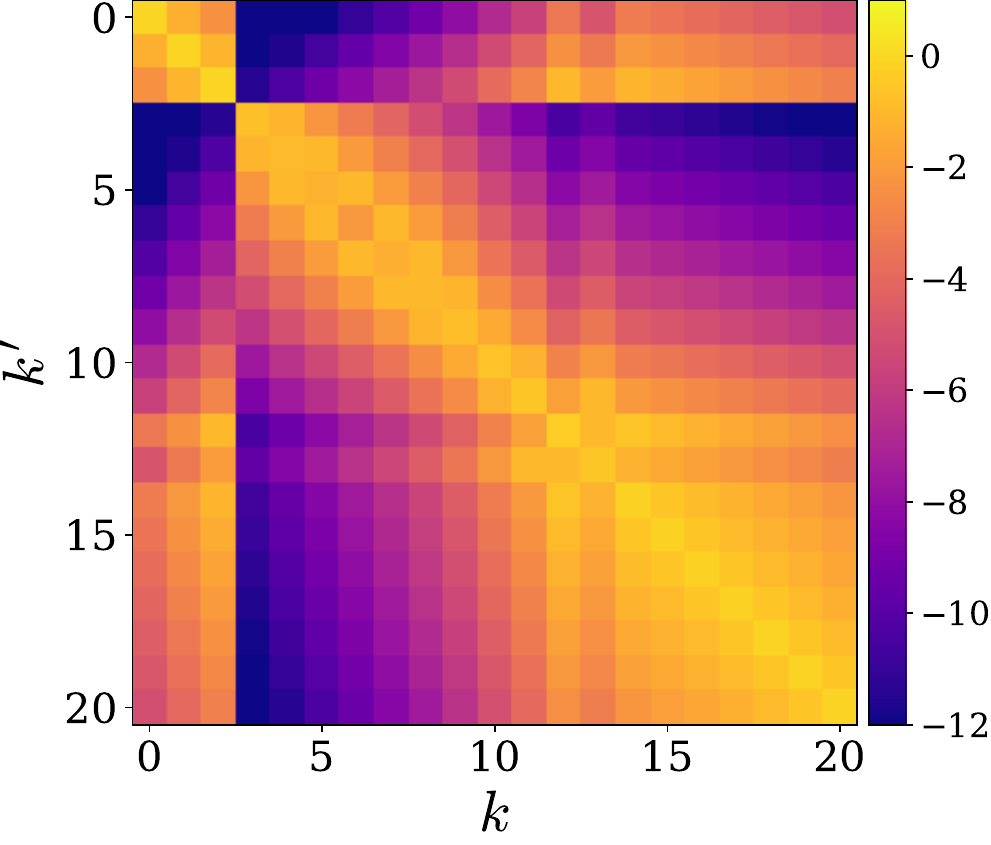}
        \\
        \hspace{-0.3cm} \includegraphics[width=0.23\textwidth]{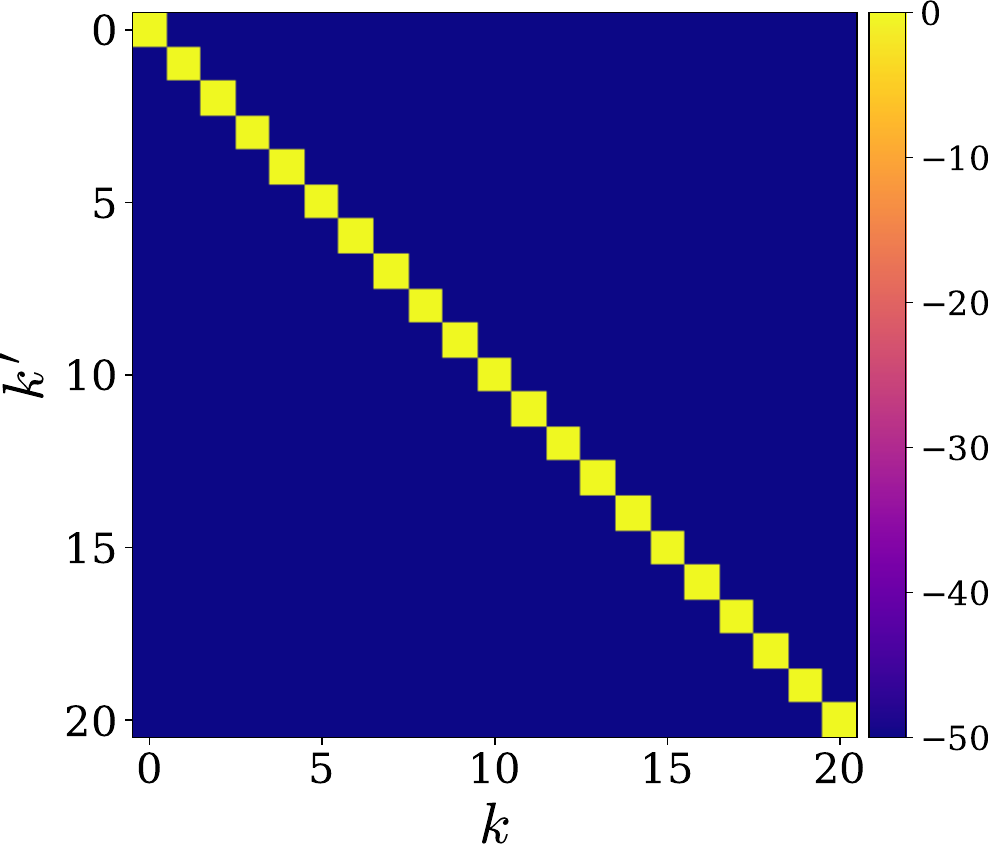}
        &
        \includegraphics[width=0.23\textwidth]{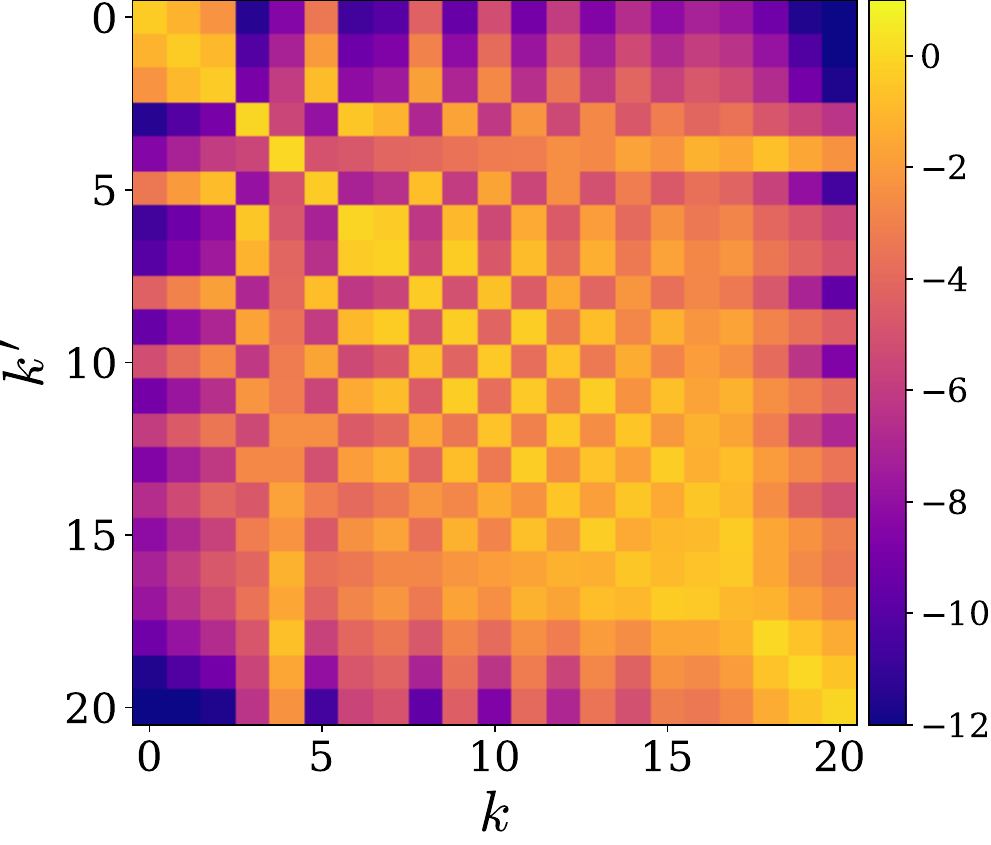}
    \end{tabular}
    
    \caption{ Gramian matrices for three \qnm{} overtone labels: $n=0$ (top row), $n=6$ (middle row), and $n=12$ (bottom row). Considered here are the order $p=20$ normalized confluent Heun polynomials with $s=2$, $a/M=0.70$ and \qnm{} frequency labels $(\ell,m)=(2,2)$. The \qnm{} frequencies for top middle and bottom rows are $M\cw=0.5326-0.0808i$, $M\cw=0.4239-1.0954i$ and $M\cw=0.4098-3.1270i$, respectively. The left column shows the absolute values of $\langle y_{pk'} | y_{pk} \rangle$; there, note that the color bar is log scaled between $10^{-50}$ and $10^0$. The right column shows absolute values of $\langle y_{pk'}| \xi | y_{pk} \rangle$; there, the color bar is logs scaled between $10^{-12}$ and $10^0$ to accentuate graph structure. 
    }
    \label{F5} 

\end{figure}

 \Fig{F5} shows examples of orthogonality at fixed polynomial order, 
\begin{align}
    p\;=\;20 \; .
\end{align}
for three cases.
At focus, is the effect of increasing \qnm{} overtone label or, equivalent, decreasing the \qnm{} decay time $\tau$, such that $1/\tau$ increases within $\cw=\omega-i/\tau$.
The $p=20$ polynomials are considered to be somewhat extreme cases: for this and all lower polynomial orders, no appreciable numerical artifacts are encountered for the numerical and physical parameters used here. 
\par The left columns of \fig{F5} show the absolute values of Gramian matrix elements, $\langle y_{pk'} | y_{pk} \rangle$, and the right columns show absolute values of $\langle y_{pk'}| \xi | y_{pk} \rangle$.
Scalar products have been evaluated using analytic continuation~(\ceqn{e20}) with $64$ significant figures of precision via the \texttt{dps} setting of \texttt{mpmath}~\cite{mpmath}.
\par Orthogonality of the polynomials at fixed order is seen through the diagonal structure of each Gramian.
The top left panel, having $n=0$, displays a slight deviation away from orthogonality (i.e. values near $10^{-40}$ rather than $10^{-50}$ and below) as $k$ nears it maximum value of $20$.
\par This effect has two causes, one technical and the other mathematical.
The technical cause of this feature is simply the value of the precision parameter, \texttt{dps}$=64$, which means that numerical results near $10^{-64}$ are more prone to round-off error. 
The mathematical cause of this feature is that, like the legendre polynomials, the confluent Heun polynomials have normalization constants that generally decrease as polynomial order increases\footnote{For the Legendre polynomials, the analog of $\eta_{pk}^2$~(see \ceqn{e47}) is $(2p+1)^{-1}$.}. 
Consequently the action of normalization effectively scales each polynomial by a very large number, exacerbating finite precision errors.
At worst, square normalization constants might be $10^{-64}$ or smaller and would thereby be corrupted by finite precision error.
In turn, this would correspond to normalization constants $10^{-32}$ or smaller, and the act of normalization would be to scale the non-normalized polynomials by a corrupted number of order $10^{32}$.
This situation can be avoided by increasing numerical precision.
\smallskip
\par The confluent Heun polynomials, while orthogonal at fixed order, do not display a feature common among classical polynomials. 
In particular, they do not possess three-term-recursion relations between polynomials of different values of $k$~\cite{chihara2011introduction,ARFKEN2013401}.
This may be understood thusly: Since $ \xi \times y_{pk}$ is a polynomial or order $p+1$, it is to be expected that it cannot be written in terms of \textit{only} polynomials of order $p$.
Consequently, the matrix whose elements are $\langle y_{pk'}| \xi | y_{pk} \rangle$ will in general have no non-zero values, and therefore will not be band-diagonal.
\par This feature of the confluent Heun polynomials is shown in the right column of \fig{F5}.
There, the structure of $|\langle y_{pk'}| \xi | y_{pk} \rangle|$ is seen to increase non-trivially with overtone label, $n$.
This may be partially understood in the context of the polynomial eigenvalues.
\smallskip
\par
\begin{figure*}[ht]
    
    \hspace{-1cm}
    \begin{tabular}{cc}
        \includegraphics[width=0.48\textwidth]{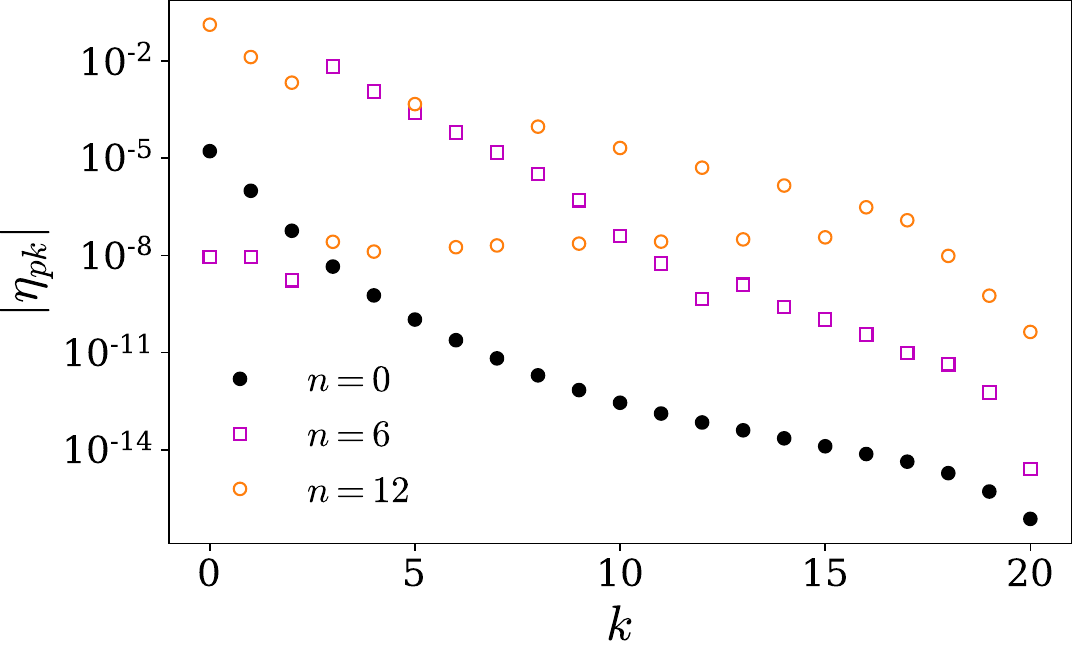}
        &
        \hspace{0.1cm}
        \includegraphics[width=0.48\textwidth]{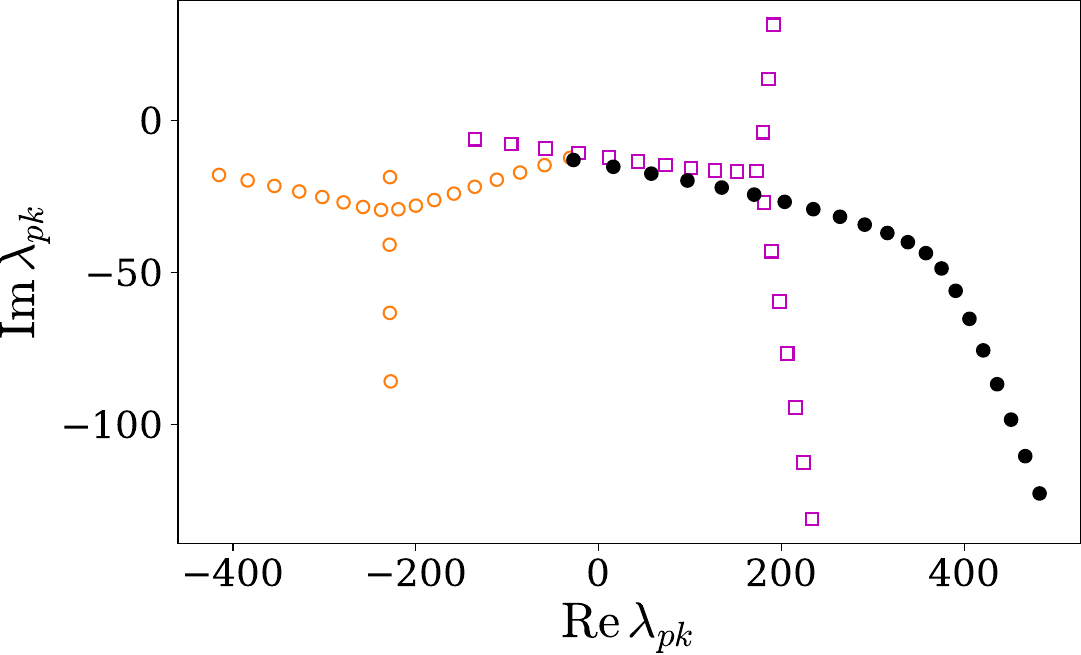} 
    \end{tabular}
    
    \caption{ 
        Distribution of normalization constants, $\eta_{pk}$, and eigenvalues, $\lambda_{pk} = \mathrm{Re}\lambda_{pk} + i\,\mathrm{Im}\lambda_{pk}$, for three different \qnm{} overtone labels: $n\in\{0,6,12\}$. For all values of $n$, the spin weight is $s=2$, the black hole dimensionless spin is $a/M=0.70$, and the \qnm{} $(\ell,m)$ are $(2,2)$. The left panel shows normalization constants, and the right panel shows eigenvalues. Black dots mark $n=0$, orange circles mark $n=6$, and magenta squares mark $n=12$.
      }
    \label{F6}
\end{figure*}

 It was previously noted that the small deviations from orthogonality are linked to small normalization constants, and that increasing structure in the distribution of $|\langle y_{pk'}| \xi | y_{pk} \rangle|$ are linked to the polynomial eigenvalues.
Polynomial normalization constants and eigenvalues are shown, respectively, in the left and right panels of \fig{F6}.
There, the same three cases seen in \fig{F5} are considered.
\par At focus in the left panel of \fig{F6} is the distribution of normalization constants. 
While normalization constants are found to generally decrease with polynomial order (not shown in figure), they are also found to generally decrease as $k$ increases, and they are found to generally increase as the imaginary part $\cw$ increases.
%
% Branching in the distribution of normalization constants is linked to branching in the respective polynomial eigenvalues.
%
\par At focus in the right panel of \fig{F6} is how the distribution of eigenvalues changes as the imaginary part of $\cw$ becomes more negative (i.e. as $n$ increases).
For $n=0$, either $\tx{Re}\,\cw$ or $\tx{Im}\,\cw$ may be used to define equivalent orders (i.e. maps between polynomials and the index $k$).
This is not the case for $n\in\{6,12\}$, where branches appear in the eigenvalue distribution. 
In these cases, the eigenvalues are still unique, so a well defined order can still be constructed.
%
%%%
\section{Implications for analytic solutions to \tk{}'s radial equation}
\label{s7}
\par We previously set out to apply two concepts to \tk{}'s radial problem.
When stated in the context of the radial equation, these two concepts are:
\begin{itemize}
    \item[(\textit{i})] That a scalar product may be developed from the structure of the radial equation~(i.e. \ref{Q1}).  
    \item[(\textit{ii})] That special functions may be developed from the structure of the radial equation, and that these functions \textit{may} be used to exactly represent analytic solutions~(i.e. \ref{Q2}).
\end{itemize}
\par Previous sections show that concept (\textit{i}) is indeed applicable to the radial equation, thus affirmatively answering \ref{Q1}. 
A radial scalar product was developed in \sec{s3}.
Methods for evaluation of the scalar product were developed in \sec{s4}.
Connections between the scalar product and concept (\textit{ii}) were discussed in \sec{s5}.
In particular, the scalar products connection with problem specific special functions (i.e. the confluent Heun polynomials) was detailed in \sec{s5b}.
Here, we take a first step towards applying confluent Heun polynomials to the representation of analytic solutions to \tk{}'s radial equation for \qnm{s}.
A non-classical feature of the confluent Heun polynomials means that, for now, \textit{only a preliminary and conceptual step} is taken.
\par Recall that \tk{}'s radial equation is a confluent Heun equation of the form,
\begin{align}
    \label{i98}
    \mcL{_\xi}  \; &= \; [\text{C}_0+\text{C}_1 (1-\xi )]+ \mathcal{D}_{\xi}\;.
\end{align}
In \eqn{i98}, the transformed radial operator, $\mcL{_\xi}$, is simply rewritten such the differential part is explicitly identified as $\mathcal{D}_{\xi}$~ (see \ceqn{e27}).
In \sec{s5b}, it was found that $\mathcal{D}_{\xi}$ generates confluent Heun polynomials, $y_{pk}(\xi)$, such that each $y_{pk}(\xi)$ satisfies a two parameter eigenvalue problem,
\begin{align}
    \label{i99}
    \mcD{_\xi} \; \ket{{y}_{pk}}  \; &= \; (\lambda_{pk} \; + \; \mu_{p} \, \xi )\; \ket{{y}_{pk}} \;,
\end{align}
where, for a given polynomial order $p$, there are $p+1$ solutions, $y_{pk}$.
This situation differs starkly from that of classical (e.g. Jacobi) polynomials, for which every \textit{single} polynomial order maps to a \textit{single} polynomial solution.
In other words, for each polynomial order, classical polynomials have a \textit{simplex} of solutions, while the confluent Heun polynomials have a \textit{multiplex}.
\par The existence of a multiplex for each polynomial order prompts a nontrivial conceptual tension:
The Jacobi polynomials' recursion relationships play a key role in the angular problem~(see \capx{Apx-3}); however, there cannot exist similar recursion relationships between the confluent Heun polynomials, $y_{pk}$.
This is because such recursion relationships require an orthogonal polynomial sequence, wherein each polynomial has a unique order (i.e. is a simplex)~\cite{chihara2011introduction}. 
Further, since each confluent Heun polynomial multiplex is composed of polynomials of like order, the collection of all such multiplexes would appear to contain \textit{too many polynomials}, meaning that the they allow for multiple equivalent representations (i.e. decompositions) of radial functions. 
\smallskip
\par This does not, however, preclude the possibility that there are special combinations of confluent Heun polynomials allow for unique representations, and classical recursion relationships. 
While this possibility is perhaps too open ended to be fully pursued here, we may \textit{assume} (somewhat adventurously) that such polynomials exist, and then contemplate implications for \tk{}'s radial equation:
\smallskip
\par \textit{Suppose that solutions to the radial equation, $\ket{f}$, can be exactly represented in terms of orthonormal polynomials, $u_n$,
\begin{subequations}
    \label{i100}
    \begin{align}
        \label{i100a}
        \ket{f} \; = \; \sum_{p=0}^{\infty} a_{p} \, \ket{u_{p}} \; ,
    \end{align}
where it is additionally supposed that
    \begin{align}
        \label{i100b}
        \ket{u_p} \; &= \; \sum_{k=0}^{p} \, \text{c}_{pk} \, \ket{{y}_{pk}} \; ,
        \\
        \label{i100c}
        \I \; &= \; \sum_{p=0}^{\infty}\,\ket{u_p}\bra{u_p}  \; ,
    \end{align}
and
    \begin{align}
        \label{i100d}
        \xi \ket{u_p} = \sigma_{p0}\ket{u_{p-1}} + \sigma_{p1}\ket{u_{p}} + \sigma_{p2}\ket{u_{p+1}} \; .
    \end{align}
\end{subequations}
}
\par In \eqns{i100a}{i100d}, despite not being classical, $u_n$ are constructed to have the essential properties of classical polynomials, namely completeness, orthonormality and three-term recursion~\cite{chihara2011introduction,Chen:2010}.
In this sense, we may think of $u_n$ as having \textit{canonical} properties. 
This motivates their being referred to as \textit{canonical confluent Heun polynomials}.
\par Further, in \eqns{i100a}{i100d}, $a_p$, $c_{pk}$, $\sigma_{p0}$, $\sigma_{p1}$ and $\sigma_{p2}$ are assumed to be generally non-zero complex valued constants.
Henceforth, $\sigma_{pj}$ will refer to complex valued constants that are assumed to be known if $u_p$ are known.
In \eqn{i100c}, $\I$ is the identity operator for the space of all confluent Heun functions of the fractional radial coordinate, $\xi$. 
\Eqn{i100c} requires that the expansion coefficients present in \eqn{i100a} are $a_p=\brak{u_p}{f}$.
The three-term-recursion relation given by \eqn{i100d} is a universal feature of all (simplexical) orthogonal polynomial sequences, and it is a direct consequence of \eqn{i100c} (i.e. completeness) along with the fact that $\brak{u_j}{\xi|u_k}=\brak{\xi u_j}{u_k}=\brak{u_j}{\xi u_k}$~\cite{chihara2011introduction}.
\par Naively, if each order $p$ polynomial multiplex is a $p+1$ dimensional orthonormal basis, then it should be capable of exactly representing any order $p$ polynomial on the same domain~\cite{Axler:2015}.
Thus it may be possible to construct $u_p$ such that \eqns{i100b}{i100d} hold.
\smallskip
\par In effect, \eqn{i100} fully defines a solution ansatz.
This solution ansatz may be used to investigate whether the scalar product and confluent Heun polynomials imply that \tk{}'s radial equation is, like the angular equation, exactly tridiagonalizable.
To this end, we may attempt to represent the radial equation in the basis defined by $\{u_p\}_{p=0}^{\infty}$.
\smallskip
\par In bra-ket notation, the transformed radial equation is
\begin{align}
    \label{i101}
    \mcL{_\xi} \, \ket{f} \; &= \; A \, \ket{f} \; .
\end{align}
Applying \eqn{i100c} such that $\mcL_{\xi}$ is represented as a matrix, $\hat{L}$, and $\ket{f}$ a vector, $\vec{f}$, gives
\begin{subequations}
    \label{i102}
    \begin{align}
        \label{i102a}
        \sum_{p,p'} \, \ket{u_{p'}}&\bra{u_{p'}} \, \mcL{_\xi} \,\ket{u_{p}} \brak{u_p}{f}\; \; 
        \\ \nonumber
        &= \; A \, \sum_{p} \, \ket{u_{p}}\brak{u_p}{f} \; ,
        \\
        \label{i102b}
        \hat{L}\,\vec{f} \; &= \; A\, \vec{f} \; .
    \end{align}
\end{subequations}
In \eqn{i102b}, the elements of $\hat{L}$ are $\bra{u_{p'}} \, \mcL{_\xi} \,\ket{u_{p}}$, and the elements of $\vec{f}$ are $\brak{u_p}{f}$.
Henceforth, $p'$ will label rows, and $p$ will label columns, meaning that $p \ge p'$ denotes the upper triangle of $\hat{L}$, and $p \le p'$ its lower.
To compute the matrix elements, $\bra{u_{p'}} \, \mcL{_\xi} \,\ket{u_{p}}$, the action of $\mcL{_\xi}$ on $\ket{u_p}$ must be considered in more detail.
\par Rewriting $\mcL{_\xi}$ in terms of $\mathcal{D}_{\xi}$~(see \ceqn{i98}) allows $\mcL{_\xi} \,\ket{u_{p}}$ to be expanded as
\begin{align}
    \label{i103}
    \mcL{_\xi} \,\ket{u_{p}} \; = \; (\text{C}_0+\text{C}_1 [1-\xi] ) \,\ket{u_{p}}+ \mathcal{D}_{\xi} \,\ket{u_{p}} \; .
\end{align}
In \eqn{i103}, $\mathcal{D}_{\xi} \,\ket{u_{p}}$ can be rewritten using the definition of $u_p$ in terms of confluent Heun polynomials~(\ceqn{i100b}), and the two parameter eigenvalue relationship~(\ceqn{i99}).
Doing so yields
\begin{subequations}
    \label{i104}
    \begin{align}
        \label{i104a}
        \mathcal{D}_{\xi} \,\ket{u_{p}} \; &= \; \sum_{k=0}^{p} \, \text{c}_{pk} \, \mathcal{D}_{\xi} \, \ket{{y}_{pk}}
        \\ 
        \label{i104b}
        &= \; \sum_{k=0}^{p} \, \text{c}_{pk} \, (\lambda_{pk}\,+\,\mu_{p}\,\xi) \, \ket{{y}_{pk}} \; ,
        \\ 
        \label{i104c}
        &= \; \ket{v_p} \, + \, \mu_p \,\xi \,\ket{u_p} \; .
    \end{align}
\end{subequations}
In going from \eqn{i104b} to \eqn{i104c}, the term dependent on the confluent Heun eigenvalues, $\lambda_{pk}$, has been used to define a new polynomial,
\begin{align}
    \label{i105}
    \ket{v_p} \; = \; \sum_{k=0}^{p} \, \text{c}_{pk} \, \lambda_{pk} \, \ket{{y}_{pk}} \; ,
\end{align}
and it has been noticed that the remaining term in \eqn{i104b} is simply proportional to $\ket{u_n}$.
\smallskip
\par It follows from \eqn{i104c} that $\mcL{_\xi} \,\ket{u_{p}}$ may be rewritten as
\begin{align}
    \label{i106}
    \mcL{_\xi} \ket{u_{p}}  =  (\text{C}_0+\text{C}_1)\ket{u_{p}} +  \ket{v_p}  +  (\mu_p-\tx{C}_1) \xi \ket{u_p} \; ,
\end{align}
and that the matrix element, $\bra{u_{p'}} \mcL{_\xi} \ket{u_{p}}$, is 
\begin{subequations}
    \label{i107}
    \begin{align}
        \label{i107a}
        \bra{u_{p'}} \mcL{_\xi} \ket{u_{p}}  \;=\;  &(\text{C}_0+\text{C}_1)\brak{u_{p'}}{u_{p}} 
        \\
        \label{i107b}
        & +  (\mu_p-\tx{C}_1) \bra{u_{p'}}\xi \ket{u_p}  
        \\ 
        \label{i107c}
        & +  \brak{u_{p'}}{v_p} \; .
    \end{align}
\end{subequations}
Since $\ket{u_p}$ are orthonormal (i.e. $\brak{u_{p'}}{u_{p}}=\delta_{p'p}$), \eqn{i107a} only contributes to diagonal matrix elements.
Since $\ket{u_p}$ possess a three term recursion relation given by \eqn{i100d}, \eqn{i107b} will only contribute when $|p'-p| \le 1$.
\smallskip
\par It happens that $\brak{u_{p'}}{v_p}$ is not only non-zero when $|p'-p| \le 1$, but also restricted to the upper-triangle (i.e. $p \ge p'$).
This follows from the fact that $\mcD_\xi$ is self-adjoint w.r.t. the scalar product, meaning that
\begin{align}
    \label{i108}
    \brak{u_{p'}}{\mcD_\xi \, u_p} \; = \; \brak{\mcD_\xi \, u_{p'}}{ u_p} \; .
\end{align}
Using \eqn{i104c} to expand both sides of \eqn{i108} yields,
\begin{align}
    \label{i109}
    \brak{u_{p'}}{v_p} - \brak{v_{p'}}{u_p} \; = \; (\mu_p - \mu_{p'})\brak{u_{p'}}{\xi | u_p} \; .
\end{align}
\Eqn{i109} may be understood thusly: 
Since $u_{p'}$ is a member of a complete orthonormal sequence, its scalar product with any order $p$ polynomial will be non-zero if $p \ge p'$ and zero otherwise~\cite{chihara2011introduction}.
In other words, any order $p$ polynomial may be expressed as a linear combination of $u_{p'}$, where $p' \le p$ is required since the order of interest is $p$. 
Thus $\brak{u_{p'}}{v_p}$ and $\brak{v_{p'}}{u_p}$ \textit{only} have non-zero values in the respective upper ($p\ge p'$) and lower ($p\le p'$) triangle of $\hat{L}$. 
Further, $\xi\,u_p$ is a linear combination of $\{u_{p-1},u_p,u_{p+1}\}$ due to \eqn{i100d}, so the matrix whose elements are $\brak{u_{p'}}{\xi | u_p}$ will be tridiagonal.
Together, these points mean that the only way that \eqn{i109} can hold, i.e. the only way that the upper and lower triangular structure of $\brak{u_{p'}}{v_p}$ and $\brak{v_{p'}}{u_p}$ can be equal to the tridiagonal structure of $\brak{u_{p'}}{\xi | u_p}$, is if $\{\brak{u_{p'}}{v_p}\}_{p'p}$ is upper-triangular \textit{and} tridiagonal.
This means that $\ket{v_p}$ must be a linear combination of $u_{p-1}$ and $u_{p}$, 
\begin{align}
    \label{i110}
    \ket{v_p} = \sigma_{p3}\ket{u_{p-1}} + \sigma_{p4}\ket{u_{p}}  \; ,
\end{align}
where $\sigma_{p3}=\brak{u_{p-1}}{v_p}$ and $\sigma_{p4}=\brak{u_p}{v_p}$ are complex valued constants which are determined by $c_{pk}$ (i.e. $u_p$) and the confluent Heun eigenvalues $\lambda_{pk}$.
\par Together, \eqn{i100d}, \eqn{i107} and \eqn{i110} allow $\mcL_\xi \ket{u_p}$ to be written as 
\begin{align}
    \label{i111}
    \mcL_\xi \ket{u_p}  \; = \;\sigma_{p5} \, \ket{u_{p-1}} 
    + \sigma_{p6} \, \ket{u_{p}} 
    + \sigma_{p7} \, \ket{u_{p+1}} \; ,
\end{align} 
where
\begin{subequations}
    \label{i112}
    \begin{align}
        \sigma_{p5} \; &= \; (\mu_p-\tx{C}_1)\sigma_{p0} + \sigma_{p3} \; ,
        \\
        \sigma_{p6} \; &= \; (\mu_p-\tx{C}_1)\sigma_{p1} + \sigma_{p4} + (\text{C}_0+\text{C}_1) \; ,
        \\
        \sigma_{p7} \; &= \; (\mu_p-\tx{C}_1)\sigma_{p2} \; .
    \end{align}
\end{subequations}
The implication of \eqns{i111}{i112} is that, if the canonical polynomials $u_n$ can be determined, then \tk{}'s radial operator is exactly tridiagonalizable.
%
%%%
\section{Discussion}
\label{s8}
\par The primary results of this article are
(\textit{i}) the introduction of a radial scalar product for \qnms~(\csec{s3}), 
(\textit{ii}) the development of \qnm{} specific confluent Heun polynomials that may be defined for arbitrary frequency parameter~(\csec{s5b}), and 
(\textit{iii}) the demonstration that, in principle, the scalar product and proposed canonical confluent Heun polynomials enable the exact tridiagonalization of \tk{'s} radial equation for \qnms{}~(\csec{s7}).
\par It is also of note that the scalar product may be evaluated by direct integration with a complex valued radius, or by more standard analytic extension using confluent hypergeometric functions~(See \csec{s5a}). 
While analytic extension was found to be the more practical route, the ability to construct a complex radial coordinate, $\xi(z)$, for evaluation of the scalar product also implies (via \eqn{p30}) that {the \qnm{s} themselves are spatially bounded on $\xi(z)$}.
Since $\xi(z)$ must be constructed to coincide with Sturm-Liouville boundary conditions, a \qnm{} on $\xi(z)$ would have physically relevant values at both the event horizon and spatial infinity. 
\par Regarding the canonical polynomials, a clear next step is to investigate their computation.
Like the Gram-Schmidt polynomials, which are unique once certain processing choices are made, one might expect that there are multiple constructions of the canonical polynomials.
This a topic of ongoing research that is near conclusion~\cite{London:202XP2}.    
\par Similarly, the tridiagonalization of \tk{}'s radial equation directly implies that its solutions (for fixed frequency parameter) are orthogonal. 
This would be directly analogous to the orthogonality of the spheroidal harmonics at fixed frequency (i.e. fixed ``oblateness'')~\cite{London:2020uva}.
It would also be compatible with the conclusions of complementary studies of linear perturbations of Kerr that have encountered a kind of \qnm{} orthogonality~\cite{Green:2022htq}.
\par An adjacent implication is that the radial functions are complete (i.e. isomorphic to canonical confluent Heun polynomials).
If the radial functions are both orthogonal and complete for given value of the frequency parameter $\cw$, then this would strongly imply the existence of a biorthogonal dual for the radial solutions determined by the \qnm{} frequencies.
Again, this would be directly analogous to the \qnm{'s} spheroidal harmonics, where each spheroidal harmonic requires input from a different \qnm{} frequency~\cite{London:2020uva}.
A consequence would be that the \qnms{} are ``\textit{quasi-complete}'', i.e. complete in the radial and angular domains, but not complete (specifically under- or over-complete) in the time domain.
Investigation of these implications is also a topic of ongoing work that is near conclusion~\cite{London:202XP3}. 
\par Lastly, these ideas imply a potentially useful direction of research for e.g. \bbh{} post-merger signals: rather than attempting to fit various \qnm{s'} within numerical data, it may be possible to \textit{directly project} out individual \qnm{s'} time series using orthogonality. 
In the context of ongoing efforts in Numerical Relativity and signal modeling for future \gw{} detectors, this may be a fruitful direction for future research.
\section{Acknowledgements}
The author is thankful to Katy Clough, Michelle Foucoin, Scott Hughes, Bernard Whiting and Melize Ferrus for useful discussions and feedback. Lionel London was supported at the University of Amsterdam by the GRAPPA Prize Postdoctoral Fellowship, and then at King's College London by the Royal Society ({URF{\textbackslash}R1{\textbackslash}211451}).
\appendix
\section{Notation}
\label{Apx-1}
This section lists explicit expressions for the main text's schematic notation. The details here are provided for convenience and completeness. 
\par In Boyer-Lindquist coordinates, \tk{}'s master differential operator, $\LMaster$, is
\begin{align}
    \LMaster \;  \; = \; \left[\frac{(r^2+a^2)^2}{\Delta(r)} - a^2\sin^2\theta\right]
       \frac{\partial^2}{\partial{t}^2}
     & \\ \nonumber
    + \frac{4Mar}{\Delta(r)}\frac{ \partial^2}{\partial{t}\partial\phi}
    + \left[\frac{a^2}{\Delta(r)} - \frac1{\sin^2\theta}\right]
       \frac{\partial^2}{\partial\phi^2}
     & \\ \nonumber
    - \Delta(r)^{-s}\frac{\partial}{\partial{r}}\left(
        \Delta(r)^{s+1}\frac{\partial}{\partial{r}}\right)
    -  \frac1{ \sin\theta}\frac{ \partial}{\partial\theta}\left(
       \sin\theta\frac{\partial}{\partial\theta}\right)
     & \\ \nonumber
    - 2s\left[\frac{a(r-M)}{\Delta(r)} +  \frac{i\cos\theta}{\sin^2\theta}\right]
         \frac{\partial}{\partial\phi}
     & \\ \nonumber
    - 2s\left[\frac{M(r^2-a^2)}{ \Delta(r)} - r - ia\cos\theta\right]
         \frac{\partial}{\partial{t}}
     & \\ \nonumber
    + (s^2\cot^2\theta - s) \; ,
\end{align}
where
\begin{subequations}
    \begin{align}
        \Delta(r) \; &= \; (r-\rp)(r-\rm) \; ,
    \end{align}
and
    \begin{align}
        r_{\pm} \; &= M \, \pm \,\sqrt{ M^2 - a^2 } \; .
    \end{align}
\end{subequations}
Upon separation via a separable ansatz or integral transform~(see \csec{s2}), \tk{}'s angular differential operator is encapsulated by the following expression, 
\begin{subequations} 

    \begin{align}
        \label{LSa}
        u \; = \; \cos(\theta) \; ,
        \\
        \label{LSb}
        \mcL_{u} = V_S(u) + \partial_{u}(1-u^2)\partial_{u} \; ,
        \\
        \label{LSc}
        V_S(u) \; = \; s+ u  a \cw (u a \cw-2s)-\frac{(m+su)^2}{1-u^2} \; .
    \end{align}

\end{subequations} 
Similarly, \tk{}'s radial differential operator is commonly written as
\begin{align}
    \label{A4}
    \mcL{_r} \; = \; &V_R(r)  \; + \;  \Delta(r)^{-s} \partial_r \left[\Delta(r)^{s+1} \partial_r \right] \; ,
\end{align} 
where
\begin{subequations}
    \begin{align}
        V_R(r) \; &= \; \frac{K(r)^2 -2is(r-M)K(r)}{\Delta(r)}  + \; 4is\cw{}{r} 
        \\ \nonumber
        & \hspace{3.4cm} - a^2\cw{}^2 + 2am\cw{}
        \\
        K(r)\; &= \;(r^2+a^2) \, \cw{} - am \; .
    \end{align}
\end{subequations}
The expressions above may also be found in Refs.~\cite{TeuPre74_3,leaver85,OSullivan:2014ywd}.
\section{Solving \tk{}'s angular equation with Jacobi polynomials}
\label{Apx-3}
\tk{}'s angular equation is
\begin{align}
    \label{d1}
    \mcL_{u} \, S(u) \; = \; -\,A S(u) \; .
\end{align}
In \eqn{d1}, 
\begin{subequations} 
\label{d2}
    \begin{align}
        \label{d2a}
        u \; &= \; \cos(\theta) \; ,
        \\
        \label{d2b}
        \mcL_{u} &= V_S(u) + \partial_{u}(1-u^2)\partial_{u} \; ,
    \end{align}
where the equation's potential, $V_S$, is
    \begin{align}
        \label{d2c}
        V_S(u) \; &= \; s+ u  \gamma (u \gamma-2s)-\frac{(m+su)^2}{1-u^2} \;
    \end{align}
with
    \begin{align}
        \label{d2d}
        \gamma \; &= \; a \cw \; .
    \end{align}
\end{subequations} 
Like \tk{}'s radial equation, \eqn{d1} is a confluent Heun equation; however, \eqn{d1} differs from the radial case in two ways of current relevance: 
\begin{itemize}
    \item[(\textit{i})] \Eqn{d1} has two regular singular points at $u\in \{-1,1\}$, and one irregular singular point at $u\rightarrow\infty$. The irregular singular point at $u\rightarrow\infty$ is not within the physical domain (i.e. there is no boundary condition defined there). The relevant asymptotic boundary conditions are essentially that $S(u)$ are regular when $u \in \{-1,1\}$.  
    \item[(\textit{ii})] The degeneration of \eqn{d1} to a hypergeometric equation occurs at a physical limit ($a\rightarrow 0$), whereas the degeneration of the radial equation to a hypergeometric one occurs at a non-physical limit for \qnm{s} ($\cw\rightarrow 0$).
\end{itemize}
Respectively, these differences mean that (\textit{i}) global\footnote{By \textit{global}, one means a solution that applies to the entire domain. This is in contrast to local solutions, such as those use in the MST method, that are only valid in the vicinity of a singular point~\cite{Mano:1996vt,ronveaux1995heun}.} solutions to the angular equation can be solved with high accuracy using non-confluent basis functions, and that (\textit{ii}) the optimal basis functions may be constructed from physical solutions at the $a\rightarrow 0$ limit. 
%`
\par The construction of analytic solutions begins by finding a similarity transformation that results in the transformed potential being regular when $u \in [-1,1]$. 
The similarity transformation is equivalent to taking the solution ansatz,
\begin{align}
    \label{d3}
    S(u) \; = \; \eta(u) \, g(u) \; ,
\end{align}
where,
\begin{align}
    \label{d4}
    \eta(u) \; = \; e^{u k_2} \; (1-u)^{k_0} \; (1+u)^{k_1}\;,
\end{align}
and then applying it to the angular equation thusly,
\begin{subequations}
    \label{d5}
    \begin{align}
        \label{d5a}
        \left[ \eta(u)^{-1} \, \mcL{_u} \, \eta(u) \right] g(u) \; &= \; -\left[\eta(u)^{-1} A \, \eta(u) \right] g(u) \;
    \end{align}
    \begin{align}
        \label{d5b}
        \mcL{_u'} \; &= \; \left[ \eta(u)^{-1} \, \mcL{_u} \, \eta(u) \right]
    \end{align}
    \begin{align}
        \label{d5c}
        \mcL{_u'} \, g(u) \; &= \; -\,A \, g(u) \; .
    \end{align}
\end{subequations}
In \eqn{d5a}, the similarity transformation of $\mcL{_u}$ is explicitly performed.
In \eqn{d5b}, the transformed operator, $\mcL{_u'}$, is defined, and in \eqn{d5c} its eigenvalue problem is stated.
\par The values of $k_0$, $k_1$ and $k_2$ that result in the potential of $\mcL{_u'}$ being regular and linear on $u\in\{-1,1\}$ are,
\begin{subequations}
    \label{d6}
    \begin{align}
        \label{d6a}
        k_0 \; &= \; \frac{1}{2} |m + s| \; ,
        \\
        \label{d6b}
        k_1 \; &= \; \frac{1}{2} |m - s| \; ,
        \\
        \label{d6c}
        k_2 \; &= \; \gamma \; .
    \end{align}
\end{subequations}
Equivalently, $k_0$, $k_1$ and $k_2$ are the singular exponents that may be found via the method of Frobenius~\cite{Courant1954,ARFKEN2013401,pinchover_rubinstein_2005}.
In effect, $k_0$ and $k_1$ remove divergences from the potential at the regular singular points, and $k_2$ removes the potential's quadratic dependence on $u$.
The appearance of the absolute values, e.g. $|m+s|$, ensures that $\eta$ is always finite when $u\in\{-1,1\}$, \textit{regardless} of the values of $m$ and $s$. This is equivalent to enforcing asymptotic boundary conditions on $S(u)$.
In \eqn{d6c}, $-\gamma$ is also a viable solution for $k_2$; however, $k_2=+\gamma$ is chosen by convention~\cite{leaver85}.
The result for $\mcL{_u'}$ has two schematic forms of relevance,
\begin{subequations}
    \label{d7}
    \begin{align}
        \label{d7a}
        \mcL_{u}' \; = \; &(\text{a}_0+\text{a}_1\, u) 
        \\  
        \label{d7b}
        &+ \; \left(\text{a}_2+\text{a}_3\, u+\text{a}_4\, [1-u^2]\right)\, \partial_u
        \\  
        \label{d7c}
        &+ \; \left(1-u^2\right)\partial^2_u \; ,
    \end{align}
\end{subequations}
and
\begin{align}
    \label{d8}
    \mcL_{u}' \; = \; &(\text{a}_0+\text{a}_1\, u) \, + \, \mathcal{D}'_u \; .
\end{align}
In \eqn{d7}, $\tx{a}_0$ through $\tx{a}_4$ are constant coefficients defined in \eqns{d9a}{d9e}.
They should not to be confused with the similarly labeled coefficients defined in the main text. 
\begin{subequations}
    \label{d9}
    \begin{align}
        \label{d9a}
        \text{a}_0 \; &= \; -k_0 (2 k_1+1)-k_1 - \frac{1}{2} [m^2+ s (s+2)]
        \\ \nonumber 
        & \;\;\;\;\;\,   +  \gamma ^2+2 \gamma  (k_1-k_0)
        \\
        \label{d9b}
        \text{a}_1 \; &= \; -2 \gamma  (k_0+k_1+s+1)
        \\
        \label{d9c}
        \text{a}_2 \; &= \; -2 (k_0-k_1)
        \\
        \label{d9d}
        \text{a}_3 \; &= \; -2 (k_0+k_1+1)
        \\
        \label{d9e}
        \text{a}_4 \; &= \; 2 \gamma
    \end{align}
\end{subequations}
\Eqn{d8} defines $\mathcal{D}'_u$ to be the differential part of \eqn{d7} (i.e. \ceqnsa{d7b}{d7c}).
The operator, $\mathcal{D}'_u$, is useful for understanding the eigenproblem~(\ceqn{d5c}) in the non-spinning limit (i.e. $\gamma=a\cw=0$), and for general spins. 
%
% \smallskip
\par In the non-spinning limit, $\mathcal{D}'_u$ is simply $\mathcal{D}^{\tx{Jac.}}_u$, the differential operator for which the Jacobi polynomials are eigenfunctions:
\begin{align}
    \label{d10}
    \mathcal{D}'_u|_{\gamma=0} \; &= \;  \mathcal{D}^{\tx{Jac.}}_u \;.
\end{align} 
The Jacobi operator is  
\begin{align}
    \label{d11}
    \mathcal{D}^{\tx{Jac.}}_u \; = \; &-(\,\alpha -\beta +(\alpha +\beta +2)\,u\,) \partial_u   
    \\ \nonumber
    &+  (1-u^2) \partial^2_u \; ,
\end{align}
where
\begin{align}
    \label{d12}
    \alpha \; &= \; 2\, k_0 \, ,
    \\
    \label{d13}
    \beta \; &= \; 2\, k_1 \, .
\end{align}
The simplest functions generated by $\mathcal{D}^{\tx{Jac.}}_u$ are the Jacobi polynomials, $p_n(u)$.
Their eigenvalue relationship is 
\begin{align}
    \label{d14}
    \mathcal{D}^{\tx{Jac.}}_u \, p_n(u) \; &= \;  -n (\alpha +\beta +n+1)\, p_n(u) \; .
\end{align}
In \eqn{d14}, the eigenvalue is determined by the requirement that the series solution for $p_n(u)$ terminates after $n$ terms.
\par The zero-spin eigenvalue, $A|_{a=0}=A|_{\gamma=0}$, can be determined by requiring equivalence between the physical eigenvalue problem (\ceqn{d5c}), and the Jacobi one~(\ceqn{d14}).
This amounts to holding the non-derivative terms in each eigenvalue relation to be equivalent in the zero spin limit,
\begin{align}
    \label{d15}
    (a_0+A)|_{\gamma\rightarrow0} \; = \; n (\alpha +\beta +n+1)
\end{align}
\Eqn{d15} can be solved for the eigenvalue,
\begin{subequations}
    \label{d16}
    \begin{align}
        \label{d16a}
        A_0  \; &= \; A|_{\gamma=0}
        \\
        \label{d16b}
        \; &= \; -\left(\ell-s\right) \left(\ell+s+1\right) \; .
    \end{align}
\end{subequations}
In \eqn{d16b}, $A_0$ is the usual spin weighted spherical harmonic eigenvalue~\cite{NP66,leaver85,Fackerell:1977}, and  index $\ell$ is related to the polynomial order, $n$, via
\begin{subequations}
    \label{d17}
    \begin{align}
        \label{d17a}
        \ell \; &= \; \ell_{\min }+n \, ,
        \\
        \label{d17b}
        \ell_{\min } \; &= \; k_0+k_1 \; .
    \end{align}
\end{subequations}
\par For non-zero spin, term by term comparison of $\mathcal{D}'_u$ and $\mathcal{D}^{\tx{Jac.}}_u$ yields  
\begin{align}
    \label{d18}
    \mathcal{D}'_u \; &= \;  \mathcal{D}^{\tx{Jac.}}_u + \tx{a}_4 (1-u^2)\partial_u \; .
\end{align}
\Eqn{d18}, in effect, casts $\mathcal{D}'_u$ as a {deformation} of the Jacobi operator. 
This implies that it is natural to consider the spheroidal functions $g(u)$ to be deformations of the Jacobi polynomials.
In particular, we will take $g(u)$ to be \textit{exactly represented} by Jacobi polynomials,
\begin{align}
    \label{d19}
    g(u) \; = \; \sum_{n=0}^{\infty} \, a_n \, p_n(u) \; .
\end{align}
\Eqn{d19} is justified not only because the Jacobi polynomials are known to be complete, but also because $g(u)$ and $p_n(u)$ satisfy the same asymptotic boundary conditions on the same domain.
\par \Eqn{d19} is an analytic solution for $g(u)$ that is motivated by the structure of \tk{}'s angular equation.
In effect, \eqn{d19} is a spectral decomposition of $g(u)$ into Jacobi moments, $a_n$.
Thus, finding the values of each $a_n$ is equivalent to finding $g(u)$.
\par To do so, the Jacobi polynomials' recursive and orthogonality properties must be used.
The relevant recursion rules for Jacobi polynomials are
\begin{subequations}
    \label{d20}
    \begin{align}
        \label{d20a}
        (1-u^2) \, \partial_u p_n \; &= \; \text{b}_0\, p_{n-1} + \text{b}_1\, p_{n} + \text{b}_2\, p_{n+1} \; ,
        \\
        \label{d20b}
        u \, p_n \; &= \; \text{b}_3\, p_{n-1} + \text{b}_4\, p_{n} + \text{b}_5\, p_{n+1} \; .
    \end{align}
\end{subequations}
Normalization of the Jacobi polynomials is defined with respect to the weighted inner product,
\begin{subequations}
    \label{d21}
    \begin{align}
        \label{d21a}
        \brak{p_n}{p_{n'}} \; &= \; \int_{-1}^1 p_n(u) \, p_{n'}(u) \; \tx{W}^{\tx{Jac.}}(u) \, \tx{d}u 
        \\
        \label{d21b}
        \; &= \; \tx{b}_6 \, \delta_{n'n} \; ,
    \end{align}
\end{subequations}
where the Jacobi weight function, $\tx{W}^{\tx{Jac.}}(u)$, is 
\begin{align}
    \label{d22}
    \tx{W}^{\tx{Jac.}} \; = \; (1-u)^\alpha (1+u)^\beta \; .
\end{align}
In \eqnsa{d20}{d21}, $\tx{b}_0$ through $\tx{b}_6$ are $n$ dependent constants defined in \eqns{d25a}{d25g}.
% %
%
\begin{subequations}
    \label{d25}
    \begin{align}
        \label{d25a}
        \tx{b}_0(n) \;&=\; -\frac{2 (\alpha +n) (\beta +n) (\alpha +\beta +n+1)}{(\alpha +\beta +2 n) (\alpha +\beta +2 n+1)}
        \\
        \label{d25b}
        \tx{b}_1(n) \;&=\; -\frac{2 n (\beta -\alpha ) (\alpha +\beta +n+1)}{(\alpha +\beta +2 n) (\alpha +\beta +2 n+2)}
    \end{align}
    \begin{align}
        \label{d25c}
        \tx{b}_2(n) \;&=\; \frac{2 n (n+1) (\alpha +\beta +n+1)}{(\alpha +\beta +2 n+1) (\alpha +\beta +2 n+2)}
        \\
        \label{d25d}
        \tx{b}_3(n) \;&=\; -\frac{2 (\alpha +n) (\beta +n)}{(\alpha +\beta +2 n) (\alpha +\beta +2 n+1)}
    \end{align}
    \begin{align}
        \label{d25e}
        \tx{b}_4(n) \;&=\; -\frac{\alpha ^2-\beta ^2}{(\alpha +\beta +2 n) (\alpha +\beta +2 n+2)}
        \\
        \label{d25f}
        \tx{b}_5(n) \;&=\; -\frac{2 (n+1) (\alpha +\beta +n+1)}{(\alpha +\beta +2 n+1) (\alpha +\beta +2 n+2)}
    \end{align}
    \begin{align}
        \label{d25g}
        \tx{b}_6(n) \;&=\; \frac{\Gamma (n+\alpha +1) \Gamma (n+\beta +1)}{(\alpha +\beta +2 n+1) \Gamma (n+1)  \Gamma (n+\alpha +\beta +1)}
    \end{align}
\end{subequations}
Observe that $\tx{W}^{\tx{Jac.}}(u)$ is very similar to the scaling function $\eta(u)$ introduced in \eqn{d4}.
This can be understood as resulting from the fact that the transformation present in \eqn{d5a} is conformal, meaning that
\begin{subequations}
    \label{d23}
    \begin{align}
        \label{d23a}
        \int_{-1}^1 p_n(u) \, p_{n'}&(u) \, \tx{W}^{\tx{Jac.}}(u) \, \tx{d}u 
        \\  \nonumber
        \; &= \; 
        \\  
        \label{d23b}
        \int_{-1}^1 ([p_n(u) \eta(u)&] \, [p_{n'}(u) \eta(u)])|_{\gamma=0}\, \tx{d}u 
    \end{align}
\end{subequations}
There is no weight function in \eqn{d23b} because \eqn{d2b} is explicitly in Sturm-Liouville form (with weight function $1$)~\cite{ARFKEN2013401}.
The integrands of \eqnsa{d23a}{d23b} must be equivalent, meaning that $\tx{W}^{\tx{Jac.}}(u)$ and $\eta(u)|_{\gamma=0}$ are related,
\begin{align}
    \label{d24}
    \tx{W}^{\tx{Jac.}}(u) \; = \; \eta(u)^2|_{\gamma=0} \; .
\end{align}
In general, it can be shown that if $f_1$ satisfies the second order linear differential equation $\mcL_1 f_1=0$, where $\mcL$ is self-adjoint with weight function $\lambda_1$, then $f_2 = \eta^{-1} f_1$ will satisfy $\mcL_2 g=0$, where $\mcL_2 = \eta^{-1}\mcL_1 \eta $ is self-adjoint with new weight function $\lambda_2 = \eta^2 \lambda_1$.
\par \Eqns{d20}{d25} can be used to determine the values of $a_n$ as follows.
Given the scalar product, \eqn{d21a}, we will use bra-ket notation to simplify to representation of various steps.
In that notation, the completeness of the Jacobi polynomials is expressed as
\begin{align}
    \label{d26}
    \sum_{n=0}^{\infty} \, \ket{p_n}\bra{p_n} \; = \; \I \; .
\end{align}
In \eqn{d26}, $\I$ is the identity operator for all functions that have the same asymptotic behavior as the Jacobi polynomials on $u\in[-1,1]$, and the polynomials $\ket{p_n}$ are implicitly normalized so that $\brak{p_n}{p_n}=1$.
The bra-ket notation form of the physical eigenvalue relation is,
\begin{align}
    \label{d27}
    \mcL{_u'} \, \ket{g} \; &= \; -\,A \, \ket{g} \; .
\end{align}
From this perspective, the Jacobi moments, $a_n$, are simply the components of $g(u)$ in the basis of Jacobi polynomials. 
This is made explicit by using \eqn{d26} to express both sides of \eqn{d27} in the Jacobi polynomial basis, 
\begin{align}
    \label{d28}
    \sum_{n,n'}\ket{p_{n'}}\bra{p_{n'}} \mcL{_u'} \ket{p_n}\brak{p_n}{g} \; &= \; -\,A \, \sum_{n} \ket{p_n}\brak{p_n}{g} \; .
\end{align}
Note that the right hand side of \eqn{d28} is equivalent to that of \eqn{d18} (i.e. $a_n = \brak{p_n}{g}$).
\Eqn{d28} may be recast as a simple matrix equation
\begin{align}
    \label{d29}
    \hat{L}\,\vec{g} \; &= \; -A\,\vec{g}\; ,
\end{align}
where $\hat{L}$ and $\vec{g}$ are defined from \eqn{d28} as
\begin{subequations}
    \label{d30}
    \begin{align}
        \label{d30a}
        \hat{L} \; &= \;  \{ \bra{p_{n'}} \mcL{_u'} \ket{p_n} \}_{n,n'}\;,
        \\
        \label{d30b}
        \vec{g} \; &= \; \{ \brak{p_n}{g} \}_n \; .
    \end{align}
\end{subequations}
Thus, taking a finite dimensional truncation of $\hat{L}$ allows $\vec{g}$~(i.e. $a_n$) to be determined using standard linear algebraic methods~\cite{Axler:2015,2020SciPy-NMeth}.
\par The essential feature of this approach is that, starting with \eqn{d29}, the recursion and orthogonality relationships, \eqns{d20}{d25}, can be used to show that $\hat{L}$ is at most \textit{tridiagonal}.
In particular, when $\mcL{_u'}$ acts on $\ket{p_n}$, only lower to upper diagonal terms contribute,
\begin{align}
    \label{d31}
    \mcL{_u'}  \ket{p_n}  \;=\;  \sigma_n \, \ket{p_{n-1}} 
    + \mu_n \, \ket{p_{n}}
    + \nu_n \, \ket{p_{n+1}} \; .
\end{align}
In \eqn{d31}, the $n$ dependent constants, $\sigma_n$, $\mu_n$, and $\nu_n$ are
\begin{subequations}
    \label{d32}
    \begin{align}
        \label{d32a}
        \sigma_n \;&=\; \frac{\tx{b}_6(n-1) }{\tx{b}_6(n)} [\tx{a}_1 \tx{b}_3(n)+\tx{a}_4 \tx{b}_0(n)]
        \\
        \label{d32b}
        \mu_n \;&=\; \tx{a}_0+\tx{a}_1 \tx{b}_4(n)+\tx{a}_4 \tx{b}_1(n)
        \\ \nonumber
        & \quad\quad\quad\quad\quad\quad -n (\alpha +\beta +n+1)
        \\
        \label{d32c}
        \nu_n \;&=\; \frac{\tx{b}_6(n+1)}{\tx{b}_6(n)} [ \tx{a}_1 \tx{b}_5(n)+\tx{a}_4 \tx{b}_2(n) ]
    \end{align}
\end{subequations}
For any $N$ dimensional truncation of $\hat{L}$, say $\hat{L}_N$, \eqn{d32} provides the matrix elements,
\begin{align}
    \label{d33}
    \hat{L}_N \; &= \; \left(\begin{array}{ccccc}
        \mu_0 & \nu_0 & 0 & \cdots & 0 \\
        \sigma_1 & \mu_1 & \nu_1 & \ddots & 0 \\
        0 & \sigma_2 & \mu_2 & \ddots & \vdots \\
        \vdots & \ddots & \ddots & \ddots & \nu_{N-1} \\
        0 & 0 & \cdots & \sigma_N & \mu_N
        \end{array}\right) \; .  
\end{align}
The use of \eqn{d33} to determine angular functions $S(u)=\eta(u)g(u)$, and their eigenvalues $A(u)$, amounts to determining the eigenvectors and eigenvalues of $\hat{L}_N$.
Since this method explicitly estimates the point spectra of $\hat{L}$, it is a purely spectral method (e.g. not a generalized spectral method) for the construction of analytic solutions to \tk{}'s angular equation.  
\par This construction facilitates the following conclusions:
Since the transformed angular equation degenerates to the Jacobi equation when $\gamma=0$, the spheroidal functions $g(u)$ may be exactly identified with Jacobi polynomials in that limit.
For general values of $\gamma$, the fact that $\hat{L}$ is tridiagonal means that for every $g(u)$ associated with index $\ell$, there always exists a unique Jacobi polynomial with that same index~\cite{London:2020uva}.
In this way, the \qnm{} index $\ell$ is inherently \textit{linked} to a Jacobi polynomial of order $n=\ell-\ell_\tx{min}$.
\section{The confluent Heun equation in natural general form}
\label{Apx-2}
\par The confluent Heun equation has many forms that differ with respect to transformations of the independent and dependent variables~\cite{ronveaux1995heun}. 
Here, we briefly outline the connection between Teukolsky's radial equation and the confluent Heun equation in \textit{natural general form}.
\par Perhaps the most referenced~(e.g. Refs.~\cite{Hortacsu:2011rr,Leaver:1986JMP,Cook:2020otn}) form of confluent Heun equation is,
\begin{subequations}
    \label{A2-1}
    \begin{align}
        \label{A2-1a}
        \mcL^{(1)}_{\text{cH}} \, y(z) \; = \; 0 \; ,
    \end{align}
where
    \begin{align}
        \label{A2-1b}
        \mcL^{(1)}_{\text{cH}} \; = \; (\alpha&  \nu  z -\sigma ) 
        \\ \nonumber
        \; + \, &\left[\gamma  (z-1)+z (\delta +\nu  (z-1)) \right] \, \partial_{z} 
        \\ \nonumber
        \; + \, &(z-1) z \, \partial^2_{z} \; .
    \end{align}
\end{subequations}
\Eqn{A2-1} may be derived from the Heun equation (which has $4$ regular singular points) by confluence of two singular points (see e.g.~\cite{ARFKEN2013401}) into a single {irregular} one.
\Eqn{A2-1a} is simply a definition of the confluent Heun equation in terms of its associated linear differential operator, $\mcL^{(1)}_{z}$.
In \eqn{A2-1b}, $\alpha$, $\nu$, $\sigma$, $\gamma$ and $\delta$ are complex valued parameters, and $z$ is nominally real valued. 
\par \Eqn{A2-1}, is the confluent Heun equation in what's called \textit{nonsymmetrical canonical form}~\cite{ronveaux1995heun,Cook:2014cta,Fiziev:2009}. 
It happens that this is actually not the most general form of the confluent Heun equation.
As a result, a naive comparison between \eqn{A2-1} and \tk{}'s radial equation may generate unnecessary confusion.
To facilitate the upcoming discussion, we will henceforth use $r$ rather than $z$, and we will discuss forms of the confluent Heun equation by referring to the related differential operators, $\mcL^{(i)}_{\text{cH}}$, such that
\begin{align}
    \label{A2-2}
    \mcL^{(i)}_{\text{cH}}\,y\;=\;0 \; .
\end{align}
\par The most general form of the confluent Heun differential operator with an irregular singular point at infinity is~(see e.g. Ref.~\cite{ronveaux1995heun})
\begin{align}
    \label{A2-3}
    \mcL^{(2)}_{\text{cH}} \; = \;& \left({\alpha_0}+{\alpha_1} r+{\alpha_2} r^2+\frac{{\alpha_4}}{r-{\rm}}+\frac{{\alpha_3}}{r-{\rp}}\right)
    \\ \nonumber
    \; &+\; \left({\alpha_5}+{\alpha_6} r+{\alpha_7} r^2\right) \, \partial_{r}
    \\ \nonumber
    \; &+\; (r-{\rm}) (r-{\rp}) \, \partial^2_{r} \; .
\end{align}
In \eqn{A2-3}, $\mcL^{(2)}_{\text{cH}}$ corresponds to the confluent Heun equation in what is called \textit{natural general form}~\cite{ronveaux1995heun}.
\par In \eqn{A2-3}, $\mcL^{(2)}_{\text{cH}}$ has been formatted to ease comparison with Teukolsky's radial differential operator.
In particular, the two regular singular points have been placed at $\rp$ and $\rm$, just as they are in the case of Teukolsky's radial differential operator~(\ceqn{A4}), which can be formatted as
\begin{align}
    \label{A2-4}
    \mcL{_r} \; = \; &\left({\text{A}_0}+{\text{A}_1}r+\text{A}_2\,r^2+\frac{{\text{A}_3}}{r-{\rm}}+\frac{{\text{A}_4}}{r-{\rp}}\right) 
    \\ \nonumber
    \; &+\; ({\text{A}_5}+{\text{A}_6} r) \, \partial_r  
    \\ \nonumber 
    \; &+\; (r-{\rm}) (r-{\rp}) \, \partial_r^2 \; .
\end{align}
For completeness, parameters $\tx{A}_1$ through $\tx{A}_6$ are defined in terms of $\rp$, $\rm$ and $s$ as 
\begin{subequations}
    \label{A2-5}
    \begin{align}
        \tx{A}_0 \; =& \; \cw{}  ({\rm}+{\rp}) (\cw{}  ({\rm}+{\rp})-i s)
        \\
        \tx{A}_1 \; =& \; \cw{}  (\cw{}  ({\rm}+{\rp})+2 i s)
        \\
        \tx{A}_2 \; =& \; \cw{} ^2
    \end{align}
    \begin{align}
        \tx{A}_3 \; =& \; \frac{{\rm}^2 \left(m^2+4 {\rm}^2 \cw{} ^2\right)}{{\rm}-{\rp}}
        \\ \nonumber 
        \;&-\;{\rm} \left(m^2+{\rm} \cw{}  (3 {\rm} \cw{} +i s)\right)
        \\ \nonumber 
        \;&+\;\frac{4 m {\rm}^2 \cw{}  \sqrt{{\rm} {\rp}}}{{\rp}-{\rm}}+i m s \sqrt{{\rm} {\rp}}
        \\ \nonumber 
        \;&+\;2 m {\rm} \cw{}  \sqrt{{\rm} {\rp}}-{\rm} {\rp} \cw{}  ({\rm} \cw{} +i s)
    \end{align}
    \begin{align}
        \tx{A}_4 \; =& \; -\frac{{\rp}^2 \left(m^2+4 {\rp}^2 \cw{} ^2\right)}{{\rm}-{\rp}}
        \\ \nonumber 
        \;&-\;{\rp} \left(m^2+{\rp} \cw{}  (3 {\rp} \cw{} +i s)\right)
        \\ \nonumber 
        \;&+\;\frac{4 m {\rp}^2 \cw{}  \sqrt{{\rm} {\rp}}}{{\rm}-{\rp}}+i m s \sqrt{{\rm} {\rp}}
        \\ \nonumber 
        \;&+\;2 m {\rp} \cw{}  \sqrt{{\rm} {\rp}}-{\rm} {\rp} \cw{}  ({\rp} \cw{} +i s)
    \end{align}
    \begin{align}
        \tx{A}_5 \; =& \; (s+1) (-({\rm}+{\rp}))
        \\
        \tx{A}_6 \; =& \; 2 (s+1) \; .
    \end{align}
\end{subequations}
\par The fact that \tk{}'s radial operator is a special case of Heun's confluent operator is clear by comparing \eqn{A2-3} to \eqn{A2-4}.
The confluent Heun nature of the operator is not changed under transformations of the dependent or independent variables, but such transformations can impact the physical nature of solutions.
For example, application of the \qnm{} boundary conditions (see e.g. \ceqns{p11}{p13}) has the effect of transforming \eqn{A2-4} into a \textit{different version of} \eqn{A2-3}. 
However, if the solution of the transformed equation (e.g. \eqn{p14}) is to be consistent with \qnm{} boundary conditions, then it cannot have asymptotic behavior that affects a transformation into yet \textit{another} version of the confluent Heun equation.
Application of a compactified coordinate (see \ceqn{p10}) has the effect of changing the coordinate position of singular points, but not their regular or irregular nature. 
\bibliography{references.bib}
\end{document}

%% file: macro.tex
% definitions

\newenvironment{aside}
    {\begin{addmargin}[1em]{2em}% 2em left, 1em right
\begin{center} \noindent\rule{0.65\paperwidth}{0.4pt} \end{center}
    }
    {
    \begin{center} \noindent\rule{0.65\paperwidth}{0.4pt} \end{center}
\end{addmargin}
    }

% CUSTOM COMMANDS
\newcommand\xquote[1]{``#1"}
\newcommand\psinr[2]{$\psi_{{#1}{#2}}^{NR}$}
\newcommand\red[1]{{\color[rgb]{0.75,0.0,0.0} #1}}
\newcommand\redoff[1]{#1}
\newcommand\greenoff[1]{#1}
\newcommand\bred[1]{{\color[rgb]{0.75,0.0,0.0} \textbf{#1}}}
\newcommand\green[1]{{\color[rgb]{0.0,0.60,0.08} #1}}
\newcommand\blue[1]{{\color[rgb]{0,0.20,0.65} #1}}
\newcommand\cyan[1]{{\color[HTML]{00c3ff} #1}}
\newcommand\bluey[1]{{\color[rgb]{0.11,0.20,0.4} #1}}
\newcommand\gray[1]{{\color[rgb]{0.7,0.70,0.7} #1}}
\newcommand\grey[1]{{\color[rgb]{0.7,0.70,0.7} #1}}
\newcommand\white[1]{{\color[rgb]{1,1,1} #1}}
\newcommand\darkgray[1]{{\color[rgb]{0.3,0.30,0.3} #1}}
\newcommand\orange[1]{{\color[rgb]{.86,0.24,0.08} #1}}
\newcommand\purple[1]{{\color[rgb]{0.45,0.10,0.45} #1}}
\newcommand\note[1]{\colorbox[rgb]{0.85,0.94,1}{\textcolor{black}{\textsc{\textsf{#1}}}}}

\newcommand*{\figfactor}{0.495}

% Math symbols
% ###################################### %
%\def\mh{{\hat{h}}} % model  for strain
\def\m#1{{\hat{#1}}} % model  for strain
\def\ma#1{{\hat{#1}_{\bm{a}}}} % model  for strain
\newcommand{\var}[1]{\mathcal{{#1}}}
\newcommand{\mlam}{{\bm{\lambda}}}
\newcommand{\lam}{{\lambda}}
\newcommand{\mLam}{{\bm{\Lambda}}}
\newcommand{\bigo}[1]{{\cal O}({#1})}
\newcommand{\braket}[2]{ {\langle {#1} \, | \, {#2} \rangle} }
\newcommand{\bra}[1]{ \langle {#1} |  }
\newcommand{\ket}[1]{ | {#1} \rangle }
\newcommand{\ketbra}[2]{ \ket{#1}\bra{#2} }
\newcommand{\tx}[1]{\text{#1}}
\def\T{\dagger}
% ###################################### %

% ###################################### %
\definecolor{lightblue}{rgb}{.82,.88,0.95}
\definecolor{lightred}{rgb}{0.95,.86,0.86}
\definecolor{yellow}{rgb}{0.95,0.95,0.86}
\definecolor{green}{rgb}{.90,1,0.95}
\definecolor{lightpurple}{rgb}{.95,0.85,0.95}
% ###################################### %

% CUSTOM DEFINITIONS
% ***************************************** %
\def\prd{Phys.Rev.D}
% ***************************************** %
\def\gt{Georgia Tech}
% ***************************************** %
\def\tk{{Teukolsky}}
% ***************************************** %
\def\mpl{multipole}
% ***************************************** %
\def\ee{Einstein's equations}
% ***************************************** %
\def\toolkit#1{NRDA--Toolkit{#1}}
% \def\toolkit#1{Data Analysis Toolkit{#1}}
% ***************************************** %
\def\wf#1{waveform#1}
% ***************************************** %
% \def\gr#1{General Relativity#1}
\def\gr#1{General Relativity#1
  (GR#1)\gdef\gr{GR}}
% ***************************************** %
\def\gwa#1{Gravitational Wave Astrophysics#1}
% ***************************************** %
\def\gwf#1{gravitational waveform#1}
% ***************************************** %
\def\gwa#1{\gw{} astronomy#1}
% ***************************************** %
\def\grad#1{gravitational radiation#1}
% ***************************************** %
\def\nht#1{No-Hair Theorem#1}
% ***************************************** %
\def\rm#1{\mathrm{#1}}
% ***************************************** %
\def\BL{Boyer-Lindquist}

% Referencing
% ***************************************** %
\def\prt#1{Part~(\ref{#1})}
% ***************************************** %
\def\Apx#1{Appendix~(\ref{#1})}
% ***************************************** %
\def\apx#1{Appx.~(\ref{#1})}
% ***************************************** %
\def\capx#1{Appx.~\ref{#1}}
% \def\ap#1{Appendix~(\ref{#1})}
% ***************************************** %
\def\ch#1{Chapter~(\ref{#1})}
% ***************************************** %
\def\cch#1{Chapter~\ref{#1}}
% ***************************************** %
\newcommand{\chs}[2]{Chapters~(\ref{#1}-\ref{#2})}
% ***************************************** %
\newcommand{\cchs}[2]{Chapters~\ref{#1}-\ref{#2}}
% ***************************************** %
\def\Sec#1{Section~\ref{#1}}
% ***************************************** %
\def\sec#1{Sec.~\ref{#1}}
% \def\sec#1{Sec.~(\ref{#1})}
% ***************************************** %
\def\csec#1{Sec.~\ref{#1}}
% ***************************************** %
\newcommand{\csecs}[2]{Secs.~\ref{#1}-\ref{#2}}
% ***************************************** %
\def\tk#1{Teukolsky#1}
% ***************************************** %
\newcommand{\secs}[2]{Secs.~\ref{#1}-\ref{#2}}
% ***************************************** %
\newcommand{\secsa}[2]{Sec.~\ref{#1} and Sec.~\ref{#2}}
% ***************************************** %
\def\Tbl#1{Table~(\ref{#1})}
% ***************************************** %
\def\tbl#1{Table~(\ref{#1})}
% ***************************************** %
\def\ctbl#1{Table~\ref{#1}}
% ***************************************** %
\def\Fig#1{Figure~\ref{#1}}
% ***************************************** %
\def\fig#1{Fig.~\ref{#1}}
% \def\fig#1{Fig.~(\ref{#1})}
% ***************************************** %
\def\cfig#1{Fig.~\ref{#1}}
% ***************************************** %
\newcommand{\figs}[2]{Figures~(\ref{#1}-\ref{#2})}
\newcommand{\Figs}[2]{Figures~(\ref{#1}-\ref{#2})}
\newcommand{\Figsa}[2]{Figures~(\ref{#1}) and (\ref{#2})}
% ***************************************** %
\def\Eqn#1{Equation~(\ref{#1})}
% ***************************************** %
\def\eqn#1{Eq.~(\ref{#1})}
% \def\eqn#1{\hyperref[#1]{Equation~\ref{#1}}} %
% ***************************************** %
\def\ceqn#1{Eq.~\ref{#1}}
% ***************************************** %
\newcommand{\Eqns}[2]{Equations~(\ref{#1}-\ref{#2})}
\newcommand{\Eqnsa}[2]{Equations~(\ref{#1}) and (\ref{#2})}
% ***************************************** %
\newcommand{\eqns}[2]{Eqs.~(\ref{#1}-\ref{#2})}
\newcommand{\eqnsa}[2]{Eqs.~(\ref{#1}) and (\ref{#2})}
% ***************************************** %
\newcommand{\ceqns}[2]{Eqs.~\ref{#1}-\ref{#2}}
\newcommand{\ceqnsa}[2]{Eqs.~\ref{#1} and \ref{#2}}
% ***************************************** %

% ***************************************** %
\def\tridiag#1{tridiagonal#1}
% ***************************************** %
\def\ply#1{{polynomial#1}}
% ***************************************** %
\def\plys{{polynomials}}
% ***************************************** %
\def\orthog{orthogonality}
% ***************************************** %
\def\da#1{data analysis#1}
% ***************************************** %
\def\sw{spin weighted}
% ***************************************** %
\def\swsh{spin weighted spherical harmonic}
% ***************************************** %
\def\lal#1{LIGO Analysis Library#1
  (LAL#1)\gdef\lal{LAL}}
% ***************************************** %
\def\nrda#1{\nr{} Data Analysis#1
  (NRDA#1)\gdef\nrda{NRDA}}
% ***************************************** %
\def\tt#1{\textit{transverse--traceless}#1
  (TT#1)\gdef\tt{TT}}
% ***************************************** %
\def\et#1{Einstein Telescope#1
  (ET#1)\gdef\et{ET}}
% ***************************************** %
\def\ce#1{Cosmic Explorer#1
  (CE#1)\gdef\ce{CE}}
% ***************************************** %
\def\ego#1{European Gravitational Observatory#1
  (EGO#1)\gdef\ego{EGO}}
% ***************************************** %
\def\lisa#1{Laser Interferometer Space Antenna#1
  (LISA#1)\gdef\lisa{LISA}}
% ***************************************** %
\def\ligo#1{Laser Interferometer Gravitational Wave Observatory#1
  (LIGO#1)\gdef\ligo{LIGO}}
% \def\ligo#1{LIGO#1}
% ***************************************** %
\def\igwn#1{International Gravitational Wave Network#1
  (IGWN)\gdef\igwn{IGWN}}
% ***************************************** %
\def\lvk#1{LIGO-Virgo-Kagra collaboration#1
  (LVK)\gdef\lvk{LVK}}
% ***************************************** %
\def\lv#1{#1
(LIGO-Virgo#1)\gdef\lv{LV}}
% ***************************************** %
\def\virgo#1{Virgo#1}
% ***************************************** %
\def\aligo#1{Advanced LIGO#1
  (Adv. LIGO#1)\gdef\aligo{Adv. LIGO}}
% ***************************************** %
% \def\snr#1{signal to noise ratio#1}
\def\snr#1{signal to noise ratio#1
  (SNR#1)\gdef\snr{SNR}}
% ***************************************** %
\def\psd#1{power spectral density#1
  (PSD#1)\gdef\psd{PSD}}
% ***************************************** %
\def\rom#1{reduced order model#1
  (ROM#1)\gdef\rom{ROM}}
% ***************************************** %
\def\gatech#1{Georgia Institute of Technology#1
  (GaTech#1)\gdef\gatech{GaTech}}
% ***************************************** %
\def\ffi#1{Fixed-Frequency Integration#1
  (FFI#1)\gdef\ffi{FFI}}
% ***************************************** %
\def\sxs#1{Simulating Extreme Spacetimes#1
  (SXS#1)\gdef\sxs{SXS}}
% ***************************************** %
\def\bam#1{Bifunctional Adaptive Mesh#1
  (BAM#1)\gdef\bam{BAM}}
% ***************************************** %
\def\adm#1{Arnowitt-Deser-Misner
	(ADM#1)\gdef\adm{ADM}}
\def\frmse#1{Fractional Root-Mean Square Error
	(FRMSE)\gdef\frmse{FRMSE}}

% ***************************************** %
\def\bh#1{black hole#1
 (BH#1)\gdef\bh{BH}}
% \def\bh#1{black hole#1}
% ***************************************** %
\def\bbh#1{binary black hole#1
 (BBH#1)\gdef\bbh{BBH}}
% \def\bbh#1{binary \bh{}#1}
% ***************************************** %
\def\bhb#1{\bh{} binary#1}
% ***************************************** %

% ***************************************** %
\def\qnm#1{Quasi-Normal Mode#1
(QNM#1)\gdef\qnm{QNM}}
\def\Qnm#1{Quasi-Normal Mode#1}
\def\Qnms{Quasi-Normal Modes}
% ***************************************** %
\def\eob#1{Effective One Body#1
  (EOB#1)\gdef\eob{EOB}}
% ***************************************** %
\def\sws#1{spheroidal harmonics of spin weight -2#1}
\def\swy#1{spherical harmonics of spin weight -2#1}
% ***************************************** %
\def\gw#1{gravitational wave#1}
% \def\gw#1{gravitational wave#1
%  (GW#1)\gdef\gw{GW}}
% ***************************************** %
\def\gwa#1{gravitational wave astronomy#1}
% ***************************************** %
% \def\pn#1{Post-Newtonian#1}
\def\pn#1{Post-Newtonian#1
 (PN#1)\gdef\pn{PN}}
\def\pnl#1{post-Newtonian-like#1
  (PN-like#1)\gdef\pnl{PN-like}}
% ***************************************** %
% \def\nr{Numerical Relativity}
\def\NR{{\text{NR}}}
\def\nr{Numerical Relativity
 (NR)\gdef\nr{NR}}
% ***************************************** %
\def\pt{\bh{} perturbation theory}
% ***************************************** %
\def\GOLS#1{\textit{greedy ordinary least-squares}#1}
% ***************************************** %
\def\rd{ringdown}
% ***************************************** %
\def\imr{inspiral-merger-ringdown}
% \def\imr#1{inspiral-merger-ringdown#1
%   (IMR#1)\gdef\imr{IMR}}
% ***************************************** %
\def\cbc#1{compact object coalescence#1}
% ***************************************** %
\def\bbc#1{\bbh{} coalescence#1}
%\def\bbc#1{binary black hole coalescence#1
%  (BBC#1)\gdef\bbc{BBC}}
% ***************************************** %
\def\pc#1{principle component#1}
% ***************************************** %
\def\pca#1{principle component analysis#1
  (PCA#1)\gdef\pca{PCA}}
% ***************************************** %
\def\svd#1{Singular Value Decomposition#1
  (SVD#1)\gdef\svd{SVD}}
% ***************************************** %
\def\gs#1{Gram-Schmidt#1}
\newcommand{\y}[2]{ {_{#1}}Y_{#2} }
\def\sylm{ \y{s}{\ell m } }
% ***************************************** %
\def\mo{\mathcal{D}}
\def\adjmo{{\mathcal{D}}^{\dagger}}
% ***************************************** %
\def\yo{\mathcal{D}_Y}
\def\adjyo{{\mathcal{D}_Y}^{\hspace{-2pt}\dagger}}
% ***************************************** %
\def\ro{\mathcal{D}_R}
\def\adjro{{\mathcal{D}_R}^{\hspace{-2pt}\dagger}}
% ***************************************** %
\def\bo{\mathcal{D}_B}
\def\adjbo{{\mathcal{D}_B}^{\hspace{-2pt}\dagger}}
% ***************************************** %
\def\so{\mathcal{D}_S}
\def\adjso{{\mathcal{D}_S}^{\hspace{-2pt}\dagger}}
% ***************************************** %
\def\to{\mathcal{D}_T}
\def\adjto{{\mathcal{D}_T}^{\hspace{-2pt}\dagger}}
% ***************************************** %
\def\adj#1{{#1}^{\dagger}}
\def\dadj#1{{#1}^{\ddag}}
% ***************************************** %
\def\ethp{\eth}
\def\ethm{{\eth'}}
\def\A{{\L m}}
\def\a{{\alpha}}
\def\mcl{\mathcal{L}}
\def\mct{{\mathcal{T}}}
\def\mcv{{\mathcal{V}}}
\def\cmcv{{\mcv^*}}
\def\cmct{{\mct^*}}
\def\tmct{{\tilde{\mct}}}
\def\tmcv{{\tilde{\mcv}}}
\def\mclo{{\mathcal{L}_o}}
\def\mcto{{\mathcal{T}_o}}
\def\mcvo{{\mathcal{V}_o}}
\def\cmcvo{{{\mcv_o^*}}}
\def\cmcto{{{\mct_o^*}}}
\def\tmcto{{\tilde{\mct}_o}}
\def\tmcvo{{\tilde{\mcv}_o}}
\def\tmcl{\tilde{\mcl}}
\def\mcp{\mathcal{P}}
\def\mcq{\mathcal{Q}}
\def\amcl{\adj{\mcl}}
\def\cmcl{{\mcl^*}}
\def\sjk{{\sigma_{\lp\ell}}}
\def\mcD{\mathcal{D}}
\def\mcL{\mathcal{L}}
\def\mcT{\mathcal{T}}
\def\mcV{\mathcal{V}}
\def\mcP{\mathcal{P}}
\def\mcQ{\mathcal{Q}}
\def\Lo{ {\mathcal{K}} }
\def\I{{\mathbb{I}}}
\def\max{\mathrm{max}}

% ***************************************** %
\def\gs{Gram-Schmidt}
\def\polys{polynomials}
% ***************************************** %

% ----------------------------------------- %
\def\lMn{{{\ell \M n}}}
\def\lmn{{{\ell m n}}}
\def\lmin{{{\ell_\mathrm{min}}}}
\def\lpmn{{{\ell' m n}}}
\def\lpmnp{{{\ell' m n'}}}
\def\lm{{{\ell m}}}
\def\lpm{{{\ell' m}}}
\def\l{{{\ell}}}
\def\n{{\bar{n}}}
\def\lmbn{{{\ell m \n}}}
\def\lpmbn{{{\ell' m \n}}}
\def\lp{{{\ell'}}}
\def\pp{p}

\mathchardef\minus = "002D
\newcommand{\swY}[4][]{{}_{{}_{#2}}\!Y^{#1}_{#3}(#4)}
\newcommand{\swSH}[5][]{{}_{{}_{#2}}S^{#1}_{#3}(#4;#5)}
\newcommand{\swS}[5][]{{}_{{}_{#2}}S^{#1}_{#3}(#4;#5)}
\newcommand{\scA}[4][]{{}_{{}_{#2}}A^{#1}_{#3}(#4)}
\newcommand{\YSH}[3][]{\mathcal{A}^{#1}_{#2}(#3)}

\def\LMaster{ \mathcal{L}_{t r \theta \phi} }
\def\LMasterB{ \mathcal{L}_{t r u \phi} }
\def\rp{ r_{+} }
\def\rm{ r_{-} }

\def\rhs{right hand side}
\def\lhs{left hand side}
\def\te{\tk{'s} equation}

\def\L{\bar{\ell}}
\def\M{\bar{m}}
\def\LM{{\L\M}}
\def\Lop{\mathcal{L}_{\k}}

% ***************************************** %
\newcommand{\brak}[2]{ \braket{#1}{#2} }
% ***************************************** %
%
% ***************************************** %
\newcommand*{\factor}{0.95} % for figure scale
\newcommand*{\rscale}{1.3}
% ***************************************** %

\newcommand{\hlgreen}[1]{\sethlcolor{green}\hl{#1}{\sethlcolor{yellow}}}
\newcommand{\hlyellow}[1]{\sethlcolor{yellow}\hl{#1}{\sethlcolor{yellow}}}
\newcommand{\hlblue}[1]{\sethlcolor{lightblue}\hl{#1}{\sethlcolor{yellow}}}
\newcommand{\hlred}[1]{\sethlcolor{lightred}\hl{#1}{\sethlcolor{yellow}}}
\newcommand{\hlpurple}[1]{\sethlcolor{lightpurple}\hl{#1}{\sethlcolor{yellow}}}

\newcommand{\qnms}{\qnm{s}}

\def\check#1{\red{#1}}
\def\new#1{\blue{#1}}
\def\changed#1{\underline{\textbf{\red{#1}}}}
\def\remove#1{\hlred{#1}}
\newcommand{\cw}{\tilde{\omega}}
\newcommand{\CW}{\tilde{\Omega}}
\newcommand{\CWr}{{\Omega}^{\mathrm{r}}}
\newcommand{\CWc}{{\Omega}^{\mathrm{c}}}
\newcommand{\SC}{\mathcal{K}}
\newcommand{\CC}{\mathcal{C}}
\newcommand{\SCr}{\mathcal{K}^{\mathrm{r}}}
\newcommand{\SCc}{\mathcal{K}^{\mathrm{c}}}
\newcommand{\lalapprox}{\texttt{MMRDNS}}
\def\jf{j_f}
\def\mf{M_f}
\newcommand{\LL}{\bar{l}}
\newcommand{\MM}{\bar{m}}
\def\gmvp#1{greedy-multivariate-polynomial#1
  (\texttt{GMVP}#1)\gdef\gmvp{\texttt{GMVP}}}
\def\gmvr#1{greedy-multivariate-rational#1
  (\texttt{GMVR}#1)\gdef\gmvr{\texttt{GMVR}}}

% Paper ref macros
\def\PaperOne{\hyperlink{cite.London:202XP1}{Paper {I}}}
\def\PaperTwo{\hyperlink{cite.London:202XP2}{Paper {II}}}
\def\FigThreeA{Fig.~\hyperref[F3]{3a}}
\def\FigThreeB{Fig.~\hyperref[F3]{3b}}
\def\FigThreeC{Fig.~\hyperref[F3]{3c}}
\def\FigThreeD{Fig.~\hyperref[F3]{3d}}
\def\FigThreeE{Fig.~\hyperref[F3]{3e}}
\def\FigThreeF{Fig.~\hyperref[F3]{3f}}
\def\FigThreeG{Fig.~\hyperref[F3]{3g}}
\def\FigFourA{Fig.~\hyperref[F4]{4a}}
\def\FigFourB{Fig.~\hyperref[F4]{4b}}
\def\FigFourC{Fig.~\hyperref[F4]{4c}}
\def\ccHp#1{canonical confluent Heun polynomial#1}
\def\CcHp#1{Canonical confluent Heun polynomial#1}
\def\cHp#1{confluent Heun polynomial#1}
\def\CHp#1{Confluent Heun polynomial#1}
\def\cp#1{canonical polynomial#1}
\def\i{(\textit{i})}
\def\ii{(\textit{ii})}
\def\iii{(\textit{iii})}
\def\ci{\textit{i}}
\def\cii{\textit{ii}}
\def\ciii{\textit{iii}}
\def\monm#1{\langle\xi^{#1}\rangle}
\def\wrt{with respect to }
\def\pspm{{p_\star^\pm} }
\def\psp{{p_\star^+} }
\def\psm{{p_\star^-} }
\def\LcH{ \hat{L}{_{p}}^{(\tx{cH})} }
\def\CZero{{\tx{C}_0} }
\def\COne{{\tx{C}_1} }
\def\CTwo{{\tx{C}_2} }
\def\CThree{{\tx{C}_3} }
\def\CFour{{\tx{C}_4} }
\def\nnint{non-negative integer}

\def\bpar#1{\smallskip\smallskip\paragraph*{\textbf{#1}}--~}
\def\bparNoSkip#1{\paragraph*{\textbf{#1}}--~}
\def\iparNoSkip#1{\paragraph*{\textit{#1}}--~}
\def\tre{\tk{}'s radial equation}
\def\trp{the radial problem}

%% file: paper.bbl
\begin{thebibliography}{163}%
\makeatletter
\providecommand \@ifxundefined [1]{%
 \@ifx{#1\undefined}
}%
\providecommand \@ifnum [1]{%
 \ifnum #1\expandafter \@firstoftwo
 \else \expandafter \@secondoftwo
 \fi
}%
\providecommand \@ifx [1]{%
 \ifx #1\expandafter \@firstoftwo
 \else \expandafter \@secondoftwo
 \fi
}%
\providecommand \natexlab [1]{#1}%
\providecommand \enquote  [1]{``#1''}%
\providecommand \bibnamefont  [1]{#1}%
\providecommand \bibfnamefont [1]{#1}%
\providecommand \citenamefont [1]{#1}%
\providecommand \href@noop [0]{\@secondoftwo}%
\providecommand \href [0]{\begingroup \@sanitize@url \@href}%
\providecommand \@href[1]{\@@startlink{#1}\@@href}%
\providecommand \@@href[1]{\endgroup#1\@@endlink}%
\providecommand \@sanitize@url [0]{\catcode `\\12\catcode `\$12\catcode `\&12\catcode `\#12\catcode `\^12\catcode `\_12\catcode `\%12\relax}%
\providecommand \@@startlink[1]{}%
\providecommand \@@endlink[0]{}%
\providecommand \url  [0]{\begingroup\@sanitize@url \@url }%
\providecommand \@url [1]{\endgroup\@href {#1}{\urlprefix }}%
\providecommand \urlprefix  [0]{URL }%
\providecommand \Eprint [0]{\href }%
\providecommand \doibase [0]{http://dx.doi.org/}%
\providecommand \selectlanguage [0]{\@gobble}%
\providecommand \bibinfo  [0]{\@secondoftwo}%
\providecommand \bibfield  [0]{\@secondoftwo}%
\providecommand \translation [1]{[#1]}%
\providecommand \BibitemOpen [0]{}%
\providecommand \bibitemStop [0]{}%
\providecommand \bibitemNoStop [0]{.\EOS\space}%
\providecommand \EOS [0]{\spacefactor3000\relax}%
\providecommand \BibitemShut  [1]{\csname bibitem#1\endcsname}%
\let\auto@bib@innerbib\@empty
%</preamble>
\bibitem [{\citenamefont {Vishveshwara}(1970)}]{Vishv:1970}%
  \BibitemOpen
  \bibfield  {author} {\bibinfo {author} {\bibfnamefont {C.~V.}\ \bibnamefont {Vishveshwara}},\ }\href {\doibase 10.1103/PhysRevD.1.2870} {\bibfield  {journal} {\bibinfo  {journal} {Phys. Rev. D}\ }\textbf {\bibinfo {volume} {1}},\ \bibinfo {pages} {2870} (\bibinfo {year} {1970})}\BibitemShut {NoStop}%
\bibitem [{\citenamefont {Abbott}\ \emph {et~al.}(2016{\natexlab{a}})\citenamefont {Abbott} \emph {et~al.}}]{LIGOScientific:2016aoc}%
  \BibitemOpen
  \bibfield  {author} {\bibinfo {author} {\bibfnamefont {B.~P.}\ \bibnamefont {Abbott}} \emph {et~al.} (\bibinfo {collaboration} {LIGO Scientific, Virgo}),\ }\href {\doibase 10.1103/PhysRevLett.116.061102} {\bibfield  {journal} {\bibinfo  {journal} {Phys. Rev. Lett.}\ }\textbf {\bibinfo {volume} {116}},\ \bibinfo {pages} {061102} (\bibinfo {year} {2016}{\natexlab{a}})},\ \Eprint {http://arxiv.org/abs/1602.03837} {arXiv:1602.03837 [gr-qc]} \BibitemShut {NoStop}%
\bibitem [{\citenamefont {Abbott}\ \emph {et~al.}(2016{\natexlab{b}})\citenamefont {Abbott} \emph {et~al.}}]{TheLIGOScientific:2016src}%
  \BibitemOpen
  \bibfield  {author} {\bibinfo {author} {\bibfnamefont {B.~P.}\ \bibnamefont {Abbott}} \emph {et~al.} (\bibinfo {collaboration} {Virgo, LIGO Scientific}),\ }\href {\doibase 10.1103/PhysRevLett.116.221101} {\bibfield  {journal} {\bibinfo  {journal} {Phys. Rev. Lett.}\ }\textbf {\bibinfo {volume} {116}},\ \bibinfo {pages} {221101} (\bibinfo {year} {2016}{\natexlab{b}})},\ \Eprint {http://arxiv.org/abs/1602.03841} {arXiv:1602.03841 [gr-qc]} \BibitemShut {NoStop}%
%%CITATION = ARXIV:1602.03841;%%
\bibitem [{\citenamefont {Abbott}\ \emph {et~al.}(2021{\natexlab{a}})\citenamefont {Abbott} \emph {et~al.}}]{LIGOScientific:2020ibl}%
  \BibitemOpen
  \bibfield  {author} {\bibinfo {author} {\bibfnamefont {R.}~\bibnamefont {Abbott}} \emph {et~al.} (\bibinfo {collaboration} {LIGO Scientific, Virgo}),\ }\href {\doibase 10.1103/PhysRevX.11.021053} {\bibfield  {journal} {\bibinfo  {journal} {Phys. Rev. X}\ }\textbf {\bibinfo {volume} {11}},\ \bibinfo {pages} {021053} (\bibinfo {year} {2021}{\natexlab{a}})},\ \Eprint {http://arxiv.org/abs/2010.14527} {arXiv:2010.14527 [gr-qc]} \BibitemShut {NoStop}%
\bibitem [{\citenamefont {Abbott}\ \emph {et~al.}(2021{\natexlab{b}})\citenamefont {Abbott} \emph {et~al.}}]{LIGOScientific:2021djp}%
  \BibitemOpen
  \bibfield  {author} {\bibinfo {author} {\bibfnamefont {R.}~\bibnamefont {Abbott}} \emph {et~al.} (\bibinfo {collaboration} {LIGO Scientific, VIRGO, KAGRA}),\ }\href@noop {} {\  (\bibinfo {year} {2021}{\natexlab{b}})},\ \Eprint {http://arxiv.org/abs/2111.03606} {arXiv:2111.03606 [gr-qc]} \BibitemShut {NoStop}%
\bibitem [{\citenamefont {Abbott}\ \emph {et~al.}(2019{\natexlab{a}})\citenamefont {Abbott} \emph {et~al.}}]{LIGOScientific:2018mvr}%
  \BibitemOpen
  \bibfield  {author} {\bibinfo {author} {\bibfnamefont {B.}~\bibnamefont {Abbott}} \emph {et~al.} (\bibinfo {collaboration} {LIGO Scientific, Virgo}),\ }\href {\doibase 10.1103/PhysRevX.9.031040} {\bibfield  {journal} {\bibinfo  {journal} {Phys. Rev. X}\ }\textbf {\bibinfo {volume} {9}},\ \bibinfo {pages} {031040} (\bibinfo {year} {2019}{\natexlab{a}})},\ \Eprint {http://arxiv.org/abs/1811.12907} {arXiv:1811.12907 [astro-ph.HE]} \BibitemShut {NoStop}%
\bibitem [{\citenamefont {Kamaretsos}\ \emph {et~al.}(2012)\citenamefont {Kamaretsos}, \citenamefont {Hannam},\ and\ \citenamefont {Sathyaprakash}}]{Kamaretsos:2012bs}%
  \BibitemOpen
  \bibfield  {author} {\bibinfo {author} {\bibfnamefont {I.}~\bibnamefont {Kamaretsos}}, \bibinfo {author} {\bibfnamefont {M.}~\bibnamefont {Hannam}}, \ and\ \bibinfo {author} {\bibfnamefont {B.}~\bibnamefont {Sathyaprakash}},\ }\href {\doibase 10.1103/PhysRevLett.109.141102} {\bibfield  {journal} {\bibinfo  {journal} {Phys.Rev.Lett.}\ }\textbf {\bibinfo {volume} {109}},\ \bibinfo {pages} {141102} (\bibinfo {year} {2012})},\ \Eprint {http://arxiv.org/abs/1207.0399} {arXiv:1207.0399 [gr-qc]} \BibitemShut {NoStop}%
%%CITATION = ARXIV:1207.0399;%%
\bibitem [{\citenamefont {London}\ \emph {et~al.}(2014)\citenamefont {London}, \citenamefont {Shoemaker},\ and\ \citenamefont {Healy}}]{London:2014cma}%
  \BibitemOpen
  \bibfield  {author} {\bibinfo {author} {\bibfnamefont {L.}~\bibnamefont {London}}, \bibinfo {author} {\bibfnamefont {D.}~\bibnamefont {Shoemaker}}, \ and\ \bibinfo {author} {\bibfnamefont {J.}~\bibnamefont {Healy}},\ }\href {\doibase 10.1103/PhysRevD.90.124032} {\bibfield  {journal} {\bibinfo  {journal} {Phys. Rev.}\ }\textbf {\bibinfo {volume} {D90}},\ \bibinfo {pages} {124032} (\bibinfo {year} {2014})},\ \Eprint {http://arxiv.org/abs/1404.3197} {arXiv:1404.3197 [gr-qc]} \BibitemShut {NoStop}%
%%CITATION = ARXIV:1404.3197;%%
\bibitem [{\citenamefont {Hughes}\ \emph {et~al.}(2019)\citenamefont {Hughes}, \citenamefont {Apte}, \citenamefont {Khanna},\ and\ \citenamefont {Lim}}]{Hughes:2019zmt}%
  \BibitemOpen
  \bibfield  {author} {\bibinfo {author} {\bibfnamefont {S.~A.}\ \bibnamefont {Hughes}}, \bibinfo {author} {\bibfnamefont {A.}~\bibnamefont {Apte}}, \bibinfo {author} {\bibfnamefont {G.}~\bibnamefont {Khanna}}, \ and\ \bibinfo {author} {\bibfnamefont {H.}~\bibnamefont {Lim}},\ }\href {\doibase 10.1103/PhysRevLett.123.161101} {\bibfield  {journal} {\bibinfo  {journal} {Phys. Rev. Lett.}\ }\textbf {\bibinfo {volume} {123}},\ \bibinfo {pages} {161101} (\bibinfo {year} {2019})},\ \Eprint {http://arxiv.org/abs/1901.05900} {arXiv:1901.05900 [gr-qc]} \BibitemShut {NoStop}%
\bibitem [{\citenamefont {Berti}\ \emph {et~al.}(2006{\natexlab{a}})\citenamefont {Berti}, \citenamefont {Cardoso},\ and\ \citenamefont {Will}}]{Berti:2005ys}%
  \BibitemOpen
  \bibfield  {author} {\bibinfo {author} {\bibfnamefont {E.}~\bibnamefont {Berti}}, \bibinfo {author} {\bibfnamefont {V.}~\bibnamefont {Cardoso}}, \ and\ \bibinfo {author} {\bibfnamefont {C.~M.}\ \bibnamefont {Will}},\ }\href {\doibase 10.1103/PhysRevD.73.064030} {\bibfield  {journal} {\bibinfo  {journal} {Phys. Rev.}\ }\textbf {\bibinfo {volume} {D73}},\ \bibinfo {pages} {064030} (\bibinfo {year} {2006}{\natexlab{a}})},\ \Eprint {http://arxiv.org/abs/gr-qc/0512160} {arXiv:gr-qc/0512160 [gr-qc]} \BibitemShut {NoStop}%
%%CITATION = GR-QC/0512160;%%
\bibitem [{\citenamefont {Berti}\ \emph {et~al.}(2009)\citenamefont {Berti}, \citenamefont {Cardoso},\ and\ \citenamefont {Starinets}}]{QNMTopicalReview09}%
  \BibitemOpen
  \bibfield  {author} {\bibinfo {author} {\bibfnamefont {E.}~\bibnamefont {Berti}}, \bibinfo {author} {\bibfnamefont {V.}~\bibnamefont {Cardoso}}, \ and\ \bibinfo {author} {\bibfnamefont {A.~O.}\ \bibnamefont {Starinets}},\ }\href {http://stacks.iop.org/0264-9381/26/i=16/a=163001} {\bibfield  {journal} {\bibinfo  {journal} {Classical and Quantum Gravity}\ }\textbf {\bibinfo {volume} {26}},\ \bibinfo {pages} {163001} (\bibinfo {year} {2009})}\BibitemShut {NoStop}%
\bibitem [{\citenamefont {Maselli}\ \emph {et~al.}(2017)\citenamefont {Maselli}, \citenamefont {Kokkotas},\ and\ \citenamefont {Laguna}}]{Maselli:2017kvl}%
  \BibitemOpen
  \bibfield  {author} {\bibinfo {author} {\bibfnamefont {A.}~\bibnamefont {Maselli}}, \bibinfo {author} {\bibfnamefont {K.}~\bibnamefont {Kokkotas}}, \ and\ \bibinfo {author} {\bibfnamefont {P.}~\bibnamefont {Laguna}},\ }\href {\doibase 10.1103/PhysRevD.95.104026} {\bibfield  {journal} {\bibinfo  {journal} {Phys. Rev.}\ }\textbf {\bibinfo {volume} {D95}},\ \bibinfo {pages} {104026} (\bibinfo {year} {2017})},\ \Eprint {http://arxiv.org/abs/1702.01110} {arXiv:1702.01110 [gr-qc]} \BibitemShut {NoStop}%
%%CITATION = ARXIV:1702.01110;%%
\bibitem [{\citenamefont {Kokkotas}\ and\ \citenamefont {Schmidt}(1999)}]{LivRevQNM}%
  \BibitemOpen
  \bibfield  {author} {\bibinfo {author} {\bibfnamefont {K.~D.}\ \bibnamefont {Kokkotas}}\ and\ \bibinfo {author} {\bibfnamefont {B.~G.}\ \bibnamefont {Schmidt}},\ }\href {http://arxiv.org/abs/gr-qc/9909058} {\bibfield  {journal} {\bibinfo  {journal} {Living Review in Relativity}\ }\textbf {\bibinfo {volume} {2}} (\bibinfo {year} {1999})}\BibitemShut {NoStop}%
\bibitem [{\citenamefont {Evstafyeva}\ \emph {et~al.}(2023)\citenamefont {Evstafyeva}, \citenamefont {Agathos},\ and\ \citenamefont {Ripley}}]{Evstafyeva:2022rve}%
  \BibitemOpen
  \bibfield  {author} {\bibinfo {author} {\bibfnamefont {T.}~\bibnamefont {Evstafyeva}}, \bibinfo {author} {\bibfnamefont {M.}~\bibnamefont {Agathos}}, \ and\ \bibinfo {author} {\bibfnamefont {J.~L.}\ \bibnamefont {Ripley}},\ }\href {\doibase 10.1103/PhysRevD.107.124010} {\bibfield  {journal} {\bibinfo  {journal} {Phys. Rev. D}\ }\textbf {\bibinfo {volume} {107}},\ \bibinfo {pages} {124010} (\bibinfo {year} {2023})},\ \Eprint {http://arxiv.org/abs/2212.11359} {arXiv:2212.11359 [gr-qc]} \BibitemShut {NoStop}%
\bibitem [{\citenamefont {Arun}\ \emph {et~al.}(2022)\citenamefont {Arun} \emph {et~al.}}]{LISA:2022kgy}%
  \BibitemOpen
  \bibfield  {author} {\bibinfo {author} {\bibfnamefont {K.~G.}\ \bibnamefont {Arun}} \emph {et~al.} (\bibinfo {collaboration} {LISA}),\ }\href {\doibase 10.1007/s41114-022-00036-9} {\bibfield  {journal} {\bibinfo  {journal} {Living Rev. Rel.}\ }\textbf {\bibinfo {volume} {25}},\ \bibinfo {pages} {4} (\bibinfo {year} {2022})},\ \Eprint {http://arxiv.org/abs/2205.01597} {arXiv:2205.01597 [gr-qc]} \BibitemShut {NoStop}%
\bibitem [{\citenamefont {Pacilio}\ and\ \citenamefont {Bhagwat}(2023)}]{Pacilio:2023mvk}%
  \BibitemOpen
  \bibfield  {author} {\bibinfo {author} {\bibfnamefont {C.}~\bibnamefont {Pacilio}}\ and\ \bibinfo {author} {\bibfnamefont {S.}~\bibnamefont {Bhagwat}},\ }\href {\doibase 10.1103/PhysRevD.107.083021} {\bibfield  {journal} {\bibinfo  {journal} {Phys. Rev. D}\ }\textbf {\bibinfo {volume} {107}},\ \bibinfo {pages} {083021} (\bibinfo {year} {2023})},\ \Eprint {http://arxiv.org/abs/2301.02267} {arXiv:2301.02267 [gr-qc]} \BibitemShut {NoStop}%
\bibitem [{\citenamefont {Bailes}\ \emph {et~al.}(2021)\citenamefont {Bailes}, \citenamefont {Berger}, \citenamefont {Brady}, \citenamefont {Branchesi}, \citenamefont {Danzmann}, \citenamefont {Evans}, \citenamefont {Holley-Bockelmann}, \citenamefont {Iyer}, \citenamefont {Kajita}, \citenamefont {Katsanevas}, \citenamefont {Kramer}, \citenamefont {Lazzarini}, \citenamefont {Lehner}, \citenamefont {Losurdo}, \citenamefont {L\"{u}ck}, \citenamefont {McClelland}, \citenamefont {McLaughlin}, \citenamefont {Punturo}, \citenamefont {Ransom}, \citenamefont {Raychaudhury}, \citenamefont {Reitze}, \citenamefont {Ricci}, \citenamefont {Rowan}, \citenamefont {Saito}, \citenamefont {Sanders}, \citenamefont {Sathyaprakash}, \citenamefont {Schutz}, \citenamefont {Sesana}, \citenamefont {Shinkai}, \citenamefont {Siemens}, \citenamefont {Shoemaker}, \citenamefont {Thorpe}, \citenamefont {van~den Brand},\ and\ \citenamefont {Vitale}}]{Bailes2021}%
  \BibitemOpen
  \bibfield  {author} {\bibinfo {author} {\bibfnamefont {M.}~\bibnamefont {Bailes}}, \bibinfo {author} {\bibfnamefont {B.~K.}\ \bibnamefont {Berger}}, \bibinfo {author} {\bibfnamefont {P.~R.}\ \bibnamefont {Brady}}, \bibinfo {author} {\bibfnamefont {M.}~\bibnamefont {Branchesi}}, \bibinfo {author} {\bibfnamefont {K.}~\bibnamefont {Danzmann}}, \bibinfo {author} {\bibfnamefont {M.}~\bibnamefont {Evans}}, \bibinfo {author} {\bibfnamefont {K.}~\bibnamefont {Holley-Bockelmann}}, \bibinfo {author} {\bibfnamefont {B.~R.}\ \bibnamefont {Iyer}}, \bibinfo {author} {\bibfnamefont {T.}~\bibnamefont {Kajita}}, \bibinfo {author} {\bibfnamefont {S.}~\bibnamefont {Katsanevas}}, \bibinfo {author} {\bibfnamefont {M.}~\bibnamefont {Kramer}}, \bibinfo {author} {\bibfnamefont {A.}~\bibnamefont {Lazzarini}}, \bibinfo {author} {\bibfnamefont {L.}~\bibnamefont {Lehner}}, \bibinfo {author} {\bibfnamefont {G.}~\bibnamefont {Losurdo}}, \bibinfo {author} {\bibfnamefont {H.}~\bibnamefont {L\"{u}ck}}, \bibinfo {author} {\bibfnamefont {D.~E.}\ \bibnamefont {McClelland}}, \bibinfo {author} {\bibfnamefont {M.~A.}\ \bibnamefont {McLaughlin}}, \bibinfo {author} {\bibfnamefont {M.}~\bibnamefont {Punturo}}, \bibinfo {author} {\bibfnamefont {S.}~\bibnamefont {Ransom}}, \bibinfo {author} {\bibfnamefont {S.}~\bibnamefont {Raychaudhury}}, \bibinfo {author} {\bibfnamefont {D.~H.}\ \bibnamefont {Reitze}}, \bibinfo {author} {\bibfnamefont {F.}~\bibnamefont {Ricci}}, \bibinfo {author} {\bibfnamefont {S.}~\bibnamefont {Rowan}}, \bibinfo {author} {\bibfnamefont {Y.}~\bibnamefont {Saito}}, \bibinfo {author} {\bibfnamefont {G.~H.}\ \bibnamefont {Sanders}}, \bibinfo {author} {\bibfnamefont {B.~S.}\ \bibnamefont {Sathyaprakash}}, \bibinfo {author} {\bibfnamefont {B.~F.}\ \bibnamefont {Schutz}}, \bibinfo {author} {\bibfnamefont {A.}~\bibnamefont {Sesana}}, \bibinfo {author} {\bibfnamefont {H.}~\bibnamefont {Shinkai}}, \bibinfo {author} {\bibfnamefont {X.}~\bibnamefont {Siemens}}, \bibinfo {author} {\bibfnamefont {D.~H.}\ \bibnamefont {Shoemaker}}, \bibinfo {author} {\bibfnamefont {J.}~\bibnamefont {Thorpe}}, \bibinfo {author} {\bibfnamefont {J.~F.~J.}\ \bibnamefont {van~den Brand}}, \ and\ \bibinfo {author} {\bibfnamefont {S.}~\bibnamefont {Vitale}},\ }\href {\doibase 10.1038/s42254-021-00303-8} {\bibfield  {journal} {\bibinfo  {journal} {Nature Reviews Physics}\ }\textbf {\bibinfo {volume} {3}},\ \bibinfo {pages} {344} (\bibinfo {year} {2021})}\BibitemShut {NoStop}%
\bibitem [{\citenamefont {Sathyaprakash}\ \emph {et~al.}(2012)\citenamefont {Sathyaprakash} \emph {et~al.}}]{Sathyaprakash:2012jk}%
  \BibitemOpen
  \bibfield  {author} {\bibinfo {author} {\bibfnamefont {B.}~\bibnamefont {Sathyaprakash}} \emph {et~al.},\ }\bibfield  {booktitle} {\emph {\bibinfo {booktitle} {{Gravitational waves. Numerical relativity - data analysis. Proceedings, 9th Edoardo Amaldi Conference, Amaldi 9, and meeting, NRDA 2011, Cardiff, UK, July 10-15, 2011}}},\ }\href {\doibase 10.1088/0264-9381/29/12/124013, 10.1088/0264-9381/30/7/079501} {\bibfield  {journal} {\bibinfo  {journal} {Class. Quant. Grav.}\ }\textbf {\bibinfo {volume} {29}},\ \bibinfo {pages} {124013} (\bibinfo {year} {2012})},\ \bibinfo {note} {[Erratum: Class. Quant. Grav.30,079501(2013)]},\ \Eprint {http://arxiv.org/abs/1206.0331} {arXiv:1206.0331 [gr-qc]} \BibitemShut {NoStop}%
%%CITATION = ARXIV:1206.0331;%%
\bibitem [{\citenamefont {Vallisneri}\ \emph {et~al.}(2015)\citenamefont {Vallisneri}, \citenamefont {Kanner}, \citenamefont {Williams}, \citenamefont {Weinstein},\ and\ \citenamefont {Stephens}}]{Vallisneri:2014vxa}%
  \BibitemOpen
  \bibfield  {author} {\bibinfo {author} {\bibfnamefont {M.}~\bibnamefont {Vallisneri}}, \bibinfo {author} {\bibfnamefont {J.}~\bibnamefont {Kanner}}, \bibinfo {author} {\bibfnamefont {R.}~\bibnamefont {Williams}}, \bibinfo {author} {\bibfnamefont {A.}~\bibnamefont {Weinstein}}, \ and\ \bibinfo {author} {\bibfnamefont {B.}~\bibnamefont {Stephens}},\ }\bibfield  {booktitle} {\emph {\bibinfo {booktitle} {{Proceedings, 10th International LISA Symposium: Gainesville, Florida, USA, May 18-23, 2014}}},\ }\href {\doibase 10.1088/1742-6596/610/1/012021} {\bibfield  {journal} {\bibinfo  {journal} {J. Phys. Conf. Ser.}\ }\textbf {\bibinfo {volume} {610}},\ \bibinfo {pages} {012021} (\bibinfo {year} {2015})},\ \Eprint {http://arxiv.org/abs/1410.4839} {arXiv:1410.4839 [gr-qc]} \BibitemShut {NoStop}%
%%CITATION = ARXIV:1410.4839;%%
\bibitem [{\citenamefont {Karnesis}\ \emph {et~al.}(2022)\citenamefont {Karnesis} \emph {et~al.}}]{Karnesis:2022vdp}%
  \BibitemOpen
  \bibfield  {author} {\bibinfo {author} {\bibfnamefont {N.}~\bibnamefont {Karnesis}} \emph {et~al.},\ }\href@noop {} {\  (\bibinfo {year} {2022})},\ \Eprint {http://arxiv.org/abs/2209.04358} {arXiv:2209.04358 [gr-qc]} \BibitemShut {NoStop}%
\bibitem [{\citenamefont {Abbott}\ \emph {et~al.}(2017)\citenamefont {Abbott} \emph {et~al.}}]{Abbott:2017oio}%
  \BibitemOpen
  \bibfield  {author} {\bibinfo {author} {\bibfnamefont {B.~P.}\ \bibnamefont {Abbott}} \emph {et~al.} (\bibinfo {collaboration} {Virgo, LIGO Scientific}),\ }\href {\doibase 10.1103/PhysRevLett.119.141101} {\bibfield  {journal} {\bibinfo  {journal} {Phys. Rev. Lett.}\ }\textbf {\bibinfo {volume} {119}},\ \bibinfo {pages} {141101} (\bibinfo {year} {2017})},\ \Eprint {http://arxiv.org/abs/1709.09660} {arXiv:1709.09660 [gr-qc]} \BibitemShut {NoStop}%
%%CITATION = ARXIV:1709.09660;%%
\bibitem [{\citenamefont {Abbott}\ \emph {et~al.}(2016{\natexlab{c}})\citenamefont {Abbott} \emph {et~al.}}]{TheLIGOScientific:2016wfe}%
  \BibitemOpen
  \bibfield  {author} {\bibinfo {author} {\bibfnamefont {B.~P.}\ \bibnamefont {Abbott}} \emph {et~al.} (\bibinfo {collaboration} {Virgo, LIGO Scientific}),\ }\href {\doibase 10.1103/PhysRevLett.116.241102} {\bibfield  {journal} {\bibinfo  {journal} {Phys. Rev. Lett.}\ }\textbf {\bibinfo {volume} {116}},\ \bibinfo {pages} {241102} (\bibinfo {year} {2016}{\natexlab{c}})},\ \Eprint {http://arxiv.org/abs/1602.03840} {arXiv:1602.03840 [gr-qc]} \BibitemShut {NoStop}%
%%CITATION = ARXIV:1602.03840;%%
\bibitem [{\citenamefont {{Misner}}\ \emph {et~al.}(1973)\citenamefont {{Misner}}, \citenamefont {{Thorne}},\ and\ \citenamefont {{Wheeler}}}]{Misner1973}%
  \BibitemOpen
  \bibfield  {author} {\bibinfo {author} {\bibfnamefont {C.~W.}\ \bibnamefont {{Misner}}}, \bibinfo {author} {\bibfnamefont {K.~S.}\ \bibnamefont {{Thorne}}}, \ and\ \bibinfo {author} {\bibfnamefont {J.~A.}\ \bibnamefont {{Wheeler}}},\ }\href@noop {} {\emph {\bibinfo {title} {San Francisco: W.H.~Freeman and Co., 1973}}},\ edited by\ \bibinfo {editor} {\bibnamefont {{Misner, C.~W., Thorne, K.~S., \& Wheeler, J.~A.}}}\ (\bibinfo {year} {1973})\BibitemShut {NoStop}%
\bibitem [{\citenamefont {Maggiore}(2007)}]{Maggiore2007}%
  \BibitemOpen
  \bibfield  {author} {\bibinfo {author} {\bibfnamefont {M.}~\bibnamefont {Maggiore}},\ }\href {\doibase 10.1093/acprof:oso/9780198570745.001.0001} {\emph {\bibinfo {title} {Gravitational Waves}}}\ (\bibinfo  {publisher} {Oxford University {PressOxford}},\ \bibinfo {year} {2007})\BibitemShut {NoStop}%
\bibitem [{\citenamefont {Newman}\ and\ \citenamefont {Penrose}(1966)}]{NP66}%
  \BibitemOpen
  \bibfield  {author} {\bibinfo {author} {\bibfnamefont {E.~T.}\ \bibnamefont {Newman}}\ and\ \bibinfo {author} {\bibfnamefont {R.}~\bibnamefont {Penrose}},\ }\href {\doibase 10.1063/1.1931221} {\bibfield  {journal} {\bibinfo  {journal} {Journal of Mathematical Physics}\ }\textbf {\bibinfo {volume} {7}},\ \bibinfo {pages} {863} (\bibinfo {year} {1966})},\ \Eprint {http://arxiv.org/abs/https://doi.org/10.1063/1.1931221} {https://doi.org/10.1063/1.1931221} \BibitemShut {NoStop}%
\bibitem [{\citenamefont {{Newman}}\ and\ \citenamefont {{Penrose}}(1962)}]{NP62}%
  \BibitemOpen
  \bibfield  {author} {\bibinfo {author} {\bibfnamefont {E.}~\bibnamefont {{Newman}}}\ and\ \bibinfo {author} {\bibfnamefont {R.}~\bibnamefont {{Penrose}}},\ }\href {\doibase 10.1063/1.1724257} {\bibfield  {journal} {\bibinfo  {journal} {Journal of Mathematical Physics}\ }\textbf {\bibinfo {volume} {3}},\ \bibinfo {pages} {566} (\bibinfo {year} {1962})}\BibitemShut {NoStop}%
\bibitem [{\citenamefont {Teukolsky}(1972)}]{PhysRevLett.29.1114}%
  \BibitemOpen
  \bibfield  {author} {\bibinfo {author} {\bibfnamefont {S.~A.}\ \bibnamefont {Teukolsky}},\ }\href {\doibase 10.1103/PhysRevLett.29.1114} {\bibfield  {journal} {\bibinfo  {journal} {Phys. Rev. Lett.}\ }\textbf {\bibinfo {volume} {29}},\ \bibinfo {pages} {1114} (\bibinfo {year} {1972})}\BibitemShut {NoStop}%
\bibitem [{\citenamefont {Whiting}(1989)}]{Whiting:1989ms}%
  \BibitemOpen
  \bibfield  {author} {\bibinfo {author} {\bibfnamefont {B.~F.}\ \bibnamefont {Whiting}},\ }\href {\doibase 10.1063/1.528308} {\bibfield  {journal} {\bibinfo  {journal} {Journal of Mathematical Physics}\ }\textbf {\bibinfo {volume} {30}},\ \bibinfo {pages} {1301} (\bibinfo {year} {1989})},\ \Eprint {http://arxiv.org/abs/https://pubs.aip.org/aip/jmp/article-pdf/30/6/1301/8158708/1301\_1\_online.pdf} {https://pubs.aip.org/aip/jmp/article-pdf/30/6/1301/8158708/1301\_1\_online.pdf} \BibitemShut {NoStop}%
\bibitem [{\citenamefont {Giesler}\ \emph {et~al.}(2019)\citenamefont {Giesler}, \citenamefont {Isi}, \citenamefont {Scheel},\ and\ \citenamefont {Teukolsky}}]{Giesler:2019uxc}%
  \BibitemOpen
  \bibfield  {author} {\bibinfo {author} {\bibfnamefont {M.}~\bibnamefont {Giesler}}, \bibinfo {author} {\bibfnamefont {M.}~\bibnamefont {Isi}}, \bibinfo {author} {\bibfnamefont {M.~A.}\ \bibnamefont {Scheel}}, \ and\ \bibinfo {author} {\bibfnamefont {S.}~\bibnamefont {Teukolsky}},\ }\href {\doibase 10.1103/PhysRevX.9.041060} {\bibfield  {journal} {\bibinfo  {journal} {Phys. Rev. X}\ }\textbf {\bibinfo {volume} {9}},\ \bibinfo {pages} {041060} (\bibinfo {year} {2019})},\ \Eprint {http://arxiv.org/abs/1903.08284} {arXiv:1903.08284 [gr-qc]} \BibitemShut {NoStop}%
\bibitem [{\citenamefont {McWilliams}(2019)}]{McWilliams:2018ztb}%
  \BibitemOpen
  \bibfield  {author} {\bibinfo {author} {\bibfnamefont {S.~T.}\ \bibnamefont {McWilliams}},\ }\href {\doibase 10.1103/PhysRevLett.122.191102} {\bibfield  {journal} {\bibinfo  {journal} {Phys. Rev. Lett.}\ }\textbf {\bibinfo {volume} {122}},\ \bibinfo {pages} {191102} (\bibinfo {year} {2019})},\ \Eprint {http://arxiv.org/abs/1810.00040} {arXiv:1810.00040 [gr-qc]} \BibitemShut {NoStop}%
\bibitem [{\citenamefont {Carullo}\ \emph {et~al.}(2019)\citenamefont {Carullo}, \citenamefont {Del~Pozzo},\ and\ \citenamefont {Veitch}}]{Carullo:2019flw}%
  \BibitemOpen
  \bibfield  {author} {\bibinfo {author} {\bibfnamefont {G.}~\bibnamefont {Carullo}}, \bibinfo {author} {\bibfnamefont {W.}~\bibnamefont {Del~Pozzo}}, \ and\ \bibinfo {author} {\bibfnamefont {J.}~\bibnamefont {Veitch}},\ }\href {\doibase 10.1103/PhysRevD.99.123029} {\bibfield  {journal} {\bibinfo  {journal} {Phys. Rev. D}\ }\textbf {\bibinfo {volume} {99}},\ \bibinfo {pages} {123029} (\bibinfo {year} {2019})},\ \bibinfo {note} {[Erratum: Phys.Rev.D 100, 089903 (2019)]},\ \Eprint {http://arxiv.org/abs/1902.07527} {arXiv:1902.07527 [gr-qc]} \BibitemShut {NoStop}%
\bibitem [{\citenamefont {Ota}\ and\ \citenamefont {Chirenti}(2020)}]{Ota:2019bzl}%
  \BibitemOpen
  \bibfield  {author} {\bibinfo {author} {\bibfnamefont {I.}~\bibnamefont {Ota}}\ and\ \bibinfo {author} {\bibfnamefont {C.}~\bibnamefont {Chirenti}},\ }\href {\doibase 10.1103/PhysRevD.101.104005} {\bibfield  {journal} {\bibinfo  {journal} {Phys. Rev. D}\ }\textbf {\bibinfo {volume} {101}},\ \bibinfo {pages} {104005} (\bibinfo {year} {2020})},\ \Eprint {http://arxiv.org/abs/1911.00440} {arXiv:1911.00440 [gr-qc]} \BibitemShut {NoStop}%
\bibitem [{\citenamefont {Bhagwat}\ \emph {et~al.}(2020)\citenamefont {Bhagwat}, \citenamefont {Forteza}, \citenamefont {Pani},\ and\ \citenamefont {Ferrari}}]{Bhagwat:2019dtm}%
  \BibitemOpen
  \bibfield  {author} {\bibinfo {author} {\bibfnamefont {S.}~\bibnamefont {Bhagwat}}, \bibinfo {author} {\bibfnamefont {X.~J.}\ \bibnamefont {Forteza}}, \bibinfo {author} {\bibfnamefont {P.}~\bibnamefont {Pani}}, \ and\ \bibinfo {author} {\bibfnamefont {V.}~\bibnamefont {Ferrari}},\ }\href {\doibase 10.1103/PhysRevD.101.044033} {\bibfield  {journal} {\bibinfo  {journal} {Phys. Rev. D}\ }\textbf {\bibinfo {volume} {101}},\ \bibinfo {pages} {044033} (\bibinfo {year} {2020})},\ \Eprint {http://arxiv.org/abs/1910.08708} {arXiv:1910.08708 [gr-qc]} \BibitemShut {NoStop}%
\bibitem [{\citenamefont {Forteza}\ \emph {et~al.}(2020)\citenamefont {Forteza}, \citenamefont {Bhagwat}, \citenamefont {Pani},\ and\ \citenamefont {Ferrari}}]{Forteza:2020hbw}%
  \BibitemOpen
  \bibfield  {author} {\bibinfo {author} {\bibfnamefont {X.~J.}\ \bibnamefont {Forteza}}, \bibinfo {author} {\bibfnamefont {S.}~\bibnamefont {Bhagwat}}, \bibinfo {author} {\bibfnamefont {P.}~\bibnamefont {Pani}}, \ and\ \bibinfo {author} {\bibfnamefont {V.}~\bibnamefont {Ferrari}},\ }\href@noop {} {\  (\bibinfo {year} {2020})},\ \Eprint {http://arxiv.org/abs/2005.03260} {arXiv:2005.03260 [gr-qc]} \BibitemShut {NoStop}%
\bibitem [{\citenamefont {Leong}\ \emph {et~al.}(2023)\citenamefont {Leong}, \citenamefont {Calder\'on~Bustillo}, \citenamefont {Gracia-Linares},\ and\ \citenamefont {Laguna}}]{Leong:2023nuk}%
  \BibitemOpen
  \bibfield  {author} {\bibinfo {author} {\bibfnamefont {S.~H.~W.}\ \bibnamefont {Leong}}, \bibinfo {author} {\bibfnamefont {J.}~\bibnamefont {Calder\'on~Bustillo}}, \bibinfo {author} {\bibfnamefont {M.}~\bibnamefont {Gracia-Linares}}, \ and\ \bibinfo {author} {\bibfnamefont {P.}~\bibnamefont {Laguna}},\ }\href@noop {} {\  (\bibinfo {year} {2023})},\ \Eprint {http://arxiv.org/abs/2308.03250} {arXiv:2308.03250 [gr-qc]} \BibitemShut {NoStop}%
\bibitem [{\citenamefont {Khera}\ \emph {et~al.}(2023)\citenamefont {Khera}, \citenamefont {Ribes~Metidieri}, \citenamefont {Bonga}, \citenamefont {Forteza}, \citenamefont {Krishnan}, \citenamefont {Poisson}, \citenamefont {Pook-Kolb}, \citenamefont {Schnetter},\ and\ \citenamefont {Yang}}]{Khera:2023lnc}%
  \BibitemOpen
  \bibfield  {author} {\bibinfo {author} {\bibfnamefont {N.}~\bibnamefont {Khera}}, \bibinfo {author} {\bibfnamefont {A.}~\bibnamefont {Ribes~Metidieri}}, \bibinfo {author} {\bibfnamefont {B.}~\bibnamefont {Bonga}}, \bibinfo {author} {\bibfnamefont {X.~J.}\ \bibnamefont {Forteza}}, \bibinfo {author} {\bibfnamefont {B.}~\bibnamefont {Krishnan}}, \bibinfo {author} {\bibfnamefont {E.}~\bibnamefont {Poisson}}, \bibinfo {author} {\bibfnamefont {D.}~\bibnamefont {Pook-Kolb}}, \bibinfo {author} {\bibfnamefont {E.}~\bibnamefont {Schnetter}}, \ and\ \bibinfo {author} {\bibfnamefont {H.}~\bibnamefont {Yang}},\ }\href@noop {} {\  (\bibinfo {year} {2023})},\ \Eprint {http://arxiv.org/abs/2306.11142} {arXiv:2306.11142 [gr-qc]} \BibitemShut {NoStop}%
\bibitem [{\citenamefont {Franchini}\ and\ \citenamefont {V\"olkel}(2023)}]{Franchini:2023eda}%
  \BibitemOpen
  \bibfield  {author} {\bibinfo {author} {\bibfnamefont {N.}~\bibnamefont {Franchini}}\ and\ \bibinfo {author} {\bibfnamefont {S.~H.}\ \bibnamefont {V\"olkel}},\ }\href@noop {} {\  (\bibinfo {year} {2023})},\ \Eprint {http://arxiv.org/abs/2305.01696} {arXiv:2305.01696 [gr-qc]} \BibitemShut {NoStop}%
\bibitem [{\citenamefont {Nee}\ \emph {et~al.}(2023)\citenamefont {Nee}, \citenamefont {V\"olkel},\ and\ \citenamefont {Pfeiffer}}]{Nee:2023osy}%
  \BibitemOpen
  \bibfield  {author} {\bibinfo {author} {\bibfnamefont {P.~J.}\ \bibnamefont {Nee}}, \bibinfo {author} {\bibfnamefont {S.~H.}\ \bibnamefont {V\"olkel}}, \ and\ \bibinfo {author} {\bibfnamefont {H.~P.}\ \bibnamefont {Pfeiffer}},\ }\href {\doibase 10.1103/PhysRevD.108.044032} {\bibfield  {journal} {\bibinfo  {journal} {Phys. Rev. D}\ }\textbf {\bibinfo {volume} {108}},\ \bibinfo {pages} {044032} (\bibinfo {year} {2023})},\ \Eprint {http://arxiv.org/abs/2302.06634} {arXiv:2302.06634 [gr-qc]} \BibitemShut {NoStop}%
\bibitem [{\citenamefont {Cotesta}\ \emph {et~al.}(2022)\citenamefont {Cotesta}, \citenamefont {Carullo}, \citenamefont {Berti},\ and\ \citenamefont {Cardoso}}]{Cotesta:2022pci}%
  \BibitemOpen
  \bibfield  {author} {\bibinfo {author} {\bibfnamefont {R.}~\bibnamefont {Cotesta}}, \bibinfo {author} {\bibfnamefont {G.}~\bibnamefont {Carullo}}, \bibinfo {author} {\bibfnamefont {E.}~\bibnamefont {Berti}}, \ and\ \bibinfo {author} {\bibfnamefont {V.}~\bibnamefont {Cardoso}},\ }\href {\doibase 10.1103/PhysRevLett.129.111102} {\bibfield  {journal} {\bibinfo  {journal} {Phys. Rev. Lett.}\ }\textbf {\bibinfo {volume} {129}},\ \bibinfo {pages} {111102} (\bibinfo {year} {2022})},\ \Eprint {http://arxiv.org/abs/2201.00822} {arXiv:2201.00822 [gr-qc]} \BibitemShut {NoStop}%
\bibitem [{\citenamefont {Baibhav}\ \emph {et~al.}(2023)\citenamefont {Baibhav}, \citenamefont {Cheung}, \citenamefont {Berti}, \citenamefont {Cardoso}, \citenamefont {Carullo}, \citenamefont {Cotesta}, \citenamefont {Del~Pozzo},\ and\ \citenamefont {Duque}}]{Baibhav:2023clw}%
  \BibitemOpen
  \bibfield  {author} {\bibinfo {author} {\bibfnamefont {V.}~\bibnamefont {Baibhav}}, \bibinfo {author} {\bibfnamefont {M.~H.-Y.}\ \bibnamefont {Cheung}}, \bibinfo {author} {\bibfnamefont {E.}~\bibnamefont {Berti}}, \bibinfo {author} {\bibfnamefont {V.}~\bibnamefont {Cardoso}}, \bibinfo {author} {\bibfnamefont {G.}~\bibnamefont {Carullo}}, \bibinfo {author} {\bibfnamefont {R.}~\bibnamefont {Cotesta}}, \bibinfo {author} {\bibfnamefont {W.}~\bibnamefont {Del~Pozzo}}, \ and\ \bibinfo {author} {\bibfnamefont {F.}~\bibnamefont {Duque}},\ }\href@noop {} {\  (\bibinfo {year} {2023})},\ \Eprint {http://arxiv.org/abs/2302.03050} {arXiv:2302.03050 [gr-qc]} \BibitemShut {NoStop}%
\bibitem [{\citenamefont {Abbott}\ \emph {et~al.}(2021{\natexlab{c}})\citenamefont {Abbott} \emph {et~al.}}]{LIGOScientific:2021usb}%
  \BibitemOpen
  \bibfield  {author} {\bibinfo {author} {\bibfnamefont {R.}~\bibnamefont {Abbott}} \emph {et~al.} (\bibinfo {collaboration} {LIGO Scientific, VIRGO}),\ }\href@noop {} {\  (\bibinfo {year} {2021}{\natexlab{c}})},\ \Eprint {http://arxiv.org/abs/2108.01045} {arXiv:2108.01045 [gr-qc]} \BibitemShut {NoStop}%
\bibitem [{\citenamefont {Abbott}\ \emph {et~al.}(2019{\natexlab{b}})\citenamefont {Abbott} \emph {et~al.}}]{LIGOScientific:2019fpa}%
  \BibitemOpen
  \bibfield  {author} {\bibinfo {author} {\bibfnamefont {B.~P.}\ \bibnamefont {Abbott}} \emph {et~al.} (\bibinfo {collaboration} {LIGO Scientific, Virgo}),\ }\href {\doibase 10.1103/PhysRevD.100.104036} {\bibfield  {journal} {\bibinfo  {journal} {Phys. Rev. D}\ }\textbf {\bibinfo {volume} {100}},\ \bibinfo {pages} {104036} (\bibinfo {year} {2019}{\natexlab{b}})},\ \Eprint {http://arxiv.org/abs/1903.04467} {arXiv:1903.04467 [gr-qc]} \BibitemShut {NoStop}%
\bibitem [{\citenamefont {Abbott}\ \emph {et~al.}(2021{\natexlab{d}})\citenamefont {Abbott} \emph {et~al.}}]{LIGOScientific:2020tif}%
  \BibitemOpen
  \bibfield  {author} {\bibinfo {author} {\bibfnamefont {R.}~\bibnamefont {Abbott}} \emph {et~al.} (\bibinfo {collaboration} {LIGO Scientific, Virgo}),\ }\href {\doibase 10.1103/PhysRevD.103.122002} {\bibfield  {journal} {\bibinfo  {journal} {Phys. Rev. D}\ }\textbf {\bibinfo {volume} {103}},\ \bibinfo {pages} {122002} (\bibinfo {year} {2021}{\natexlab{d}})},\ \Eprint {http://arxiv.org/abs/2010.14529} {arXiv:2010.14529 [gr-qc]} \BibitemShut {NoStop}%
\bibitem [{\citenamefont {Abbott}\ \emph {et~al.}(2021{\natexlab{e}})\citenamefont {Abbott} \emph {et~al.}}]{LIGOScientific:2021sio}%
  \BibitemOpen
  \bibfield  {author} {\bibinfo {author} {\bibfnamefont {R.}~\bibnamefont {Abbott}} \emph {et~al.} (\bibinfo {collaboration} {LIGO Scientific, VIRGO, KAGRA}),\ }\href@noop {} {\  (\bibinfo {year} {2021}{\natexlab{e}})},\ \Eprint {http://arxiv.org/abs/2112.06861} {arXiv:2112.06861 [gr-qc]} \BibitemShut {NoStop}%
\bibitem [{\citenamefont {Berti}\ \emph {et~al.}(2016)\citenamefont {Berti}, \citenamefont {Sesana}, \citenamefont {Barausse}, \citenamefont {Cardoso},\ and\ \citenamefont {Belczynski}}]{Berti:2016lat}%
  \BibitemOpen
  \bibfield  {author} {\bibinfo {author} {\bibfnamefont {E.}~\bibnamefont {Berti}}, \bibinfo {author} {\bibfnamefont {A.}~\bibnamefont {Sesana}}, \bibinfo {author} {\bibfnamefont {E.}~\bibnamefont {Barausse}}, \bibinfo {author} {\bibfnamefont {V.}~\bibnamefont {Cardoso}}, \ and\ \bibinfo {author} {\bibfnamefont {K.}~\bibnamefont {Belczynski}},\ }\href {\doibase 10.1103/PhysRevLett.117.101102} {\bibfield  {journal} {\bibinfo  {journal} {Phys. Rev. Lett.}\ }\textbf {\bibinfo {volume} {117}},\ \bibinfo {pages} {101102} (\bibinfo {year} {2016})},\ \Eprint {http://arxiv.org/abs/1605.09286} {arXiv:1605.09286 [gr-qc]} \BibitemShut {NoStop}%
%%CITATION = ARXIV:1605.09286;%%
\bibitem [{\citenamefont {Mishra}\ \emph {et~al.}(2010)\citenamefont {Mishra}, \citenamefont {Arun}, \citenamefont {Iyer},\ and\ \citenamefont {Sathyaprakash}}]{Mishra:2010tp}%
  \BibitemOpen
  \bibfield  {author} {\bibinfo {author} {\bibfnamefont {C.~K.}\ \bibnamefont {Mishra}}, \bibinfo {author} {\bibfnamefont {K.~G.}\ \bibnamefont {Arun}}, \bibinfo {author} {\bibfnamefont {B.~R.}\ \bibnamefont {Iyer}}, \ and\ \bibinfo {author} {\bibfnamefont {B.~S.}\ \bibnamefont {Sathyaprakash}},\ }\href {\doibase 10.1103/PhysRevD.82.064010} {\bibfield  {journal} {\bibinfo  {journal} {Phys. Rev. D}\ }\textbf {\bibinfo {volume} {82}},\ \bibinfo {pages} {064010} (\bibinfo {year} {2010})},\ \Eprint {http://arxiv.org/abs/1005.0304} {arXiv:1005.0304 [gr-qc]} \BibitemShut {NoStop}%
\bibitem [{\citenamefont {Robson}\ \emph {et~al.}(2019)\citenamefont {Robson}, \citenamefont {Cornish},\ and\ \citenamefont {Liu}}]{Robson:2018ifk}%
  \BibitemOpen
  \bibfield  {author} {\bibinfo {author} {\bibfnamefont {T.}~\bibnamefont {Robson}}, \bibinfo {author} {\bibfnamefont {N.~J.}\ \bibnamefont {Cornish}}, \ and\ \bibinfo {author} {\bibfnamefont {C.}~\bibnamefont {Liu}},\ }\href {\doibase 10.1088/1361-6382/ab1101} {\bibfield  {journal} {\bibinfo  {journal} {Class. Quant. Grav.}\ }\textbf {\bibinfo {volume} {36}},\ \bibinfo {pages} {105011} (\bibinfo {year} {2019})},\ \Eprint {http://arxiv.org/abs/1803.01944} {arXiv:1803.01944 [astro-ph.HE]} \BibitemShut {NoStop}%
\bibitem [{\citenamefont {Klein}\ \emph {et~al.}(2016)\citenamefont {Klein} \emph {et~al.}}]{Klein:2015hvg}%
  \BibitemOpen
  \bibfield  {author} {\bibinfo {author} {\bibfnamefont {A.}~\bibnamefont {Klein}} \emph {et~al.},\ }\href {\doibase 10.1103/PhysRevD.93.024003} {\bibfield  {journal} {\bibinfo  {journal} {Phys. Rev. D}\ }\textbf {\bibinfo {volume} {93}},\ \bibinfo {pages} {024003} (\bibinfo {year} {2016})},\ \Eprint {http://arxiv.org/abs/1511.05581} {arXiv:1511.05581 [gr-qc]} \BibitemShut {NoStop}%
\bibitem [{\citenamefont {Macedo}\ \emph {et~al.}(2013)\citenamefont {Macedo}, \citenamefont {Pani}, \citenamefont {Cardoso},\ and\ \citenamefont {Crispino}}]{Macedo_2013}%
  \BibitemOpen
  \bibfield  {author} {\bibinfo {author} {\bibfnamefont {C.~F.~B.}\ \bibnamefont {Macedo}}, \bibinfo {author} {\bibfnamefont {P.}~\bibnamefont {Pani}}, \bibinfo {author} {\bibfnamefont {V.}~\bibnamefont {Cardoso}}, \ and\ \bibinfo {author} {\bibfnamefont {L.~C.~B.}\ \bibnamefont {Crispino}},\ }\href {\doibase 10.1088/0004-637X/774/1/48} {\bibfield  {journal} {\bibinfo  {journal} {The Astrophysical Journal}\ }\textbf {\bibinfo {volume} {774}},\ \bibinfo {pages} {48} (\bibinfo {year} {2013})}\BibitemShut {NoStop}%
\bibitem [{\citenamefont {Barausse}\ \emph {et~al.}(2015)\citenamefont {Barausse}, \citenamefont {Cardoso},\ and\ \citenamefont {Pani}}]{Barausse:2014pra}%
  \BibitemOpen
  \bibfield  {author} {\bibinfo {author} {\bibfnamefont {E.}~\bibnamefont {Barausse}}, \bibinfo {author} {\bibfnamefont {V.}~\bibnamefont {Cardoso}}, \ and\ \bibinfo {author} {\bibfnamefont {P.}~\bibnamefont {Pani}},\ }\href {\doibase 10.1088/1742-6596/610/1/012044} {\bibfield  {journal} {\bibinfo  {journal} {J. Phys. Conf. Ser.}\ }\textbf {\bibinfo {volume} {610}},\ \bibinfo {pages} {012044} (\bibinfo {year} {2015})},\ \Eprint {http://arxiv.org/abs/1404.7140} {arXiv:1404.7140 [astro-ph.CO]} \BibitemShut {NoStop}%
\bibitem [{\citenamefont {Zhao}\ \emph {et~al.}(2023)\citenamefont {Zhao}, \citenamefont {Sun}, \citenamefont {Cao}, \citenamefont {Lin},\ and\ \citenamefont {Qian}}]{Zhao:2023itk}%
  \BibitemOpen
  \bibfield  {author} {\bibinfo {author} {\bibfnamefont {Y.}~\bibnamefont {Zhao}}, \bibinfo {author} {\bibfnamefont {B.}~\bibnamefont {Sun}}, \bibinfo {author} {\bibfnamefont {Z.}~\bibnamefont {Cao}}, \bibinfo {author} {\bibfnamefont {K.}~\bibnamefont {Lin}}, \ and\ \bibinfo {author} {\bibfnamefont {W.-L.}\ \bibnamefont {Qian}},\ }\href@noop {} {\  (\bibinfo {year} {2023})},\ \Eprint {http://arxiv.org/abs/2308.15371} {arXiv:2308.15371 [gr-qc]} \BibitemShut {NoStop}%
\bibitem [{\citenamefont {Berti}\ \emph {et~al.}(2003)\citenamefont {Berti}, \citenamefont {Cardoso}, \citenamefont {Kokkotas},\ and\ \citenamefont {Onozawa}}]{Berti:2003jh}%
  \BibitemOpen
  \bibfield  {author} {\bibinfo {author} {\bibfnamefont {E.}~\bibnamefont {Berti}}, \bibinfo {author} {\bibfnamefont {V.}~\bibnamefont {Cardoso}}, \bibinfo {author} {\bibfnamefont {K.~D.}\ \bibnamefont {Kokkotas}}, \ and\ \bibinfo {author} {\bibfnamefont {H.}~\bibnamefont {Onozawa}},\ }\href {\doibase 10.1103/PhysRevD.68.124018} {\bibfield  {journal} {\bibinfo  {journal} {Phys. Rev. D}\ }\textbf {\bibinfo {volume} {68}},\ \bibinfo {pages} {124018} (\bibinfo {year} {2003})},\ \Eprint {http://arxiv.org/abs/hep-th/0307013} {arXiv:hep-th/0307013} \BibitemShut {NoStop}%
\bibitem [{\citenamefont {Berti}\ and\ \citenamefont {Klein}(2014)}]{Berti:2014fga}%
  \BibitemOpen
  \bibfield  {author} {\bibinfo {author} {\bibfnamefont {E.}~\bibnamefont {Berti}}\ and\ \bibinfo {author} {\bibfnamefont {A.}~\bibnamefont {Klein}},\ }\href {\doibase 10.1103/PhysRevD.90.064012} {\bibfield  {journal} {\bibinfo  {journal} {Phys. Rev. D}\ }\textbf {\bibinfo {volume} {90}},\ \bibinfo {pages} {064012} (\bibinfo {year} {2014})},\ \Eprint {http://arxiv.org/abs/1408.1860} {arXiv:1408.1860 [gr-qc]} \BibitemShut {NoStop}%
\bibitem [{\citenamefont {Berti}\ \emph {et~al.}(2006{\natexlab{b}})\citenamefont {Berti}, \citenamefont {Cardoso},\ and\ \citenamefont {Casals}}]{Berti:2005gp}%
  \BibitemOpen
  \bibfield  {author} {\bibinfo {author} {\bibfnamefont {E.}~\bibnamefont {Berti}}, \bibinfo {author} {\bibfnamefont {V.}~\bibnamefont {Cardoso}}, \ and\ \bibinfo {author} {\bibfnamefont {M.}~\bibnamefont {Casals}},\ }\href {\doibase 10.1103/PhysRevD.73.109902, 10.1103/PhysRevD.73.024013} {\bibfield  {journal} {\bibinfo  {journal} {Phys. Rev.}\ }\textbf {\bibinfo {volume} {D73}},\ \bibinfo {pages} {024013} (\bibinfo {year} {2006}{\natexlab{b}})},\ \bibinfo {note} {[Erratum: Phys. Rev.D73,109902(2006)]},\ \Eprint {http://arxiv.org/abs/gr-qc/0511111} {arXiv:gr-qc/0511111 [gr-qc]} \BibitemShut {NoStop}%
%%CITATION = GR-QC/0511111;%%
\bibitem [{\citenamefont {London}(2020)}]{London:2018gaq}%
  \BibitemOpen
  \bibfield  {author} {\bibinfo {author} {\bibfnamefont {L.}~\bibnamefont {London}},\ }\href {\doibase 10.1103/PhysRevD.102.084052} {\enquote {\bibinfo {title} {{Modeling ringdown. II. Aligned-spin binary black holes, implications for data analysis and fundamental theory}},}\ } (\bibinfo {year} {2020}),\ \Eprint {http://arxiv.org/abs/1801.08208} {arXiv:1801.08208 [gr-qc]} \BibitemShut {NoStop}%
\bibitem [{\citenamefont {London}(2023)}]{London:2020uva}%
  \BibitemOpen
  \bibfield  {author} {\bibinfo {author} {\bibfnamefont {L.~T.}\ \bibnamefont {London}},\ }\href {\doibase 10.1103/PhysRevD.107.044056} {\bibfield  {journal} {\bibinfo  {journal} {Phys. Rev. D}\ }\textbf {\bibinfo {volume} {107}},\ \bibinfo {pages} {044056} (\bibinfo {year} {2023})},\ \Eprint {http://arxiv.org/abs/2006.11449} {arXiv:2006.11449 [gr-qc]} \BibitemShut {NoStop}%
\bibitem [{\citenamefont {London}\ and\ \citenamefont {Fauchon-Jones}(2019)}]{London:2018nxs}%
  \BibitemOpen
  \bibfield  {author} {\bibinfo {author} {\bibfnamefont {L.}~\bibnamefont {London}}\ and\ \bibinfo {author} {\bibfnamefont {E.}~\bibnamefont {Fauchon-Jones}},\ }\href {\doibase 10.1088/1361-6382/ab2f11} {\bibfield  {journal} {\bibinfo  {journal} {Class. Quant. Grav.}\ }\textbf {\bibinfo {volume} {36}},\ \bibinfo {pages} {235015} (\bibinfo {year} {2019})},\ \Eprint {http://arxiv.org/abs/1810.03550} {arXiv:1810.03550 [gr-qc]} \BibitemShut {NoStop}%
\bibitem [{\citenamefont {Husa}\ \emph {et~al.}(2016)\citenamefont {Husa}, \citenamefont {Khan}, \citenamefont {Hannam}, \citenamefont {Pürrer}, \citenamefont {Ohme}, \citenamefont {Jiménez~Forteza},\ and\ \citenamefont {Bohé}}]{Husa:2015iqa}%
  \BibitemOpen
  \bibfield  {author} {\bibinfo {author} {\bibfnamefont {S.}~\bibnamefont {Husa}}, \bibinfo {author} {\bibfnamefont {S.}~\bibnamefont {Khan}}, \bibinfo {author} {\bibfnamefont {M.}~\bibnamefont {Hannam}}, \bibinfo {author} {\bibfnamefont {M.}~\bibnamefont {Pürrer}}, \bibinfo {author} {\bibfnamefont {F.}~\bibnamefont {Ohme}}, \bibinfo {author} {\bibfnamefont {X.}~\bibnamefont {Jiménez~Forteza}}, \ and\ \bibinfo {author} {\bibfnamefont {A.}~\bibnamefont {Bohé}},\ }\href {\doibase 10.1103/PhysRevD.93.044006} {\bibfield  {journal} {\bibinfo  {journal} {Phys. Rev.}\ }\textbf {\bibinfo {volume} {D93}},\ \bibinfo {pages} {044006} (\bibinfo {year} {2016})},\ \Eprint {http://arxiv.org/abs/1508.07250} {arXiv:1508.07250 [gr-qc]} \BibitemShut {NoStop}%
%%CITATION = ARXIV:1508.07250;%%
\bibitem [{\citenamefont {Hamilton}\ \emph {et~al.}(2021)\citenamefont {Hamilton}, \citenamefont {London}, \citenamefont {Thompson}, \citenamefont {Fauchon-Jones}, \citenamefont {Hannam}, \citenamefont {Kalaghatgi}, \citenamefont {Khan}, \citenamefont {Pannarale},\ and\ \citenamefont {Vano-Vinuales}}]{Hamilton:2021pkf}%
  \BibitemOpen
  \bibfield  {author} {\bibinfo {author} {\bibfnamefont {E.}~\bibnamefont {Hamilton}}, \bibinfo {author} {\bibfnamefont {L.}~\bibnamefont {London}}, \bibinfo {author} {\bibfnamefont {J.~E.}\ \bibnamefont {Thompson}}, \bibinfo {author} {\bibfnamefont {E.}~\bibnamefont {Fauchon-Jones}}, \bibinfo {author} {\bibfnamefont {M.}~\bibnamefont {Hannam}}, \bibinfo {author} {\bibfnamefont {C.}~\bibnamefont {Kalaghatgi}}, \bibinfo {author} {\bibfnamefont {S.}~\bibnamefont {Khan}}, \bibinfo {author} {\bibfnamefont {F.}~\bibnamefont {Pannarale}}, \ and\ \bibinfo {author} {\bibfnamefont {A.}~\bibnamefont {Vano-Vinuales}},\ }\href@noop {} {\  (\bibinfo {year} {2021})},\ \Eprint {http://arxiv.org/abs/2107.08876} {arXiv:2107.08876 [gr-qc]} \BibitemShut {NoStop}%
\bibitem [{\citenamefont {Hamilton}\ \emph {et~al.}(2023)\citenamefont {Hamilton}, \citenamefont {London},\ and\ \citenamefont {Hannam}}]{Hamilton:2023znn}%
  \BibitemOpen
  \bibfield  {author} {\bibinfo {author} {\bibfnamefont {E.}~\bibnamefont {Hamilton}}, \bibinfo {author} {\bibfnamefont {L.}~\bibnamefont {London}}, \ and\ \bibinfo {author} {\bibfnamefont {M.}~\bibnamefont {Hannam}},\ }\href {\doibase 10.1103/PhysRevD.107.104035} {\bibfield  {journal} {\bibinfo  {journal} {Phys. Rev. D}\ }\textbf {\bibinfo {volume} {107}},\ \bibinfo {pages} {104035} (\bibinfo {year} {2023})},\ \Eprint {http://arxiv.org/abs/2301.06558} {arXiv:2301.06558 [gr-qc]} \BibitemShut {NoStop}%
\bibitem [{\citenamefont {Cotesta}\ \emph {et~al.}(2020)\citenamefont {Cotesta}, \citenamefont {Marsat},\ and\ \citenamefont {P\"urrer}}]{Cotesta:2020qhw}%
  \BibitemOpen
  \bibfield  {author} {\bibinfo {author} {\bibfnamefont {R.}~\bibnamefont {Cotesta}}, \bibinfo {author} {\bibfnamefont {S.}~\bibnamefont {Marsat}}, \ and\ \bibinfo {author} {\bibfnamefont {M.}~\bibnamefont {P\"urrer}},\ }\href {\doibase 10.1103/PhysRevD.101.124040} {\bibfield  {journal} {\bibinfo  {journal} {Phys. Rev. D}\ }\textbf {\bibinfo {volume} {101}},\ \bibinfo {pages} {124040} (\bibinfo {year} {2020})},\ \Eprint {http://arxiv.org/abs/2003.12079} {arXiv:2003.12079 [gr-qc]} \BibitemShut {NoStop}%
\bibitem [{\citenamefont {Pompili}\ \emph {et~al.}(2023)\citenamefont {Pompili} \emph {et~al.}}]{Pompili:2023tna}%
  \BibitemOpen
  \bibfield  {author} {\bibinfo {author} {\bibfnamefont {L.}~\bibnamefont {Pompili}} \emph {et~al.},\ }\href@noop {} {\  (\bibinfo {year} {2023})},\ \Eprint {http://arxiv.org/abs/2303.18039} {arXiv:2303.18039 [gr-qc]} \BibitemShut {NoStop}%
\bibitem [{\citenamefont {Ossokine}\ \emph {et~al.}(2020)\citenamefont {Ossokine} \emph {et~al.}}]{Ossokine:2020kjp}%
  \BibitemOpen
  \bibfield  {author} {\bibinfo {author} {\bibfnamefont {S.}~\bibnamefont {Ossokine}} \emph {et~al.},\ }\href {\doibase 10.1103/PhysRevD.102.044055} {\bibfield  {journal} {\bibinfo  {journal} {Phys. Rev. D}\ }\textbf {\bibinfo {volume} {102}},\ \bibinfo {pages} {044055} (\bibinfo {year} {2020})},\ \Eprint {http://arxiv.org/abs/2004.09442} {arXiv:2004.09442 [gr-qc]} \BibitemShut {NoStop}%
\bibitem [{\citenamefont {Lagos}\ and\ \citenamefont {Hui}(2022)}]{Lagos:2022otp}%
  \BibitemOpen
  \bibfield  {author} {\bibinfo {author} {\bibfnamefont {M.}~\bibnamefont {Lagos}}\ and\ \bibinfo {author} {\bibfnamefont {L.}~\bibnamefont {Hui}},\ }\href@noop {} {\  (\bibinfo {year} {2022})},\ \Eprint {http://arxiv.org/abs/2208.07379} {arXiv:2208.07379 [gr-qc]} \BibitemShut {NoStop}%
\bibitem [{\citenamefont {Press}(1971)}]{Press:1971ApJ}%
  \BibitemOpen
  \bibfield  {author} {\bibinfo {author} {\bibfnamefont {W.~H.}\ \bibnamefont {Press}},\ }\href {\doibase 10.1086/180849} {\bibfield  {journal} {\bibinfo  {journal} {Astrophys. J. Lett.}\ }\textbf {\bibinfo {volume} {170}},\ \bibinfo {pages} {L105} (\bibinfo {year} {1971})}\BibitemShut {NoStop}%
\bibitem [{\citenamefont {Leaver}(1985)}]{leaver85}%
  \BibitemOpen
  \bibfield  {author} {\bibinfo {author} {\bibfnamefont {E.}~\bibnamefont {Leaver}},\ }\href {\doibase 10.1098/rspa.1985.0119} {\bibfield  {journal} {\bibinfo  {journal} {Proc. Roy. Soc. Lond. A}\ }\textbf {\bibinfo {volume} {A402}},\ \bibinfo {pages} {285} (\bibinfo {year} {1985})}\BibitemShut {NoStop}%
\bibitem [{\citenamefont {Press}\ and\ \citenamefont {Teukolsky}(1973)}]{Press:1973zz}%
  \BibitemOpen
  \bibfield  {author} {\bibinfo {author} {\bibfnamefont {W.~H.}\ \bibnamefont {Press}}\ and\ \bibinfo {author} {\bibfnamefont {S.~A.}\ \bibnamefont {Teukolsky}},\ }\href {\doibase 10.1086/152445} {\bibfield  {journal} {\bibinfo  {journal} {Astrophys. J.}\ }\textbf {\bibinfo {volume} {185}},\ \bibinfo {pages} {649} (\bibinfo {year} {1973})}\BibitemShut {NoStop}%
\bibitem [{\citenamefont {Ghosh}\ \emph {et~al.}(2018)\citenamefont {Ghosh}, \citenamefont {Johnson-Mcdaniel}, \citenamefont {Ghosh}, \citenamefont {Mishra}, \citenamefont {Ajith}, \citenamefont {Del~Pozzo}, \citenamefont {Berry}, \citenamefont {Nielsen},\ and\ \citenamefont {London}}]{Ghosh:2017gfp}%
  \BibitemOpen
  \bibfield  {author} {\bibinfo {author} {\bibfnamefont {A.}~\bibnamefont {Ghosh}}, \bibinfo {author} {\bibfnamefont {N.~K.}\ \bibnamefont {Johnson-Mcdaniel}}, \bibinfo {author} {\bibfnamefont {A.}~\bibnamefont {Ghosh}}, \bibinfo {author} {\bibfnamefont {C.~K.}\ \bibnamefont {Mishra}}, \bibinfo {author} {\bibfnamefont {P.}~\bibnamefont {Ajith}}, \bibinfo {author} {\bibfnamefont {W.}~\bibnamefont {Del~Pozzo}}, \bibinfo {author} {\bibfnamefont {C.~P.~L.}\ \bibnamefont {Berry}}, \bibinfo {author} {\bibfnamefont {A.~B.}\ \bibnamefont {Nielsen}}, \ and\ \bibinfo {author} {\bibfnamefont {L.}~\bibnamefont {London}},\ }\href {\doibase 10.1088/1361-6382/aa972e} {\bibfield  {journal} {\bibinfo  {journal} {Class. Quant. Grav.}\ }\textbf {\bibinfo {volume} {35}},\ \bibinfo {pages} {014002} (\bibinfo {year} {2018})},\ \Eprint {http://arxiv.org/abs/1704.06784} {arXiv:1704.06784 [gr-qc]} \BibitemShut {NoStop}%
\bibitem [{\citenamefont {Yang}\ \emph {et~al.}(2012)\citenamefont {Yang}, \citenamefont {Nichols}, \citenamefont {Zhang}, \citenamefont {Zimmerman}, \citenamefont {Zhang},\ and\ \citenamefont {Chen}}]{Yang:2012he}%
  \BibitemOpen
  \bibfield  {author} {\bibinfo {author} {\bibfnamefont {H.}~\bibnamefont {Yang}}, \bibinfo {author} {\bibfnamefont {D.~A.}\ \bibnamefont {Nichols}}, \bibinfo {author} {\bibfnamefont {F.}~\bibnamefont {Zhang}}, \bibinfo {author} {\bibfnamefont {A.}~\bibnamefont {Zimmerman}}, \bibinfo {author} {\bibfnamefont {Z.}~\bibnamefont {Zhang}}, \ and\ \bibinfo {author} {\bibfnamefont {Y.}~\bibnamefont {Chen}},\ }\href {\doibase 10.1103/PhysRevD.86.104006} {\bibfield  {journal} {\bibinfo  {journal} {Phys. Rev. D}\ }\textbf {\bibinfo {volume} {86}},\ \bibinfo {pages} {104006} (\bibinfo {year} {2012})},\ \Eprint {http://arxiv.org/abs/1207.4253} {arXiv:1207.4253 [gr-qc]} \BibitemShut {NoStop}%
\bibitem [{\citenamefont {Yang}\ \emph {et~al.}(2013)\citenamefont {Yang}, \citenamefont {Zhang}, \citenamefont {Zimmerman}, \citenamefont {Nichols}, \citenamefont {Berti},\ and\ \citenamefont {Chen}}]{Yang:2012pj}%
  \BibitemOpen
  \bibfield  {author} {\bibinfo {author} {\bibfnamefont {H.}~\bibnamefont {Yang}}, \bibinfo {author} {\bibfnamefont {F.}~\bibnamefont {Zhang}}, \bibinfo {author} {\bibfnamefont {A.}~\bibnamefont {Zimmerman}}, \bibinfo {author} {\bibfnamefont {D.~A.}\ \bibnamefont {Nichols}}, \bibinfo {author} {\bibfnamefont {E.}~\bibnamefont {Berti}}, \ and\ \bibinfo {author} {\bibfnamefont {Y.}~\bibnamefont {Chen}},\ }\href {\doibase 10.1103/PhysRevD.87.041502} {\bibfield  {journal} {\bibinfo  {journal} {Phys. Rev.}\ }\textbf {\bibinfo {volume} {D87}},\ \bibinfo {pages} {041502} (\bibinfo {year} {2013})},\ \Eprint {http://arxiv.org/abs/1212.3271} {arXiv:1212.3271 [gr-qc]} \BibitemShut {NoStop}%
%%CITATION = ARXIV:1212.3271;%%
\bibitem [{\citenamefont {Zimmerman}\ and\ \citenamefont {Mark}(2016)}]{Zimmerman:2015trm}%
  \BibitemOpen
  \bibfield  {author} {\bibinfo {author} {\bibfnamefont {A.}~\bibnamefont {Zimmerman}}\ and\ \bibinfo {author} {\bibfnamefont {Z.}~\bibnamefont {Mark}},\ }\href {\doibase 10.1103/PhysRevD.93.044033, 10.1103/PhysRevD.93.089905} {\bibfield  {journal} {\bibinfo  {journal} {Phys. Rev.}\ }\textbf {\bibinfo {volume} {D93}},\ \bibinfo {pages} {044033} (\bibinfo {year} {2016})},\ \bibinfo {note} {[Erratum: Phys. Rev.D93,no.8,089905(2016)]},\ \Eprint {http://arxiv.org/abs/1512.02247} {arXiv:1512.02247 [gr-qc]} \BibitemShut {NoStop}%
%%CITATION = ARXIV:1512.02247;%%
\bibitem [{\citenamefont {Dariescu}\ \emph {et~al.}(2021)\citenamefont {Dariescu}, \citenamefont {Dariescu},\ and\ \citenamefont {Stelea}}]{Dariescu:2021zve}%
  \BibitemOpen
  \bibfield  {author} {\bibinfo {author} {\bibfnamefont {C.}~\bibnamefont {Dariescu}}, \bibinfo {author} {\bibfnamefont {M.-A.}\ \bibnamefont {Dariescu}}, \ and\ \bibinfo {author} {\bibfnamefont {C.}~\bibnamefont {Stelea}},\ }\href {\doibase 10.1155/2021/5512735} {\bibfield  {journal} {\bibinfo  {journal} {Adv. High Energy Phys.}\ }\textbf {\bibinfo {volume} {2021}},\ \bibinfo {pages} {5512735} (\bibinfo {year} {2021})},\ \Eprint {http://arxiv.org/abs/2102.03850} {arXiv:2102.03850 [hep-th]} \BibitemShut {NoStop}%
\bibitem [{\citenamefont {Fiziev}(2010)}]{Fiziev:2010}%
  \BibitemOpen
  \bibfield  {author} {\bibinfo {author} {\bibfnamefont {P.~P.}\ \bibnamefont {Fiziev}},\ }\href {\doibase 10.1088/0264-9381/27/13/135001} {\bibfield  {journal} {\bibinfo  {journal} {Classical and Quantum Gravity}\ }\textbf {\bibinfo {volume} {27}},\ \bibinfo {pages} {135001} (\bibinfo {year} {2010})}\BibitemShut {NoStop}%
\bibitem [{\citenamefont {Hatsuda}(2021)}]{Hatsuda:2020iql}%
  \BibitemOpen
  \bibfield  {author} {\bibinfo {author} {\bibfnamefont {Y.}~\bibnamefont {Hatsuda}},\ }\href {\doibase 10.1007/s10714-021-02866-4} {\bibfield  {journal} {\bibinfo  {journal} {Gen. Rel. Grav.}\ }\textbf {\bibinfo {volume} {53}},\ \bibinfo {pages} {93} (\bibinfo {year} {2021})},\ \Eprint {http://arxiv.org/abs/2007.07906} {arXiv:2007.07906 [gr-qc]} \BibitemShut {NoStop}%
\bibitem [{\citenamefont {Brauer}(1964)}]{brauer1964}%
  \BibitemOpen
  \bibfield  {author} {\bibinfo {author} {\bibfnamefont {F.}~\bibnamefont {Brauer}},\ }\href {\doibase 10.1307/mmj/1028999193} {\bibfield  {journal} {\bibinfo  {journal} {Michigan Math. J.}\ }\textbf {\bibinfo {volume} {11}},\ \bibinfo {pages} {379} (\bibinfo {year} {1964})}\BibitemShut {NoStop}%
\bibitem [{\citenamefont {Christensen}(2003)}]{Christensen2003}%
  \BibitemOpen
  \bibfield  {author} {\bibinfo {author} {\bibfnamefont {O.}~\bibnamefont {Christensen}},\ }\href {\doibase 10.1007/978-0-8176-8224-8} {\emph {\bibinfo {title} {An Introduction to Frames and Riesz Bases}}}\ (\bibinfo  {publisher} {Birkh\"{a}user Boston},\ \bibinfo {year} {2003})\BibitemShut {NoStop}%
\bibitem [{\citenamefont {Fackerell}\ and\ \citenamefont {Crossman}(2008)}]{Fackerell:1977}%
  \BibitemOpen
  \bibfield  {author} {\bibinfo {author} {\bibfnamefont {E.~D.}\ \bibnamefont {Fackerell}}\ and\ \bibinfo {author} {\bibfnamefont {R.~G.}\ \bibnamefont {Crossman}},\ }\href {\doibase 10.1063/1.523499} {\bibfield  {journal} {\bibinfo  {journal} {Journal of Mathematical Physics}\ }\textbf {\bibinfo {volume} {18}},\ \bibinfo {pages} {1849} (\bibinfo {year} {2008})},\ \Eprint {http://arxiv.org/abs/https://pubs.aip.org/aip/jmp/article-pdf/18/9/1849/11184786/1849\_1\_online.pdf} {https://pubs.aip.org/aip/jmp/article-pdf/18/9/1849/11184786/1849\_1\_online.pdf} \BibitemShut {NoStop}%
\bibitem [{\citenamefont {Goldberg}\ \emph {et~al.}(1967)\citenamefont {Goldberg}, \citenamefont {MacFarlane}, \citenamefont {Newman}, \citenamefont {Rohrlich},\ and\ \citenamefont {Sudarshan}}]{Goldberg:1966uu}%
  \BibitemOpen
  \bibfield  {author} {\bibinfo {author} {\bibfnamefont {J.~N.}\ \bibnamefont {Goldberg}}, \bibinfo {author} {\bibfnamefont {A.~J.}\ \bibnamefont {MacFarlane}}, \bibinfo {author} {\bibfnamefont {E.~T.}\ \bibnamefont {Newman}}, \bibinfo {author} {\bibfnamefont {F.}~\bibnamefont {Rohrlich}}, \ and\ \bibinfo {author} {\bibfnamefont {E.~C.~G.}\ \bibnamefont {Sudarshan}},\ }\href {\doibase 10.1063/1.1705135} {\bibfield  {journal} {\bibinfo  {journal} {J. Math. Phys.}\ }\textbf {\bibinfo {volume} {8}},\ \bibinfo {pages} {2155} (\bibinfo {year} {1967})}\BibitemShut {NoStop}%
%%CITATION = JMAPA,8,2155;%%
\bibitem [{\citenamefont {Blanchet}(2014)}]{Blanchet:2013haa}%
  \BibitemOpen
  \bibfield  {author} {\bibinfo {author} {\bibfnamefont {L.}~\bibnamefont {Blanchet}},\ }\href {\doibase 10.12942/lrr-2014-2} {\bibfield  {journal} {\bibinfo  {journal} {Living Rev. Rel.}\ }\textbf {\bibinfo {volume} {17}},\ \bibinfo {pages} {2} (\bibinfo {year} {2014})},\ \Eprint {http://arxiv.org/abs/1310.1528} {arXiv:1310.1528 [gr-qc]} \BibitemShut {NoStop}%
\bibitem [{\citenamefont {Abbott}\ \emph {et~al.}(2016{\natexlab{d}})\citenamefont {Abbott} \emph {et~al.}}]{TheLIGOScientific:2016pea}%
  \BibitemOpen
  \bibfield  {author} {\bibinfo {author} {\bibfnamefont {B.~P.}\ \bibnamefont {Abbott}} \emph {et~al.} (\bibinfo {collaboration} {Virgo, LIGO Scientific}),\ }\href {\doibase 10.1103/PhysRevX.6.041015} {\bibfield  {journal} {\bibinfo  {journal} {Phys. Rev.}\ }\textbf {\bibinfo {volume} {X6}},\ \bibinfo {pages} {041015} (\bibinfo {year} {2016}{\natexlab{d}})},\ \Eprint {http://arxiv.org/abs/1606.04856} {arXiv:1606.04856 [gr-qc]} \BibitemShut {NoStop}%
%%CITATION = ARXIV:1606.04856;%%
\bibitem [{\citenamefont {O'Sullivan}\ and\ \citenamefont {Hughes}(2014)}]{OSullivan:2014ywd}%
  \BibitemOpen
  \bibfield  {author} {\bibinfo {author} {\bibfnamefont {S.}~\bibnamefont {O'Sullivan}}\ and\ \bibinfo {author} {\bibfnamefont {S.~A.}\ \bibnamefont {Hughes}},\ }\href {\doibase 10.1103/PhysRevD.91.109901} {\bibfield  {journal} {\bibinfo  {journal} {Phys. Rev. D}\ }\textbf {\bibinfo {volume} {90}},\ \bibinfo {pages} {124039} (\bibinfo {year} {2014})},\ \bibinfo {note} {[Erratum: Phys.Rev.D 91, 109901 (2015)]},\ \Eprint {http://arxiv.org/abs/1407.6983} {arXiv:1407.6983 [gr-qc]} \BibitemShut {NoStop}%
\bibitem [{\citenamefont {Ruiz}\ \emph {et~al.}(2008)\citenamefont {Ruiz}, \citenamefont {Takahashi}, \citenamefont {Alcubierre},\ and\ \citenamefont {Nunez}}]{Ruiz:2007yx}%
  \BibitemOpen
  \bibfield  {author} {\bibinfo {author} {\bibfnamefont {M.}~\bibnamefont {Ruiz}}, \bibinfo {author} {\bibfnamefont {R.}~\bibnamefont {Takahashi}}, \bibinfo {author} {\bibfnamefont {M.}~\bibnamefont {Alcubierre}}, \ and\ \bibinfo {author} {\bibfnamefont {D.}~\bibnamefont {Nunez}},\ }\href {\doibase 10.1007/s10714-007-0570-8} {\bibfield  {journal} {\bibinfo  {journal} {Gen. Rel. Grav.}\ }\textbf {\bibinfo {volume} {40}},\ \bibinfo {pages} {2467} (\bibinfo {year} {2008})},\ \Eprint {http://arxiv.org/abs/0707.4654} {arXiv:0707.4654 [gr-qc]} \BibitemShut {NoStop}%
\bibitem [{\citenamefont {Breuer}\ \emph {et~al.}(1977)\citenamefont {Breuer}, \citenamefont {Ryan},\ and\ \citenamefont {Waller}}]{Breuer77}%
  \BibitemOpen
  \bibfield  {author} {\bibinfo {author} {\bibfnamefont {R.~A.}\ \bibnamefont {Breuer}}, \bibinfo {author} {\bibfnamefont {M.~P.}\ \bibnamefont {Ryan}}, \ and\ \bibinfo {author} {\bibfnamefont {S.}~\bibnamefont {Waller}},\ }\href {http://www.jstor.org/stable/79369} {\bibfield  {journal} {\bibinfo  {journal} {Proceedings of the Royal Society of London. Series A, Mathematical and Physical Sciences}\ }\textbf {\bibinfo {volume} {358}},\ \bibinfo {pages} {71} (\bibinfo {year} {1977})}\BibitemShut {NoStop}%
\bibitem [{\citenamefont {London}\ and\ \citenamefont {Hughes}(2022)}]{London:2021P2}%
  \BibitemOpen
  \bibfield  {author} {\bibinfo {author} {\bibfnamefont {L.}~\bibnamefont {London}}\ and\ \bibinfo {author} {\bibfnamefont {S.~A.}\ \bibnamefont {Hughes}},\ }\href@noop {} {\  (\bibinfo {year} {2022})},\ \Eprint {http://arxiv.org/abs/2206.15246} {arXiv:2206.15246 [gr-qc]} \BibitemShut {NoStop}%
\bibitem [{\citenamefont {Green}\ \emph {et~al.}(2023)\citenamefont {Green}, \citenamefont {Hollands}, \citenamefont {Sberna}, \citenamefont {Toomani},\ and\ \citenamefont {Zimmerman}}]{Green:2022htq}%
  \BibitemOpen
  \bibfield  {author} {\bibinfo {author} {\bibfnamefont {S.~R.}\ \bibnamefont {Green}}, \bibinfo {author} {\bibfnamefont {S.}~\bibnamefont {Hollands}}, \bibinfo {author} {\bibfnamefont {L.}~\bibnamefont {Sberna}}, \bibinfo {author} {\bibfnamefont {V.}~\bibnamefont {Toomani}}, \ and\ \bibinfo {author} {\bibfnamefont {P.}~\bibnamefont {Zimmerman}},\ }\href {\doibase 10.1103/PhysRevD.107.064030} {\bibfield  {journal} {\bibinfo  {journal} {Phys. Rev. D}\ }\textbf {\bibinfo {volume} {107}},\ \bibinfo {pages} {064030} (\bibinfo {year} {2023})},\ \Eprint {http://arxiv.org/abs/2210.15935} {arXiv:2210.15935 [gr-qc]} \BibitemShut {NoStop}%
\bibitem [{\citenamefont {Leaver}(1986)}]{Leaver86c}%
  \BibitemOpen
  \bibfield  {author} {\bibinfo {author} {\bibfnamefont {E.~W.}\ \bibnamefont {Leaver}},\ }\href {\doibase 10.1103/PhysRevD.34.384} {\bibfield  {journal} {\bibinfo  {journal} {Phys. Rev. D}\ }\textbf {\bibinfo {volume} {34}},\ \bibinfo {pages} {384} (\bibinfo {year} {1986})}\BibitemShut {NoStop}%
\bibitem [{\citenamefont {Teukolsky}(1973)}]{Teukolsky:1973ha}%
  \BibitemOpen
  \bibfield  {author} {\bibinfo {author} {\bibfnamefont {S.~A.}\ \bibnamefont {Teukolsky}},\ }\href {\doibase 10.1086/152444} {\bibfield  {journal} {\bibinfo  {journal} {Astrophys. J.}\ }\textbf {\bibinfo {volume} {185}},\ \bibinfo {pages} {635} (\bibinfo {year} {1973})}\BibitemShut {NoStop}%
\bibitem [{\citenamefont {Hughes}(2000{\natexlab{a}})}]{Hughes:1999bq}%
  \BibitemOpen
  \bibfield  {author} {\bibinfo {author} {\bibfnamefont {S.~A.}\ \bibnamefont {Hughes}},\ }\href {\doibase 10.1103/PhysRevD.65.069902} {\bibfield  {journal} {\bibinfo  {journal} {Phys. Rev. D}\ }\textbf {\bibinfo {volume} {61}},\ \bibinfo {pages} {084004} (\bibinfo {year} {2000}{\natexlab{a}})},\ \bibinfo {note} {[Erratum: Phys.Rev.D 63, 049902 (2001), Erratum: Phys.Rev.D 65, 069902 (2002), Erratum: Phys.Rev.D 67, 089901 (2003), Erratum: Phys.Rev.D 78, 109902 (2008), Erratum: Phys.Rev.D 90, 109904 (2014)]},\ \Eprint {http://arxiv.org/abs/gr-qc/9910091} {arXiv:gr-qc/9910091} \BibitemShut {NoStop}%
\bibitem [{\citenamefont {Taracchini}\ \emph {et~al.}(2014)\citenamefont {Taracchini}, \citenamefont {Buonanno}, \citenamefont {Khanna},\ and\ \citenamefont {Hughes}}]{Taracchini:2014zpa}%
  \BibitemOpen
  \bibfield  {author} {\bibinfo {author} {\bibfnamefont {A.}~\bibnamefont {Taracchini}}, \bibinfo {author} {\bibfnamefont {A.}~\bibnamefont {Buonanno}}, \bibinfo {author} {\bibfnamefont {G.}~\bibnamefont {Khanna}}, \ and\ \bibinfo {author} {\bibfnamefont {S.~A.}\ \bibnamefont {Hughes}},\ }\href {\doibase 10.1103/PhysRevD.90.084025} {\bibfield  {journal} {\bibinfo  {journal} {Phys. Rev. D}\ }\textbf {\bibinfo {volume} {90}},\ \bibinfo {pages} {084025} (\bibinfo {year} {2014})},\ \Eprint {http://arxiv.org/abs/1404.1819} {arXiv:1404.1819 [gr-qc]} \BibitemShut {NoStop}%
\bibitem [{\citenamefont {Islam}\ \emph {et~al.}(2022)\citenamefont {Islam}, \citenamefont {Field}, \citenamefont {Hughes}, \citenamefont {Khanna}, \citenamefont {Varma}, \citenamefont {Giesler}, \citenamefont {Scheel}, \citenamefont {Kidder},\ and\ \citenamefont {Pfeiffer}}]{Islam:2022laz}%
  \BibitemOpen
  \bibfield  {author} {\bibinfo {author} {\bibfnamefont {T.}~\bibnamefont {Islam}}, \bibinfo {author} {\bibfnamefont {S.~E.}\ \bibnamefont {Field}}, \bibinfo {author} {\bibfnamefont {S.~A.}\ \bibnamefont {Hughes}}, \bibinfo {author} {\bibfnamefont {G.}~\bibnamefont {Khanna}}, \bibinfo {author} {\bibfnamefont {V.}~\bibnamefont {Varma}}, \bibinfo {author} {\bibfnamefont {M.}~\bibnamefont {Giesler}}, \bibinfo {author} {\bibfnamefont {M.~A.}\ \bibnamefont {Scheel}}, \bibinfo {author} {\bibfnamefont {L.~E.}\ \bibnamefont {Kidder}}, \ and\ \bibinfo {author} {\bibfnamefont {H.~P.}\ \bibnamefont {Pfeiffer}},\ }\href {\doibase 10.1103/PhysRevD.106.104025} {\bibfield  {journal} {\bibinfo  {journal} {Phys. Rev. D}\ }\textbf {\bibinfo {volume} {106}},\ \bibinfo {pages} {104025} (\bibinfo {year} {2022})},\ \Eprint {http://arxiv.org/abs/2204.01972} {arXiv:2204.01972 [gr-qc]} \BibitemShut {NoStop}%
\bibitem [{\citenamefont {Rifat}\ \emph {et~al.}(2019)\citenamefont {Rifat}, \citenamefont {Khanna},\ and\ \citenamefont {Burko}}]{Rifat:2019fkt}%
  \BibitemOpen
  \bibfield  {author} {\bibinfo {author} {\bibfnamefont {N.~E.~M.}\ \bibnamefont {Rifat}}, \bibinfo {author} {\bibfnamefont {G.}~\bibnamefont {Khanna}}, \ and\ \bibinfo {author} {\bibfnamefont {L.~M.}\ \bibnamefont {Burko}},\ }\href {\doibase 10.1103/PhysRevResearch.1.033150} {\bibfield  {journal} {\bibinfo  {journal} {Phys. Rev. Research.}\ }\textbf {\bibinfo {volume} {1}},\ \bibinfo {pages} {033150} (\bibinfo {year} {2019})},\ \Eprint {http://arxiv.org/abs/1910.03462} {arXiv:1910.03462 [gr-qc]} \BibitemShut {NoStop}%
\bibitem [{\citenamefont {Campanelli}\ \emph {et~al.}(2001)\citenamefont {Campanelli}, \citenamefont {Khanna}, \citenamefont {Laguna}, \citenamefont {Pullin},\ and\ \citenamefont {Ryan}}]{Campanelli:2000nc}%
  \BibitemOpen
  \bibfield  {author} {\bibinfo {author} {\bibfnamefont {M.}~\bibnamefont {Campanelli}}, \bibinfo {author} {\bibfnamefont {G.}~\bibnamefont {Khanna}}, \bibinfo {author} {\bibfnamefont {P.}~\bibnamefont {Laguna}}, \bibinfo {author} {\bibfnamefont {J.}~\bibnamefont {Pullin}}, \ and\ \bibinfo {author} {\bibfnamefont {M.~P.}\ \bibnamefont {Ryan}},\ }\href {\doibase 10.1088/0264-9381/18/8/310} {\bibfield  {journal} {\bibinfo  {journal} {Class. Quant. Grav.}\ }\textbf {\bibinfo {volume} {18}},\ \bibinfo {pages} {1543} (\bibinfo {year} {2001})},\ \Eprint {http://arxiv.org/abs/gr-qc/0010034} {arXiv:gr-qc/0010034} \BibitemShut {NoStop}%
\bibitem [{\citenamefont {Krivan}\ \emph {et~al.}(1996)\citenamefont {Krivan}, \citenamefont {Laguna},\ and\ \citenamefont {Papadopoulos}}]{Krivan:1996da}%
  \BibitemOpen
  \bibfield  {author} {\bibinfo {author} {\bibfnamefont {W.}~\bibnamefont {Krivan}}, \bibinfo {author} {\bibfnamefont {P.}~\bibnamefont {Laguna}}, \ and\ \bibinfo {author} {\bibfnamefont {P.}~\bibnamefont {Papadopoulos}},\ }\href {\doibase 10.1103/PhysRevD.54.4728} {\bibfield  {journal} {\bibinfo  {journal} {Phys. Rev. D}\ }\textbf {\bibinfo {volume} {54}},\ \bibinfo {pages} {4728} (\bibinfo {year} {1996})},\ \Eprint {http://arxiv.org/abs/gr-qc/9606003} {arXiv:gr-qc/9606003} \BibitemShut {NoStop}%
\bibitem [{\citenamefont {Vishal}\ \emph {et~al.}(2023)\citenamefont {Vishal}, \citenamefont {Field}, \citenamefont {Rink}, \citenamefont {Gottlieb},\ and\ \citenamefont {Khanna}}]{Vishal:2023fye}%
  \BibitemOpen
  \bibfield  {author} {\bibinfo {author} {\bibfnamefont {M.}~\bibnamefont {Vishal}}, \bibinfo {author} {\bibfnamefont {S.~E.}\ \bibnamefont {Field}}, \bibinfo {author} {\bibfnamefont {K.}~\bibnamefont {Rink}}, \bibinfo {author} {\bibfnamefont {S.}~\bibnamefont {Gottlieb}}, \ and\ \bibinfo {author} {\bibfnamefont {G.}~\bibnamefont {Khanna}},\ }\href@noop {} {\  (\bibinfo {year} {2023})},\ \Eprint {http://arxiv.org/abs/2307.01349} {arXiv:2307.01349 [gr-qc]} \BibitemShut {NoStop}%
\bibitem [{\citenamefont {{Leaver}}(1986)}]{Leaver:1986JMP}%
  \BibitemOpen
  \bibfield  {author} {\bibinfo {author} {\bibfnamefont {E.~W.}\ \bibnamefont {{Leaver}}},\ }\href {\doibase 10.1063/1.527130} {\bibfield  {journal} {\bibinfo  {journal} {Journal of Mathematical Physics}\ }\textbf {\bibinfo {volume} {27}},\ \bibinfo {pages} {1238} (\bibinfo {year} {1986})}\BibitemShut {NoStop}%
\bibitem [{\citenamefont {Isoyama}\ \emph {et~al.}(2022)\citenamefont {Isoyama}, \citenamefont {Fujita}, \citenamefont {Chua}, \citenamefont {Nakano}, \citenamefont {Pound},\ and\ \citenamefont {Sago}}]{Isoyama:2021jjd}%
  \BibitemOpen
  \bibfield  {author} {\bibinfo {author} {\bibfnamefont {S.}~\bibnamefont {Isoyama}}, \bibinfo {author} {\bibfnamefont {R.}~\bibnamefont {Fujita}}, \bibinfo {author} {\bibfnamefont {A.~J.~K.}\ \bibnamefont {Chua}}, \bibinfo {author} {\bibfnamefont {H.}~\bibnamefont {Nakano}}, \bibinfo {author} {\bibfnamefont {A.}~\bibnamefont {Pound}}, \ and\ \bibinfo {author} {\bibfnamefont {N.}~\bibnamefont {Sago}},\ }\href {\doibase 10.1103/PhysRevLett.128.231101} {\bibfield  {journal} {\bibinfo  {journal} {Phys. Rev. Lett.}\ }\textbf {\bibinfo {volume} {128}},\ \bibinfo {pages} {231101} (\bibinfo {year} {2022})},\ \Eprint {http://arxiv.org/abs/2111.05288} {arXiv:2111.05288 [gr-qc]} \BibitemShut {NoStop}%
\bibitem [{\citenamefont {Casals}\ and\ \citenamefont {Zimmerman}(2019)}]{Casals:2018eev}%
  \BibitemOpen
  \bibfield  {author} {\bibinfo {author} {\bibfnamefont {M.}~\bibnamefont {Casals}}\ and\ \bibinfo {author} {\bibfnamefont {P.}~\bibnamefont {Zimmerman}},\ }\href {\doibase 10.1103/PhysRevD.100.124027} {\bibfield  {journal} {\bibinfo  {journal} {Phys. Rev. D}\ }\textbf {\bibinfo {volume} {100}},\ \bibinfo {pages} {124027} (\bibinfo {year} {2019})},\ \Eprint {http://arxiv.org/abs/1801.05830} {arXiv:1801.05830 [gr-qc]} \BibitemShut {NoStop}%
\bibitem [{\citenamefont {Sasaki}\ and\ \citenamefont {Tagoshi}(2003)}]{Sasaki:2003xr}%
  \BibitemOpen
  \bibfield  {author} {\bibinfo {author} {\bibfnamefont {M.}~\bibnamefont {Sasaki}}\ and\ \bibinfo {author} {\bibfnamefont {H.}~\bibnamefont {Tagoshi}},\ }\href {\doibase 10.12942/lrr-2003-6} {\bibfield  {journal} {\bibinfo  {journal} {Living Rev. Rel.}\ }\textbf {\bibinfo {volume} {6}},\ \bibinfo {pages} {6} (\bibinfo {year} {2003})},\ \Eprint {http://arxiv.org/abs/gr-qc/0306120} {arXiv:gr-qc/0306120} \BibitemShut {NoStop}%
\bibitem [{\citenamefont {Fujita}\ and\ \citenamefont {Tagoshi}(2004)}]{Fujita:2004rb}%
  \BibitemOpen
  \bibfield  {author} {\bibinfo {author} {\bibfnamefont {R.}~\bibnamefont {Fujita}}\ and\ \bibinfo {author} {\bibfnamefont {H.}~\bibnamefont {Tagoshi}},\ }\href {\doibase 10.1143/PTP.112.415} {\bibfield  {journal} {\bibinfo  {journal} {Prog. Theor. Phys.}\ }\textbf {\bibinfo {volume} {112}},\ \bibinfo {pages} {415} (\bibinfo {year} {2004})},\ \Eprint {http://arxiv.org/abs/gr-qc/0410018} {arXiv:gr-qc/0410018} \BibitemShut {NoStop}%
\bibitem [{\citenamefont {Ronveaux}\ and\ \citenamefont {Arscott}(1995)}]{ronveaux1995heun}%
  \BibitemOpen
  \bibfield  {author} {\bibinfo {author} {\bibfnamefont {A.}~\bibnamefont {Ronveaux}}\ and\ \bibinfo {author} {\bibfnamefont {F.}~\bibnamefont {Arscott}},\ }\href {https://books.google.co.uk/books?id=5p65FD8caCgC} {\emph {\bibinfo {title} {Heun's Differential Equations}}},\ Oxford science publications\ (\bibinfo  {publisher} {Oxford University Press},\ \bibinfo {year} {1995})\BibitemShut {NoStop}%
\bibitem [{\citenamefont {Chen}\ and\ \citenamefont {Jing}(2023)}]{Chen:2023ese}%
  \BibitemOpen
  \bibfield  {author} {\bibinfo {author} {\bibfnamefont {C.}~\bibnamefont {Chen}}\ and\ \bibinfo {author} {\bibfnamefont {J.}~\bibnamefont {Jing}},\ }\href@noop {} {\  (\bibinfo {year} {2023})},\ \Eprint {http://arxiv.org/abs/2307.14616} {arXiv:2307.14616 [gr-qc]} \BibitemShut {NoStop}%
\bibitem [{\citenamefont {Fiziev}(2009{\natexlab{a}})}]{Fiziev:2009ud}%
  \BibitemOpen
  \bibfield  {author} {\bibinfo {author} {\bibfnamefont {P.~P.}\ \bibnamefont {Fiziev}},\ }\href {\doibase 10.1103/PhysRevD.80.124001} {\bibfield  {journal} {\bibinfo  {journal} {Phys. Rev. D}\ }\textbf {\bibinfo {volume} {80}},\ \bibinfo {pages} {124001} (\bibinfo {year} {2009}{\natexlab{a}})},\ \Eprint {http://arxiv.org/abs/0906.5108} {arXiv:0906.5108 [gr-qc]} \BibitemShut {NoStop}%
\bibitem [{\citenamefont {Zarrinkamar}\ and\ \citenamefont {Horta{\c c}su}(2018)}]{Hortacsu:2011rr}%
  \BibitemOpen
  \bibfield  {author} {\bibinfo {author} {\bibfnamefont {S.}~\bibnamefont {Zarrinkamar}}\ and\ \bibinfo {author} {\bibfnamefont {M.}~\bibnamefont {Horta{\c c}su}},\ }\href {\doibase 10.1155/2018/8621573} {\bibfield  {journal} {\bibinfo  {journal} {Advances in High Energy Physics}\ }\textbf {\bibinfo {volume} {2018}},\ \bibinfo {pages} {8621573} (\bibinfo {year} {2018})}\BibitemShut {NoStop}%
\bibitem [{\citenamefont {Magnus}\ \emph {et~al.}(2021)\citenamefont {Magnus}, \citenamefont {Ndayiragije},\ and\ \citenamefont {Ronveaux}}]{MAGNUS2021105522}%
  \BibitemOpen
  \bibfield  {author} {\bibinfo {author} {\bibfnamefont {A.~P.}\ \bibnamefont {Magnus}}, \bibinfo {author} {\bibfnamefont {F.}~\bibnamefont {Ndayiragije}}, \ and\ \bibinfo {author} {\bibfnamefont {A.}~\bibnamefont {Ronveaux}},\ }\href {\doibase https://doi.org/10.1016/j.jat.2020.105522} {\bibfield  {journal} {\bibinfo  {journal} {Journal of Approximation Theory}\ }\textbf {\bibinfo {volume} {263}},\ \bibinfo {pages} {105522} (\bibinfo {year} {2021})}\BibitemShut {NoStop}%
\bibitem [{\citenamefont {Bonelli}\ \emph {et~al.}(2022)\citenamefont {Bonelli}, \citenamefont {Iossa}, \citenamefont {Lichtig},\ and\ \citenamefont {Tanzini}}]{Bonelli:2021uvf}%
  \BibitemOpen
  \bibfield  {author} {\bibinfo {author} {\bibfnamefont {G.}~\bibnamefont {Bonelli}}, \bibinfo {author} {\bibfnamefont {C.}~\bibnamefont {Iossa}}, \bibinfo {author} {\bibfnamefont {D.~P.}\ \bibnamefont {Lichtig}}, \ and\ \bibinfo {author} {\bibfnamefont {A.}~\bibnamefont {Tanzini}},\ }\href {\doibase 10.1103/PhysRevD.105.044047} {\bibfield  {journal} {\bibinfo  {journal} {Phys. Rev. D}\ }\textbf {\bibinfo {volume} {105}},\ \bibinfo {pages} {044047} (\bibinfo {year} {2022})},\ \Eprint {http://arxiv.org/abs/2105.04483} {arXiv:2105.04483 [hep-th]} \BibitemShut {NoStop}%
\bibitem [{\citenamefont {Arfken}\ \emph {et~al.}(2013)\citenamefont {Arfken}, \citenamefont {Weber},\ and\ \citenamefont {Harris}}]{ARFKEN2013401}%
  \BibitemOpen
  \bibfield  {author} {\bibinfo {author} {\bibfnamefont {G.~B.}\ \bibnamefont {Arfken}}, \bibinfo {author} {\bibfnamefont {H.~J.}\ \bibnamefont {Weber}}, \ and\ \bibinfo {author} {\bibfnamefont {F.~E.}\ \bibnamefont {Harris}},\ }in\ \href {\doibase https://doi.org/10.1016/B978-0-12-384654-9.00009-8} {\emph {\bibinfo {booktitle} {Mathematical Methods for Physicists (Seventh Edition)}}},\ \bibinfo {editor} {edited by\ \bibinfo {editor} {\bibfnamefont {G.~B.}\ \bibnamefont {Arfken}}, \bibinfo {editor} {\bibfnamefont {H.~J.}\ \bibnamefont {Weber}}, \ and\ \bibinfo {editor} {\bibfnamefont {F.~E.}\ \bibnamefont {Harris}}}\ (\bibinfo  {publisher} {Academic Press},\ \bibinfo {address} {Boston},\ \bibinfo {year} {2013})\ \bibinfo {edition} {seventh edition}\ ed.,\ pp.\ \bibinfo {pages} {401--445}\BibitemShut {NoStop}%
\bibitem [{\citenamefont {Kristensson}(2010)}]{Kristensson:2010}%
  \BibitemOpen
  \bibfield  {author} {\bibinfo {author} {\bibfnamefont {G.}~\bibnamefont {Kristensson}},\ }\href@noop {} {\emph {\bibinfo {title} {Second order differential equations - special functions and their classification}}}\ (\bibinfo  {publisher} {Springer},\ \bibinfo {address} {Germany},\ \bibinfo {year} {2010})\BibitemShut {NoStop}%
\bibitem [{\citenamefont {Courant}\ and\ \citenamefont {Hilbert}(1989)}]{Courant1954}%
  \BibitemOpen
  \bibfield  {author} {\bibinfo {author} {\bibfnamefont {R.}~\bibnamefont {Courant}}\ and\ \bibinfo {author} {\bibfnamefont {D.}~\bibnamefont {Hilbert}},\ }\href {\doibase 10.1002/9783527617210} {\  (\bibinfo {year} {1989}),\ 10.1002/9783527617210}\BibitemShut {NoStop}%
\bibitem [{\citenamefont {Adkins}\ and\ \citenamefont {Weintraub}(2012)}]{adkins2012algebra}%
  \BibitemOpen
  \bibfield  {author} {\bibinfo {author} {\bibfnamefont {W.}~\bibnamefont {Adkins}}\ and\ \bibinfo {author} {\bibfnamefont {S.}~\bibnamefont {Weintraub}},\ }\href {https://books.google.co.uk/books?id=RFzdBwAAQBAJ} {\emph {\bibinfo {title} {Algebra: An Approach via Module Theory}}},\ Graduate Texts in Mathematics\ (\bibinfo  {publisher} {Springer New York},\ \bibinfo {year} {2012})\BibitemShut {NoStop}%
\bibitem [{\citenamefont {Abramowitz}\ and\ \citenamefont {Stegun}(1964)}]{abramowitz+stegun}%
  \BibitemOpen
  \bibfield  {author} {\bibinfo {author} {\bibfnamefont {M.}~\bibnamefont {Abramowitz}}\ and\ \bibinfo {author} {\bibfnamefont {I.~A.}\ \bibnamefont {Stegun}},\ }\href@noop {} {\emph {\bibinfo {title} {Handbook of Mathematical Functions with Formulas, Graphs, and Mathematical Tables}}},\ \bibinfo {edition} {ninth dover printing, tenth gpo printing}\ ed.\ (\bibinfo  {publisher} {Dover},\ \bibinfo {address} {New York},\ \bibinfo {year} {1964})\BibitemShut {NoStop}%
\bibitem [{\citenamefont {Motl}\ and\ \citenamefont {Neitzke}(2003)}]{Motl:2003cd}%
  \BibitemOpen
  \bibfield  {author} {\bibinfo {author} {\bibfnamefont {L.}~\bibnamefont {Motl}}\ and\ \bibinfo {author} {\bibfnamefont {A.}~\bibnamefont {Neitzke}},\ }\href {\doibase 10.4310/ATMP.2003.v7.n2.a4} {\bibfield  {journal} {\bibinfo  {journal} {Adv. Theor. Math. Phys.}\ }\textbf {\bibinfo {volume} {7}},\ \bibinfo {pages} {307} (\bibinfo {year} {2003})},\ \Eprint {http://arxiv.org/abs/hep-th/0301173} {arXiv:hep-th/0301173} \BibitemShut {NoStop}%
\bibitem [{\citenamefont {Andersson}(1997)}]{Andersson:1996cm}%
  \BibitemOpen
  \bibfield  {author} {\bibinfo {author} {\bibfnamefont {N.}~\bibnamefont {Andersson}},\ }\href {\doibase 10.1103/PhysRevD.55.468} {\bibfield  {journal} {\bibinfo  {journal} {Phys. Rev. D}\ }\textbf {\bibinfo {volume} {55}},\ \bibinfo {pages} {468} (\bibinfo {year} {1997})},\ \Eprint {http://arxiv.org/abs/gr-qc/9607064} {arXiv:gr-qc/9607064} \BibitemShut {NoStop}%
\bibitem [{\citenamefont {Jaramillo}\ \emph {et~al.}(2021)\citenamefont {Jaramillo}, \citenamefont {Panosso~Macedo},\ and\ \citenamefont {Al~Sheikh}}]{Jaramillo:2020tuu}%
  \BibitemOpen
  \bibfield  {author} {\bibinfo {author} {\bibfnamefont {J.~L.}\ \bibnamefont {Jaramillo}}, \bibinfo {author} {\bibfnamefont {R.}~\bibnamefont {Panosso~Macedo}}, \ and\ \bibinfo {author} {\bibfnamefont {L.}~\bibnamefont {Al~Sheikh}},\ }\href {\doibase 10.1103/PhysRevX.11.031003} {\bibfield  {journal} {\bibinfo  {journal} {Phys. Rev. X}\ }\textbf {\bibinfo {volume} {11}},\ \bibinfo {pages} {031003} (\bibinfo {year} {2021})},\ \Eprint {http://arxiv.org/abs/2004.06434} {arXiv:2004.06434 [gr-qc]} \BibitemShut {NoStop}%
\bibitem [{\citenamefont {Nollert}\ and\ \citenamefont {Price}(1999{\natexlab{a}})}]{Price1999}%
  \BibitemOpen
  \bibfield  {author} {\bibinfo {author} {\bibfnamefont {H.-P.}\ \bibnamefont {Nollert}}\ and\ \bibinfo {author} {\bibfnamefont {R.~H.}\ \bibnamefont {Price}},\ }\href {http://arxiv.org/abs/gr-qc/9810074} {\bibfield  {journal} {\bibinfo  {journal} {J.Math.Phys.}\ }\textbf {\bibinfo {volume} {40}},\ \bibinfo {pages} {980} (\bibinfo {year} {1999}{\natexlab{a}})}\BibitemShut {NoStop}%
\bibitem [{\citenamefont {Teukolsky}\ and\ \citenamefont {Press}(1974)}]{TeuPre74_3}%
  \BibitemOpen
  \bibfield  {author} {\bibinfo {author} {\bibfnamefont {S.~A.}\ \bibnamefont {Teukolsky}}\ and\ \bibinfo {author} {\bibfnamefont {W.~H.}\ \bibnamefont {Press}},\ }\href {http://articles.adsabs.harvard.edu/full/1974ApJ...193..443T} {\bibfield  {journal} {\bibinfo  {journal} {Astrophysical Journal}\ }\textbf {\bibinfo {volume} {193}},\ \bibinfo {pages} {443} (\bibinfo {year} {1974})}\BibitemShut {NoStop}%
\bibitem [{\citenamefont {{Wald}}(1979)}]{Wald:1979a}%
  \BibitemOpen
  \bibfield  {author} {\bibinfo {author} {\bibfnamefont {R.~M.}\ \bibnamefont {{Wald}}},\ }\href {\doibase 10.1007/BF00759272} {\bibfield  {journal} {\bibinfo  {journal} {General Relativity and Gravitation}\ }\textbf {\bibinfo {volume} {11}},\ \bibinfo {pages} {321} (\bibinfo {year} {1979})}\BibitemShut {NoStop}%
\bibitem [{\citenamefont {Dias}\ \emph {et~al.}(2022)\citenamefont {Dias}, \citenamefont {Godazgar},\ and\ \citenamefont {Santos}}]{Dias:2022oqm}%
  \BibitemOpen
  \bibfield  {author} {\bibinfo {author} {\bibfnamefont {O.~J.~C.}\ \bibnamefont {Dias}}, \bibinfo {author} {\bibfnamefont {M.}~\bibnamefont {Godazgar}}, \ and\ \bibinfo {author} {\bibfnamefont {J.~E.}\ \bibnamefont {Santos}},\ }\href {\doibase 10.1007/JHEP07(2022)076} {\bibfield  {journal} {\bibinfo  {journal} {JHEP}\ }\textbf {\bibinfo {volume} {07}},\ \bibinfo {pages} {076} (\bibinfo {year} {2022})},\ \Eprint {http://arxiv.org/abs/2205.13072} {arXiv:2205.13072 [gr-qc]} \BibitemShut {NoStop}%
\bibitem [{\citenamefont {Chung}\ \emph {et~al.}(2023{\natexlab{a}})\citenamefont {Chung}, \citenamefont {Wagle},\ and\ \citenamefont {Yunes}}]{Chung:2023zdq}%
  \BibitemOpen
  \bibfield  {author} {\bibinfo {author} {\bibfnamefont {A.~K.-W.}\ \bibnamefont {Chung}}, \bibinfo {author} {\bibfnamefont {P.}~\bibnamefont {Wagle}}, \ and\ \bibinfo {author} {\bibfnamefont {N.}~\bibnamefont {Yunes}},\ }\href {\doibase 10.1103/PhysRevD.107.124032} {\bibfield  {journal} {\bibinfo  {journal} {Phys. Rev. D}\ }\textbf {\bibinfo {volume} {107}},\ \bibinfo {pages} {124032} (\bibinfo {year} {2023}{\natexlab{a}})},\ \Eprint {http://arxiv.org/abs/2302.11624} {arXiv:2302.11624 [gr-qc]} \BibitemShut {NoStop}%
\bibitem [{\citenamefont {Chung}\ \emph {et~al.}(2023{\natexlab{b}})\citenamefont {Chung}, \citenamefont {Wagle},\ and\ \citenamefont {Yunes}}]{Chung:2023wkd}%
  \BibitemOpen
  \bibfield  {author} {\bibinfo {author} {\bibfnamefont {A.~K.-W.}\ \bibnamefont {Chung}}, \bibinfo {author} {\bibfnamefont {P.}~\bibnamefont {Wagle}}, \ and\ \bibinfo {author} {\bibfnamefont {N.}~\bibnamefont {Yunes}},\ }\href@noop {} {\  (\bibinfo {year} {2023}{\natexlab{b}})},\ \Eprint {http://arxiv.org/abs/2312.08435} {arXiv:2312.08435 [gr-qc]} \BibitemShut {NoStop}%
\bibitem [{\citenamefont {Ghojogh}\ \emph {et~al.}(2023)\citenamefont {Ghojogh}, \citenamefont {Karray},\ and\ \citenamefont {Crowley}}]{ghojogh2023eigenvalue}%
  \BibitemOpen
  \bibfield  {author} {\bibinfo {author} {\bibfnamefont {B.}~\bibnamefont {Ghojogh}}, \bibinfo {author} {\bibfnamefont {F.}~\bibnamefont {Karray}}, \ and\ \bibinfo {author} {\bibfnamefont {M.}~\bibnamefont {Crowley}},\ }\href@noop {} {\enquote {\bibinfo {title} {Eigenvalue and generalized eigenvalue problems: Tutorial},}\ } (\bibinfo {year} {2023}),\ \Eprint {http://arxiv.org/abs/1903.11240} {arXiv:1903.11240 [stat.ML]} \BibitemShut {NoStop}%
\bibitem [{\citenamefont {Cook}\ and\ \citenamefont {Zalutskiy}(2014)}]{Cook:2014cta}%
  \BibitemOpen
  \bibfield  {author} {\bibinfo {author} {\bibfnamefont {G.~B.}\ \bibnamefont {Cook}}\ and\ \bibinfo {author} {\bibfnamefont {M.}~\bibnamefont {Zalutskiy}},\ }\href {\doibase 10.1103/PhysRevD.90.124021} {\bibfield  {journal} {\bibinfo  {journal} {Phys. Rev. D}\ }\textbf {\bibinfo {volume} {90}},\ \bibinfo {pages} {124021} (\bibinfo {year} {2014})},\ \Eprint {http://arxiv.org/abs/1410.7698} {arXiv:1410.7698 [gr-qc]} \BibitemShut {NoStop}%
\bibitem [{\citenamefont {Nollert}(1999)}]{Nollert:1999ji}%
  \BibitemOpen
  \bibfield  {author} {\bibinfo {author} {\bibfnamefont {H.-P.}\ \bibnamefont {Nollert}},\ }\href {\doibase 10.1088/0264-9381/16/12/201} {\bibfield  {journal} {\bibinfo  {journal} {Class. Quant. Grav.}\ }\textbf {\bibinfo {volume} {16}},\ \bibinfo {pages} {R159} (\bibinfo {year} {1999})}\BibitemShut {NoStop}%
\bibitem [{\citenamefont {Nollert}\ and\ \citenamefont {Price}(1999{\natexlab{b}})}]{Nollert:1998ys}%
  \BibitemOpen
  \bibfield  {author} {\bibinfo {author} {\bibfnamefont {H.-P.}\ \bibnamefont {Nollert}}\ and\ \bibinfo {author} {\bibfnamefont {R.~H.}\ \bibnamefont {Price}},\ }\href {\doibase 10.1063/1.532698} {\bibfield  {journal} {\bibinfo  {journal} {J. Math. Phys.}\ }\textbf {\bibinfo {volume} {40}},\ \bibinfo {pages} {980} (\bibinfo {year} {1999}{\natexlab{b}})},\ \Eprint {http://arxiv.org/abs/gr-qc/9810074} {arXiv:gr-qc/9810074} \BibitemShut {NoStop}%
\bibitem [{\citenamefont {London}\ \emph {et~al.}(2020)\citenamefont {London}, \citenamefont {Fauchon},\ and\ \citenamefont {Hamilton}}]{positive:2020}%
  \BibitemOpen
  \bibfield  {author} {\bibinfo {author} {\bibfnamefont {L.}~\bibnamefont {London}}, \bibinfo {author} {\bibfnamefont {E.}~\bibnamefont {Fauchon}}, \ and\ \bibinfo {author} {\bibfnamefont {E.}~\bibnamefont {Hamilton}},\ }\href {\doibase 10.5281/zenodo.3901856} {\enquote {\bibinfo {title} {llondon6/positive: map},}\ } (\bibinfo {year} {2020})\BibitemShut {NoStop}%
\bibitem [{\citenamefont {Stein}(2019)}]{Stein:2019mop}%
  \BibitemOpen
  \bibfield  {author} {\bibinfo {author} {\bibfnamefont {L.~C.}\ \bibnamefont {Stein}},\ }\href {\doibase 10.21105/joss.01683} {\bibfield  {journal} {\bibinfo  {journal} {J. Open Source Softw.}\ }\textbf {\bibinfo {volume} {4}},\ \bibinfo {pages} {1683} (\bibinfo {year} {2019})},\ \Eprint {http://arxiv.org/abs/1908.10377} {arXiv:1908.10377 [gr-qc]} \BibitemShut {NoStop}%
\bibitem [{\citenamefont {mpmath~development team}(2023)}]{mpmath}%
  \BibitemOpen
  \bibfield  {author} {\bibinfo {author} {\bibfnamefont {T.}~\bibnamefont {mpmath~development team}},\ }\href@noop {} {\emph {\bibinfo {title} {mpmath: a {P}ython library for arbitrary-precision floating-point arithmetic (version 1.3.0)}}} (\bibinfo {year} {2023}),\ \bibinfo {note} {{\tt http://mpmath.org/}}\BibitemShut {NoStop}%
\bibitem [{\citenamefont {Cook}(2020)}]{Cook:2020otn}%
  \BibitemOpen
  \bibfield  {author} {\bibinfo {author} {\bibfnamefont {G.~B.}\ \bibnamefont {Cook}},\ }\href@noop {} {\  (\bibinfo {year} {2020})},\ \Eprint {http://arxiv.org/abs/2004.08347} {arXiv:2004.08347 [gr-qc]} \BibitemShut {NoStop}%
\bibitem [{\citenamefont {Pinchover}\ and\ \citenamefont {Rubinstein}(2005)}]{pinchover_rubinstein_2005}%
  \BibitemOpen
  \bibfield  {author} {\bibinfo {author} {\bibfnamefont {Y.}~\bibnamefont {Pinchover}}\ and\ \bibinfo {author} {\bibfnamefont {J.}~\bibnamefont {Rubinstein}},\ }\enquote {\bibinfo {title} {Equations in high dimensions},}\ in\ \href {\doibase 10.1017/CBO9780511801228.010} {\emph {\bibinfo {booktitle} {An Introduction to Partial Differential Equations}}}\ (\bibinfo  {publisher} {Cambridge University Press},\ \bibinfo {year} {2005})\ p.\ \bibinfo {pages} {226–281}\BibitemShut {NoStop}%
\bibitem [{\citenamefont {Thorne}(1980)}]{Thorne:1980}%
  \BibitemOpen
  \bibfield  {author} {\bibinfo {author} {\bibfnamefont {K.~S.}\ \bibnamefont {Thorne}},\ }\href {\doibase 10.1103/RevModPhys.52.299} {\bibfield  {journal} {\bibinfo  {journal} {Rev. Mod. Phys.}\ }\textbf {\bibinfo {volume} {52}},\ \bibinfo {pages} {299} (\bibinfo {year} {1980})}\BibitemShut {NoStop}%
\bibitem [{\citenamefont {Boyer}\ and\ \citenamefont {Lindquist}(2004)}]{BoyerLindquist:1967}%
  \BibitemOpen
  \bibfield  {author} {\bibinfo {author} {\bibfnamefont {R.~H.}\ \bibnamefont {Boyer}}\ and\ \bibinfo {author} {\bibfnamefont {R.~W.}\ \bibnamefont {Lindquist}},\ }\href {\doibase 10.1063/1.1705193} {\bibfield  {journal} {\bibinfo  {journal} {Journal of Mathematical Physics}\ }\textbf {\bibinfo {volume} {8}},\ \bibinfo {pages} {265} (\bibinfo {year} {2004})},\ \Eprint {http://arxiv.org/abs/https://pubs.aip.org/aip/jmp/article-pdf/8/2/265/11041897/265\_1\_online.pdf} {https://pubs.aip.org/aip/jmp/article-pdf/8/2/265/11041897/265\_1\_online.pdf} \BibitemShut {NoStop}%
\bibitem [{\citenamefont {Hughes}(2000{\natexlab{b}})}]{Hughes:2000pf}%
  \BibitemOpen
  \bibfield  {author} {\bibinfo {author} {\bibfnamefont {S.~A.}\ \bibnamefont {Hughes}},\ }\href {\doibase 10.1103/PhysRevD.62.044029} {\bibfield  {journal} {\bibinfo  {journal} {Phys. Rev. D}\ }\textbf {\bibinfo {volume} {62}},\ \bibinfo {pages} {044029} (\bibinfo {year} {2000}{\natexlab{b}})},\ \bibinfo {note} {[Erratum: Phys.Rev.D 67, 089902 (2003)]},\ \Eprint {http://arxiv.org/abs/gr-qc/0002043} {arXiv:gr-qc/0002043} \BibitemShut {NoStop}%
\bibitem [{\citenamefont {Mino}\ \emph {et~al.}(1997)\citenamefont {Mino}, \citenamefont {Sasaki}, \citenamefont {Shibata}, \citenamefont {Tagoshi},\ and\ \citenamefont {Tanaka}}]{Mino:1997bx}%
  \BibitemOpen
  \bibfield  {author} {\bibinfo {author} {\bibfnamefont {Y.}~\bibnamefont {Mino}}, \bibinfo {author} {\bibfnamefont {M.}~\bibnamefont {Sasaki}}, \bibinfo {author} {\bibfnamefont {M.}~\bibnamefont {Shibata}}, \bibinfo {author} {\bibfnamefont {H.}~\bibnamefont {Tagoshi}}, \ and\ \bibinfo {author} {\bibfnamefont {T.}~\bibnamefont {Tanaka}},\ }\href {\doibase 10.1143/PTPS.128.1} {\bibfield  {journal} {\bibinfo  {journal} {Progress of Theoretical Physics Supplement}\ }\textbf {\bibinfo {volume} {128}},\ \bibinfo {pages} {1} (\bibinfo {year} {1997})},\ \Eprint {http://arxiv.org/abs/http://oup.prod.sis.lan/ptps/article-pdf/doi/10.1143/PTPS.128.1/5438984/128-1.pdf} {http://oup.prod.sis.lan/ptps/article-pdf/doi/10.1143/PTPS.128.1/5438984/128-1.pdf} \BibitemShut {NoStop}%
\bibitem [{\citenamefont {Helfer}\ \emph {et~al.}(2017)\citenamefont {Helfer}, \citenamefont {Marsh}, \citenamefont {Clough}, \citenamefont {Fairbairn}, \citenamefont {Lim},\ and\ \citenamefont {Becerril}}]{Helfer:2016ljl}%
  \BibitemOpen
  \bibfield  {author} {\bibinfo {author} {\bibfnamefont {T.}~\bibnamefont {Helfer}}, \bibinfo {author} {\bibfnamefont {D.~J.~E.}\ \bibnamefont {Marsh}}, \bibinfo {author} {\bibfnamefont {K.}~\bibnamefont {Clough}}, \bibinfo {author} {\bibfnamefont {M.}~\bibnamefont {Fairbairn}}, \bibinfo {author} {\bibfnamefont {E.~A.}\ \bibnamefont {Lim}}, \ and\ \bibinfo {author} {\bibfnamefont {R.}~\bibnamefont {Becerril}},\ }\href {\doibase 10.1088/1475-7516/2017/03/055} {\bibfield  {journal} {\bibinfo  {journal} {JCAP}\ }\textbf {\bibinfo {volume} {03}},\ \bibinfo {pages} {055} (\bibinfo {year} {2017})},\ \Eprint {http://arxiv.org/abs/1609.04724} {arXiv:1609.04724 [astro-ph.CO]} \BibitemShut {NoStop}%
\bibitem [{\citenamefont {Baker}\ \emph {et~al.}(2006)\citenamefont {Baker}, \citenamefont {Centrella}, \citenamefont {Choi}, \citenamefont {Koppitz},\ and\ \citenamefont {van Meter}}]{Baker:2005vv}%
  \BibitemOpen
  \bibfield  {author} {\bibinfo {author} {\bibfnamefont {J.~G.}\ \bibnamefont {Baker}}, \bibinfo {author} {\bibfnamefont {J.}~\bibnamefont {Centrella}}, \bibinfo {author} {\bibfnamefont {D.-I.}\ \bibnamefont {Choi}}, \bibinfo {author} {\bibfnamefont {M.}~\bibnamefont {Koppitz}}, \ and\ \bibinfo {author} {\bibfnamefont {J.}~\bibnamefont {van Meter}},\ }\href {\doibase 10.1103/PhysRevLett.96.111102} {\bibfield  {journal} {\bibinfo  {journal} {Phys. Rev. Lett.}\ }\textbf {\bibinfo {volume} {96}},\ \bibinfo {pages} {111102} (\bibinfo {year} {2006})},\ \Eprint {http://arxiv.org/abs/gr-qc/0511103} {arXiv:gr-qc/0511103} \BibitemShut {NoStop}%
\bibitem [{\citenamefont {Berti}\ \emph {et~al.}(2007)\citenamefont {Berti}, \citenamefont {Cardoso}, \citenamefont {Cardoso},\ and\ \citenamefont {Cavaglia}}]{Berti:2007zu}%
  \BibitemOpen
  \bibfield  {author} {\bibinfo {author} {\bibfnamefont {E.}~\bibnamefont {Berti}}, \bibinfo {author} {\bibfnamefont {J.}~\bibnamefont {Cardoso}}, \bibinfo {author} {\bibfnamefont {V.}~\bibnamefont {Cardoso}}, \ and\ \bibinfo {author} {\bibfnamefont {M.}~\bibnamefont {Cavaglia}},\ }\href {\doibase 10.1103/PhysRevD.76.104044} {\bibfield  {journal} {\bibinfo  {journal} {Phys.Rev.}\ }\textbf {\bibinfo {volume} {D76}},\ \bibinfo {pages} {104044} (\bibinfo {year} {2007})},\ \Eprint {http://arxiv.org/abs/0707.1202} {arXiv:0707.1202 [gr-qc]} \BibitemShut {NoStop}%
%%CITATION = ARXIV:0707.1202;%%
\bibitem [{\citenamefont {Berti}\ and\ \citenamefont {Cardoso}(2006)}]{Berti:2006:ExFacs}%
  \BibitemOpen
  \bibfield  {author} {\bibinfo {author} {\bibfnamefont {E.}~\bibnamefont {Berti}}\ and\ \bibinfo {author} {\bibfnamefont {V.}~\bibnamefont {Cardoso}},\ }\href {\doibase 10.1103/physrevd.74.104020} {\bibfield  {journal} {\bibinfo  {journal} {Physical Review D}\ }\textbf {\bibinfo {volume} {74}} (\bibinfo {year} {2006}),\ 10.1103/physrevd.74.104020}\BibitemShut {NoStop}%
\bibitem [{\citenamefont {Cho}\ \emph {et~al.}(2009)\citenamefont {Cho}, \citenamefont {Cornell}, \citenamefont {Doukas},\ and\ \citenamefont {Naylor}}]{Cho:2009wf}%
  \BibitemOpen
  \bibfield  {author} {\bibinfo {author} {\bibfnamefont {H.~T.}\ \bibnamefont {Cho}}, \bibinfo {author} {\bibfnamefont {A.~S.}\ \bibnamefont {Cornell}}, \bibinfo {author} {\bibfnamefont {J.}~\bibnamefont {Doukas}}, \ and\ \bibinfo {author} {\bibfnamefont {W.}~\bibnamefont {Naylor}},\ }\href {\doibase 10.1103/PhysRevD.80.064022} {\bibfield  {journal} {\bibinfo  {journal} {Phys. Rev. D}\ }\textbf {\bibinfo {volume} {80}},\ \bibinfo {pages} {064022} (\bibinfo {year} {2009})},\ \Eprint {http://arxiv.org/abs/0904.1867} {arXiv:0904.1867 [gr-qc]} \BibitemShut {NoStop}%
\bibitem [{\citenamefont {Regge}\ and\ \citenamefont {Wheeler}(1957)}]{Regge:1957td}%
  \BibitemOpen
  \bibfield  {author} {\bibinfo {author} {\bibfnamefont {T.}~\bibnamefont {Regge}}\ and\ \bibinfo {author} {\bibfnamefont {J.~A.}\ \bibnamefont {Wheeler}},\ }\href {\doibase 10.1103/PhysRev.108.1063} {\bibfield  {journal} {\bibinfo  {journal} {Phys. Rev.}\ }\textbf {\bibinfo {volume} {108}},\ \bibinfo {pages} {1063} (\bibinfo {year} {1957})}\BibitemShut {NoStop}%
%%CITATION = PHRVA,108,1063;%%
\bibitem [{\citenamefont {Axler}(2015)}]{Axler:2015}%
  \BibitemOpen
  \bibfield  {author} {\bibinfo {author} {\bibfnamefont {S.}~\bibnamefont {Axler}},\ }\href {https://doi.org/10.1007/978-3-319-11080-6} {\emph {\bibinfo {title} {Linear algebra done right (eBook)}}},\ \bibinfo {edition} {3rd}\ ed.\ (\bibinfo  {publisher} {Springer},\ \bibinfo {address} {Cham},\ \bibinfo {year} {2015})\BibitemShut {NoStop}%
\bibitem [{\citenamefont {Lax}(2002)}]{lax2002functional}%
  \BibitemOpen
  \bibfield  {author} {\bibinfo {author} {\bibfnamefont {P.}~\bibnamefont {Lax}},\ }\href@noop {} {\emph {\bibinfo {title} {Functional analysis}}}\ (\bibinfo  {publisher} {Wiley},\ \bibinfo {address} {New York},\ \bibinfo {year} {2002})\BibitemShut {NoStop}%
\bibitem [{\citenamefont {Cho}(2020)}]{Cho:2020tzx}%
  \BibitemOpen
  \bibfield  {author} {\bibinfo {author} {\bibfnamefont {G.}~\bibnamefont {Cho}},\ }\href@noop {} {\  (\bibinfo {year} {2020})},\ \Eprint {http://arxiv.org/abs/2008.12526} {arXiv:2008.12526 [gr-qc]} \BibitemShut {NoStop}%
\bibitem [{\citenamefont {Mano}\ \emph {et~al.}(1996)\citenamefont {Mano}, \citenamefont {Suzuki},\ and\ \citenamefont {Takasugi}}]{Mano:1996vt}%
  \BibitemOpen
  \bibfield  {author} {\bibinfo {author} {\bibfnamefont {S.}~\bibnamefont {Mano}}, \bibinfo {author} {\bibfnamefont {H.}~\bibnamefont {Suzuki}}, \ and\ \bibinfo {author} {\bibfnamefont {E.}~\bibnamefont {Takasugi}},\ }\href {\doibase 10.1143/PTP.95.1079} {\bibfield  {journal} {\bibinfo  {journal} {Prog. Theor. Phys.}\ }\textbf {\bibinfo {volume} {95}},\ \bibinfo {pages} {1079} (\bibinfo {year} {1996})},\ \Eprint {http://arxiv.org/abs/gr-qc/9603020} {arXiv:gr-qc/9603020} \BibitemShut {NoStop}%
\bibitem [{\citenamefont {Ripley}(2022)}]{Ripley:2022ypi}%
  \BibitemOpen
  \bibfield  {author} {\bibinfo {author} {\bibfnamefont {J.~L.}\ \bibnamefont {Ripley}},\ }\href {\doibase 10.1088/1361-6382/ac776d} {\bibfield  {journal} {\bibinfo  {journal} {Class. Quant. Grav.}\ }\textbf {\bibinfo {volume} {39}},\ \bibinfo {pages} {145009} (\bibinfo {year} {2022})},\ \Eprint {http://arxiv.org/abs/2202.03837} {arXiv:2202.03837 [gr-qc]} \BibitemShut {NoStop}%
\bibitem [{\citenamefont {Lin}\ and\ \citenamefont {Qian}(2023)}]{Lin:2022ynv}%
  \BibitemOpen
  \bibfield  {author} {\bibinfo {author} {\bibfnamefont {K.}~\bibnamefont {Lin}}\ and\ \bibinfo {author} {\bibfnamefont {W.-L.}\ \bibnamefont {Qian}},\ }\href {\doibase 10.1088/1361-6382/acc50f} {\bibfield  {journal} {\bibinfo  {journal} {Class. Quant. Grav.}\ }\textbf {\bibinfo {volume} {40}},\ \bibinfo {pages} {085019} (\bibinfo {year} {2023})},\ \Eprint {http://arxiv.org/abs/2209.11612} {arXiv:2209.11612 [gr-qc]} \BibitemShut {NoStop}%
\bibitem [{\citenamefont {Zimmerman}\ \emph {et~al.}(2015)\citenamefont {Zimmerman}, \citenamefont {Yang}, \citenamefont {Mark}, \citenamefont {Chen},\ and\ \citenamefont {Lehner}}]{Zimmerman:2014aha}%
  \BibitemOpen
  \bibfield  {author} {\bibinfo {author} {\bibfnamefont {A.}~\bibnamefont {Zimmerman}}, \bibinfo {author} {\bibfnamefont {H.}~\bibnamefont {Yang}}, \bibinfo {author} {\bibfnamefont {Z.}~\bibnamefont {Mark}}, \bibinfo {author} {\bibfnamefont {Y.}~\bibnamefont {Chen}}, \ and\ \bibinfo {author} {\bibfnamefont {L.}~\bibnamefont {Lehner}},\ }\href {\doibase 10.1007/978-3-319-10488-1_19} {\bibfield  {journal} {\bibinfo  {journal} {Astrophys. Space Sci. Proc.}\ }\textbf {\bibinfo {volume} {40}},\ \bibinfo {pages} {217} (\bibinfo {year} {2015})},\ \Eprint {http://arxiv.org/abs/1406.4206} {arXiv:1406.4206 [gr-qc]} \BibitemShut {NoStop}%
\bibitem [{\citenamefont {Seidel}(1989)}]{Seidel:1988ue}%
  \BibitemOpen
  \bibfield  {author} {\bibinfo {author} {\bibfnamefont {E.}~\bibnamefont {Seidel}},\ }\href {\doibase 10.1088/0264-9381/6/7/012} {\bibfield  {journal} {\bibinfo  {journal} {Class. Quant. Grav.}\ }\textbf {\bibinfo {volume} {6}},\ \bibinfo {pages} {1057} (\bibinfo {year} {1989})}\BibitemShut {NoStop}%
\bibitem [{\citenamefont {Gohberg}\ \emph {et~al.}(2006)\citenamefont {Gohberg}, \citenamefont {Lancaster},\ and\ \citenamefont {Rodman}}]{gohberg2006indefinite}%
  \BibitemOpen
  \bibfield  {author} {\bibinfo {author} {\bibfnamefont {I.}~\bibnamefont {Gohberg}}, \bibinfo {author} {\bibfnamefont {P.}~\bibnamefont {Lancaster}}, \ and\ \bibinfo {author} {\bibfnamefont {L.}~\bibnamefont {Rodman}},\ }\href {https://books.google.co.uk/books?id=Q1-ZJjY8Q9UC} {\emph {\bibinfo {title} {Indefinite Linear Algebra and Applications}}}\ (\bibinfo  {publisher} {Birkh{\"a}user Basel},\ \bibinfo {year} {2006})\BibitemShut {NoStop}%
\bibitem [{\citenamefont {Chen}\ and\ \citenamefont {Dai}(2010)}]{Chen:2010}%
  \BibitemOpen
  \bibfield  {author} {\bibinfo {author} {\bibfnamefont {Y.}~\bibnamefont {Chen}}\ and\ \bibinfo {author} {\bibfnamefont {D.}~\bibnamefont {Dai}},\ }\href {\doibase 10.1016/j.jat.2010.07.005} {\bibfield  {journal} {\bibinfo  {journal} {Journal of Approximation Theory}\ }\textbf {\bibinfo {volume} {162}},\ \bibinfo {pages} {2149} (\bibinfo {year} {2010})}\BibitemShut {NoStop}%
\bibitem [{\citenamefont {Virtanen}\ \emph {et~al.}(2020)\citenamefont {Virtanen}, \citenamefont {Gommers}, \citenamefont {Oliphant}, \citenamefont {Haberland}, \citenamefont {Reddy}, \citenamefont {Cournapeau}, \citenamefont {Burovski}, \citenamefont {Peterson}, \citenamefont {Weckesser}, \citenamefont {Bright}, \citenamefont {{van der Walt}}, \citenamefont {Brett}, \citenamefont {Wilson}, \citenamefont {Millman}, \citenamefont {Mayorov}, \citenamefont {Nelson}, \citenamefont {Jones}, \citenamefont {Kern}, \citenamefont {Larson}, \citenamefont {Carey}, \citenamefont {Polat}, \citenamefont {Feng}, \citenamefont {Moore}, \citenamefont {{VanderPlas}}, \citenamefont {Laxalde}, \citenamefont {Perktold}, \citenamefont {Cimrman}, \citenamefont {Henriksen}, \citenamefont {Quintero}, \citenamefont {Harris}, \citenamefont {Archibald}, \citenamefont {Ribeiro}, \citenamefont {Pedregosa}, \citenamefont {{van Mulbregt}},\ and\ \citenamefont {{SciPy 1.0 Contributors}}}]{2020SciPy-NMeth}%
  \BibitemOpen
  \bibfield  {author} {\bibinfo {author} {\bibfnamefont {P.}~\bibnamefont {Virtanen}}, \bibinfo {author} {\bibfnamefont {R.}~\bibnamefont {Gommers}}, \bibinfo {author} {\bibfnamefont {T.~E.}\ \bibnamefont {Oliphant}}, \bibinfo {author} {\bibfnamefont {M.}~\bibnamefont {Haberland}}, \bibinfo {author} {\bibfnamefont {T.}~\bibnamefont {Reddy}}, \bibinfo {author} {\bibfnamefont {D.}~\bibnamefont {Cournapeau}}, \bibinfo {author} {\bibfnamefont {E.}~\bibnamefont {Burovski}}, \bibinfo {author} {\bibfnamefont {P.}~\bibnamefont {Peterson}}, \bibinfo {author} {\bibfnamefont {W.}~\bibnamefont {Weckesser}}, \bibinfo {author} {\bibfnamefont {J.}~\bibnamefont {Bright}}, \bibinfo {author} {\bibfnamefont {S.~J.}\ \bibnamefont {{van der Walt}}}, \bibinfo {author} {\bibfnamefont {M.}~\bibnamefont {Brett}}, \bibinfo {author} {\bibfnamefont {J.}~\bibnamefont {Wilson}}, \bibinfo {author} {\bibfnamefont {K.~J.}\ \bibnamefont {Millman}}, \bibinfo {author} {\bibfnamefont {N.}~\bibnamefont {Mayorov}}, \bibinfo {author} {\bibfnamefont {A.~R.~J.}\ \bibnamefont {Nelson}}, \bibinfo {author} {\bibfnamefont {E.}~\bibnamefont {Jones}}, \bibinfo {author} {\bibfnamefont {R.}~\bibnamefont {Kern}}, \bibinfo {author} {\bibfnamefont {E.}~\bibnamefont {Larson}}, \bibinfo {author} {\bibfnamefont {C.~J.}\ \bibnamefont {Carey}}, \bibinfo {author} {\bibfnamefont {{\.I}.}~\bibnamefont {Polat}}, \bibinfo {author} {\bibfnamefont {Y.}~\bibnamefont {Feng}}, \bibinfo {author} {\bibfnamefont {E.~W.}\ \bibnamefont {Moore}}, \bibinfo {author} {\bibfnamefont {J.}~\bibnamefont {{VanderPlas}}}, \bibinfo {author} {\bibfnamefont {D.}~\bibnamefont {Laxalde}}, \bibinfo {author} {\bibfnamefont {J.}~\bibnamefont {Perktold}}, \bibinfo {author} {\bibfnamefont {R.}~\bibnamefont {Cimrman}}, \bibinfo {author} {\bibfnamefont {I.}~\bibnamefont {Henriksen}}, \bibinfo {author} {\bibfnamefont {E.~A.}\ \bibnamefont {Quintero}}, \bibinfo {author} {\bibfnamefont {C.~R.}\ \bibnamefont {Harris}}, \bibinfo {author} {\bibfnamefont {A.~M.}\ \bibnamefont {Archibald}}, \bibinfo {author} {\bibfnamefont {A.~H.}\ \bibnamefont {Ribeiro}}, \bibinfo {author} {\bibfnamefont {F.}~\bibnamefont {Pedregosa}}, \bibinfo {author} {\bibfnamefont {P.}~\bibnamefont {{van Mulbregt}}}, \ and\ \bibinfo {author} {\bibnamefont {{SciPy 1.0 Contributors}}},\ }\href {\doibase 10.1038/s41592-019-0686-2} {\bibfield  {journal} {\bibinfo  {journal} {Nature Methods}\ }\textbf {\bibinfo {volume} {17}},\ \bibinfo {pages} {261} (\bibinfo {year} {2020})}\BibitemShut {NoStop}%
\bibitem [{\citenamefont {Kuijlaars}\ \emph {et~al.}(2003)\citenamefont {Kuijlaars}, \citenamefont {Martinez-Finkelshtein},\ and\ \citenamefont {Orive}}]{kuijlaars2003orthogonality}%
  \BibitemOpen
  \bibfield  {author} {\bibinfo {author} {\bibfnamefont {A.~B.~J.}\ \bibnamefont {Kuijlaars}}, \bibinfo {author} {\bibfnamefont {A.}~\bibnamefont {Martinez-Finkelshtein}}, \ and\ \bibinfo {author} {\bibfnamefont {R.}~\bibnamefont {Orive}},\ }\href@noop {} {\enquote {\bibinfo {title} {Orthogonality of jacobi polynomials with general parameters},}\ } (\bibinfo {year} {2003}),\ \Eprint {http://arxiv.org/abs/math/0301037} {arXiv:math/0301037 [math.CA]} \BibitemShut {NoStop}%
\bibitem [{\citenamefont {Ahlfors}()}]{Ahlfors1966}%
  \BibitemOpen
  \bibfield  {author} {\bibinfo {author} {\bibfnamefont {L.~V.}\ \bibnamefont {Ahlfors}},\ }\href@noop {} {\emph {\bibinfo {title} {Complex Analysis}}},\ \bibinfo {edition} {2nd}\ ed.\ (\bibinfo  {publisher} {McGraw-Hill Book Company})\BibitemShut {NoStop}%
\bibitem [{{\relax DLMF}({\natexlab{a}})}]{NIST:DLMF}%
  \BibitemOpen
  {\relax DLMF},\ \href {https://dlmf.nist.gov/} {\enquote {\bibinfo {title} {{\it NIST Digital Library of Mathematical Functions}},}\ }\bibinfo {howpublished} {\url{https://dlmf.nist.gov/}, Release 1.1.10 of 2023-06-15} ({\natexlab{a}}),\ \bibinfo {note} {f.~W.~J. Olver, A.~B. {Olde Daalhuis}, D.~W. Lozier, B.~I. Schneider, R.~F. Boisvert, C.~W. Clark, B.~R. Miller, B.~V. Saunders, H.~S. Cohl, and M.~A. McClain, eds.}\BibitemShut {Stop}%
\bibitem [{\citenamefont {Whittaker}\ and\ \citenamefont {Watson}(1996)}]{whittaker_watson_1996}%
  \BibitemOpen
  \bibfield  {author} {\bibinfo {author} {\bibfnamefont {E.~T.}\ \bibnamefont {Whittaker}}\ and\ \bibinfo {author} {\bibfnamefont {G.~N.}\ \bibnamefont {Watson}},\ }\href {\doibase 10.1017/CBO9780511608759} {\emph {\bibinfo {title} {A Course of Modern Analysis}}},\ \bibinfo {edition} {4th}\ ed.,\ Cambridge Mathematical Library\ (\bibinfo  {publisher} {Cambridge University Press},\ \bibinfo {year} {1996})\BibitemShut {NoStop}%
\bibitem [{\citenamefont {Andersson}(1994)}]{Andersson:1994tt}%
  \BibitemOpen
  \bibfield  {author} {\bibinfo {author} {\bibfnamefont {N.}~\bibnamefont {Andersson}},\ }\href {\doibase 10.1088/0264-9381/11/3/001} {\bibfield  {journal} {\bibinfo  {journal} {Class. Quant. Grav.}\ }\textbf {\bibinfo {volume} {11}},\ \bibinfo {pages} {L39} (\bibinfo {year} {1994})}\BibitemShut {NoStop}%
\bibitem [{\citenamefont {Dias}\ and\ \citenamefont {Reall}(2013)}]{Dias:2013hn}%
  \BibitemOpen
  \bibfield  {author} {\bibinfo {author} {\bibfnamefont {O.~J.~C.}\ \bibnamefont {Dias}}\ and\ \bibinfo {author} {\bibfnamefont {H.~S.}\ \bibnamefont {Reall}},\ }\href {\doibase 10.1088/0264-9381/30/9/095003} {\bibfield  {journal} {\bibinfo  {journal} {Class. Quant. Grav.}\ }\textbf {\bibinfo {volume} {30}},\ \bibinfo {pages} {095003} (\bibinfo {year} {2013})},\ \Eprint {http://arxiv.org/abs/1301.7068} {arXiv:1301.7068 [gr-qc]} \BibitemShut {NoStop}%
\bibitem [{\citenamefont {Chihara}(2011)}]{chihara2011introduction}%
  \BibitemOpen
  \bibfield  {author} {\bibinfo {author} {\bibfnamefont {T.}~\bibnamefont {Chihara}},\ }\href {https://books.google.co.uk/books?id=71CVAwAAQBAJ} {\emph {\bibinfo {title} {An Introduction to Orthogonal Polynomials}}},\ Dover Books on Mathematics\ (\bibinfo  {publisher} {Dover Publications},\ \bibinfo {year} {2011})\BibitemShut {NoStop}%
\bibitem [{\citenamefont {Fiziev}(2009{\natexlab{b}})}]{Fiziev:2009}%
  \BibitemOpen
  \bibfield  {author} {\bibinfo {author} {\bibfnamefont {P.~P.}\ \bibnamefont {Fiziev}},\ }\href {\doibase 10.1088/1751-8113/43/3/035203} {\bibfield  {journal} {\bibinfo  {journal} {Journal of Physics A: Mathematical and Theoretical}\ }\textbf {\bibinfo {volume} {43}},\ \bibinfo {pages} {035203} (\bibinfo {year} {2009}{\natexlab{b}})}\BibitemShut {NoStop}%
\bibitem [{{\relax DLMF}({\natexlab{b}})}]{NIST:DLMF:ConfHeunPoly}%
  \BibitemOpen
  {\relax DLMF},\ \href {https://dlmf.nist.gov/31.5/} {\enquote {\bibinfo {title} {{\it Solutions Analytic at Three Singularities: Heun Polynomials}},}\ }\bibinfo {howpublished} {\url{https://dlmf.nist.gov/}, Release 1.1.10 of 2023-06-15} ({\natexlab{b}}),\ \bibinfo {note} {f.~W.~J. Olver, A.~B. {Olde Daalhuis}, D.~W. Lozier, B.~I. Schneider, R.~F. Boisvert, C.~W. Clark, B.~R. Miller, B.~V. Saunders, H.~S. Cohl, and M.~A. McClain, eds.}\BibitemShut {Stop}%
\bibitem [{\citenamefont {Atkinson}\ and\ \citenamefont {Mingarelli}(2011)}]{Atkinson:2011mpe}%
  \BibitemOpen
  \bibfield  {author} {\bibinfo {author} {\bibfnamefont {F.}~\bibnamefont {Atkinson}}\ and\ \bibinfo {author} {\bibfnamefont {A.}~\bibnamefont {Mingarelli}},\ }\href {\doibase 10.1201/b10511} {\emph {\bibinfo {title} {Multiparameter Eigenvalue Problems: Sturm-Liouville Theory}}}\ (\bibinfo {year} {2011})\BibitemShut {NoStop}%
\bibitem [{\citenamefont {Horn}\ and\ \citenamefont {Johnson}(1990)}]{Horm:1999ma}%
  \BibitemOpen
  \bibfield  {author} {\bibinfo {author} {\bibfnamefont {R.~A.}\ \bibnamefont {Horn}}\ and\ \bibinfo {author} {\bibfnamefont {C.~R.}\ \bibnamefont {Johnson}},\ }\href {http://www.amazon.com/Matrix-Analysis-Roger-Horn/dp/0521386322%3FSubscriptionId%3D192BW6DQ43CK9FN0ZGG2%26tag%3Dws%26linkCode%3Dxm2%26camp%3D2025%26creative%3D165953%26creativeASIN%3D0521386322} {\emph {\bibinfo {title} {Matrix Analysis}}}\ (\bibinfo  {publisher} {Cambridge University Press},\ \bibinfo {year} {1990})\BibitemShut {NoStop}%
\bibitem [{\citenamefont {Harris}\ \emph {et~al.}(2020)\citenamefont {Harris}, \citenamefont {Millman}, \citenamefont {van~der Walt}, \citenamefont {Gommers}, \citenamefont {Virtanen}, \citenamefont {Cournapeau}, \citenamefont {Wieser}, \citenamefont {Taylor}, \citenamefont {Berg}, \citenamefont {Smith}, \citenamefont {Kern}, \citenamefont {Picus}, \citenamefont {Hoyer}, \citenamefont {van Kerkwijk}, \citenamefont {Brett}, \citenamefont {Haldane}, \citenamefont {Fernández~del Río}, \citenamefont {Wiebe}, \citenamefont {Peterson}, \citenamefont {Gérard-Marchant}, \citenamefont {Sheppard}, \citenamefont {Reddy}, \citenamefont {Weckesser}, \citenamefont {Abbasi}, \citenamefont {Gohlke},\ and\ \citenamefont {Oliphant}}]{2020NumPy-Array}%
  \BibitemOpen
  \bibfield  {author} {\bibinfo {author} {\bibfnamefont {C.~R.}\ \bibnamefont {Harris}}, \bibinfo {author} {\bibfnamefont {K.~J.}\ \bibnamefont {Millman}}, \bibinfo {author} {\bibfnamefont {S.~J.}\ \bibnamefont {van~der Walt}}, \bibinfo {author} {\bibfnamefont {R.}~\bibnamefont {Gommers}}, \bibinfo {author} {\bibfnamefont {P.}~\bibnamefont {Virtanen}}, \bibinfo {author} {\bibfnamefont {D.}~\bibnamefont {Cournapeau}}, \bibinfo {author} {\bibfnamefont {E.}~\bibnamefont {Wieser}}, \bibinfo {author} {\bibfnamefont {J.}~\bibnamefont {Taylor}}, \bibinfo {author} {\bibfnamefont {S.}~\bibnamefont {Berg}}, \bibinfo {author} {\bibfnamefont {N.~J.}\ \bibnamefont {Smith}}, \bibinfo {author} {\bibfnamefont {R.}~\bibnamefont {Kern}}, \bibinfo {author} {\bibfnamefont {M.}~\bibnamefont {Picus}}, \bibinfo {author} {\bibfnamefont {S.}~\bibnamefont {Hoyer}}, \bibinfo {author} {\bibfnamefont {M.~H.}\ \bibnamefont {van Kerkwijk}}, \bibinfo {author} {\bibfnamefont {M.}~\bibnamefont {Brett}}, \bibinfo {author} {\bibfnamefont {A.}~\bibnamefont {Haldane}}, \bibinfo {author} {\bibfnamefont {J.}~\bibnamefont {Fernández~del Río}}, \bibinfo {author} {\bibfnamefont {M.}~\bibnamefont {Wiebe}}, \bibinfo {author} {\bibfnamefont {P.}~\bibnamefont {Peterson}}, \bibinfo {author} {\bibfnamefont {P.}~\bibnamefont {Gérard-Marchant}}, \bibinfo {author} {\bibfnamefont {K.}~\bibnamefont {Sheppard}}, \bibinfo {author} {\bibfnamefont {T.}~\bibnamefont {Reddy}}, \bibinfo {author} {\bibfnamefont {W.}~\bibnamefont {Weckesser}}, \bibinfo {author} {\bibfnamefont {H.}~\bibnamefont {Abbasi}}, \bibinfo {author} {\bibfnamefont {C.}~\bibnamefont {Gohlke}}, \ and\ \bibinfo {author} {\bibfnamefont {T.~E.}\ \bibnamefont {Oliphant}},\ }\href {\doibase 10.1038/s41586-020-2649-2} {\bibfield  {journal} {\bibinfo  {journal} {Nature}\ }\textbf {\bibinfo {volume} {585}},\ \bibinfo {pages} {357–362} (\bibinfo {year} {2020})}\BibitemShut {NoStop}%
\bibitem [{\citenamefont {London}\ and\ \citenamefont {Gurevich}(2023{\natexlab{a}})}]{London:202XP2}%
  \BibitemOpen
  \bibfield  {author} {\bibinfo {author} {\bibfnamefont {L.}~\bibnamefont {London}}\ and\ \bibinfo {author} {\bibfnamefont {M.}~\bibnamefont {Gurevich}},\ }\href@noop {} {\enquote {\bibinfo {title} {Natural polynomials for kerr quasi-normal modes},}\ } (\bibinfo {year} {2023}{\natexlab{a}}),\ \Eprint {http://arxiv.org/abs/2312.17680} {arXiv:2312.17680 [gr-qc]} \BibitemShut {NoStop}%
\bibitem [{\citenamefont {London}\ and\ \citenamefont {Gurevich}(2023{\natexlab{b}})}]{London:202XP3}%
  \BibitemOpen
  \bibfield  {author} {\bibinfo {author} {\bibfnamefont {L.}~\bibnamefont {London}}\ and\ \bibinfo {author} {\bibfnamefont {M.}~\bibnamefont {Gurevich}},\ }\href@noop {} {\  (\bibinfo {year} {2023}{\natexlab{b}})},\ \bibinfo {note} {in Prep.}\BibitemShut {Stop}%
\end{thebibliography}%
